\documentclass{article}[12pt]
\RequirePackage{amsthm,amsmath,amsfonts,amssymb}
\RequirePackage[authoryear]{natbib}
\RequirePackage[colorlinks,citecolor=blue,urlcolor=blue]{hyperref}
\usepackage[margin=1in]{geometry}
\usepackage{setspace}
\theoremstyle{plain}

\newtheorem{assumption}{Assumption}[section]

\usepackage[utf8]{inputenc}
\usepackage{bbm}
\usepackage{graphicx}
\usepackage{siunitx} 

\graphicspath{ {images/} }
\usepackage{caption}
\usepackage{subcaption}
\usepackage{multirow}
\usepackage{enumerate}
\usepackage{float}
\usepackage{listings}
\usepackage{xcolor}
\usepackage{bigints}
\usepackage{algorithm}
\usepackage{algpseudocode}

\def\Minus{\texttt{-}}
\newcommand{\argmax}[1]{\underset{#1}{\operatorname{arg}\,\operatorname{max}}\;} 
\newcommand{\argmin}[1]{\underset{#1}{\operatorname{arg}\,\operatorname{min}}\;} 

\providecommand{\keywords}[1]{%
  {\small \textbf{\textit{Keywords---}}} #1
}

\counterwithin*{figure}{subsubsection}
\counterwithin*{figure}{subsection}
\counterwithin*{figure}{section}

\renewcommand{\thefigure}{%
  \ifnum\value{subsubsection}>0
    \thesubsubsection.\Roman{figure}
  \else
    \ifnum\value{subsection}>0
      \thesubsection.\Roman{figure}
    \else
      \thesection.\Roman{figure}
    \fi
  \fi
}

\title{Change point analysis of high-dimensional data \\
using random projections}
\author{Yi Xu, Yeonwoo Rho}

\numberwithin{equation}{section}
\begin{document}

\begin{center} {\bf\Large Change point analysis of high-dimensional data \\
using random projections}\end{center}
	\centerline{\textsc{Yi Xu and Yeonwoo Rho}\footnote{
	 Author of Correspondence: Y. Rho (yrho@mtu.edu)}}	
		\bigskip
	\centerline {Department of Mathematical Sciences, Michigan Technological University}
    \bigskip
    \centerline{\today}
	\bigskip

\begin{abstract} 
This paper develops a novel change point identification method for high-dimensional data using random projections. By projecting high-dimensional time series into a one-dimensional space, we are able to leverage the rich literature for univariate time series. We propose applying random projections multiple times and then combining the univariate test results using existing multiple comparison methods. Simulation results suggest that the proposed method tends to have better size and power, with more accurate location estimation. At the same time, random projections may introduce variability in the estimated locations. To enhance stability in practice,  we recommend repeating the procedure, and using the mode of the estimated locations as a guide for the final change point estimate. An application to an Australian temperature dataset is presented. This study, though limited to the single change point setting, demonstrates the usefulness of random projections in change point analysis. 
\end{abstract}

\section{Introduction}\label{sec:Introduction}
Change point detection plays a crucial role in science by identifying shifts in the underlying dynamics of a system. While change point analysis has been extensively studied in low-dimensional settings (e.g., \cite{csorgo1997limit,aue2013structural,horvath2014extensions,aminikhanghahi2017survey}), modern applications increasingly generate high-dimensional vectors or functional observations. Examples arise in neuroscience through functional magnetic resonance imaging \citep{aston2012evaluating}, in finance via panels of S\&P500 stock prices \citep{horvath2014testing,jirak2015uniform}, and in climate studies through long term temperature trajectories \citep{aue2018detecting}. Detecting the precise timing of dynamic shifts in such high-dimensional or functional data has increasingly attracted attention in the literature.

Current change point detection methods for high-dimensional or functional data can be broadly divided into three categories: projection-based, fully functional, and manipulations of multivariate statistics. Projection-based approaches involve some form of dimension reduction and rely on classical univariate change point tools, such as cumulative sum (CUSUM) tests \citep{page1954continuous, page1955test}. For instance, functional principal component (FPC) analysis projects the high-dimensional or functional data onto a few leading directions that capture the majority of variance \citep{berkes2009detecting, aue2009estimation, aston2012detecting}. \cite{wang2018high} seeks a single projection direction that aligns the most with the change. The CUSUM-type statistics are applied to the projected data afterwards. While these approaches leverage well-established univariate tools, reducing the data to one or a few dimensions may loose 
information if the projection does not fully capture the structure of the change. For instance, the FPC-based technique cannot consistently detect the change if the mean break function is orthogonal to the direction of FPCs 
\citep{aue2018detecting}. 
Fully-functional methods \citep{horvath2014testing, aue2018detecting, 
dette2020testing} are relatively free of such projection-induced limitations. However, they 
assume the underlying signal is sufficiently smooth, which may limit their applicability to certain types of data. These methods require a smoothing step since observations are often only on discrete grids, and improper smoothing can negatively affect results. For example, using too few basis functions for densely observed data may remove high-frequency variations \citep{jiao2022enhanced}, while using too many basis functions can produce excessively large covariance matrices, leading to slow computation.
The last category is based on multivariate statistics, often in a CUSUM form, which are then aggregated using suitable norms or other manipulations \citep{bai2010common,horvath2012change, chan2013darling,jirak2015uniform, cho2016change,enikeeva2019high, liu2020unified, yu2021finite, 
wang2022inference,wang2023computationally}.  See \cite{liu2022high} for an extensive review of existing change point detection methods in high-dimensional vector settings.
The method proposed in this paper can be viewed as a combination of the projection-based method and the last category. Our method also aggregates a multivariate CUSUM statistic, but unlike existing methods, this statistic is constructed from projections rather than the original high-dimensional vectors.

{Our aim is to develop a new change point inference framework for high-dimensional or functional data using random projections. We limit the scope to the at most one change point (AMOC) case as the first study of this kind. Our idea is to leverage the popular dimension reduction tool, random projections \citep{bingham2001random, dasgupta2013experiments}. By projecting the original data using randomly generated directions, we can reduce dimensionality while approximately preserving important geometric information in the data. The theoretical foundation of random projection is closely related to the Johnson-Lindenstrauss lemma \cite{Johnson1984ExtensionsOL}, which shows that pairwise distances among high-dimensional observations can be approximately preserved after projection into a lower-dimensional space. Compared with many data-dependent dimension-reduction methods, random projection is computationally simple and does not require estimating projection directions from the data. By transforming the multivariate problem into a univariate one, this framework eliminates the need to estimate large covariance matrices as required in traditional multivariate CUSUM approaches, while simultaneously bypassing the need for complex tools such as self-normalization or bootstrapping. Notably, no independence assumptions across coordinates are required either.

Random projection has also appeared in the change point literature. In particular, \cite{aston2018high} consider only one random projection, which is proposed as the lower benchmark when measuring the efficiency of change point detection methods. Our use of random projection is different. Rather than relying on a single projected series, we aggregate across multiple random projections. The motivation for using multiple projections is that the ability of a single random direction to detect a mean change depends on whether it aligns with the underlying change direction. One random projection is unlikely to capture the underlying structure of the original space if it is nearly orthogonal to the change, however, other random directions may preserve stronger information of the change and increase the opportunity to capture the change signal.

While our idea sounds simple -- apply multiple random projections, run univariate change-point tests, and combine their results -- its implementation involves several critical choices: how many random projections are required to guarantee valid inference, which univariate change-point tests to apply to the projected data, and how to aggregate the resulting test statistics. This paper provides a guideline of such choices} through extensive simulations, where size and power of global tests as well as accuracy of location estimates are explored. 
 
The advantages of the proposed method can be summarized as follows. First, using random projections, our method is easy to implement, taking advantage of the rich literature and computational ease of univariate time series. Second, compared to other projection-based methods, which heavily rely on the quality of selected directions, our method eliminates the need for exploration of optimal projection direction and the possible information loss due to dimension reduction. Third, our method does not rely on the functional framework, which requires smooth functions.

The remainder of the paper is organized as follows. We introduce the proposed method in Section \ref{sec: mean method}. Section \ref{sec:simulation-mean } presents simulation results: parameter choices are discussed in Subsection \ref{subsection: tuning},  and comparison with other existing methods are presented in Subsection \ref{subsec:comparison}. Subsection \ref{subsec:repeat RP method on one data} demonstrates the ability to correctly capture change locations despite the variability introduced by random projections, when our method is repeatedly applied. {Subsections \ref{subsec: sparse functional setting}--\ref{subsec: high dimensional vector setting} provide additional simulation results in settings that might be of empirical interest, such as a sparse functional setting, departures from the AMOC assumption, and a high-dimensional vector setting.} Section \ref{sec:application} illustrates the applications to an Australian temperature data. Section \ref{conclusion} is the conclusion. 

\section{Random projection change point detection method}\label{sec: mean method}

Let $\bold{x}_t=(X_{t1},\dots,X_{tp})^{T} $ be a $p$-dimensional time series. 
Our interest is to detect a change point in the mean vectors such that $E(\bold{x}_1)= E(\bold{x}_2)= \dots =E(\bold{x}_{z^{*}}) \neq  E(\bold{x}_{z^{*}+1})=\dots 
= E(\bold{x}_{n})$, where $z^{*}$ denotes a possible (unknown) change point location, and 
$z^{*}=\lfloor \theta n \rfloor$, {$\theta \in [\frac{1}{n}, 1).$}
The time series $\bold{x}_1,\dots,\bold{x}_n$ are modeled by:
\begin{equation} \label{eq: model}
\bold{x}_{t}=\boldsymbol{\mu}+\boldsymbol{\delta} \mathbbm{1}\{t > z^{*}\}+\boldsymbol{\varepsilon}_{t}, \qquad 1\leq t \leq n,
\end{equation}
where $\boldsymbol{\mu} \in \mathbb{R}^{p} $ is a baseline mean vector,
$\boldsymbol{\delta} \in \mathbb{R}^{p}$ is the change in the mean vector after $z^{*}$. We denote $\mathbb{R}^{p}$ as a space of $p$-dimensional vectors of real numbers.
$\mathbbm{1}\{t > z^{*}\}$ is an indicator function which maps $t$ to one if $t > z^{*}$ and to zero if $t \leq z^{*}$. 
We are interested in a hypothesis testing regarding the break vector $\boldsymbol{\delta}$:
\begin{equation}
\begin{aligned} 
\label{hypo:2.2}
& H_0: \boldsymbol{\delta}= \boldsymbol{0} \quad 
versus \quad & H_A: \boldsymbol{\delta} \neq \boldsymbol{0}. \quad 
\end{aligned}
\end{equation}
The $p$-dimensional errors 
$\boldsymbol{\varepsilon}_{t}=({\varepsilon}_{t1},\dots,{\varepsilon}_{tp})^{T}$ {are assumed to be weakly stationary. Here, we present a linear process assumption, following Assumption $\mathcal{A}$.1 in \cite{aston2018high}.}
\begin{assumption}
\label{assump linear error process}
The error sequence $\boldsymbol{\varepsilon}_t =({\varepsilon}_{t1},\dots,{\varepsilon}_{tp})^{T}$ is a linear process of the form {$\boldsymbol{\varepsilon}_t = \sum_{l=0}^{\infty}  \boldsymbol{\psi}_l e_{t-l},$}
where $\{e_t\}_{t\in\mathbb{Z}}$ is an independent and identically distributed (i.i.d.) sequence with
$\mathbb{E} (e_t) = 0$, $\operatorname{Var}(e_t) = 1$ 
and $\mathbb{E}(|e_t|^{2+\delta}) < \infty$ for some $\delta > 0$.
The coefficients $\boldsymbol{\psi}_l =(\psi_{1,l},\dots,\psi_{p,l})^T$ satisfy {$\sum_{l=0}^{\infty} \psi_{j,l}^2 < \infty,$ $j = 1, \dots, p.$}
\end{assumption}
{Under Assumption \ref{assump linear error process}, the coordinate-wise dependencies of the error vector $\boldsymbol{\varepsilon}_t$ are induced through the common innovation sequence $\{e_{t}\}$. The dependencies across coordinates are governed by the aggregated effect of $\boldsymbol{\psi}_{l}\boldsymbol{\psi}_{l}^T$ across $l$, while the time dependence is controlled by the lag structure of $\boldsymbol{\psi}_{l}$.

\cite{aston2018high} prove that, under Assumption \ref{assump linear error process}, the centered partial sum process of a random projection of high-dimensional data converges to a Brownian bridge limit, as in the univariate case. This justifies random projection as a valid dimension reduction tool for high-dimensional settings. However, the framework in \cite{aston2018high} relies on a single projection direction, treating random projection merely as a lower benchmark to evaluate the high-dimensional efficiency of other change-point tests. In this paper, we leverage random projections not only as a dimension reduction technique but also as a core mechanism in a powerful change-point analysis tool for high-dimensional and functional data. Through extensive simulations, we demonstrate that by employing a sufficiently large number of random directions, our proposed procedure achieve higher test power while maintaining accurate size control and competitive precision in the estimation of change-point locations.}

Our method {consists of} two steps, the random projections (RP) step and the univariate change point tests ({for example,} CUSUM) and their combination step.



\subsection{Step 1: RP Step}
\label{subsec: mean method step1}
The first step is to apply RP to a $p$-dimensional time series  $ X = (\bold{x}_1,\dots,\bold{x}_n)^{T} \in \mathbb{R}^{n \times p}$ 
to obtain $k$ univariate time series $ Y = \frac{1}{\sqrt{k}}XD =(\bold{y}_1,\dots,\bold{y}_k)  \in \mathbb{R}^{n \times k}$, where $D = (\bold{d}_1, \ldots, \bold{d}_k) \in \mathbb{R}^{p \times k}$ contains the $k$ random directions. Here, the $r^{th}$ direction is denoted by $\bold{d}_r=(d_{1,r}, \ldots, d_{p,r})^{T} \in \mathbb{R}^{p}$, $ 1\leq r \leq k$, and the $r^{th}$ projected time series by $\bold{y}_r = (y_{1,r},\dots, y_{n,r})^T \in \mathbb{R}^{n}$,  where $y_{t,r}=\frac{1}{\sqrt{k}}\bold{x}_{t}^{T} \bold{d}_r$,  $1\leq t \leq n$. 
The RP preserves pairwise distance to certain accuracy \citep{Johnson1984ExtensionsOL,arriaga1999algorithmic}. Intuitively, the projected time series also preserve useful information about the mean change and the stationarity of the original time series if $\boldsymbol{\delta}^T \bold{d_r} \neq \boldsymbol{0}$ because
\begin{equation*}
\begin{aligned}
    E(y_{t,r}|\bold{d}_r) = E(\frac{1}{\sqrt{k}}\bold{x}_{t}^{T} \bold{d}_r | \bold{d}_r) = \frac{1}{\sqrt{k}} E(\bold{x}_{t}^{T})\bold{d}_r  =   
    \begin{cases}
 \frac{1}{\sqrt{k}} \boldsymbol{\mu}^T  \bold{d}_r, & t \leq z^{*},  \\
 \frac{1}{\sqrt{k}} (\boldsymbol{\mu}+\boldsymbol{\delta})^T  \bold{d}_r, & t > z^{*}. \\
      \end{cases}     
\end{aligned}
\end{equation*}

There are two choices to make in this first step: the way to generate random directions and the number of projections. For the random directions, several approaches have been proposed to generate the entries $d_{j,r}$ for $ 1 \leq j \leq p$, $ 1\leq r \leq k$ in the direction matrix $D$. \cite{arriaga1999algorithmic} propose two methods. 
One option is selecting the entries $d_{j,r}$ independently from a standard normal distribution. Another option is selecting elements independently from a discrete distribution, where $d_{j,r} =  1 $ with probability $\frac{1}{2}$ and $d_{j,r} = \Minus 1 $ with probability $\frac{1}{2}$. In fact, \cite{arriaga1999algorithmic} prove  that the elements $d_{j,r}$ can be drawn independently from any distribution with a mean of zero and a bounded fourth moment.
\cite{achlioptas2003database} argues that having sparse $\bold{d}$ is also acceptable, proposing ``sparse random projections," where the entries $d_{j,r}$ are independent random variables following a probability distribution: 
\begin{equation} \label{random matrix elements}
  d_{j,r}= \sqrt{3} 
    \begin{cases}
      $1$ & \text{ \textit{with probability} $1/6$}, \\
      0 & \text{ \textit{with probability} $2/3$},\\
      $\Minus 1$ & \text{ \textit{with probability} $1/6$}.
    \end{cases}       
\end{equation}
As \cite{achlioptas2003database} explains, a ``threefold speedup" can be achieved using this sparse $\bold{d}$, which randomly drops approximately two-thirds of the original data.
To further reduce computational cost,
\cite{li2006very} propose ``very sparse random projections" which requires only $1/b$ of the original data to be processed, where $b\in \mathbb{Z}$ and $b \gg 3$. While a larger $b$ speeds the computation up significantly, \cite{li2006very} notice that higher sparsity increases the variance of the pairwise distances of projected data, pointing out that using $b<3$ can achieve smaller variability than a large $b$.
In our approach, we follow \cite{achlioptas2003database}, setting $b=3$, to balance between computational cost and stability. For the number of projections, a theoretical bound is provided in the Johnson-Lindenstrauss (JL) lemma (\cite{Johnson1984ExtensionsOL}), which aims to preserve pairwise distances. However, \cite{bingham2001random} achieve satisfactory empirical performance using far fewer projections than the JL bound suggests. Our simulation results in Subsection \ref{subsection: tuning} with sample size $n=50$ also reveal that using a relatively small number of projections, $k=200$, can already achieve high accuracy. The larger the signal-to-noise ratio is, the smaller the number of random projections is needed.

\subsection{Step 2: CUSUM and Combination Step}\label{subsec: mean method step2}
The second step is to apply an existing univariate change point test on each projected time series $\bold{y}_r = (y_{1,r},\dots, y_{n,r})^T$, $r=1,\ldots,k$,
and then combine the $k$ tests with a $p$-value combination method. 
There are two choices to make in the second step: the type of univariate change point test and the type of $p$-value combination method.

We first discuss the type of univariate change point test. Theorem 3.1 of \cite{aston2018high} proves that, given a random direction $\bold{d}_r$ independent of $\{ \boldsymbol{\varepsilon}_{t} : 1 \le t \le n\}$, under Assumption \ref{assump linear error process} and $H_0$,
\[ \Big \{ \frac{\mathcal{Z}_{n,r}(x)}{\hat{\sigma}} : 0\le x\le 1   | \bold{d}_r \Big \}
\ \xrightarrow{D[0,1]} \Big \{B(x):0\le x\le 1  \Big \} \qquad a.s., \]
where $\mathcal{Z}_{n,r}(x)=\frac{1}{\sqrt{n}}  \Bigl|  \sum_{t=1}^{\lfloor nx \rfloor} y_{t,r} - \frac{\lfloor nx \rfloor }{n} \sum_{t=1}^{n} y_{t,r} \Bigr|$, $\hat{\sigma}$ is a consistent estimator of the long-run variance of the projected series, and $B(x)$ is a standard Brownian bridge. This theorem justifies the use of classical change point detection methods for univariate time series, such as the standard cumulative sum (CUSUM) test  \citep{page1954continuous} and its variants, to the RP-projected data. A weighted version of the CUSUM \citep{horvath1993weighted,csorgo1997limit} applies a weight function to stabilize the variance, often attaining higher power in detecting early or late change points as explained in \cite{aue2013structural}. Other CUSUM-based tests can also be considered, we refer readers to reviews in \cite{aue2013structural} and \cite{horvath2014extensions}.  
Our simulations in Subsection \ref{subsection: tuning} suggests that the standard or weighted CUSUM perform the best, when combined with RP, in terms of size and power.
We write the standard and weighted CUSUM statistics as $T_{z,r}^{s} = \frac{1}{\hat{\sigma}_z}\mathcal{Z}_{n,r}\left(\frac{z}{n}\right)$ and 
\begin{equation} \label{weight CUSUM stats}
  \begin{split}
  T_{z,r}^w = \frac{1}{\hat{\sigma}_z \sqrt{n}}
  \left( \frac{z(n-z)}{n^2} \right)^{-\frac{1}{2}} 
  \Biggl|  \sum_{t=1}^{z} y_{t,r} - \frac{z}{n} \sum_{t=1}^{n} y_{t,r} \Biggr|
     &=
    \frac{1}{\hat{\sigma}_z }  
    \sqrt{ \frac{n}{z(n-z) }}   
    \Biggl|  \sum_{t=1}^{z} y_{t,r} - 
    \frac{z}{n} \sum_{t=1}^{n} y_{t,r} \Biggr|,
\end{split}
\end{equation}
respectively. The details on  $\hat{\sigma}_z$ will be discussed in Subsection \ref{subsection: tuning}. In what follows, $T_{z,r}$ shall denote either $T_{z,r}^s$ or $T_{z,r}^w$.

For the $p$-value combination method, two classical methods are illustrated here: the Bonferroni (Bonf) \citep{bonferroni1936teoria} and \cite{benjamini1995controlling} (BH) methods. Denote $p_{r}$, $r=1, \dots, k$, by the $p$-values of the univariate change point tests of the projected data, and $p_{(1)},\dots, p_{(k)}$ by their ordered version from smallest to largest. The Bonferroni- and BH-adjusted $p$-values are $p^{\textit{Bonf}}_{adj(r)}=k p_{r},$ and $p^{\textit{BH}}_{adj(r)} =\min\{1,\min_{r \leq h \leq k} \{\frac{k p_{(h)}}{h}\}\}$, respectively. The global null is rejected when $p_{adj(r)}$ is less than the significance level.

While other $p$-value combination methods can also be considered, such as \cite{wilson2019harmonic,liu2020cauchy}, but the advantage of Bonf and BH is that the identification of significant test is straightforward. Change locations are identified as follows. We first identify the projected data that leads to the largest change, or equivalently, smallest adjusted $p$-value: $\Tilde{r} =\argmin{1 \leq r \leq k} p_{adj(r)}.$ The change location is estimated as
\begin{equation} \label{randpro cp location}
  \begin{split}
    \hat{z}^{*}
    &=  \argmax {\lfloor n \tau \rfloor \leq z \leq n-\lfloor n \tau \rfloor}  T_{z,\Tilde{r}},
\end{split}
\end{equation}
where $\tau \in (0,\frac{1}{2})$ is a trimming parameter for the weighted CUSUM. We set $\tau=1/n$ for the standard CUSUM. {The proposed framework is outlined in Algorithm \ref{algorithm1}.} 
{
\begin{algorithm}[hbt!] 
\caption{RP method for a single mean change point detection}\label{algorithm1}
\begin{algorithmic} 
\State \textbf{Input:} Original data: $X \in \mathbb{R}^{n \times p}$,\\
\hspace{1.15cm}  Number of random projections: $k$,
\State \textbf{Output:}  Estimated change point location: $\hat{z}^{*} \in [1,n] $ , \\
\hspace{1.45cm} Combined $p$-value: $p_{comb}$.

\State \textbf{Steps:} 
\State 1. [RP]
Perform random projections { $Y = \frac{1}{\sqrt{k}}XD =(\bold{y}_1,\dots,\bold{y}_k)  \in \mathbb{R}^{n \times k}$ }, where elements of $D$ are drawn from distribution (\ref{random matrix elements}).

\State 2. [CUSUM and Combination]
Apply a univariate CUSUM test to $\bold{y}_r = (y_{1,r},\dots, y_{n,r})^T$, $1 \leq r\leq k$. Compute the adjusted $p$-values $p_{adj(r)}$, and the combined $p$-value $p_{comb}=\min_{1 \leq r \leq k} p_{adj(r)}$.\\
Let $\hat{z}^{*}=\arg \max_
    {\lfloor n \tau \rfloor \leq z \leq n-\lfloor n \tau \rfloor} 
     T_{z,\Tilde{r}} $ in equation (\ref{randpro cp location}), where $\Tilde{r}=\arg \min_{1 \leq r \leq k} p_{adj(r)}$.
\end{algorithmic}
\label{alg 1}
\end{algorithm}
}

The validity of this location estimate is anchored through the RP's ability to capture the pairwise distance \citep{lee2005metric,ghojogh2021johnson}, even with the $L_\infty$ norm, as long as the number $k$ of projections is large. Recall that under the null, the random behavior of a projected data $\bold{y}_r$ using a random projection is similar to that of a component of the high-dimensional data (Theorem 3.1 of \cite{aston2018high}). When $\Tilde{r}$ is chosen with $p^{\textit{Bonf}}_{adj(r)}$, \cite{yu2021finite} provide the rate of convergence of $\hat{z}^{*}$ to the true change location $z^*$ when weighted CUSUM $T_{z,r}^w$ (Theorem 3.4) or standard CUSUM $T_{z,r}^s$ (Theorem 3.5) is used.

\section{Simulation}
\label{sec:simulation-mean }

This simulation section has three goals. We first provide a guide on the parameter choices involved in our method through extensive simulations in Subsection \ref{subsection: tuning}, and identify reasonable RP-CUSUM procedures. Comparisons of our RP-CUSUM with other change point methods for high-dimension or functional data follow in Subsection \ref{subsec:comparison}. Subsection \ref{subsec:repeat RP method on one data} demonstrates that while RP-CUSUM may exhibit variability in a single run, it tends to correctly identify the change location with repetitions. {Subsections \ref{subsec: sparse functional setting}--\ref{subsec: high dimensional vector setting} examine three additional data-generating settings relevant for empirical applications.}

In the simulations in {Subsections \ref{subsection: tuning}--\ref{subsec:repeat RP method on one data}}, data are generated similarly to the functional setting described in \cite{aue2018detecting}. 
We first formulate our change point model presented in (\ref{eq: model}) in a functional framework:
$X_t(\frac{j}{p})=X_{tj}$, $j=1,\ldots,p$, where $X_{1},\dots, X_{n}$ denote functional time series on a unit interval $[0,1]$. A single change point model for the functional time series is
\begin{equation} \label{eq:model for fd}
X_{t}(s)=\mu(s)+\delta(s) \mathbbm{1}\{t > z^{*}\}+\varepsilon_{t}(s), \qquad 1\leq t \leq n, \quad s \in [0,1],
\end{equation}
where $\mu$, $\delta$, and $\varepsilon_{t}$ are functions defined on the unit interval $[0,1]$.
Let $\mu(s)=0$ and
\begin{equation} \label{eq:data generating process}
\varepsilon_{t}(s) =\sum_{g=1}^{D}A_{t,g}v_{g}(s),
\qquad 1\leq t \leq n, \quad s \in [0,1],
\end{equation}
with Fourier basis functions $v_{1}(s),\ldots,v_{D}(s)$. The basis coefficients $A_{t,g}$ are drawn independently from a normal distribution with mean zero and standard deviation $\sigma_{g}$ of the following three settings:
\begin{enumerate}[\empty]
   \item \textit{Setting 1}: $\sigma_{g}=1$ for $g=1,2,3$ and $\sigma_{g}=0$ for $g=4,\ldots,D$.
    \item \textit{Setting 2}: $\sigma_{g}=3^{-g}$ for $g=1,\ldots,D$.
    \item \textit{Setting 3}: $\sigma_{g}=g^{-1}$ for $g=1,\ldots,D$.
\end{enumerate}
The three settings of $\sigma_{g}$ model different forms in the decay of the ordered eigenvalues of the covariance matrix in the functional data setting. 
\textit{Setting 1} concerns the situation that only three basis functions are included in the functional data.
\textit{Setting 2} involves the situation in which the eigenvalues of the covariance matrix decay fast, while \textit{Setting 3} relates to the situation in which the eigenvalues decay slowly. 

When the eigenvalues of the covariance matrix decay faster, fewer eigenfunctions are needed to explain most of the variation in the functional data samples. Specifically, in \textit{Setting 1}, the majority of variation is explained by the first three eigenfunctions, with each contributing relatively evenly. In \textit{Setting 2}, the first two eigenfunctions tend to explain about $95\%$ of total variation, with the first eigenfunction explaining much more than the second eigenfunction. In \textit{Setting 3}, the proportion of variation explained by eigenfunctions decreases gradually, requiring more than three eigenfunctions to capture most of the variation.

The break function $\delta(s)$ in model (\ref{eq:model for fd}) is formulated with the first $m$ basis functions $v_{1}(s),...,v_{m}(s)$, where $m\leq D$. The magnitude of the break function is scaled by the signal-to-noise ratio ($SNR$) as follows:
{\begin{equation} \label{break function}
\delta(s)=\delta(s;m,c)
=\sqrt{c}*\frac{1}{\sqrt{m}}\sum_{g=1}^{m}v_{g}(s),\hspace{40pt}
c=SNR*\frac{tr(C_{\varepsilon})}{\theta(1-\theta)},
\end{equation}}
where $\theta $ is the scaled location of the change point in the interval $(0,1)$, and $C_{\varepsilon}$ is the long-run variance of the errors. Note that $m=1$ corresponds to a constant mean break. When $m>1$, the break is spread over a larger number of basis functions, resulting in a more complex form. Following the simulation codes from \cite{aue2018detecting}, the term $tr(C_{\varepsilon})$ is {estimated as the normalized} trace of the sample covariance matrix for the {simulated} basis coefficients $A_{t,g}$ {after centering}.

Following \cite{aue2018detecting}'s setting, $D=21$ and $n=50$. The values of $SNR$ are 0, 0.1, 0.2, 0.3, 0.5, 1, and 1.5, with $SNR=0$ indicating $\delta(s) = 0$ ($H_0$), while $SNR > 0$ represent alternative cases. The break function $\delta(s)$ is constructed with $m=1$, $5$, or $20$. {To mimic a real-data scenario, the functional data are first evaluated at $p=101$ equally spaced grid points $\frac{j}{p} \in [0,1]$, $j=1,\ldots,p$. For functional methods, data re-smoothing uses 21 Fourier basis functions.} All results are based on 1000 simulations unless otherwise specified.

\subsection{Parameter Choices for the RP method}\label{subsection: tuning}
Our RP methods require three choices: the types of change point tests, $p$-value combination methods, and the number of random projections. In this subsection, we provide an empirical guide on such choices through simulations. 

For the change point tests, there is extensive literature on detecting and locating change points. \cite{horvath2014extensions} provide a survey of some widely used change point methods. For those applicable to univariate time series, they include the standard CUSUM test \citep{page1954continuous}, the weighted CUSUM \citep{horvath1993weighted,csorgo1997limit}, Darling-Erd\H{o}s (DE) test \citep{darling1956limit}, Andrews test \citep{andrews1993tests}, and Hidalgo and Seo (HS) test \citep{hidalgo2013testing}. Furthermore, \cite{horvath2020new} compare the above tests with their newly developed R\'{e}nyi-type (HR) test, and mention that the Andrews test statistics have a similar performance to that of the DE test. Our simulation focuses on five tests:  the standard CUSUM, the weighted CUSUM, DE, HS, and HR tests.

When implementing the weighted CUSUM test, the change point estimation in (\ref{randpro cp location}) is restricted to the trimmed interval $[\lfloor n \tau \rfloor,  n-\lfloor n \tau \rfloor]$, where $\tau \in (0,\frac{1}{2})$. We consider choices of $\lfloor n \tau \rfloor$ including $1$, $\lfloor n^{0.25} \rfloor$,$\lfloor log(n) \rfloor$,$\lfloor n^{0.5} \rfloor$ following \cite{horvath2020new}. To simulate the limiting distribution of the weighted CUSUM, we generate 100,000 replications and use 10,000 increments for Brownian bridge. When estimating $\sigma ^2 _z$ in (\ref{weight CUSUM stats}), we consider the heteroskedasticity and autocorrelation consistent (HAC) estimation with the Bartlett kernel and the bandwidth based on \cite{andrews1991heteroskedasticity}. Alternative combinations of kernel function and bandwidth are also compared but leads to inferior results, which are not presented. We also consider a variance estimator as mentioned in \cite{horvath2020new}. Using $y_t$ to represent $y_{t,r}$ for simplicity, for any $r=1,\ldots,k$, $\hat{\sigma}^2_z= \frac{1}{n} [(\sum_{t=1}^{z} (y_t-\Bar{y}_z)^2 + \sum_{t=z+1}^{n} (y_t-\Bar{y}_{n-z})^2 )],$ where $\Bar{y}_z=\frac{1}{z} \sum_{t=1}^{z}y_t $, and
$\Bar{y}_{n-z}=\frac{1}{n-z} \sum_{t=z+1}^{n}y_t$. The overall conclusions of comparison using this variance estimator remain consistent with those using the HAC estimator. When the time series is known to be uncorrelated, the variance estimators $\hat{\sigma}^2_z$  from \cite{horvath2020new} is recommended as it is computationally faster than the HAC estimators.
We present the results using the variance estimator from \cite{horvath2020new} in the main paper, and the results obtained using the HAC estimator in the Supplementary Material \ref{subsec: Result of using HAC estimator}.

For the $p$-value combination methods, there are a variety of approaches that control the family-wise error rate (FWER) or false discovery rate (FDR) among multiple tests. In addition to the Bonf and BH methods, mentioned in Subsection \ref{subsec: mean method step2}, we also consider the Cauchy combination test (CCT) from \cite{liu2020cauchy} and the Harmonic mean $p$-value \citep{wilson2019harmonic} {for their popularity in the recent $p$-value combination literature. They tend to have better FWER control and improved power when tests are dependent. There are two variants of the harmonic mean $p$-value method. The first is the simple harmonic mean of the $p$-values, which we denote as HMP-raw. Because HMP-raw fails to control the type I error rate at larger significance levels when the underlying tests are independent, \cite{wilson2019harmonic} propose an alternative approach that utilizes a Landau distribution to adjust the aggregated $p$-value. This version, which incorporates the Landau adjustment, is referred to simply as HMP throughout the remainder of this paper.}
The Benjamini-Yekutieli (BY) method \citep{benjamini2001control}, an extension of BH method to handle dependent tests, is also considered but not presented since the BH method has better performance than the BY method when combined with random projections based on our simulation results. We further use BH and Bonf methods for change point location identification based on the smallest adjusted $p$-value.

For the number $k$ of random projections, a theoretical bound from \cite{dasgupta2003elementary} is $4(a^2/2-a^3/3)^{-1}\ln n$, where $0<a<1$ and $(1-a)$ relates to the accuracy of random projections in preserving pairwise distances. This bound suggests that approximately 900 projections are required for the data settings we consider, with $a=0.2$ and $n=50$. Given this theoretical bound and the empirical findings from \cite{bingham2001random} that a much lower number of random projections than the bound gives a comparable performance, we investigate selected numbers ranging from 10 to 1000. Note that increasing the number of random projections may result in longer computation time. We intend to select a smaller number of random projections to ensure computational efficiency while maintaining accuracy in change point location identification, as well as ensuring acceptable sizes and powers.

{ 
In the size-power comparisons of our simulation study, we explore three quantities: empirical size, raw power, and size-adjusted power. The empirical size is calculated as the proportion of rejections under the null hypothesis, where no change point is present. The raw power is calculated as the proportion of rejections under the alternative hypothesis, using the nominal significance level $0.05$. Since different methods may have different finite-sample sizes, we also report size-adjusted power for a fair comparison. The size-adjusted power is computed by first determining, from the null simulations, an empirical rejection threshold for the $p$-value such that the empirical size matches the nominal level. This threshold is then applied to the $p$-values obtained under the alternative model, and the resulting rejection proportion is reported as the size-adjusted power. In the size-adjusted power plots, the empirical size under the null is still presented to ensure this point remains informative; otherwise, the adjusted rejection rate under the null would trivially equal the nominal level of 0.05.
}

\begin{figure} [h!]
         \centering
            \hspace{-1 cm}  
            \includegraphics[width=5.5cm, height=5cm]{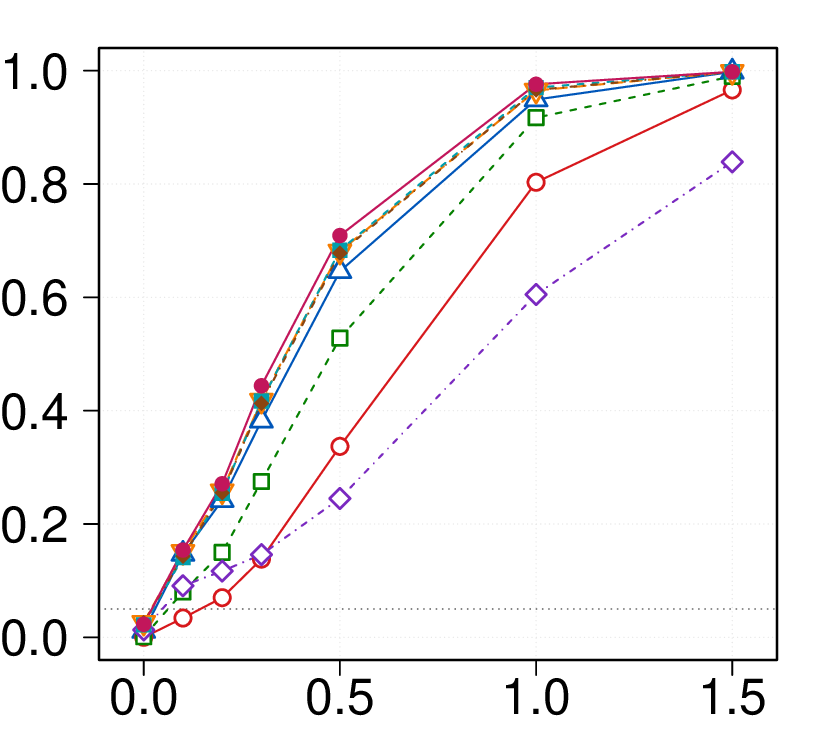}
            \hspace{-1 cm}  
            \includegraphics[width=5.5cm, height=5cm]{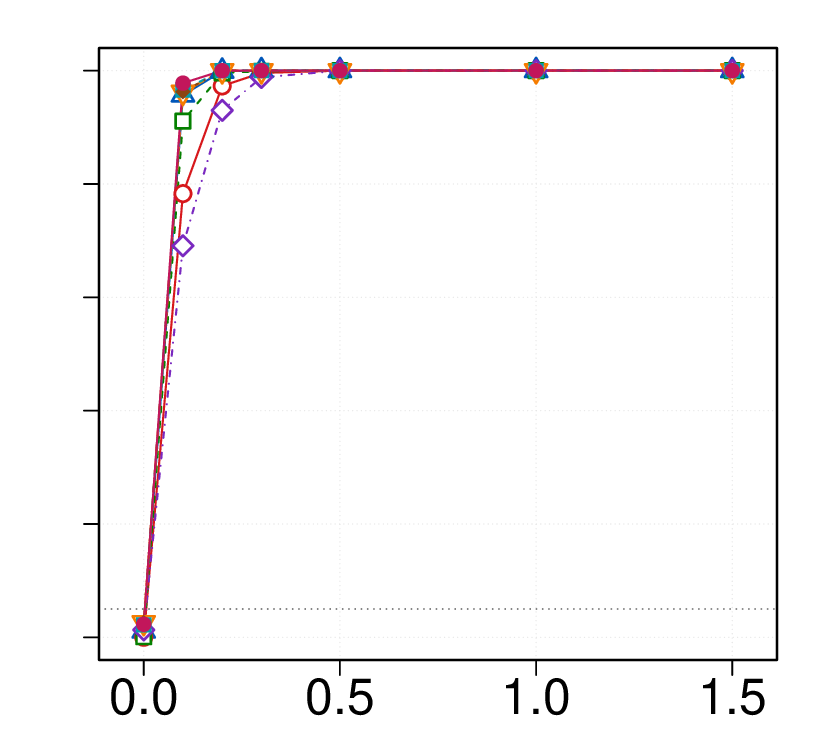}
            \hspace{-1 cm}  
            \includegraphics[width=5.5cm, height=5cm]{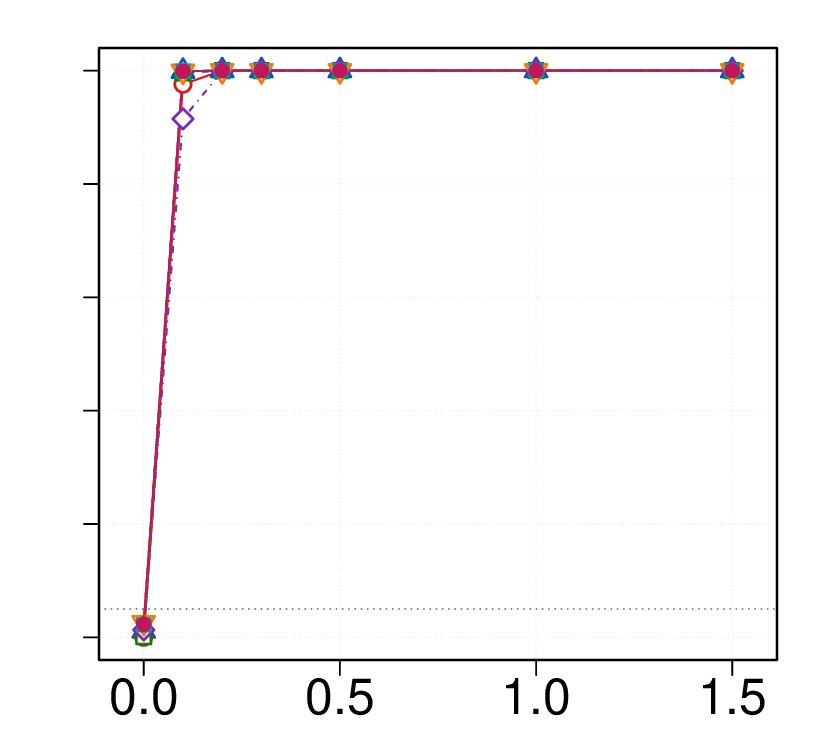}
            \hspace{-1 cm}  
            \vspace{-0.2in}
            \caption*{\small{(a) \textit{Setting 1}, with $m=1,5,20$} }

            \centering
            \hspace{-1 cm}                 \includegraphics[width=5.5cm, height=5cm]{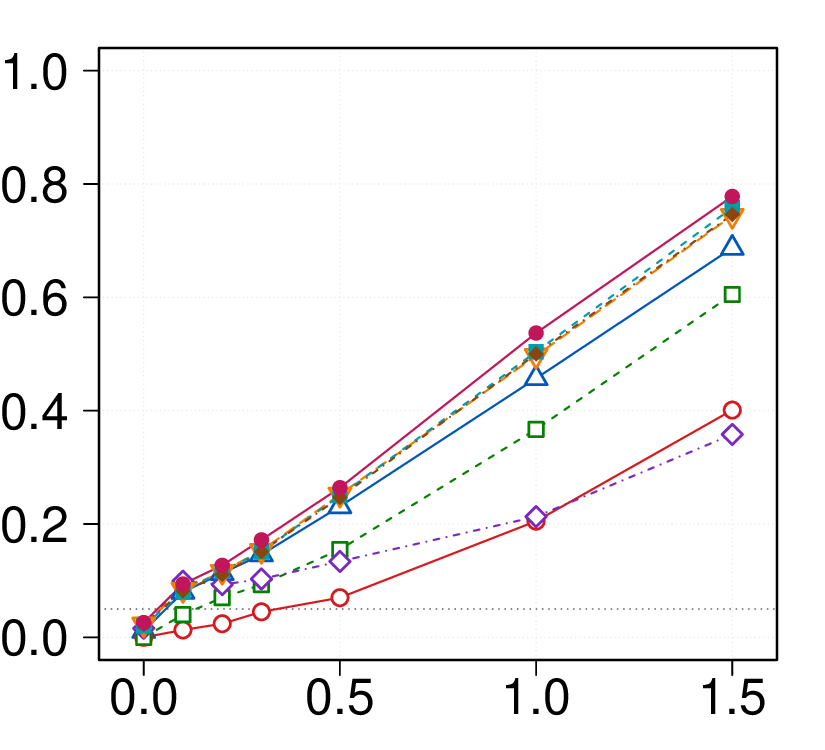}
            \hspace{-1 cm}  
            \includegraphics[width=5.5cm, height=5cm]{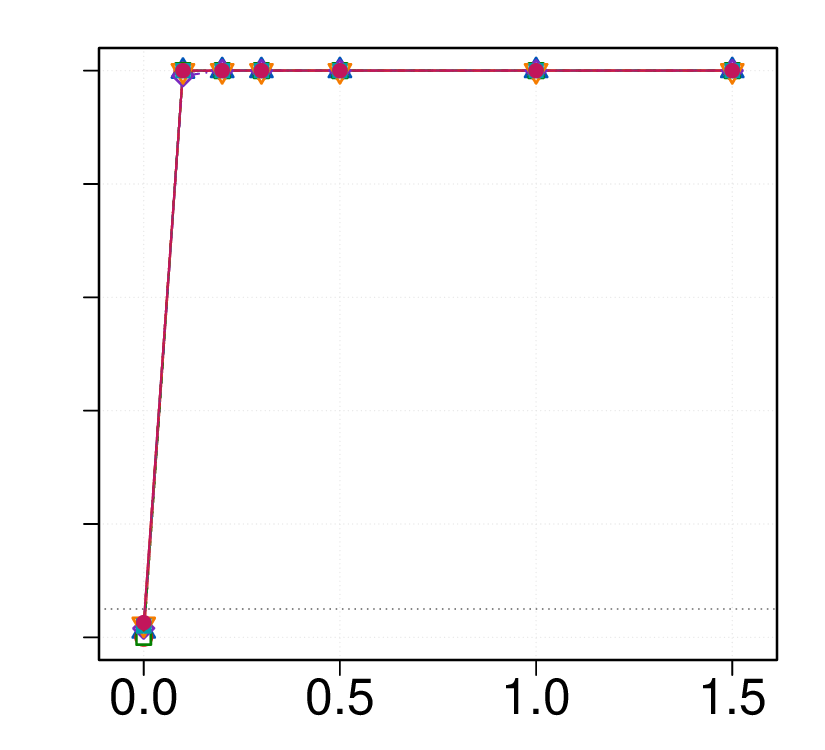}
            \hspace{-1 cm}  
            \includegraphics[width=5.5cm, height=5cm]{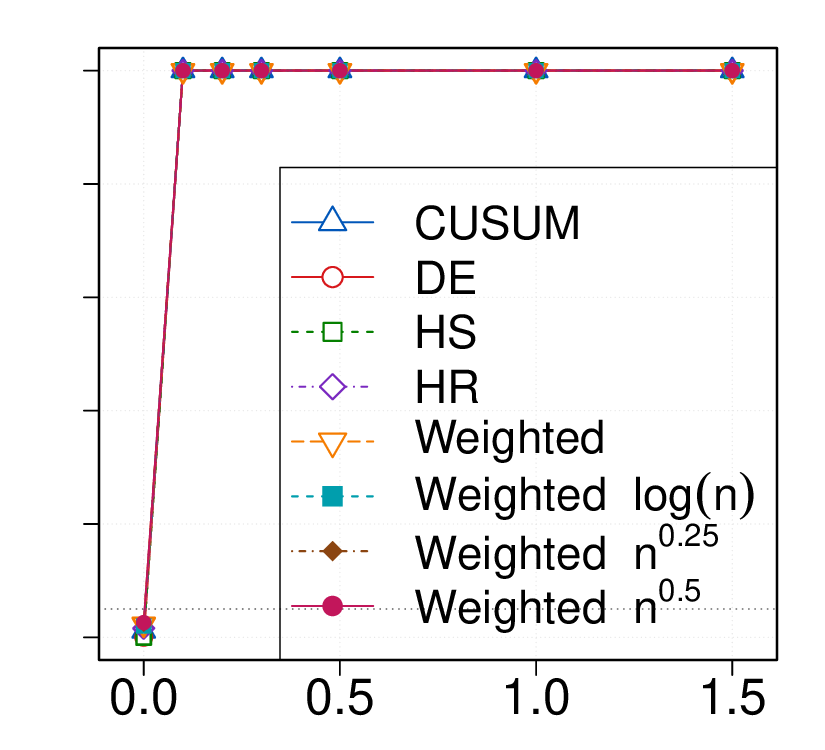}
            \hspace{-1 cm}  
            \vspace{-0.2in}
            \caption*{\small{(b) \textit{Setting 2}, with $m=1,5,20$} }

              \centering
            \hspace{-1 cm}  \includegraphics[width=5.5cm, height=5cm]{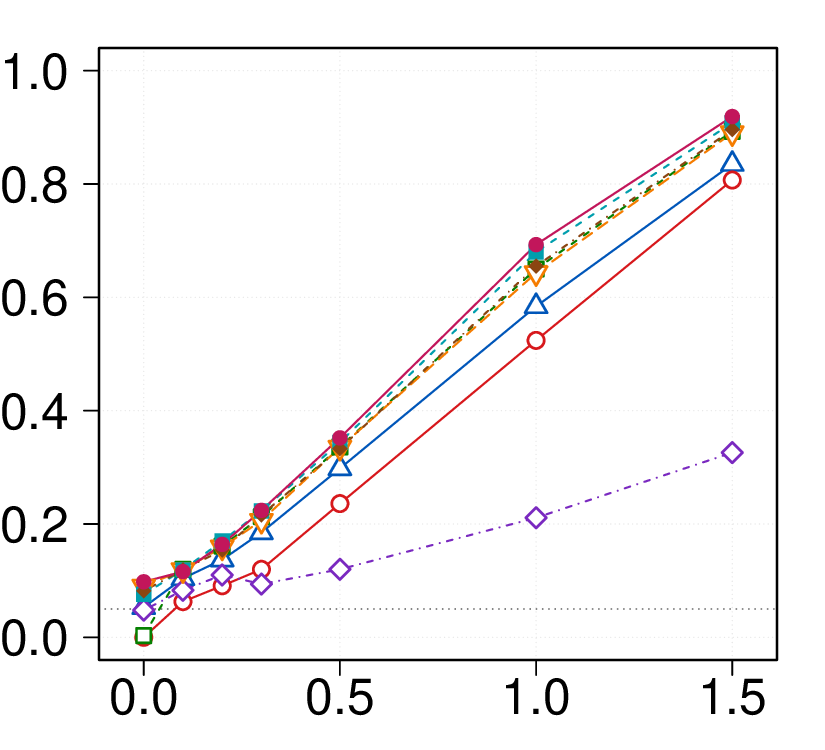}
            \hspace{-1 cm}  
            \includegraphics[width=5.5cm, height=5cm]{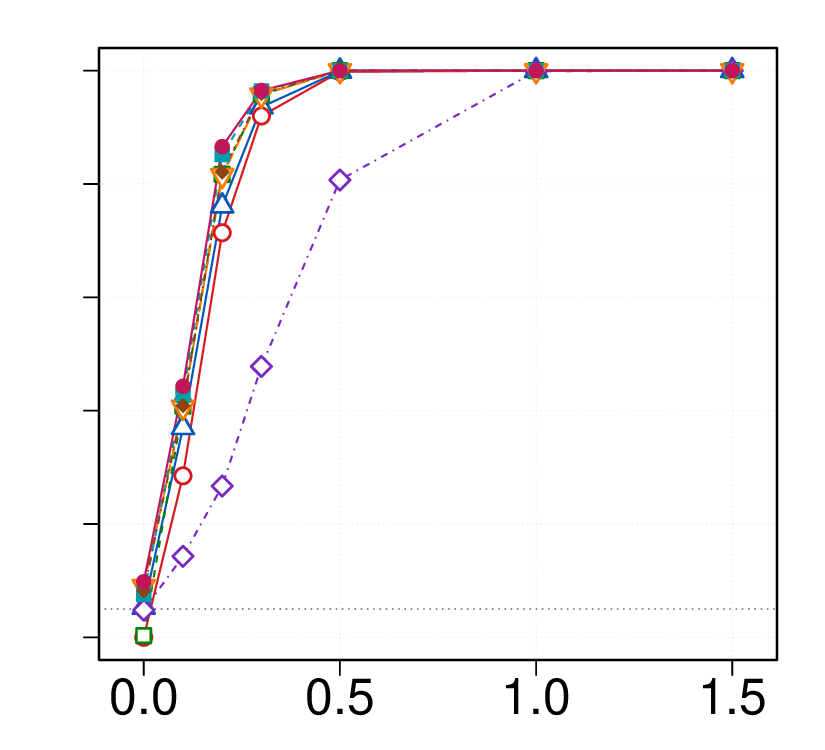}
            \hspace{-1 cm}  
            \includegraphics[width=5.5cm, height=5cm]{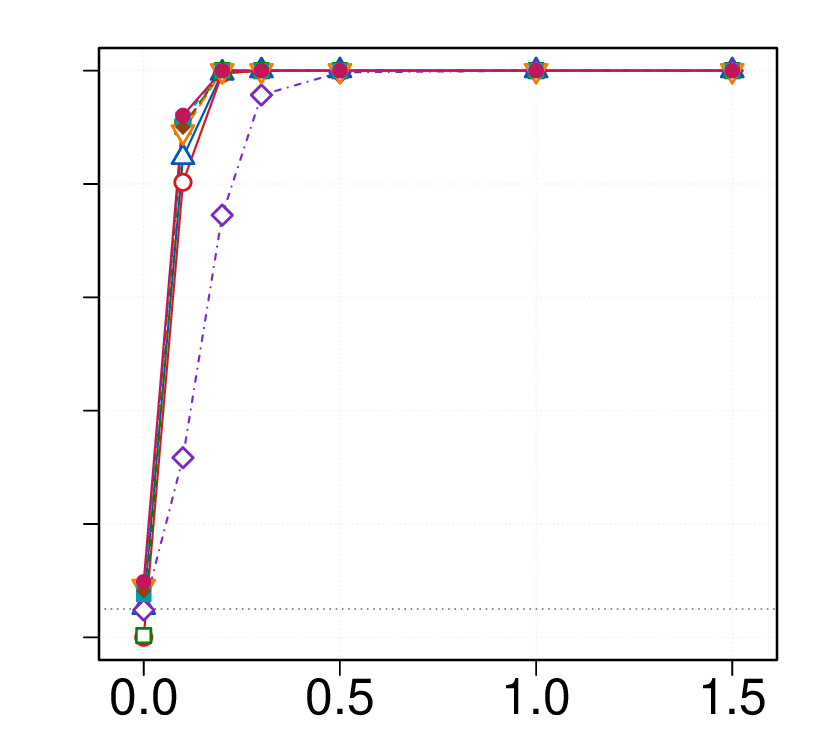}
            \hspace{-1 cm}  
            \vspace{-0.2in}
            \caption*{\small{(c) \textit{Setting 3}, with $m=1,5,20$ } }
               
            \caption{\small{Adjusted empirical rejection rates of the RP methods for various values of $SNR$ in the x-axis. The RP method performs 200 random projections and applies different change point tests (CUSUM, Weighted, DE, HS, HR) and the Bonf combination method. The data-generating process follows (\ref{eq:model for fd}), (\ref{eq:data generating process}), and (\ref{break function}), where the standard deviation $\sigma_{g}$ follows \textit{Settings 1--3}.
            The change point location is set at $\theta=0.25$.
            The empirical rejection rate is based on 1000 simulations.
            }}
            \label{fig: tuning cp test Bonf.adj}
    \end{figure}

First, we compare the effect of different change point tests when combined with RP. Figure \ref{fig: tuning cp test Bonf.adj} presents the sizes and size-adjusted powers of the RP methods using different change point tests with $k=200$ and Bonf. The corresponding raw empirical rejection rates are provided in Figure \ref{fig: tuning cp test Bonf} in the Supplementary Material \ref{subsec: Result of using IID estimator}. The change point location is set at $\theta = 0.25$. From Figure \ref{fig: tuning cp test Bonf.adj}, among the different trimming choices of the weighted CUSUM tests, the trimming choice of $\lfloor {n^{0.5}} \rfloor$ leads to slightly higher size-adjusted powers but over-rejects. The trimming choice of $\lfloor log(n) \rfloor$ has better sizes and similar size-adjusted powers as $\lfloor {n^{0.5}} \rfloor$ in most cases. We compare the weighted CUSUM test with trimming choice of $\lfloor log(n) \rfloor$ and other tests. In the case of non-constant shift ($m=5$ and $m=20$), both the weighted CUSUM and the standard CUSUM have manifestly higher size-adjusted powers in \textit{Settings 1--2}, followed by the HS test. The HR test exhibits the least powers. The weighted CUSUM exhibits slightly better size-adjusted powers in \textit{Setting 3}, closely followed by the standard CUSUM, with the standard CUSUM controlling the sizes better. In the case of constant shift ($m=1$), all the tests display low size-adjusted powers, while the weighted CUSUM and standard CUSUM still outperform. From Figure \ref{fig: tuning cp test Bonf}, the weighted CUSUM has the best powers, followed by the standard CUSUM, with the standard CUSUM controlling the sizes better. The DE test exhibits the least powers in most cases. Overall, Figure \ref{fig: tuning cp test Bonf.adj} and Figure \ref{fig: tuning cp test Bonf} suggest that the weighted CUSUM and the standard CUSUM test outperform the other change point tests in terms of size control and power. When the change point location is in the middle ($\theta=0.5$) as presented in Figure \ref{fig: tuning cp test Bonf-theta=0.5} and Figure \ref{fig: tuning cp test Bonf.adj-theta=0.5} in Supplementary Material \ref{subsec: Result of using IID estimator}, the standard CUSUM test exhibits even superior performance, acquiring higher raw powers and size-adjusted powers in all the alternative cases than the weighted CUSUM test and other tests.
While more extreme change point locations may change the results, as suggested in \cite{horvath2020new}, it appears that as long as the change points are well within the observation boundaries, the standard CUSUM is a reasonable choice to combine with random projections. More results of the sizes and size-adjusted powers of the RP methods with BH, HMP, and CCT are presented in Figure \ref {fig: tuning cp test BH.adj}, Figure \ref {fig: tuning cp test HMP.adj} and Figure \ref {fig: tuning cp test CCT.adj}, and the corresponding sizes and raw powers in Figure \ref {fig: tuning cp test BH}, Figure \ref {fig: tuning cp test HMP}, and Figure \ref {fig: tuning cp test CCT} in Supplementary Material \ref{subsec: Result of using IID estimator}.

\begin{figure} [h!]
{
        \centering
        \hspace{-1 cm}
            \includegraphics[width=5.5cm, height=5cm]{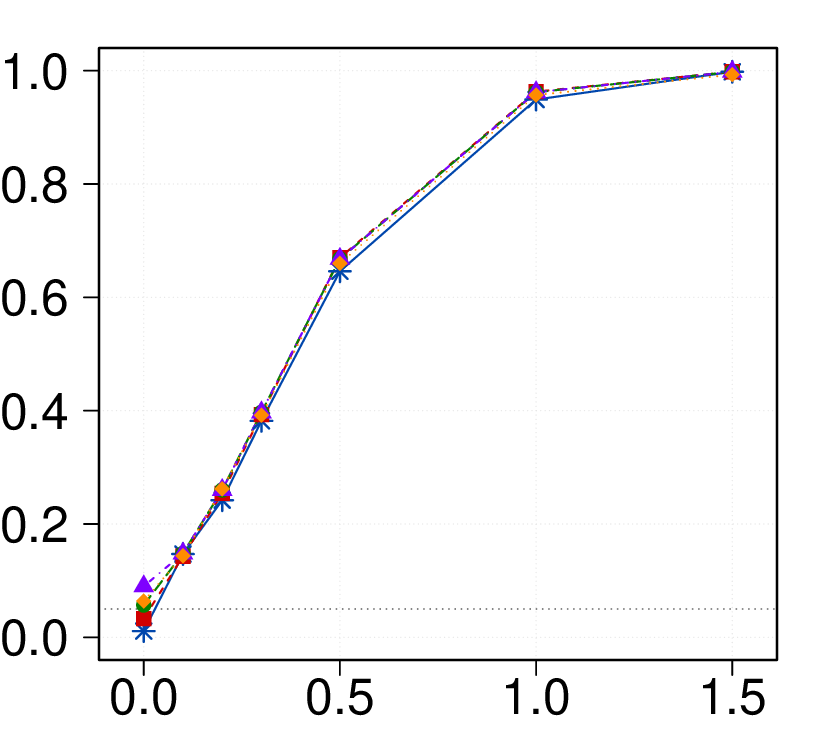}
         \hspace{-1 cm}
            \includegraphics[width=5.5cm, height=5cm]{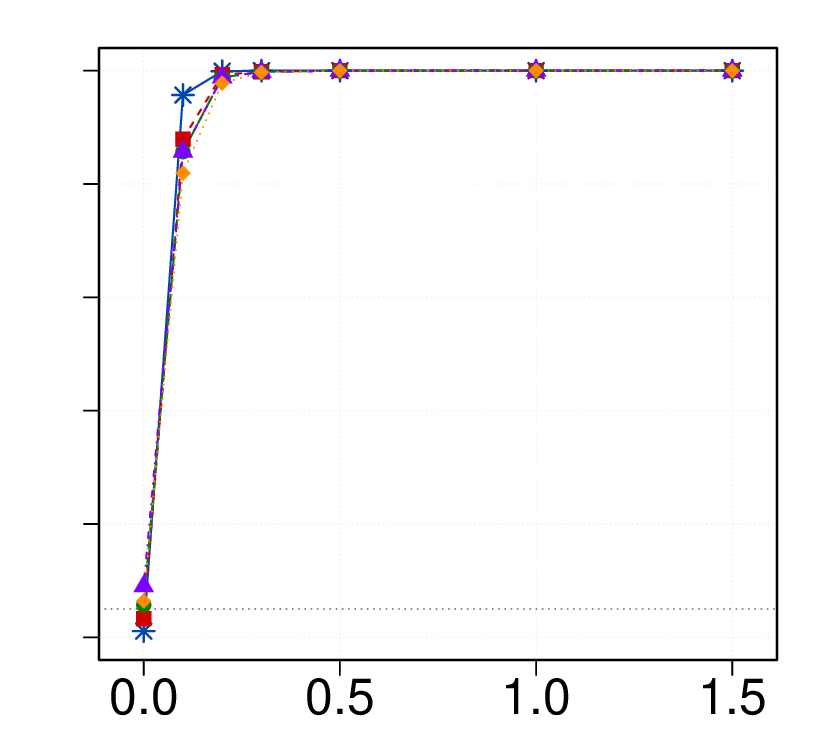}
         \hspace{-1 cm}
            \includegraphics[width=5.5cm, height=5cm]{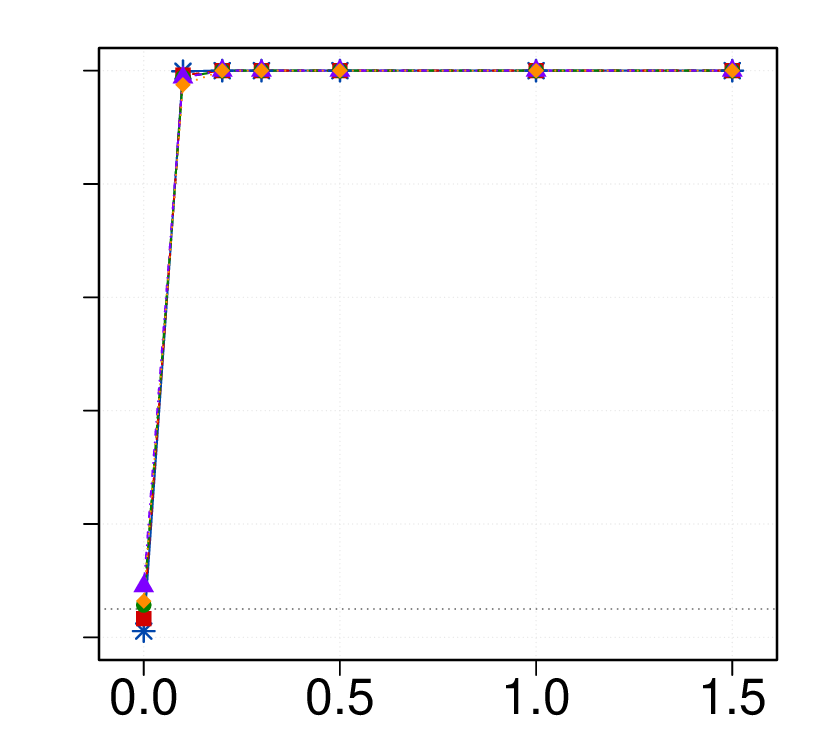}
         \hspace{-1 cm} 
         \vspace{-0.2in}
            \caption*{(a) \small{\textit{Setting 1}, with $m=1,5,20$ } }

            \centering
            \hspace{-1 cm}
            \includegraphics[width=5.5cm, height=5cm]{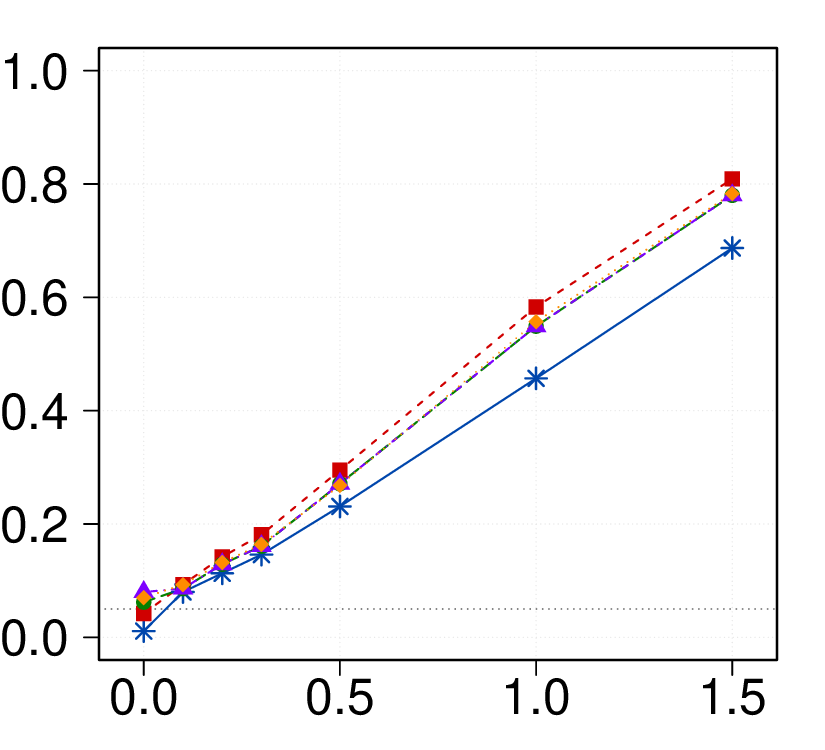}  
         \hspace{-1 cm}
            \includegraphics[width=5.5cm, height=5cm]{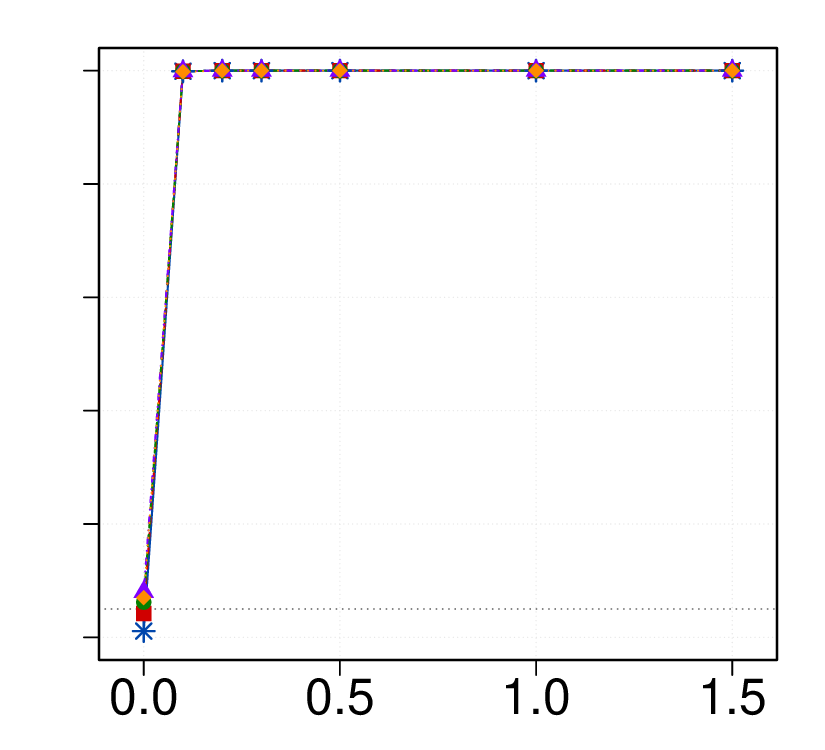}   \hspace{-1 cm}
            \includegraphics[width=5.5cm, height=5cm]{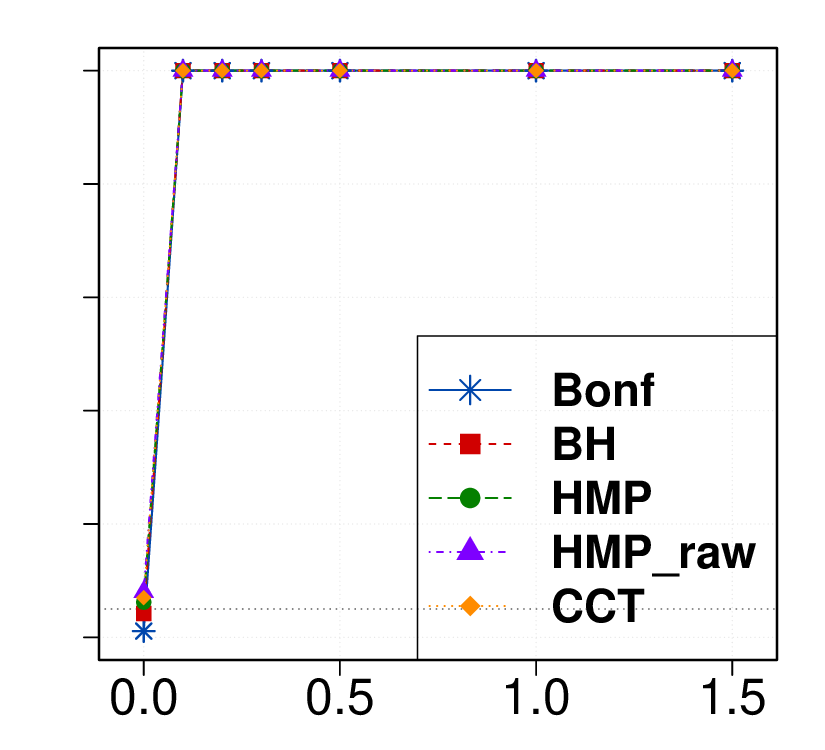}  \hspace{-1 cm} 
            \vspace{-0.2in}
            \caption*{(b) \small{\textit{Setting 2}, with $m=1,5,20$} }

            \centering
       \hspace{-1 cm}
            \includegraphics[width=5.5cm, height=5cm]{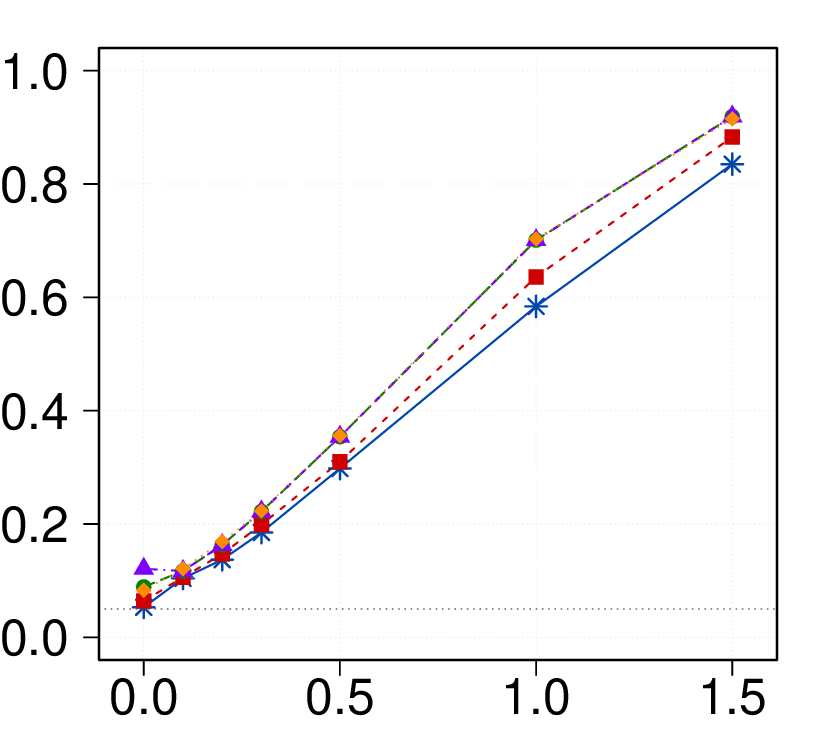} 
         \hspace{-1 cm}
            \includegraphics[width=5.5cm, height=5cm]{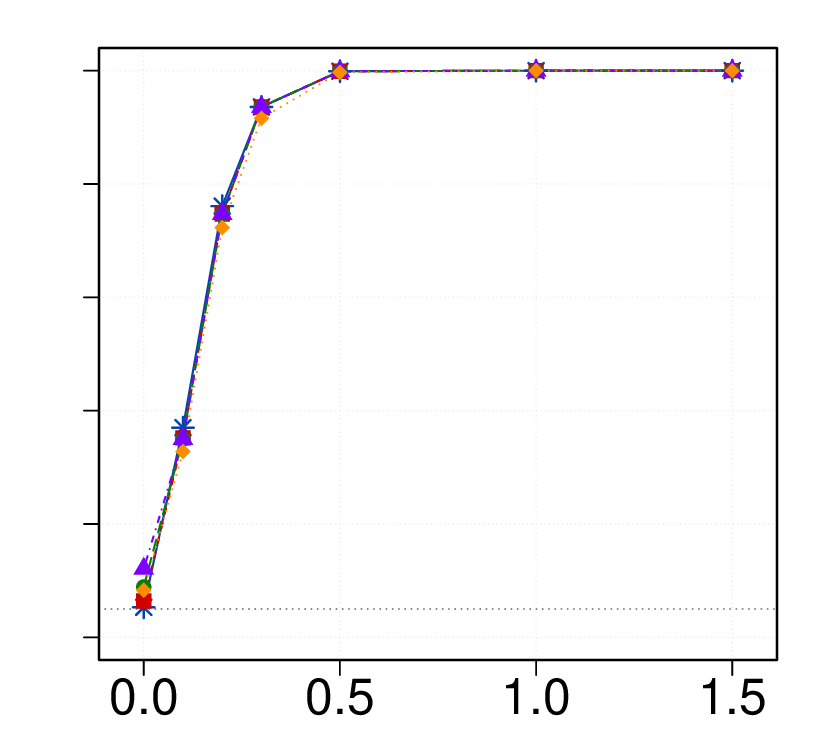}  
         \hspace{-1 cm}
            \includegraphics[width=5.5cm, height=5cm]{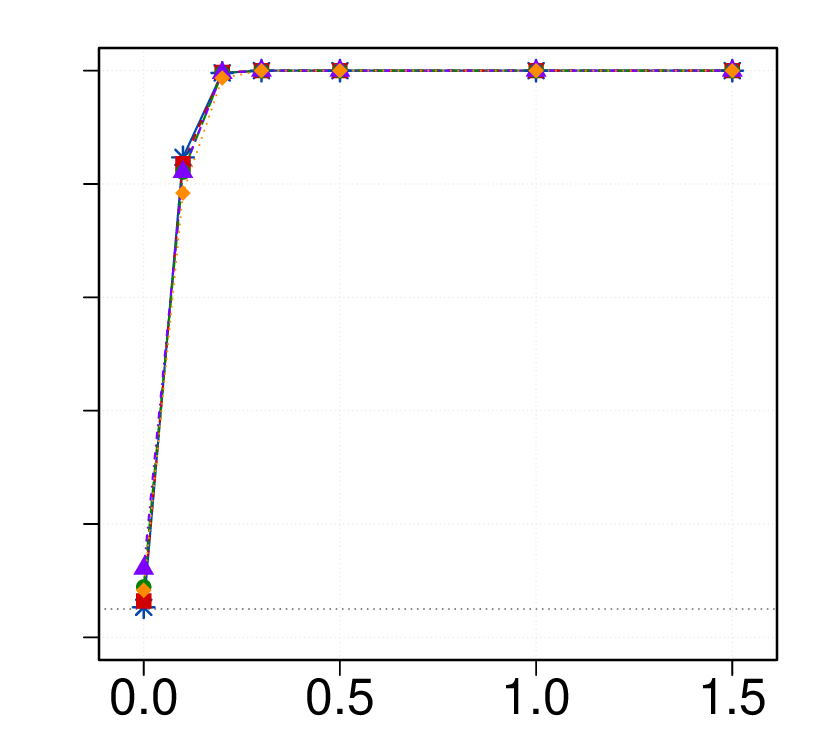} 
         \hspace{-1 cm} 
            \vspace{-0.2in}
            \caption*{(c) \small{\textit{Setting 3}, with $m=1,5,20$} }
}           
            \caption{\small{
             Adjusted empirical rejection rates of the RP-Bonf, RP-BH, RP-HMP, RP-HMP-raw, and RP-CCT methods with the standard CUSUM test for various values of $SNR$ in the x-axis. The RP method performs 200 random projections. The data-generating process follows (\ref{eq:model for fd}), (\ref{eq:data generating process}), and (\ref{break function}), where the standard deviation $\sigma_{g}$ follows \textit{Settings 1--3}.
            The change point location is set at $\theta=0.25$.
            The empirical rejection rate is based on 1000 simulations.
            }}
            \label{fig: tuning Pvalue-comb(adj)}
    \end{figure}

Secondly, we compare the effect of different $p$-value combination methods. Figure \ref{fig: tuning Pvalue-comb(adj)} presents the sizes and size-adjusted powers of the RP methods using different $p$-value combination methods with $k=200$ and the standard CUSUM test. The corresponding raw empirical rejection rates are provided in Figure \ref{fig: tuning Pvalue-comb} in Supplementary Material \ref{subsec: Result of using IID estimator}. Denote RP-Bonf, RP-BH, RP-HMP, {RP-HMP-raw}, and RP-CCT by RP methods with $p$-value combination methods Bonf, BH, HMP, {HMP-raw}, and CCT, respectively. The change point location is set at $\theta=0.25$. From Figure \ref{fig: tuning Pvalue-comb(adj)}, {all RP-based methods have similar size-adjusted power in \textit{Setting 1} or when $m=5$ or 20, with RP-Bonf having slightly better size-adjusted power in \textit{Setting 1} with $m=5$. In \textit{Settings 2--3} with $m=1$, RP-CCT and RP-HMP have the highest size-adjusted power. Regarding raw power (Figure \ref{fig: tuning Pvalue-comb} in Supplementary Material \ref{subsec: Result of using IID estimator}), performance differences are most pronounced when $m=1$, with RP-CCT, RP-HMP, and RP-HMP-raw attaining higher raw power compared to RP-Bonf. While these plots illustrate empirical sizes visually Table \ref{tab:tuning Pvalue-comb} presents exact empirical rejection rates under the null (for $k=200$ and the CUSUM test) for precise numerical comparison. Additional results using alternative CUSUM tests are provided in Table \ref{tab:IID_sizes} of Supplementary Material \ref{subsec: Result of using IID estimator}.
These tables indicate that RP-CCT, RP-HMP, and RP-HMP-raw can lose size control, with RP-HMP-raw exhibiting the most severe size distortions.} It is because these methods are valid only when the significance level is small enough, especially when dependence is weak. Conditioning on the data in the original dimension $X$, each projected univariate time series using RP are independent. For HMP{, HMP-raw,} and CCT, significance level of 0.05 may still be too large under independence, as \cite{rho2024heavy} points out. In fact, considering the independence among RPs given the high dimensional data, Bonf and BH would work reasonably well. The RP-BH maintains appropriate rejection rates under the null in \textit{Settings 1--2} and the RP-Bonf shows the best performance in controlling the sizes in \textit{Setting 3}. 
Overall, the RP-Bonf and RP-BH methods have {reasonable} size-adjusted power in most cases and control the sizes well. Considering that Bonf and BH can also naturally identify significant indicividual tests, unlike the sum-based methods such as HMP and CCT, we recommend using Bonf or BH with RP.

\begin{table}[H]
\centering
\begin{tabular}{ 
|p{1.5cm}||p{1.5cm}|p{1.5cm}|p{1.5cm}|p{1.5cm}|p{1.5cm}| }
 \hline
 & Bonf  & BH & HMP & HMP-raw & CCT \\
 \hline
 {\textit{Setting 1}} 
 & 0.011 & 0.033 & 0.057 & 0.090 & 0.064\\

 \hline
 {\textit{Setting 2}} 
 & 0.011 & 0.042 & 0.062 & 0.080 & 0.070\\

 \hline
 {\textit{Setting 3}} 
 & 0.053 & 0.064 & 0.089 & 0.121 & 0.083 \\
 
 \hline
\end{tabular}
\caption{\small{Empirical rejection rates under the null of the RP method using different $p$-value combination methods with $k=200$ and the standard CUSUM test, at significance level $0.05$.}}

\label{tab:tuning Pvalue-comb}
\end{table}

Thirdly, we compare the effect of the RP methods with different numbers of random projections.
Following the selected choices from Figures \ref{fig: tuning cp test Bonf.adj}, we limit the change point test to standard CUSUM.  
We compare the sizes, raw powers, size-adjusted powers of our methods as well as root mean squared errors (RMSE) of estimated change point locations.
The choices of the number $k$ of random projections range from 10 to 100 in increments of 10 and then from 100 to 1000 in increments of 50.

  \begin{figure} [h!]
  {
        \centering
            \includegraphics[width=5.5cm, height=5cm]{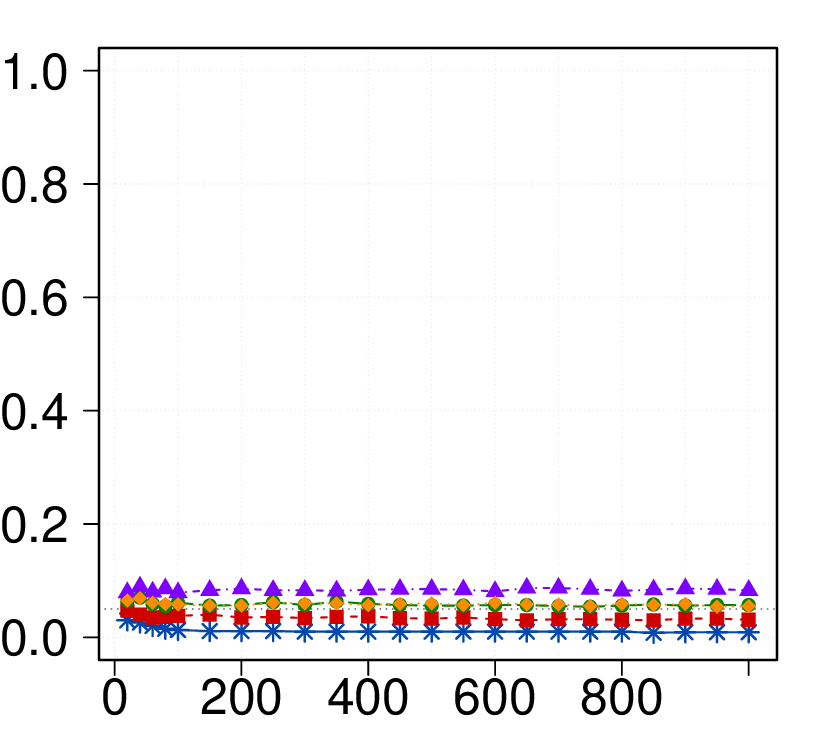}
        \vspace{-0.2in}
            \caption*{\small{(a) $SNR=0$} }

        \hspace{-1 cm}
            \includegraphics[width=5.5cm, height=5cm]{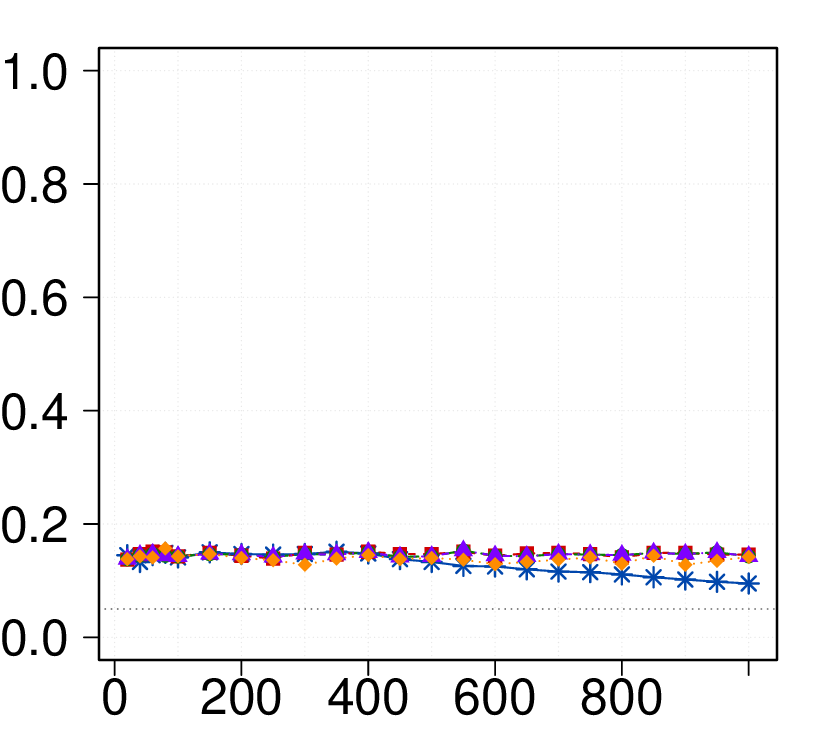}
        \hspace{-1 cm}
            \includegraphics[width=5.5cm, height=5cm]{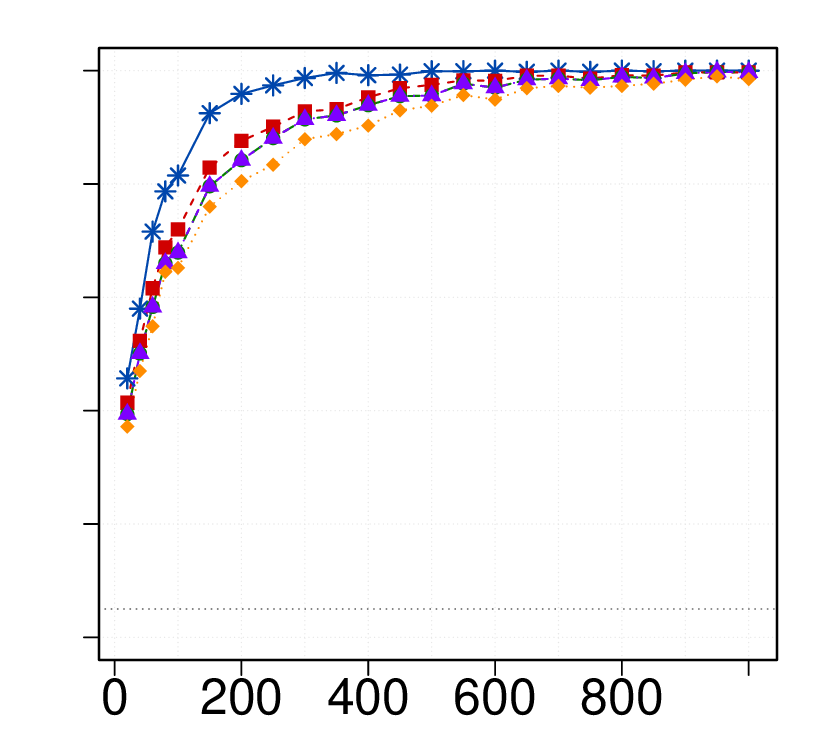}
        \hspace{-1 cm}
            \includegraphics[width=5.5cm, height=5cm]    {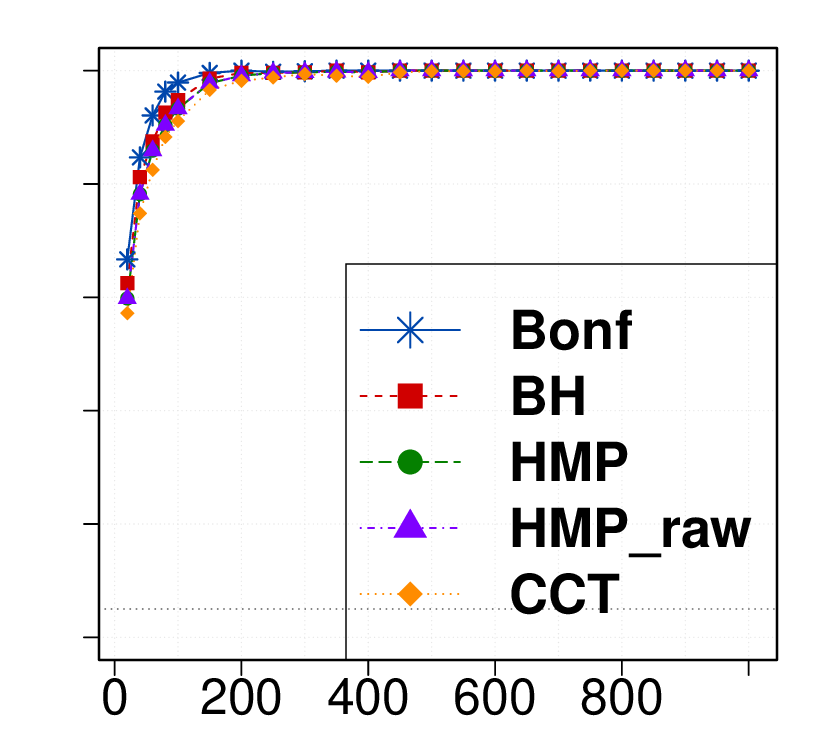}
        \hspace{-1 cm}
        \vspace{-0.2in}
            \caption*{\small{(b) $SNR=0.5$, with $m=1,5,20$ } }
            
        \hspace{-1 cm}    
             \includegraphics[width=5.5cm, height=5cm]{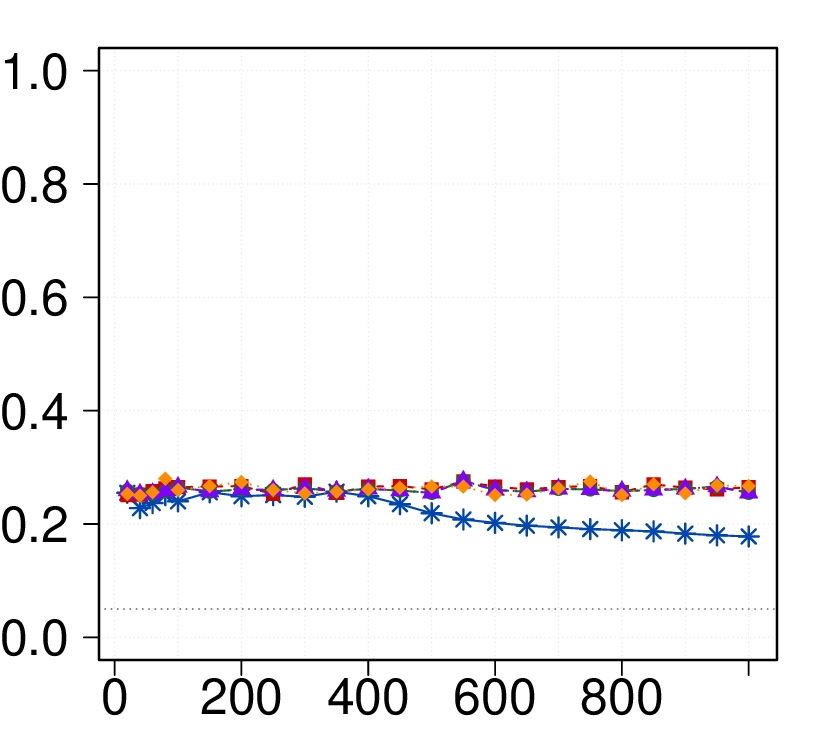}
        \hspace{-1 cm}
            \includegraphics[width=5.5cm, height=5cm]{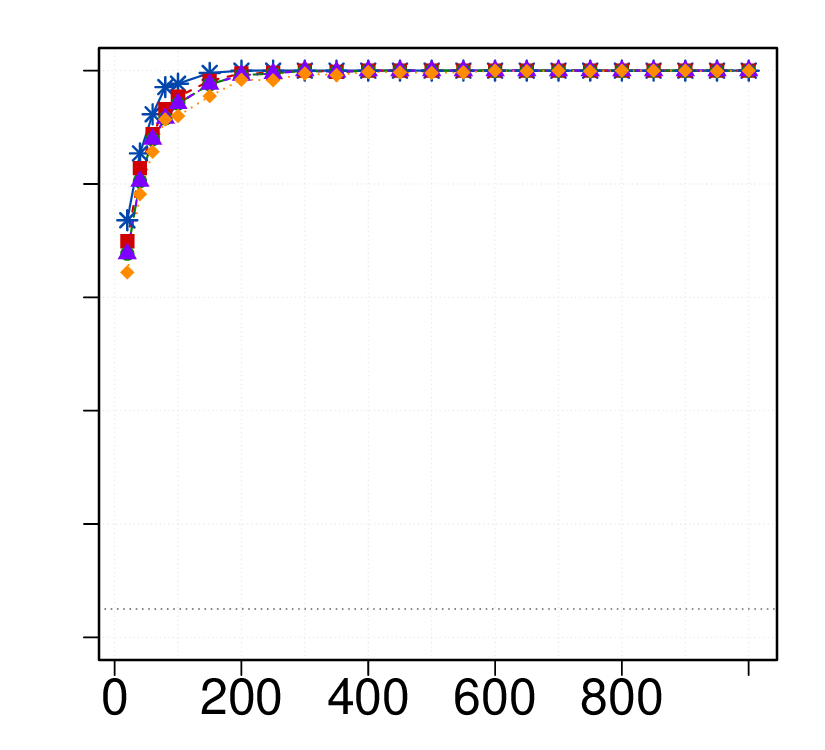}
        \hspace{-1 cm}
            \includegraphics[width=5.5cm, height=5cm]    {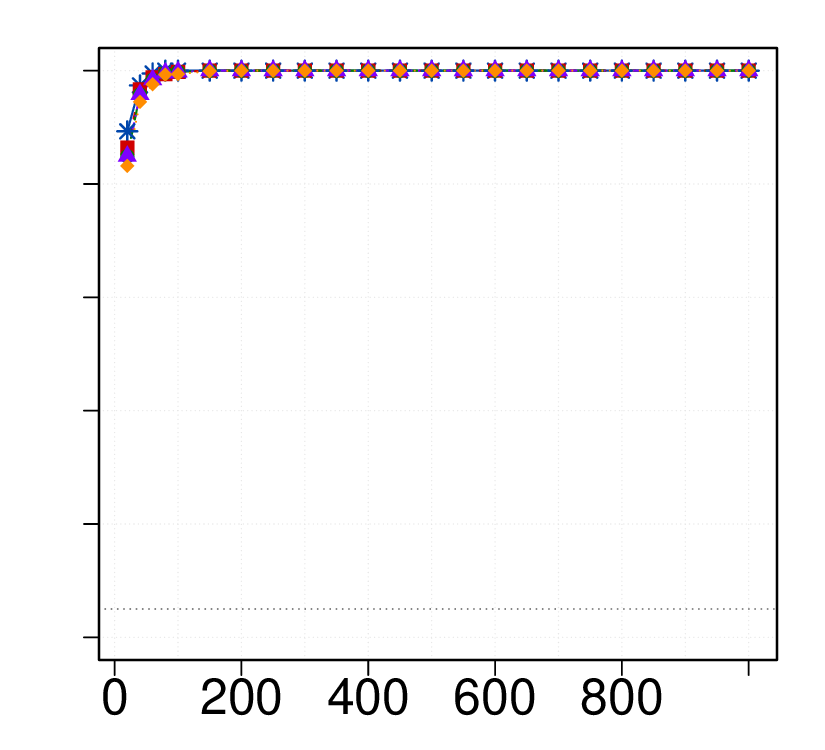}
        \hspace{-1 cm}
            \vspace{-0.2in}
            \caption*{\small{(c) $SNR=1.5$, with $m=1,5,20$ } }
 }           
            \caption{\small{
             Adjusted empirical rejection rates of the RP-Bonf, RP-BH, RP-HMP, RP-HMP-raw, and RP-CCT methods with the standard CUSUM test for various choices of number $k$ of random projections in the x-axis. 
            The data-generating process follows (\ref{eq:model for fd}), (\ref{eq:data generating process}), and (\ref{break function}), where the standard deviation $\sigma_{g}$ follows \textit{Setting 1}.
            The change point location is set at $\theta=0.25$.
            The empirical rejection rate is based on 1000 simulations.
            }}
            \label{fig:  tuning num_rps_sigma1(adj)}
    \end{figure}

Figure \ref{fig:  tuning num_rps_sigma1(adj)} displays the sizes and size-adjusted powers of the RP methods in \textit{Setting 1}. The results in \textit{Settings 2} and \textit{3} are provided in Figures \ref{fig:  tuning num_rps_sigma2(adj)} and \ref{fig:  tuning num_rps_sigma3(adj)} in Supplementary Material \ref{subsec: Result of using IID estimator}. The raw empirical rejection rates in \textit{Settings 1--3} are presented in Figures \ref{fig:  tuning num_rps_sigma1}, \ref{fig:  tuning num_rps_sigma2}, and \ref{fig:  tuning num_rps_sigma3}.
The change point location is set at $\theta=0.25$.
In Figure \ref{fig:  tuning num_rps_sigma1(adj)}, under the null ($SNR=0$), 
the RP-BH method can control the sizes well across different numbers of random projections. 
The RP-HMP{-raw} method tends to over-reject, similarly to what we observed in Table \ref{tab:tuning Pvalue-comb}.
Under the alternatives with $SNR=0.5$ and $SNR=1.5$, 
when $m=5$ and $m=20$, size-adjusted powers of the RP methods are improved significantly as the number $k$ of random projections rises. Moreover, the size-adjusted powers remain {relatively constant past} $k=200$. However, when $m=1$, the size-adjusted powers fluctuates without an ascending trend as $k$ increases.
Similar trends can be found in the raw powers.

\begin{figure} [h!]
{
            \centering
            \hspace{-1 cm}    
             \includegraphics[width=5.5cm, height=5cm]{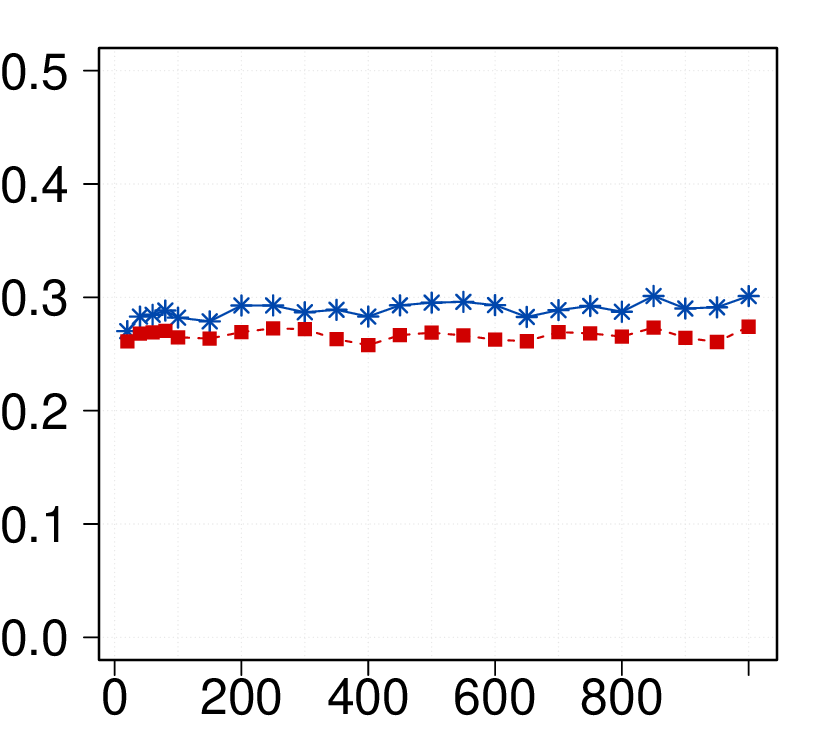}
            \hspace{-1 cm}    
             \includegraphics[width=5.5cm, height=5cm]{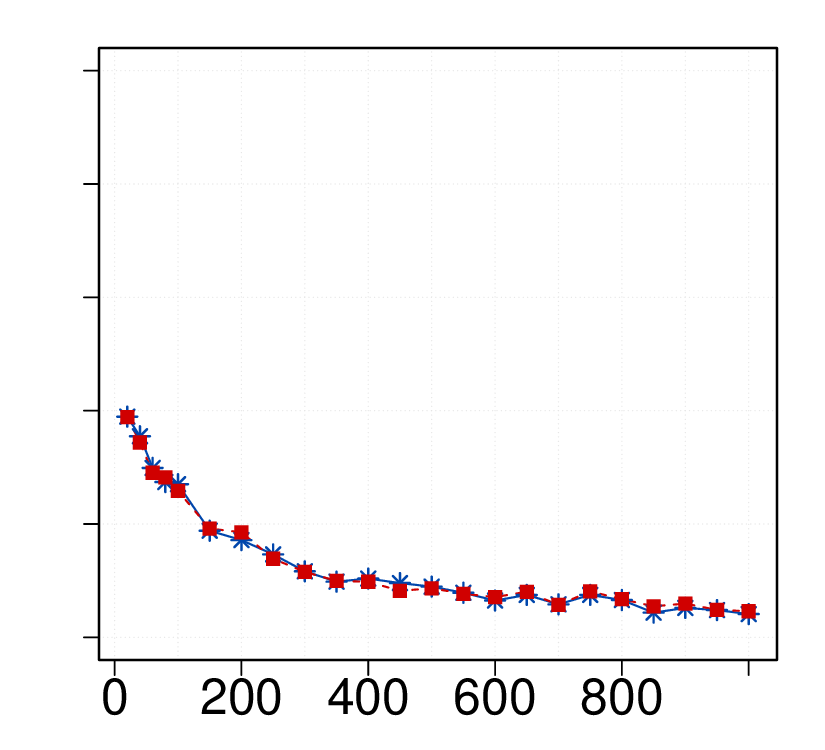}
           \hspace{-1 cm}    
             \includegraphics[width=5.5cm, height=5cm]{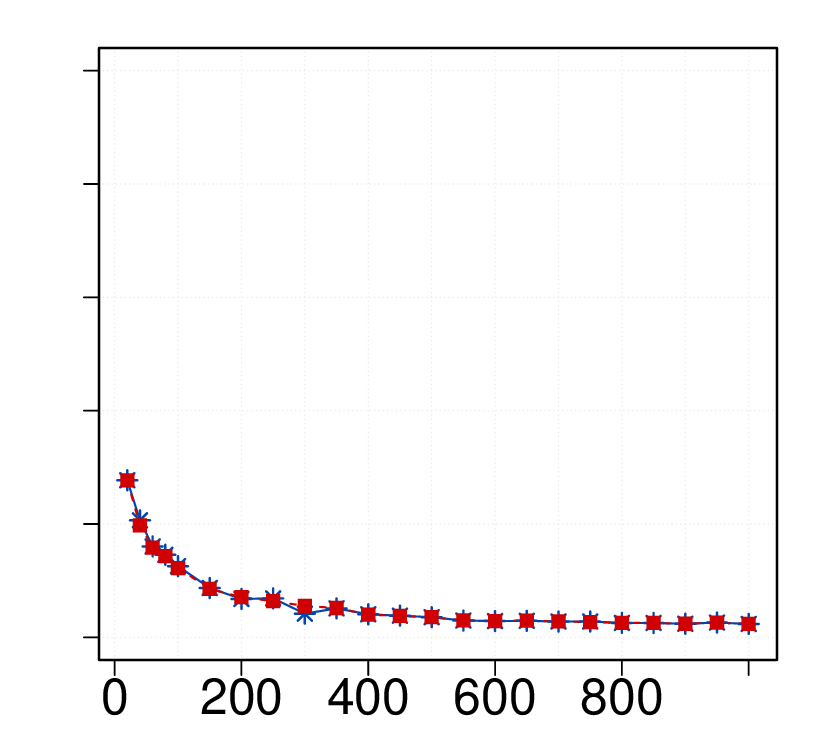}
           \hspace{-1 cm}    
            \vspace{-0.2in}
            \caption*{\small{(a) \textit{Setting 1}, with $m=1,5,20$ } }

            \centering
            \hspace{-1 cm}    
             \includegraphics[width=5.5cm, height=5cm]{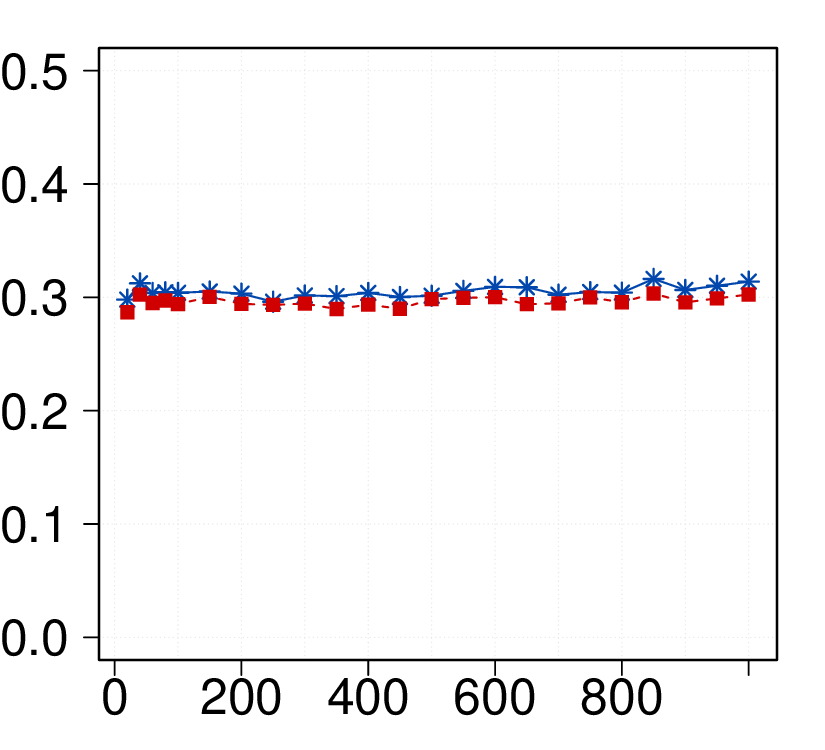}
           \hspace{-1 cm}    
             \includegraphics[width=5.5cm, height=5cm]{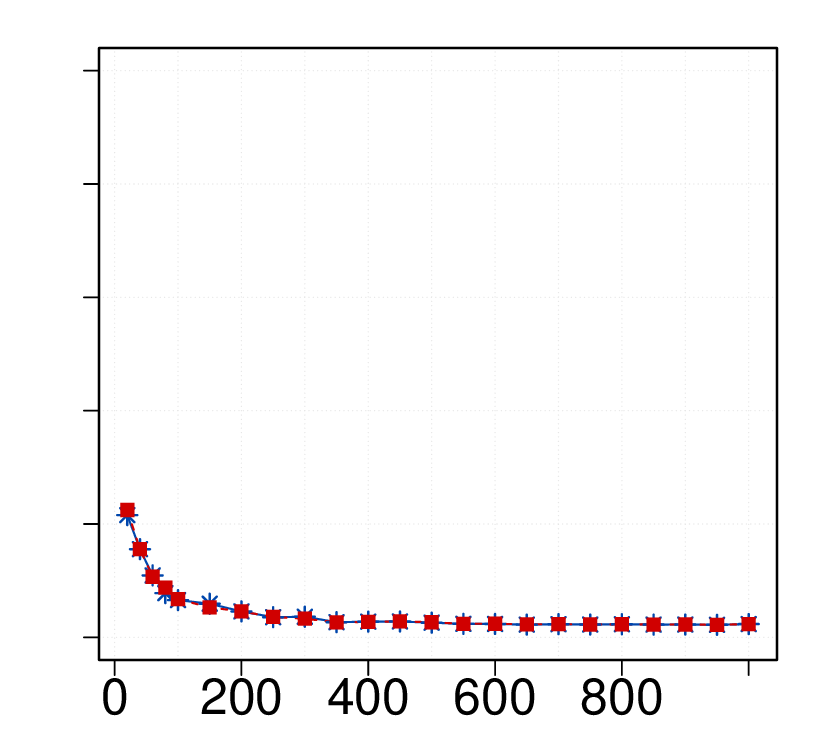}
           \hspace{-1 cm}    
             \includegraphics[width=5.5cm, height=5cm]{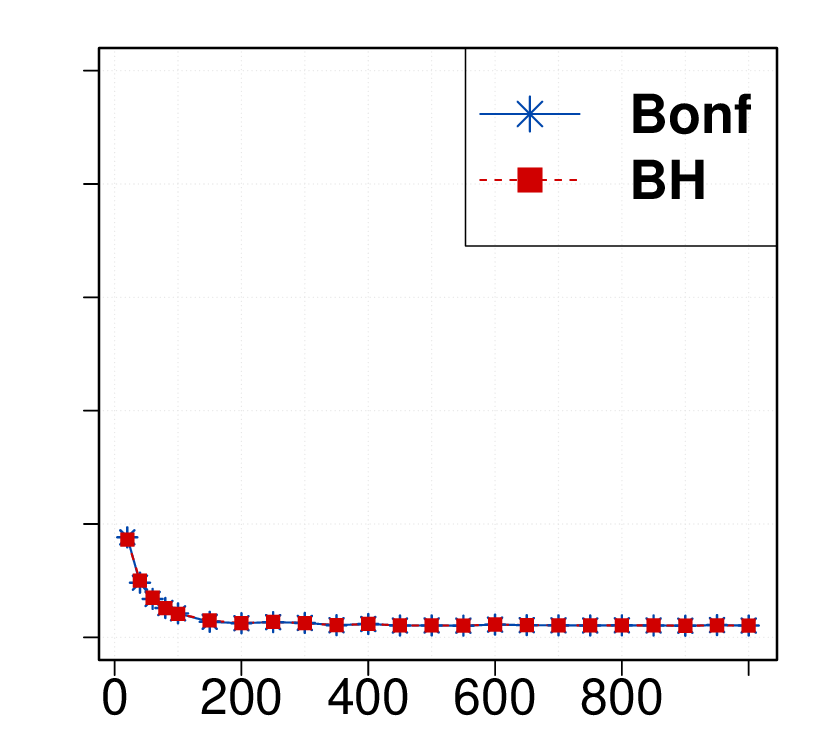}
           \hspace{-1 cm}    
            \vspace{-0.2in}
            \caption*{\small{(b) \textit{Setting 2}, with $m=1,5,20$ } }

            \centering
          \hspace{-1 cm}    
             \includegraphics[width=5.5cm, height=5cm]{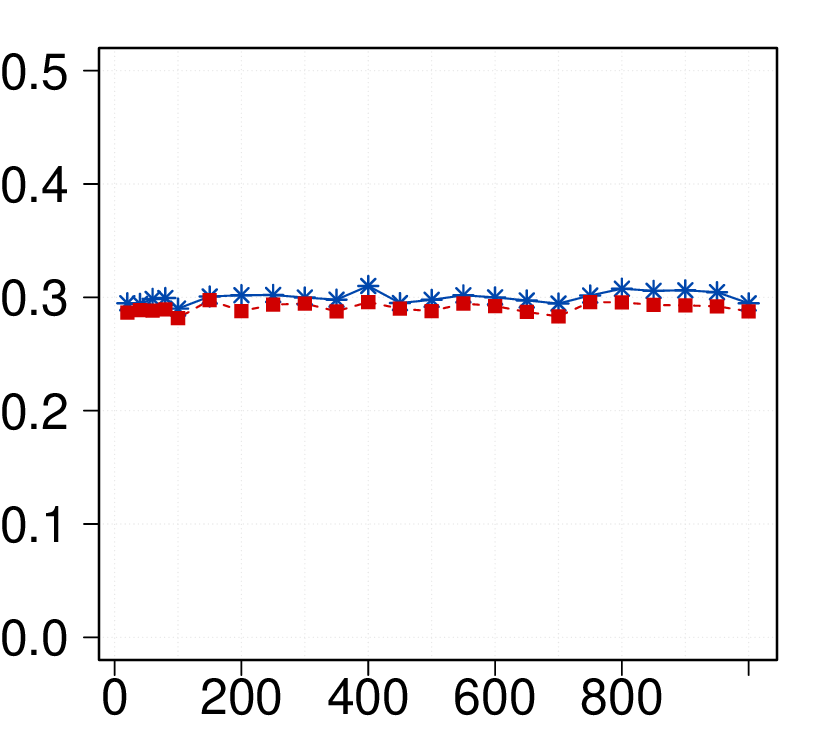}
           \hspace{-1 cm}    
             \includegraphics[width=5.5cm, height=5cm]{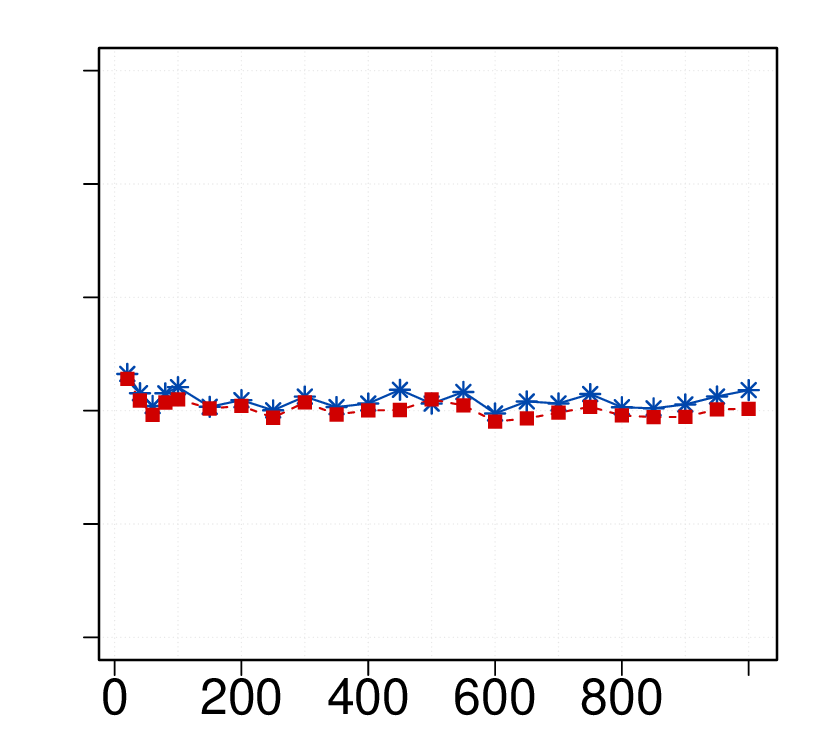}
           \hspace{-1 cm}    
             \includegraphics[width=5.5cm, height=5cm]{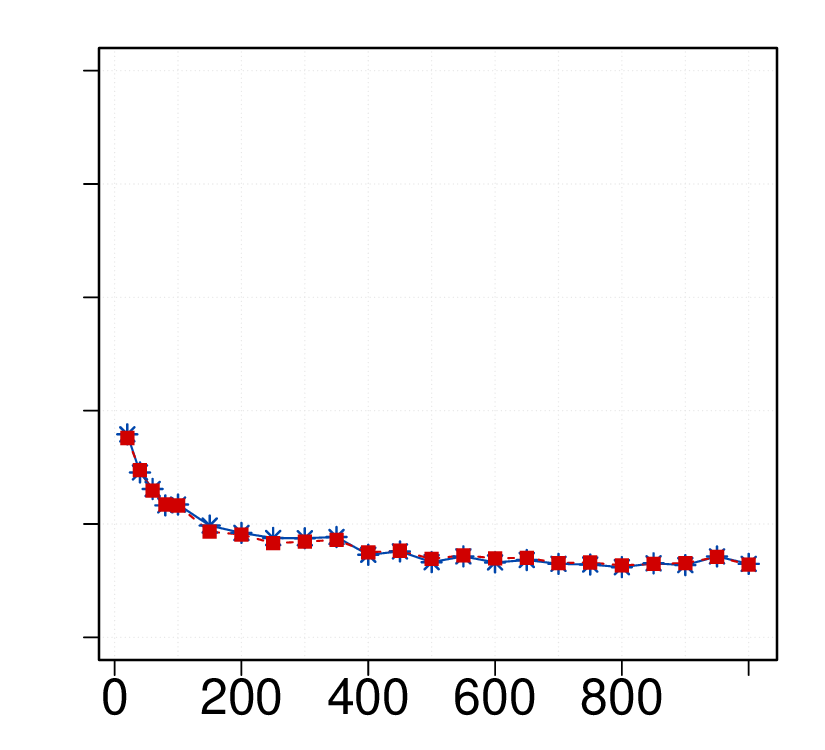}
           \hspace{-1 cm}    

            \vspace{-0.2in}
            \caption*{\small{(c) \textit{Setting 3}, with $m=1,5,20$ } }
            
}            
            \caption{\small{
            RMSE of estimated change point locations detected by the RP-Bonf and RP-BH methods with the standard CUSUM test for various choices of number $k$ of random projections in the x-axis. The data-generating process follows (\ref{eq:model for fd}), (\ref{eq:data generating process}), and (\ref{break function}), where the standard deviation $\sigma_{g}$ follows \textit{Settings 1--3}. The change point location is set at $\theta=0.25$. 
            The RMSE is based on 1000 simulations.
            }}
            \label{fig:  tuning num_rps_RMSE}
    \end{figure}

Figure \ref{fig:  tuning num_rps_RMSE} displays RMSE of all the estimated change point locations detected by the RP methods with an increasing number of random projections. The change point location is set at $\theta=0.25$. In Figure \ref{fig:  tuning num_rps_RMSE}, when $m=5$ and 20, RMSE exhibits a declining trend, decreasing rapidly at the beginning as the number $k$ of random projections increases. It seems that $k=200$ is enough in most cases. Beyond $k=200$, further increases in the number of random projections do not result in a remarkable gain in RMSE. However, when $m=1$, the RMSE is higher than that in $m=5$ and 20, and does not decline as $k$ increases. The difficulty observed when $m=1$ also arises in other comparable methods, as shown in Figure \ref{fig:  size and power Auedata} in Subsection \ref{subsec:comparison}. This may be because, when $m=1$, the break function is constant and the change is uniform across components with a small magnitude. In contrast, when $m=5$ or $m=20$, the break function is non-constant, and the change is different across components. Some components exhibit larger changes, which appear to be easier to detect. Therefore, for the data settings we consider, performing $k=200$ random projections appears to be a reasonable choice for the RP methods, as it offers a simpler computation while maintaining similar performance as with much larger $k$. The additional random projections beyond $k=200$ undermine the computational efficiency of the RP method.

\subsection{Comparison of change point methods}
\label{subsec:comparison}

In this subsection, we compare the performance of the RP methods and other existing single change point detection methods for functional or high-dimensional data, including projection-based methods and functional methods without dimension reduction. Projection-based methods include FPC-based methods from \cite{aue2009estimation} and \cite{berkes2009detecting} with different desired proportion of total variation explained (TVE): 0.85, 0.90, and 0.95 (FPC-0.85, FPC-0.90, FPC-0.95), and a method from \cite{wang2018high}, where  one optimal direction that closely aligns with the direction of the change is identified. Methods without dimension reduction include a fully functional (FF) method from \cite{aue2018detecting} and a method from \cite{dette2020testing}.

In the process of change point detection, the RP methods, the FPC-based methods, and the FF method use a standard CUSUM process, while the methods in \cite{wang2018high} and \cite{dette2020testing} and use a weighted CUSUM process.
We modify the methods in \cite{wang2018high} and \cite{dette2020testing} 
to use the standard CUSUM process and label them as WS and DKV, respectively.
The original versions of the two methods that use weighted CUSUM are labeled as WS-weighted and DKV-weighted. We present only the results of WS and DKV using the standard CUSUM process due to better size control in most cases in the considered settings than WS-weighted and DKV-weighted. The results of using weighted CUSUM are presented in the Figure \ref{fig:  CP locations Auedata} in Subsection  \ref{subsec: weightedCUSUM} in the Supplementary Material.
For \cite{wang2018high}, we also consider their simulation-based approach to find a critical value. In our results, both WS and WS-CUSUM indicate \cite{wang2018high}'s method with standard CUSUM but with two different ways to compute critical values: WS chooses the $95\%$ quantile of maximized standard CUSUM statistics of 1000 i.i.d. samples with no change point nor any temporal dependence, while the WS-CUSUM method rely on the simulation of Brownian bridges, similarly to the one described in the beginning of Subsection \ref{subsection: tuning} for the weighted CUSUM.

\begin{figure} [h!]
            \centering
            \hspace{-1 cm}  
            \includegraphics[width=5.5cm, height=5cm]{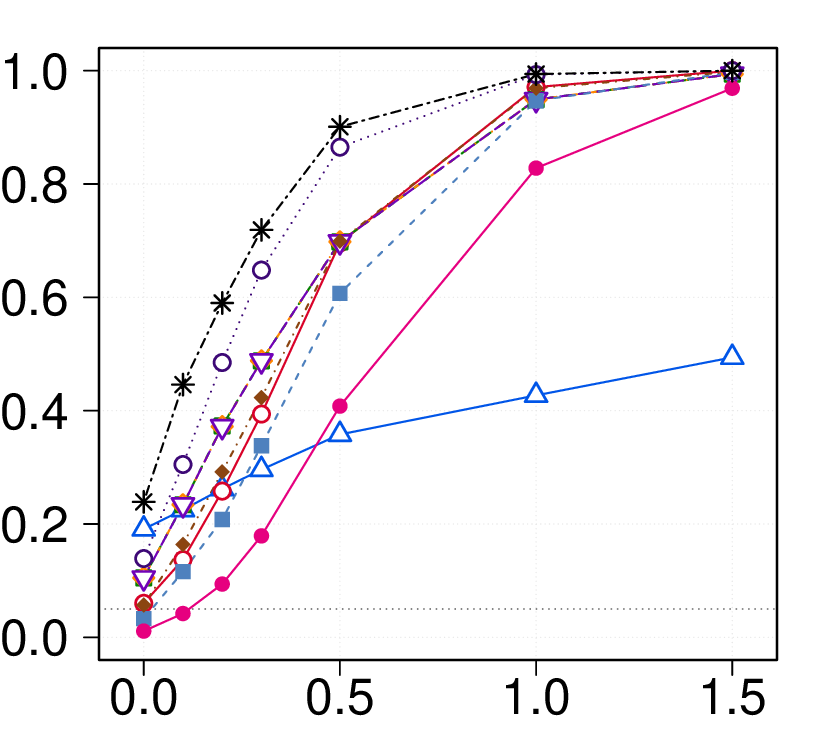}
            \hspace{-1 cm}  
            \includegraphics[width=5.5cm, height=5cm]{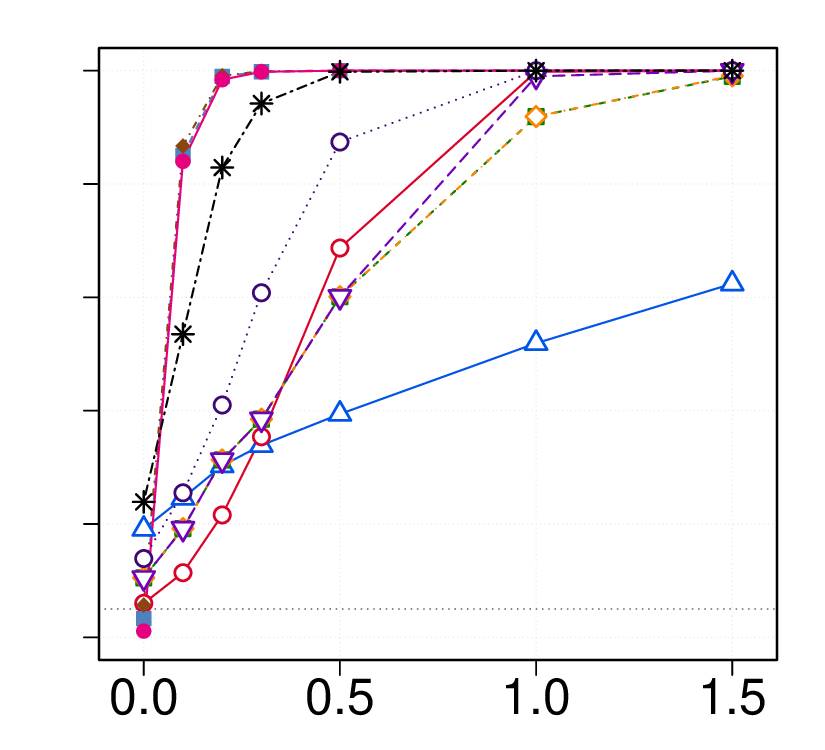}
            \hspace{-1 cm}  
            \includegraphics[width=5.5cm, height=5cm]{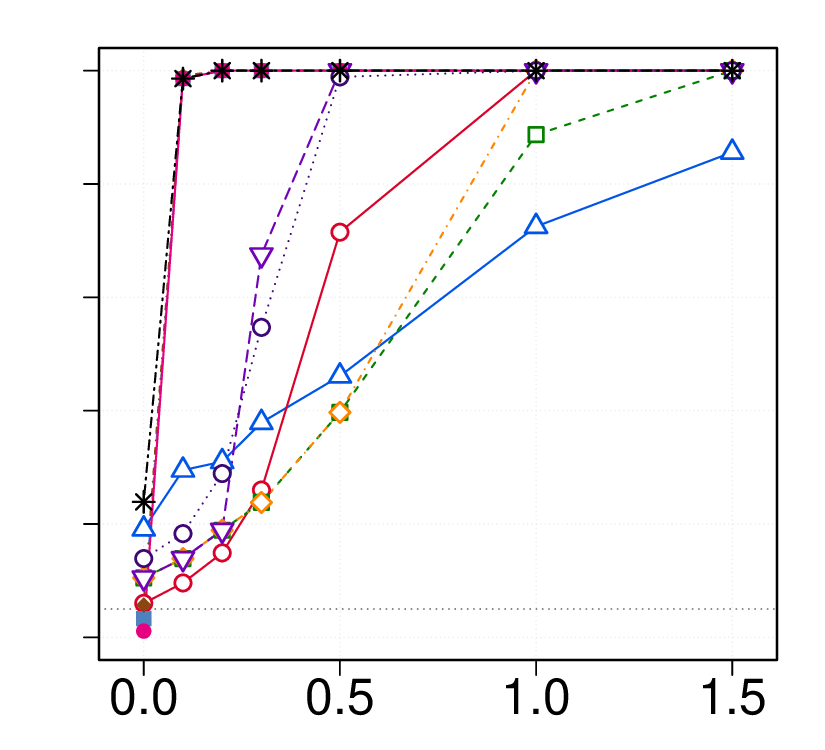}
            \hspace{-1 cm}  
            \vspace{-0.2in}
            \caption*{ \small{(a) \textit{Setting 1}, with $m=1,5,20$ } }

            \centering
            \hspace{-1 cm}  
            \includegraphics[width=5.5cm, height=5cm]{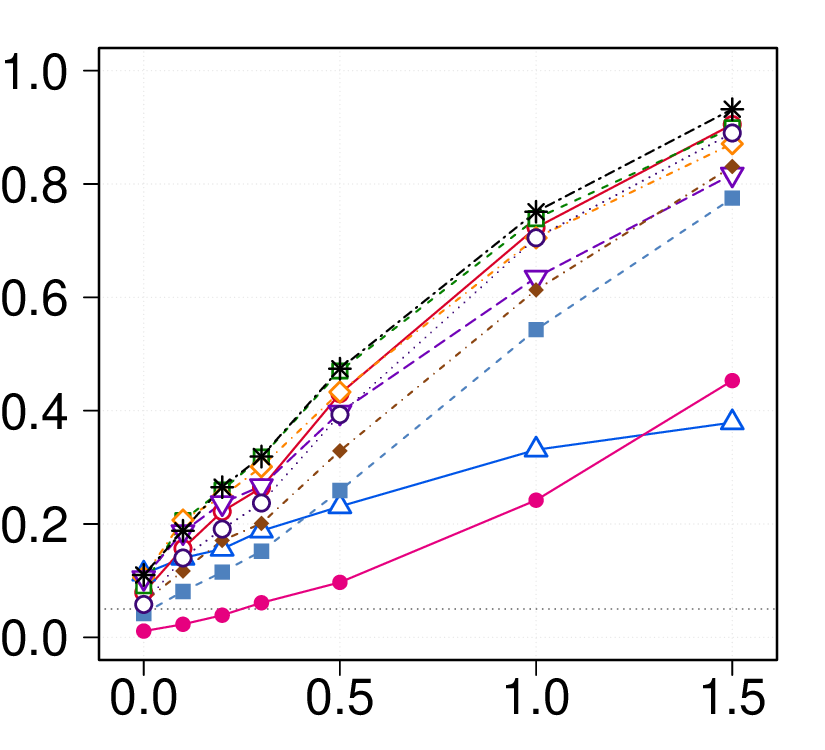}
            \hspace{-1 cm}  
            \includegraphics[width=5.5cm, height=5cm]{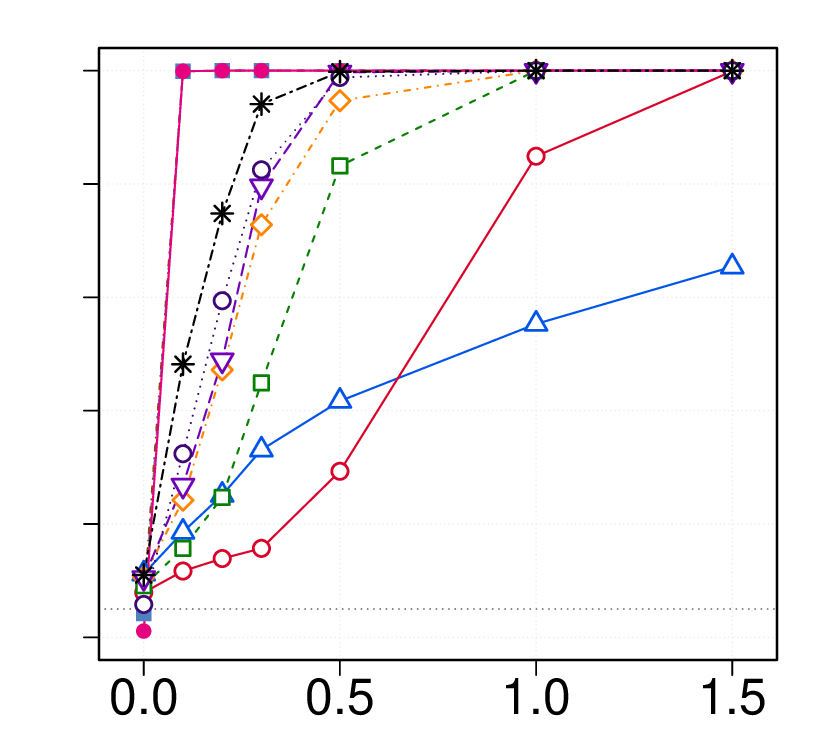}
            \hspace{-1 cm}
            \includegraphics[width=5.5cm, height=5cm]{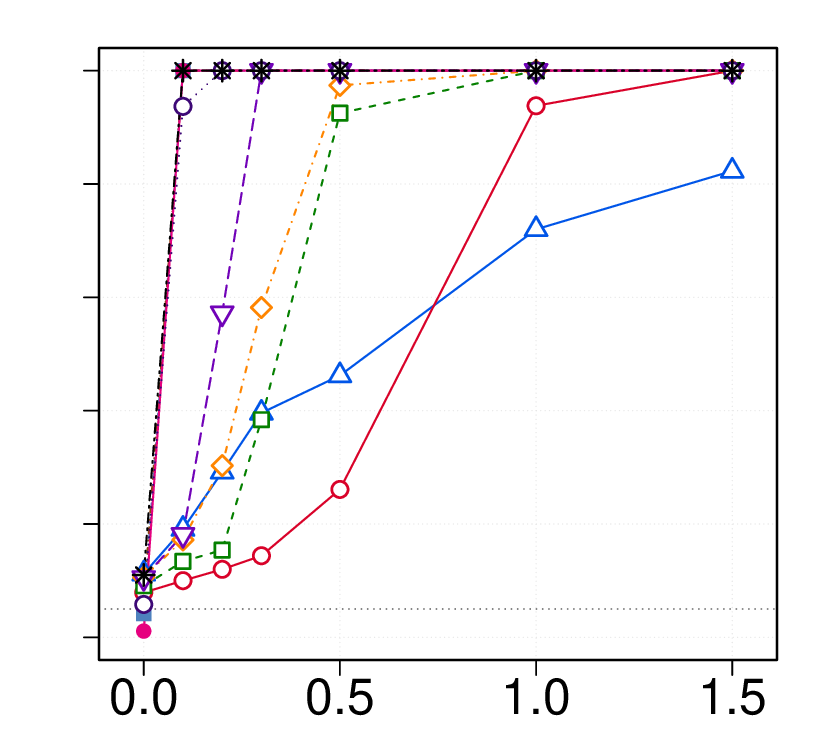}
            \hspace{-1 cm}
            \vspace{-0.2in}
            \caption*{\small{(b) \textit{Setting 2}, with $m=1,5,20$ } }

            \centering
            \hspace{-1 cm}
            \includegraphics[width=5.5cm, height=5cm]{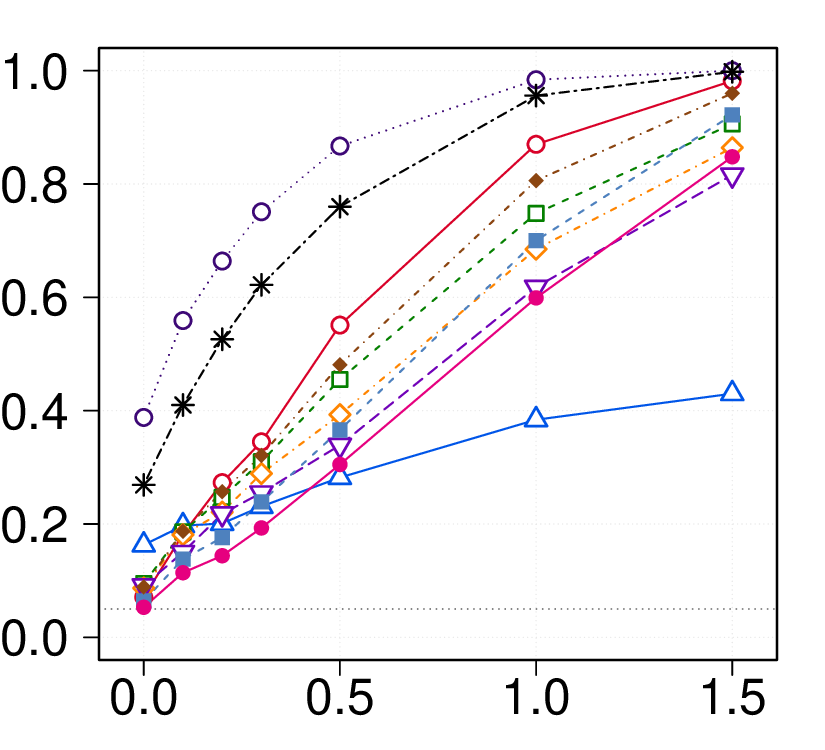}
            \hspace{-1 cm}  
            \includegraphics[width=5.5cm, height=5cm]{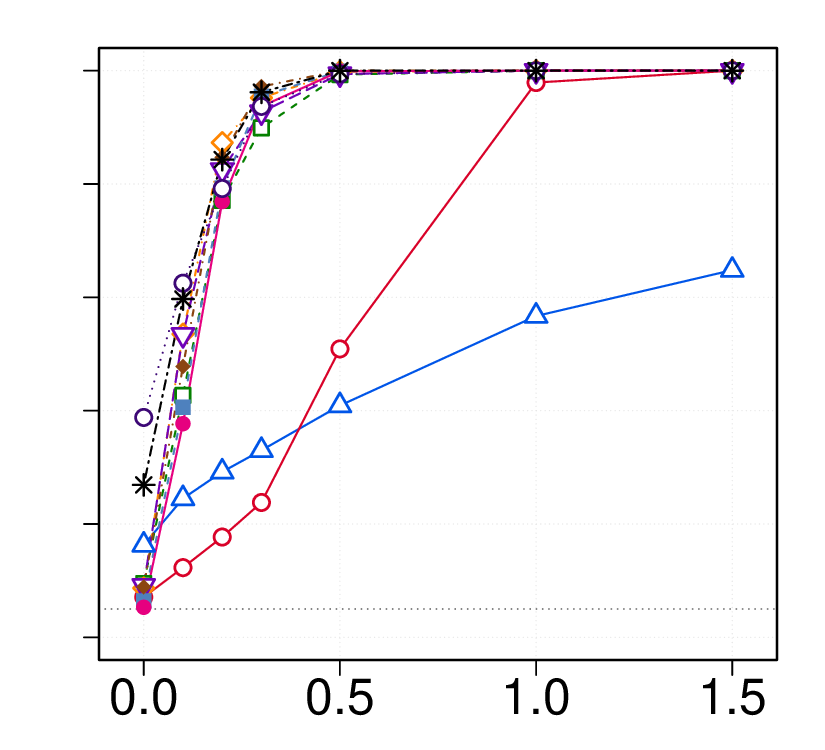}
            \hspace{-1 cm}
            \includegraphics[width=5.5cm, height=5cm]{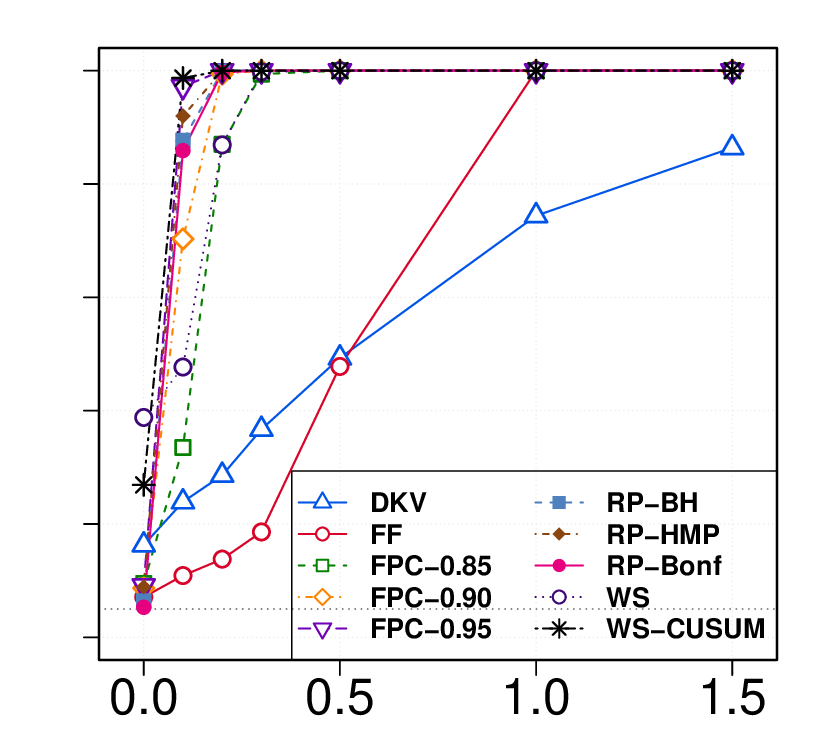}
            \hspace{-1 cm}
            \vspace{-0.2in}
            \caption*{\small{(c) \textit{Setting 3}, with $m=1,5,20$ } }

            \caption{\small{Raw empirical rejection rates of various change point detection methods, including projection-based and fully functional methods, for various values of $SNR$ in the $x$-axis.
            The data-generating process follows (\ref{eq:model for fd}), (\ref{eq:data generating process}), and (\ref{break function}), where the standard deviation $\sigma_{g}$ follows \textit{Settings 1--3}. 
            The change point location is set at $\theta =0.25$. The rejection rate is based on 1000 simulations.
            }}
           \label{fig:  size and power Auedata}
    \end{figure}

We compare the aforementioned methods in terms of sizes, raw powers, and the accuracy of estimated change point locations. Figure \ref{fig:  size and power Auedata} displays the empirical rejection rates of the RP method using 200 random projections and the other change point detection methods for high-dimensional or functional data. {In Section \ref{subsection: tuning}, we found that HMP-raw have a relatively strong size-distortion, and HMP and CCT yield almost identical size-adjusted powers. Among these three methods, we present only HMP to avoid visual clutter in Figure \ref{fig:  size and power Auedata}.} The change point location is set at $\theta=0.25$. Under the null, our RP-BH method maintains an acceptable empirical rejection rate close to a significance level of 0.05 in \textit{Settings 1--2}, while RP-Bonf method is conservative and RP-HMP is slightly over-rejecting. In \textit{Setting 3}, our RP-Bonf shows the best performance in controlling the sizes compared to RP-BH and RP-HMP. For other methods, the WS method also controls the sizes well, comparable to our RP-BH in \textit{Settings 1--2} and our RP-Bonf in \textit{Setting 3}, followed by the FF method, while other methods fail to control the sizes.
{Note that the WS-based methods tend to severely overreject under the null. Such behavior is expected, as the WS method was not developed as a global testing tool, but rather to locate changes when high-dimensional data are already known to contain structural breaks. Their method searches for the direction with the largest change, which can easily lead to false rejections.}

Under the alternatives, when $m=5$ and $20$, our RP methods have the best power in {\textit{Settings 1--3}, among those that maintain reasonable size control. FPC-based methods achieve comparable power to RP-based methods in \textit{Setting 3}, but exhibit substantially lower power in \textit{Settings 1--2}. The DKV method consistently underperform, with severe size distortions and low powers. The FF method is not as much powerful with $m=5$ or $m=20$. However, when $m=1$, FF is the most powerful method keeping the size under control. The RP-HMP and the FPC-based methods have comparable sizes to the FF method in \textit{Setting 1}, but in the other two \textit{Settings}, FF method can detect smaller changes most efficiently. RP-Bonf and RP-BH are less powerful than FF, RP-HMP, or FPC-based methods in this case ($m=1$).
Overall, if detecting the existence of a change point is the primary goal, our RP methods, particularly RP-HMP, are recommended when analyzing data with potentially non-constant break functions. The FF method is also a strong candidate, but risks losing substantial power when conditions deviate from its ideal setting.}

\begin{figure} [h!]
            \centering
            \hspace{-1 cm} 
            \includegraphics[width=5.5cm, height=5cm]{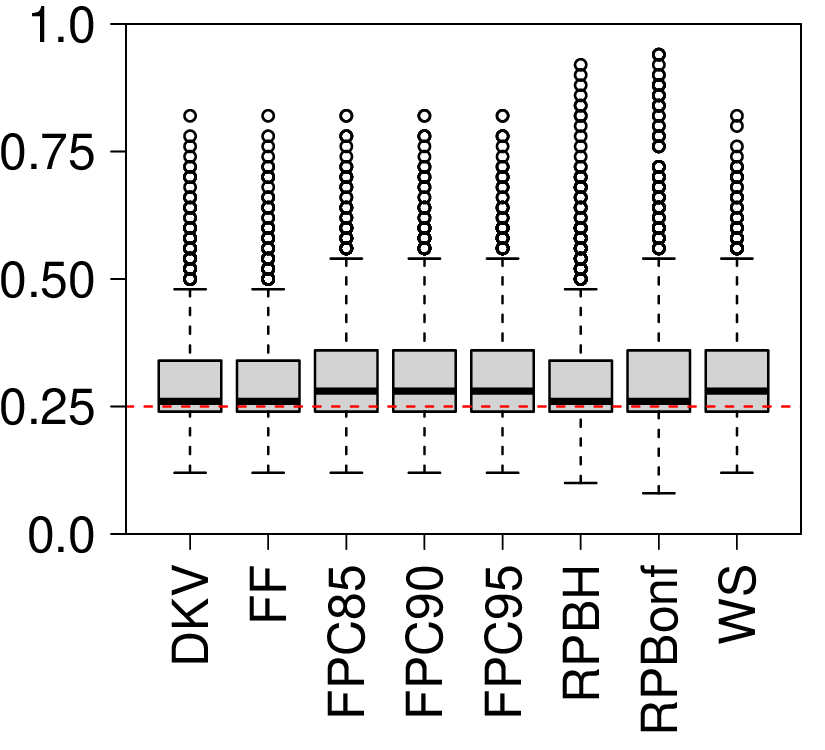}
            \hspace{-1 cm} 
            \includegraphics[width=5.5cm, height=5cm]{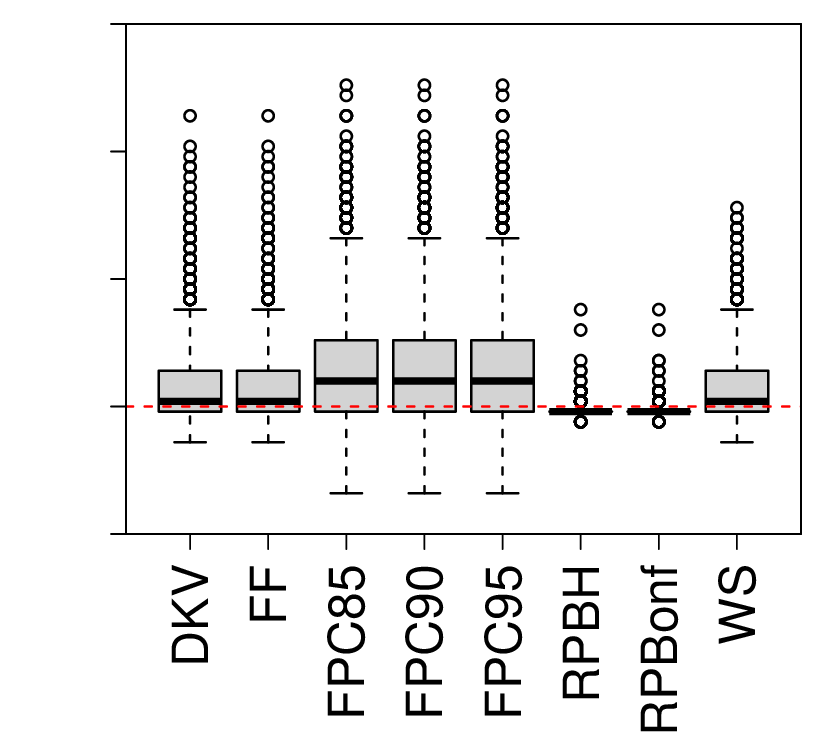}
            \hspace{-1 cm} 
            \includegraphics[width=5.5cm, height=5cm]{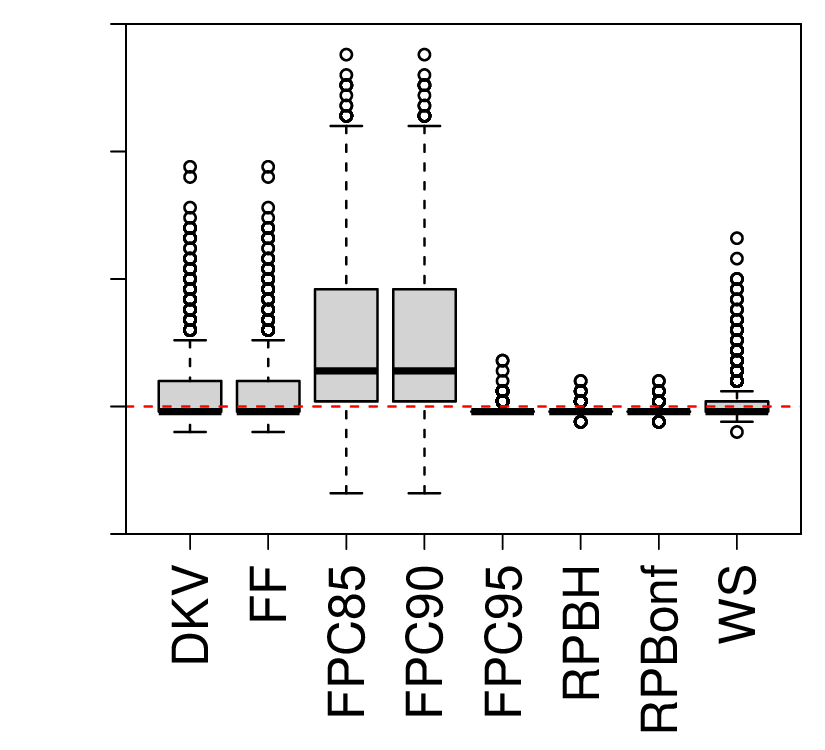}
            \hspace{-1 cm} 
            \vspace{-0.2in}
            \caption*{\small{(a) \textit{Setting 1}, with $m=1,5,20$ } }

            \centering
            \hspace{-1 cm} 
            \includegraphics[width=5.5cm, height=5cm]{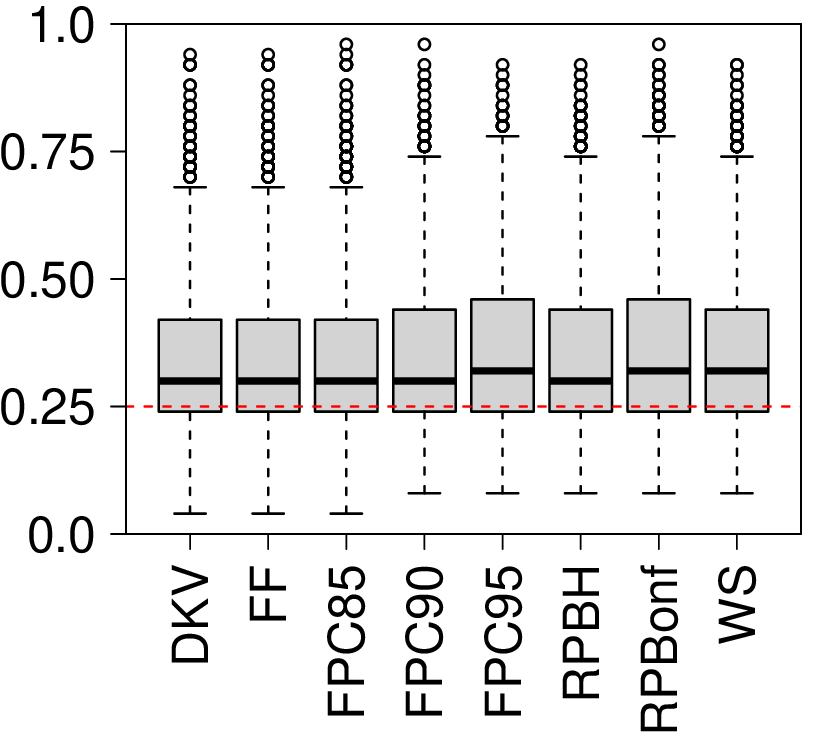}
            \hspace{-1 cm} 
            \includegraphics[width=5.5cm, height=5cm]{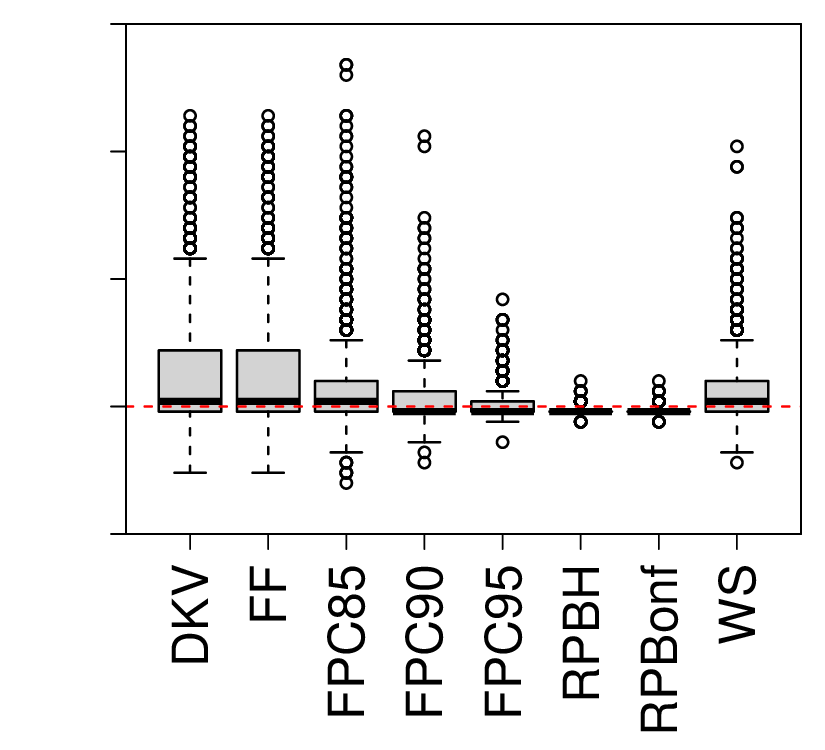}
            \hspace{-1 cm} 
            \includegraphics[width=5.5cm, height=5cm]{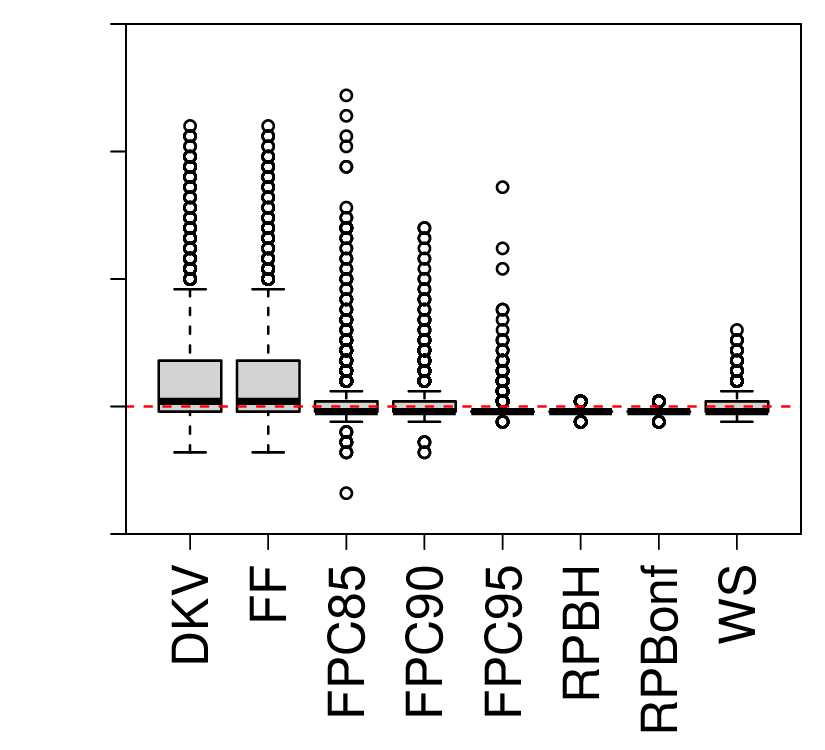}
            \hspace{-1 cm} 
            \vspace{-0.2in}
            \caption*{\small{(b) \textit{Setting 2}, with $m=1,5,20$ } }

            \centering
            \hspace{-1 cm} 
            \includegraphics[width=5.5cm, height=5cm]{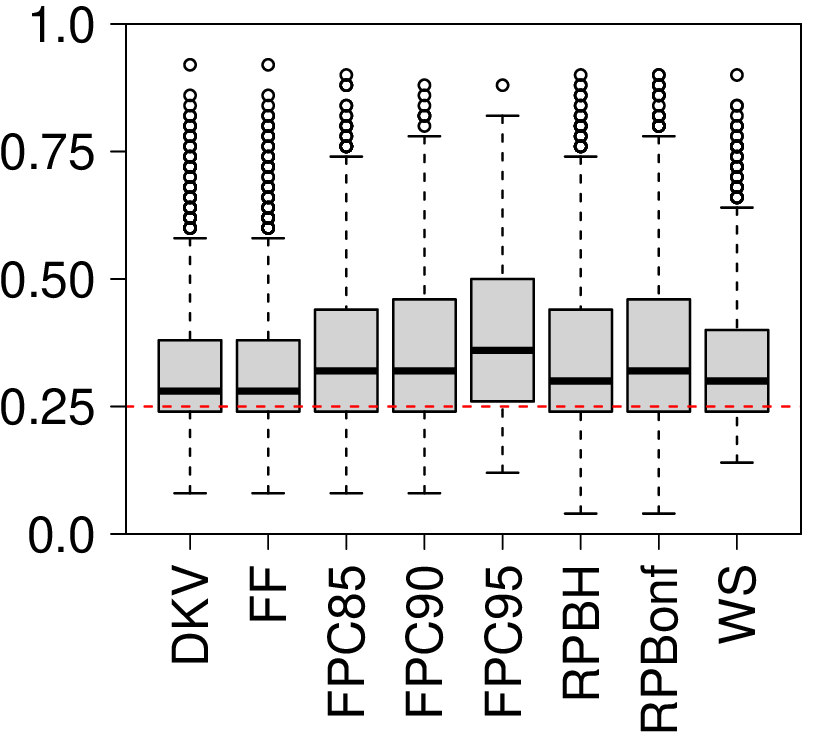}
            \hspace{-1 cm} 
            \includegraphics[width=5.5cm, height=5cm]{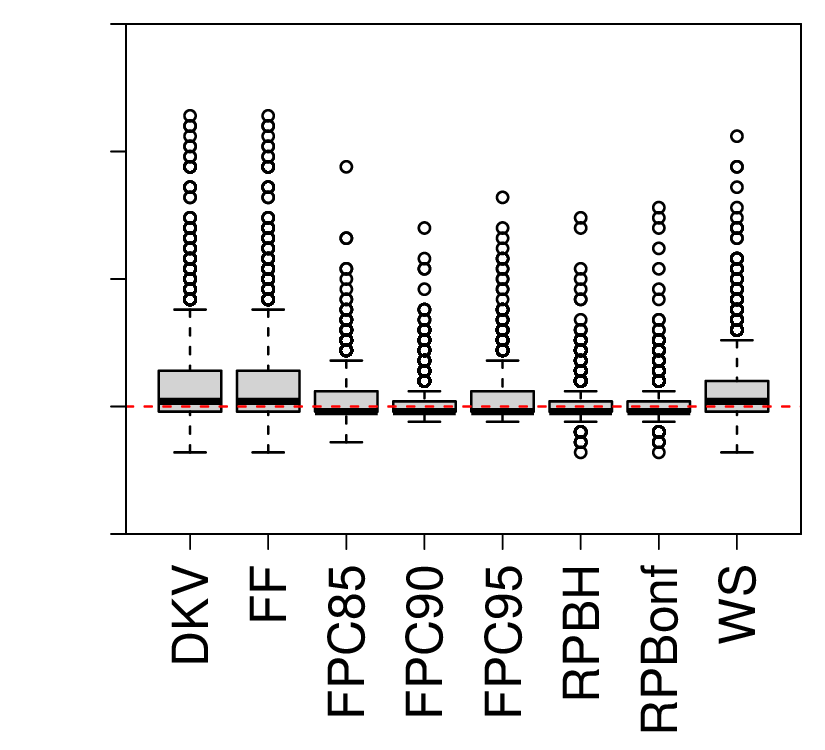}
            \hspace{-1 cm} 
            \includegraphics[width=5.5cm, height=5cm]{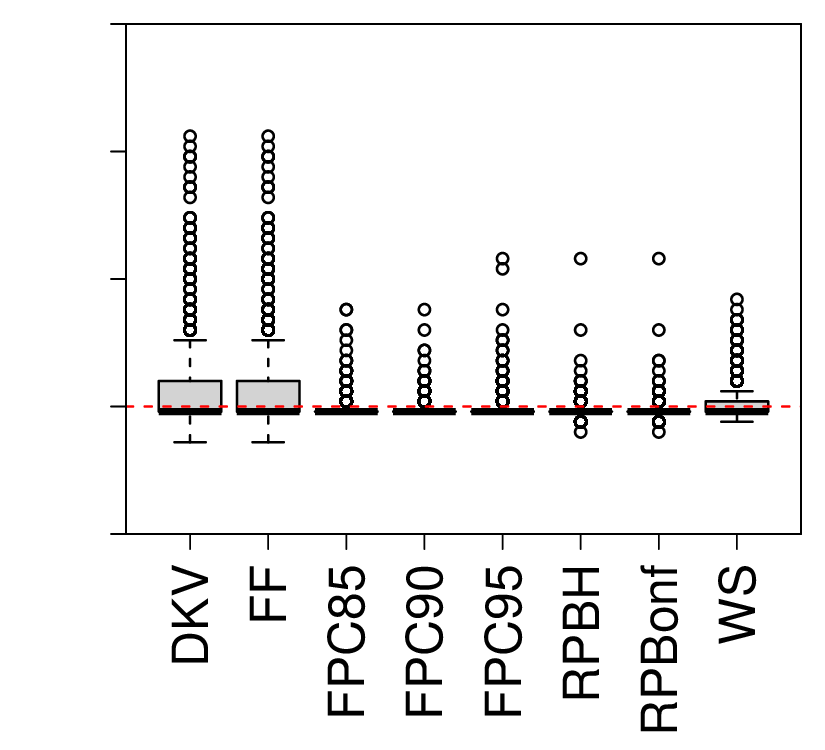}
            \hspace{-1 cm} 
            \vspace{-0.2in}
            \caption*{\small{(c) \textit{Setting 3}, with $m=1,5,20$ } }
            
            \caption{Estimated change point locations detected by different methods on 1000 simulations. The data-generating process follows (\ref{eq:model for fd}), (\ref{eq:data generating process}), and (\ref{break function}), where the standard deviation $\sigma_{g}$ follows \textit{Settings 1--3}.  The change point location is set at $\theta = 0.25$. The magnitude of the break function is scaled by $SNR=0.5$.}
            \label{fig:  CP locations Auedata standcusum}
    \end{figure}

Figure \ref{fig:  CP locations Auedata standcusum} presents the boxplots of estimated change points detected by the aforementioned methods. Our RP methods use $k=200$ random projections.
The change point location is set at $\theta = 0.25$. The magnitude of the break function is scaled by $SNR=0.5$. 
When $m = 5$ and $20$, 
in \textit{Settings 1--2}, the RP methods can detect the change point most accurately, as the median is closely aligned with the true location 0.25, followed by the WS method {or FPC 0.95 which have} a wider spread. In \textit{Setting 3}, the spread of our estimates is similar to that of {WS method or FPC-based methods} with the median of our estimates being closer to the true location. When $m=1$, the RP methods do not have distinct advantages. The RP methods are similar to the WS method in both median and spread. However, when $m=1$, methods without dimension reduction (DKV, FF) show advantages in \textit{Setting 3}, as their medians are closer to the true location. Overall, given the high accuracy and precision, RP methods are well-suited for applications with non-constant break function. In the case of constant break function, methods without dimension reduction {such as FF} may be preferable.

\subsection{
Location estimates of the RP method in practice} \label{subsec:repeat RP method on one data}

The location estimates based on RP methods may be unstable due to the randomness of directions. In practice, we propose to repeat RP-BH (or RP-Bonf) method multiple times to achieve stability. We compare the quality of location estimates of RP-based methods with those of existing methods in simulations: fix a single simulated dataset, and repeat each RP-BH (or RP-Bonf) method 1000 times. We consider $k=200$ random projections and the standard CUSUM. The change point location is set at $\theta = 0.25$. The magnitude of the break function is scaled by $SNR=0.5$.

\begin{figure}[p]
    \centering
    \includegraphics[width=\textwidth]{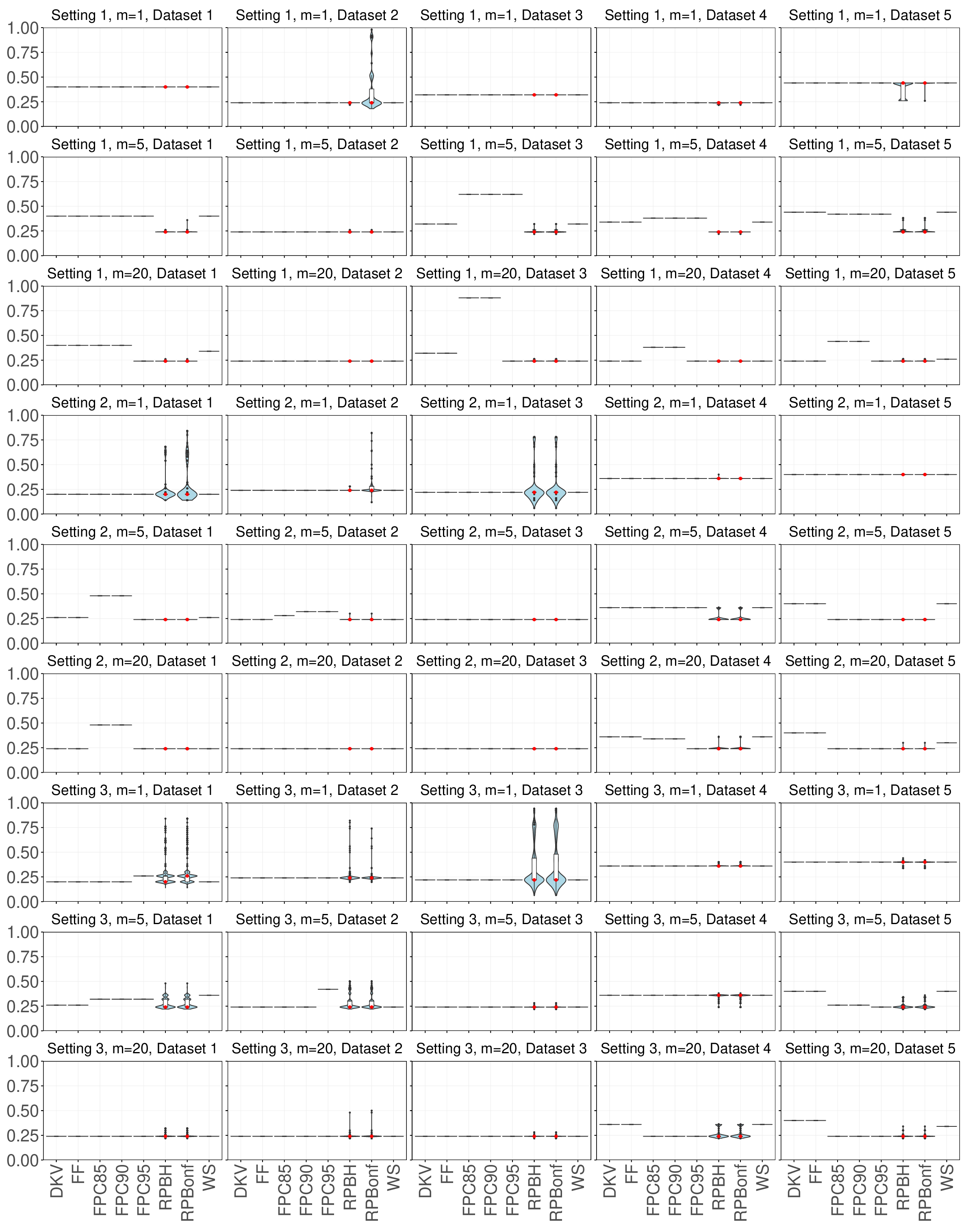}
    \caption{Estimated change point locations detected by repeating the RP-based methods on one dataset (Dataset 1 through Dataset 5) for 1000 times. For the RP methods, the mode of the estimated locations across the 1000 repetitions is marked by a red dot. 
            The data-generating process follows (\ref{eq:model for fd}), (\ref{eq:data generating process}), and (\ref{break function}), where the standard deviation $\sigma_{g}$ follows \textit{Settings 1--3}.  The change point location is set at $\theta = 0.25$. The magnitude of the break function is scaled by $SNR=0.5$.}
    \label{fig:violin_grid_data0}
\end{figure}

\begin{figure}[p]
    \centering
    \includegraphics[width=\textwidth]{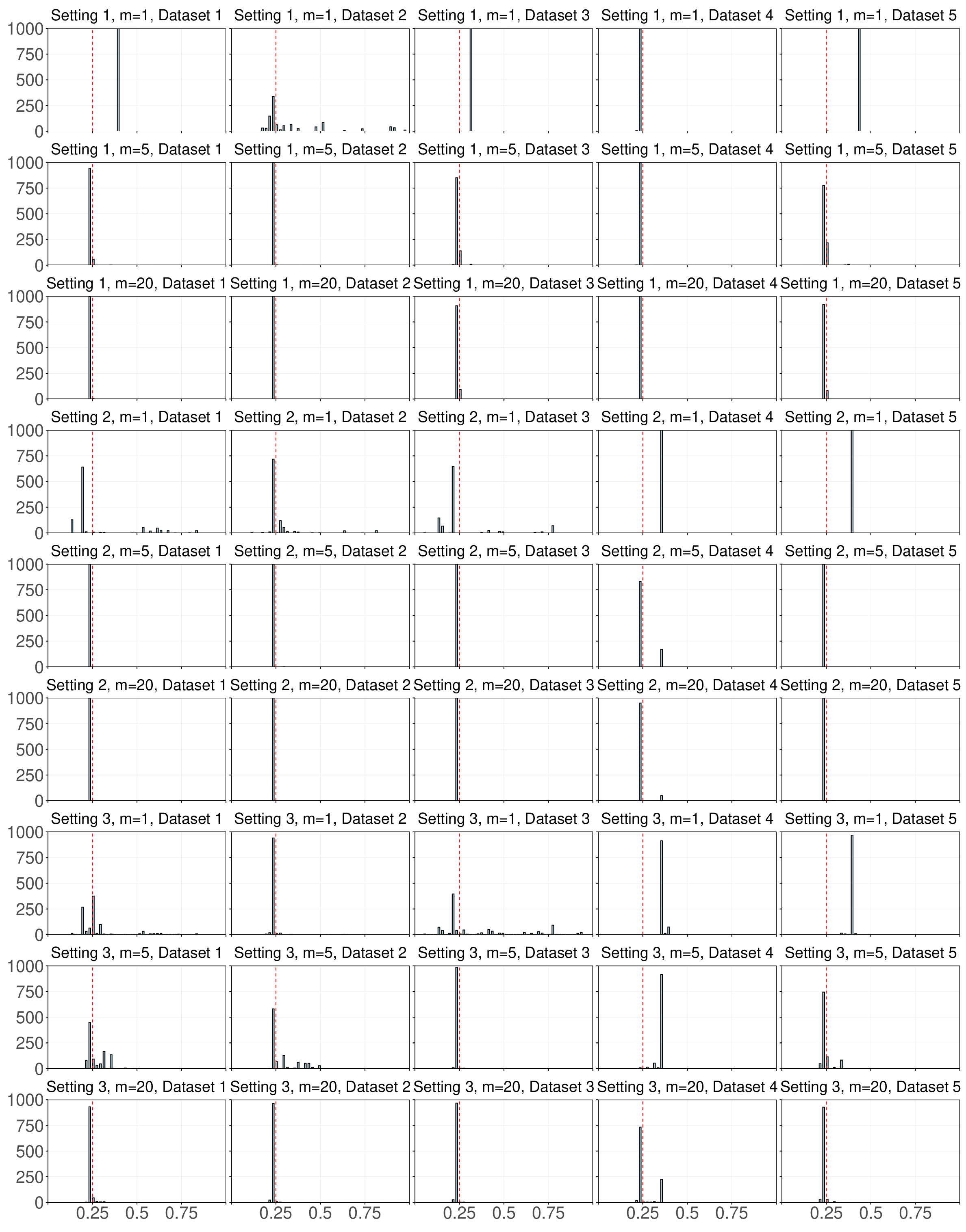}
    \caption{Frequency of estimated locations from repeating the RP-Bonf method on one dataset (Dataset 1 through Dataset 5) for 1000 times.         
            The data-generating process follows (\ref{eq:model for fd}), (\ref{eq:data generating process}), and (\ref{break function}), where the standard deviation $\sigma_{g}$ follows \textit{Settings 1--3}.  The change point location is set at $\theta = 0.25$. The magnitude of the break function is scaled by $SNR=0.5$.}
    \label{fig:hist_bonf:data0}
\end{figure}

Figure \ref{fig:violin_grid_data0} summarizes the resulting empirical distributions of the 1000 estimated change point locations using violin plots, with the red dot indicating the mode across all 1000 repetitions for RP-BH and RP-Bonf. {Five different datasets (Dataset 1 through 5) are considered. To investigate the distribution of estimated locations more precisely, Figure \ref{fig:hist_bonf:data0} presents the histograms of estimated locations from RP-Bonf for Datasets 1 through 5. Figure \ref{fig:hist_bh:data0} in Supplementary Material \ref{subsec:Repeat methods for data1-5} presents the same figure for RP-BH.} In most cases, the RP methods exhibit less variability when $m=5$ and $m=20$ than $m=1$, especially in \textit{Settings 1--2}. When $m=5$ and $m=20$, though repeating the RP method can yield variable estimated locations, the mode of the RP repetitions is accurate in most cases, tending to be closer to the true location than other methods. When $m=1$, the mode of the RP repetitions is similar to the locations obtained by other methods in most datasets. 

In practice, we recommend repeating the RP method. If the detected change points across repeated RP methods are consistent, a small number of repetition would be okay. However, if they vary widely, we recommend repeating the RP method many times, for example, 1000 times, and choosing an aggregate estimate, for example, the mode. The mode across repetitions can be reported as the final change point estimate because it targets the most frequently reproducible change point location, and is less influenced by outlier estimates. When the distribution has multi peaks, it might be an indication that there exist multiple change points, or the mean change is more gradual rather than abrupt. This related topic of detecting multiple change points will be discussed further in separate work.

{
\subsection{Sparse Functional Setting}\label{subsec: sparse functional setting}
The functional data setting in Sections \ref{subsection: tuning}--\ref{subsec:repeat RP method on one data} assumes a dense change scenario, where all coordinates experience a simultaneous structural break. However, such global changes may not be realistic for many real-world applications where structural breaks are inherently localized or sparse. To evaluate performance under more realistic conditions, this subsection compares different change-point tests under a sparse functional framework. The functional data are generated following equations (\ref{eq:model for fd}) and (\ref{eq:data generating process}) but with a different change function. While the setup in equation (\ref{break function}) constructs change functions using a global Fourier basis, using Fourier functions makes it difficult to mimic sparse structural breaks due to their infinite support. We instead adapt a B-spline basis, taking advantage of its local support property to model localized, roughly disjoint segments. A total of 23 basis functions $\tilde{v}_1(s), \dots, \tilde{v}_{23}(s)$ is considered for the cubic B-spline with 19 internal knots. The change function is constructed using the first $m$ elements of this sequence:
\begin{equation}\label{bspline break}  
\delta(s) = \delta(s; m, c) = \frac{\sqrt{c}}{\Vert{}\sum_{g=1}^m \tilde{v}_g\Vert{}} \sum_{g=1}^m \tilde{v}_g(s),
\end{equation}
where $c$ is from equation (\ref{break function}), and $\Vert{}\cdot\Vert{}$ denotes the standard $L_{2}$ norm. This scaling factor $\Vert{}\sum_{g=1}^m \tilde{v}_g\Vert{} = \left( \int \left[ \sum_{g=1}^m \tilde{v}_g(s) \right]^2 ds \right)^{1/2}$ ensures $\delta(s)/\sqrt{c}$ maintains unit signal energy across different $m$ values, aligning with the simulation framework in \cite{aue2018detecting}. Unlike the Fourier basis that are orthonormal, this explicit computation is required for the B-spline basis functions. In our simulations, $m=5,10,23$ are considered, representing roughly 25\%, 50\%, 100\% density of the impacted subregions. 

Figure \ref{fig:  size and power Auedata: data0 bspline} presents the size-power plots. We first note that as the change function becomes denser (that is, for larger $m$), all methods struggle more to achieve high power, which may seem counterintuitive. However, this is simply an artifact of maintaining the same total amount of change for $\delta(s)/\sqrt{c}$ across different values of $m$. When $m$ is smaller, fewer coordinates undergo changes, thereby inflating the per-coordinate signal strength.

In the dense scenario with weaker signals ($m=23$), functional methods except for DKV (namely, FF or FPC-based methods) perform the best, showing particularly strong performance under \textit{Setting 3}. The three RP-based methods also exhibit differing power, with RP-HMP and RP-BH outperforming RP-Bonf by a large margin. With fewer but stronger signals ($m=5$ or $m=10$), the RP-based methods generally provide strong size control and power. In this case, the three RP-based approaches yield almost identical rejection rates, boasting the best size control and power. FPC-based methods match this performance in \textit{Setting 3}, but in \textit{Settings 1} and \textit{2}, the RP-based methods are clear winners. As discussed in Section \ref{subsec:comparison}, WS-based methods may fail to maintain size control -- this size distortion is most pronounced in \textit{Settings 1} and \textit{3}. 

Figure \ref{fig:  CP locations Auedata standcusum: data0 bspline} in Supplementary Material \ref{supp:sec:sparse functional} presents boxplots of location estimates across the 1000 simulations. Consistent with the size-power plots, the RP-based methods yield more accurate location estimates in the sparse cases ($m=5, 10$) and achieve performance comparable to other methods in the dense case ($m=23$). Violin plots and histograms of location estimates across five datasets are presented in Figures \ref{fig:violin_grid:data0 bspline}, \ref{fig:hist_bonf:data0} and \ref{fig:hist_bh:data0}. As in Section \ref{subsec:repeat RP method on one data}, for a fixed dataset, the RP methods are repeated 1000 times to generate these violin plots and histograms. While the mode may not always hit the true change location $\theta=0.25$ exactly, it frequently aligns with it, validating the strategy introduced in Section \ref{subsec:repeat RP method on one data} under the sparse functional setting.

\begin{figure} [h!]
            \centering
            \hspace{-1 cm}  
            \includegraphics[width=5.5cm, height=5cm]{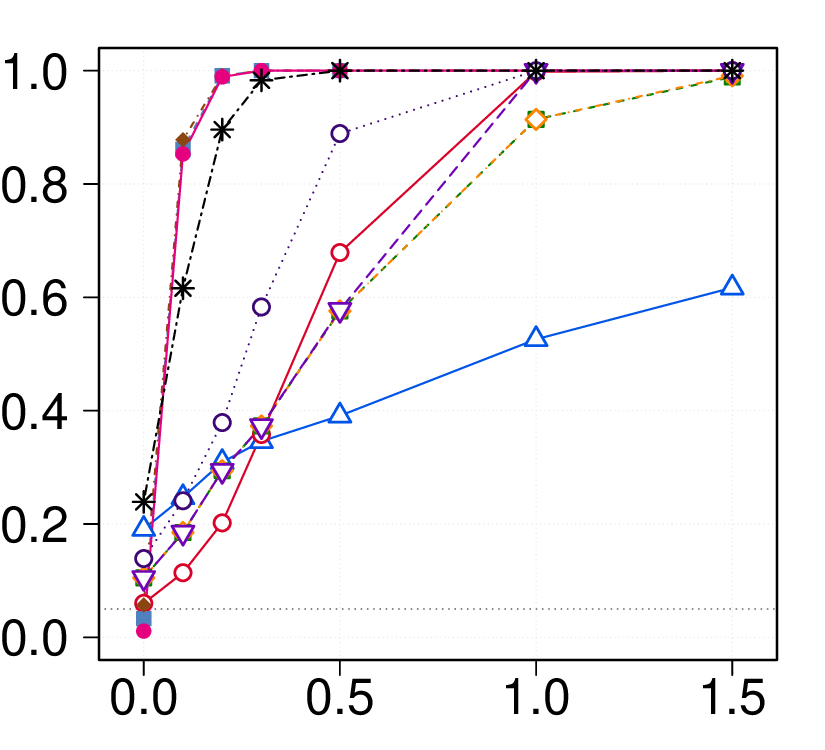}
            \hspace{-1 cm}  
            \includegraphics[width=5.5cm, height=5cm]{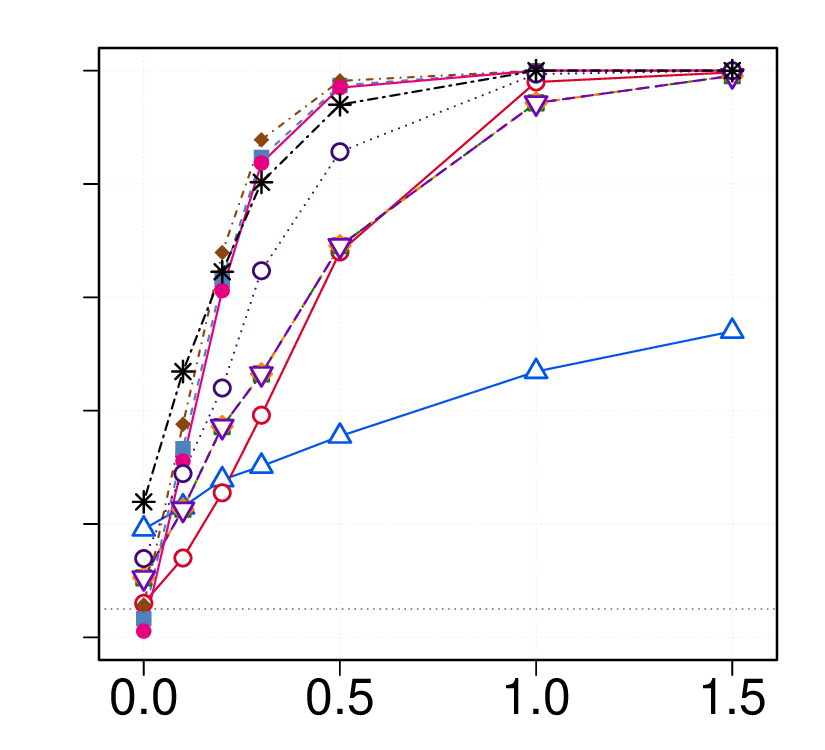}
            \hspace{-1 cm}  
            \includegraphics[width=5.5cm, height=5cm]{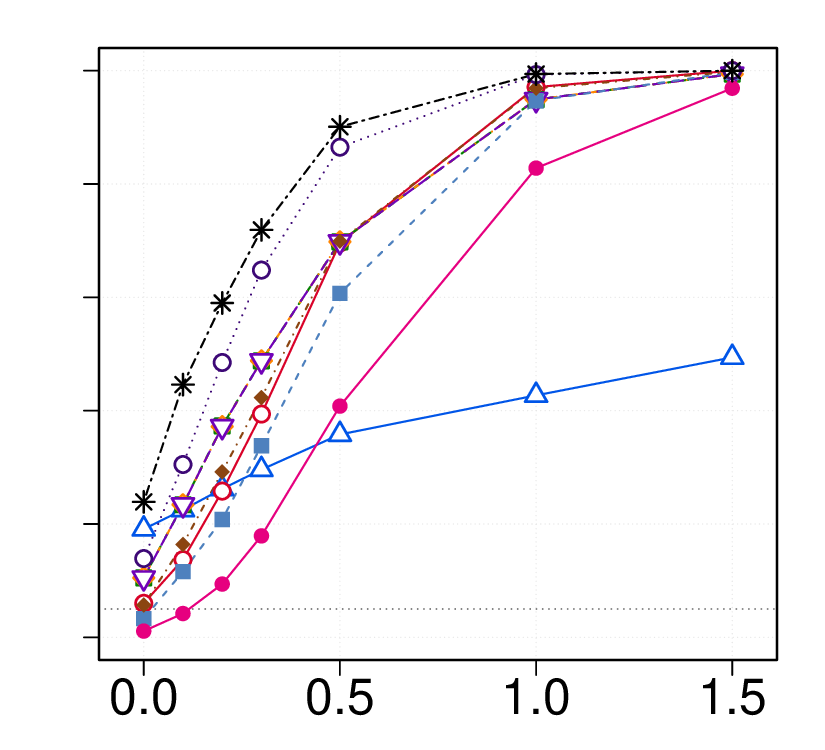}
            \hspace{-1 cm}  
            \vspace{-0.2in}
            \caption*{\small{(a) \textit{Setting 1}, with $m=5,10,23$ } }

            \centering
            \hspace{-1 cm}  
            \includegraphics[width=5.5cm, height=5cm]{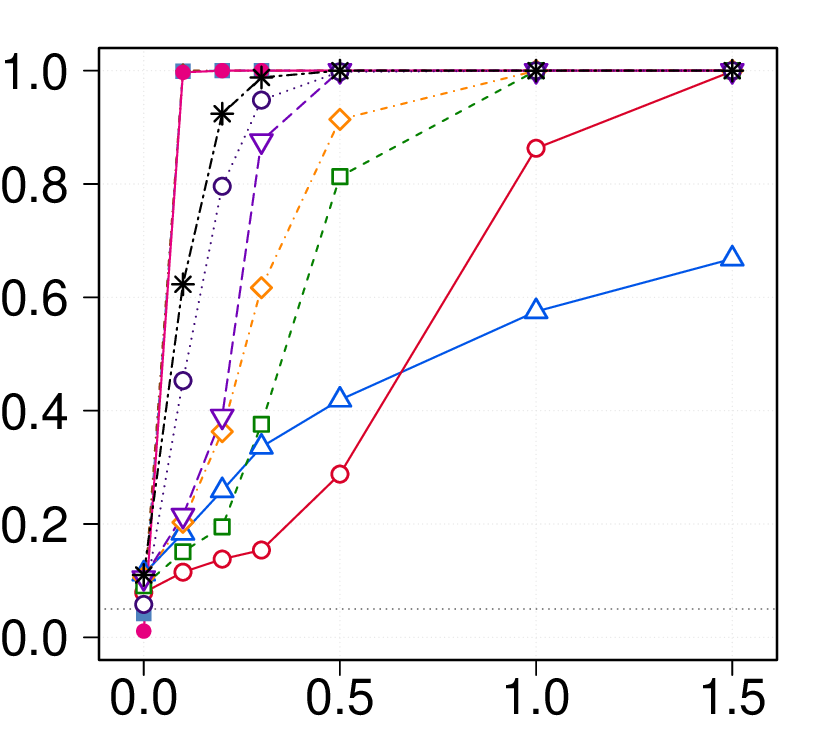}
            \hspace{-1 cm}  
            \includegraphics[width=5.5cm, height=5cm]{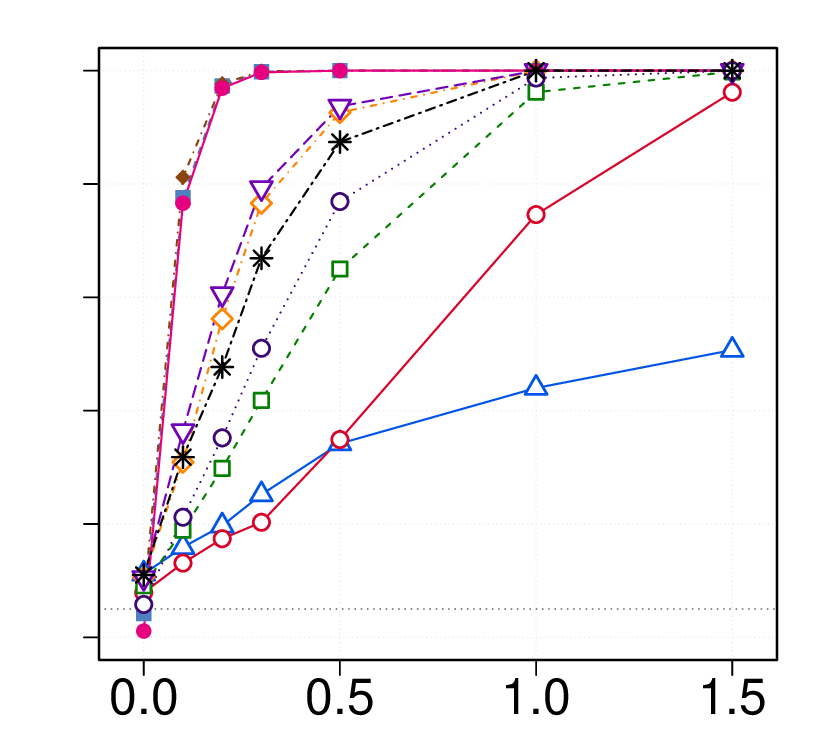}
            \hspace{-1 cm}  
            \includegraphics[width=5.5cm, height=5cm]{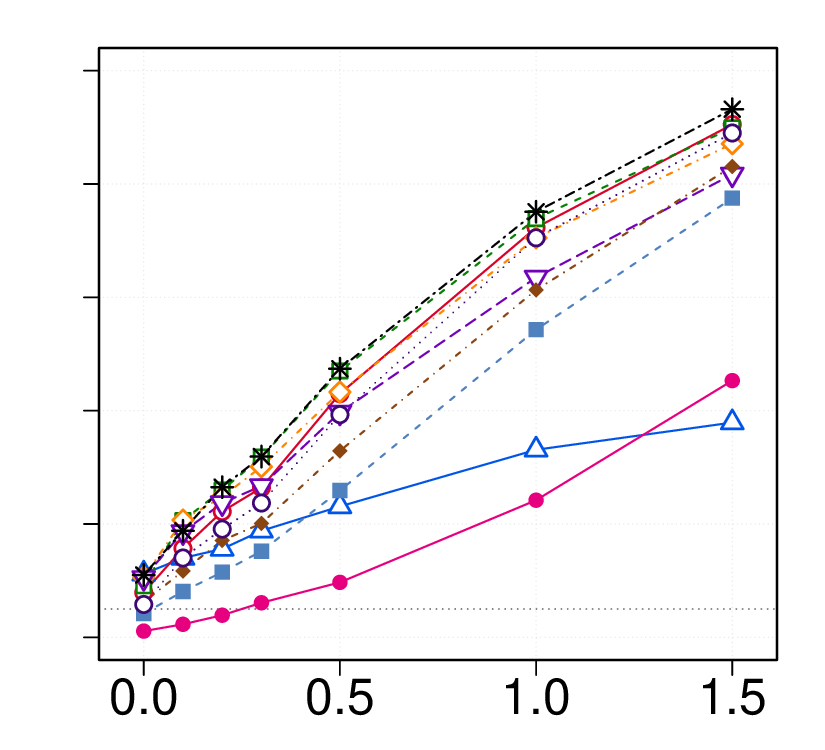}
            \hspace{-1 cm}  
            \vspace{-0.2in}
            \caption*{\small{(b) \textit{Setting 2}, with $m=5,10,23$ } }

            \centering
            \hspace{-1 cm}  
            \includegraphics[width=5.5cm, height=5cm]{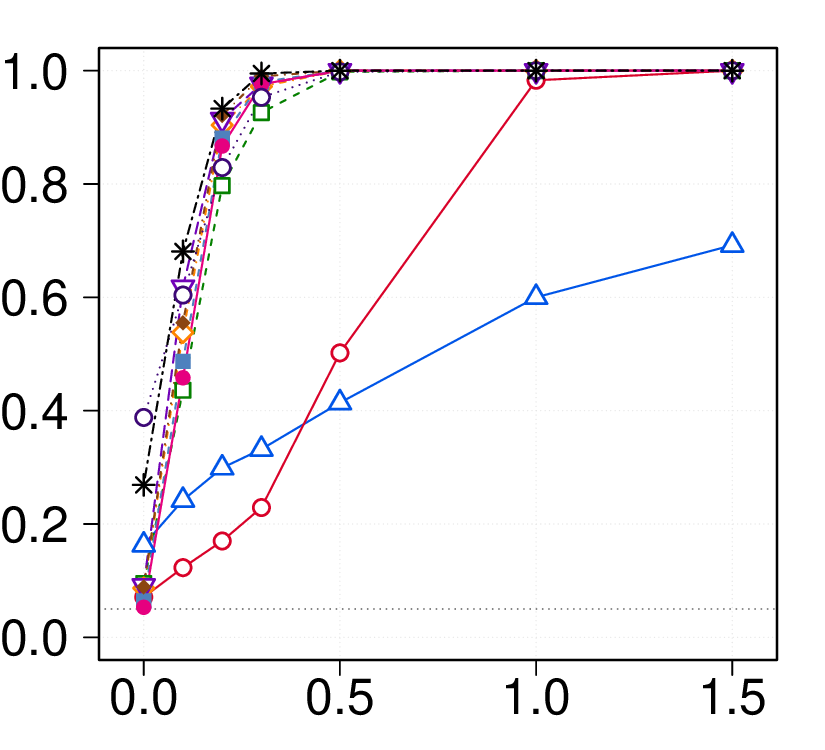}
            \hspace{-1 cm}  
            \includegraphics[width=5.5cm, height=5cm]{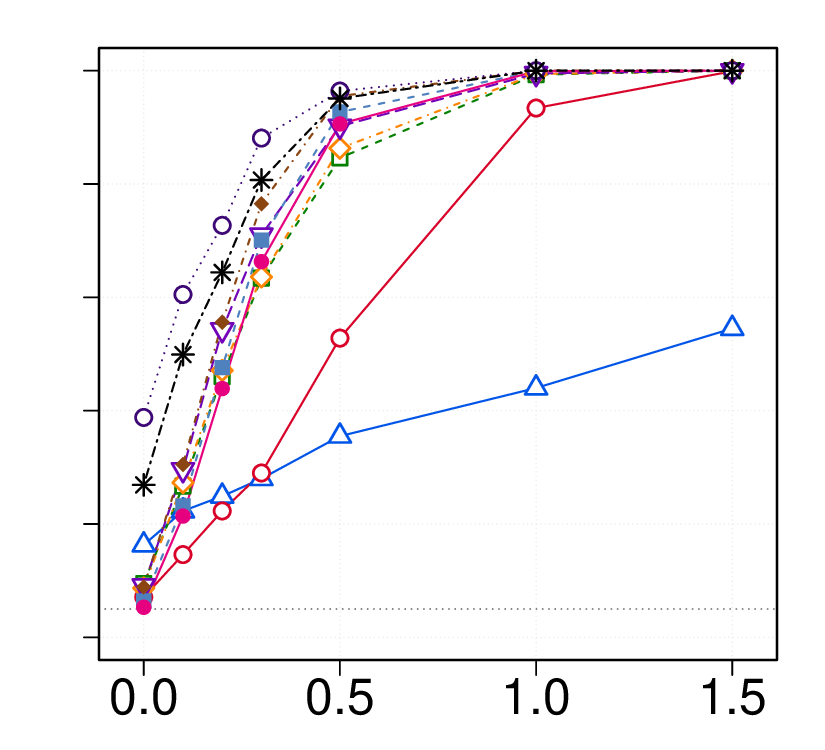}
            \hspace{-1 cm}  
            \includegraphics[width=5.5cm, height=5cm]{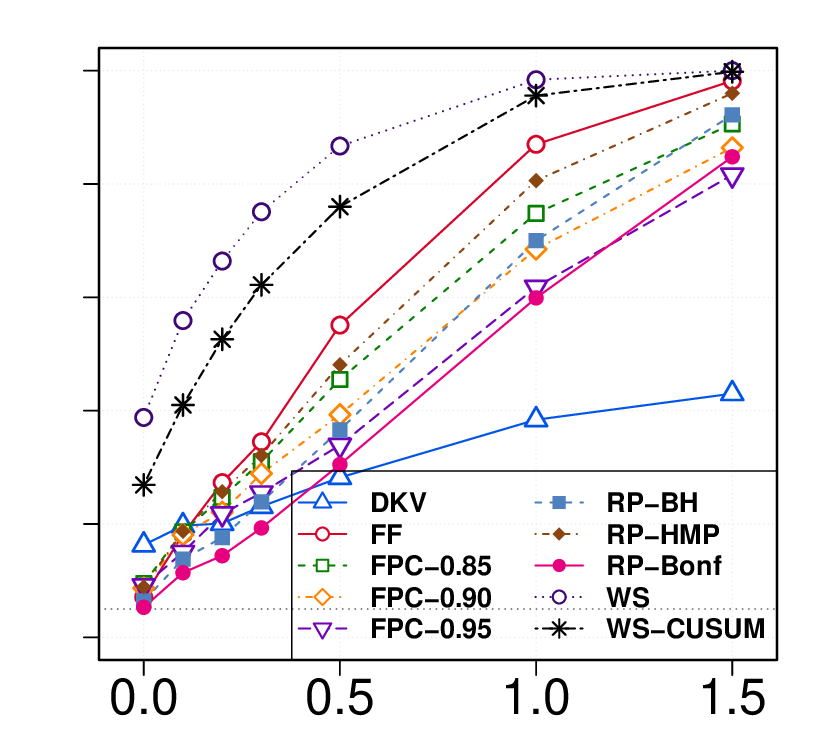}
            \hspace{-1 cm}  
            \vspace{-0.2in}
            \caption*{\small{(c) \textit{Setting 3}, with $m=5,10,23$ } }

            \caption{\small{Sparse functional setting in Section \ref{subsec: sparse functional setting}. Raw empirical rejection rates of various change point detection methods, including projection-based and fully functional methods, for various values of $SNR$ in the $x$-axis.
            The data-generating process follows (\ref{eq:model for fd}), (\ref{eq:data generating process}), and (\ref{bspline break}), where the standard deviation $\sigma_{g}$ follows \textit{Settings 1--3}. 
            The change point location is set at $\theta =0.25$. The rejection rate is based on 1000 simulations.
            }}
           \label{fig:  size and power Auedata: data0 bspline}
    \end{figure}

\subsection{Departures from the AMOC Assumption}\label{subsec: deviations from AMOC}
Although the scope of our paper is primarily focused on the AMOC setting, real data often feature multiple change points. In high-dimensional or functional settings, different components may experience changes at different times. This subsection explores how RP-based and other high-dimensional or functional change-point methods perform when the AMOC assumption is violated. 

The functional time series $X_t(s)$ is generated from
\begin{equation}\label{eq:dgp-2cpts}
    X_t(s)=\mu(s)+\delta_1(s)\mathbbm{1}\{t>z_1^*\}+\delta_2(s)\mathbbm{1}\{t>z_2^*\}+\varepsilon_t(s), \quad 1\leq t\leq n, \quad s\in[0,1],
\end{equation}
where $\mu(s)=0$, $z_1^*=\lfloor\theta_1 n\rfloor$, and $z_{2}^*=\lfloor\theta_2 n\rfloor$. The functional error sequence $\varepsilon_t(s)$ is generated as in (\ref{eq:model for fd}) under \textit{Settings 1--3}. The two change functions $\delta_1(s)$ and $\delta_2(s)$ are constructed using the B-spline basis functions $\tilde{v}_1(s), \dots, \tilde{v}_{23}(s)$ as in Section \ref{subsec: sparse functional setting},
\begin{equation}\label{bspline break 2cpts}  
\delta_1(s) = \frac{\sqrt{c}}{\Vert{}\sum_{g=1}^m \tilde{v}_g\Vert{}} \sum_{g=1}^m \tilde{v}_g(s) \quad \text{and} \quad \delta_2(s) = \frac{\sqrt{c}}{\Vert{}\sum_{g=24-m}^{23} \tilde{v}_g\Vert{}} \sum_{g=24-m}^{23} \tilde{v}_g(s),
\end{equation}
with $c$ as defined in (\ref{break function}).

The two change functions take the first (or last) $m$ elements of the 23 basis functions. In our simulations, $\theta_1=0.25$, $\theta_2=0.75$, and $m=5,10,15$ are considered, covering roughly the first (or last) 25\%, 50\%, and 75\% subregions. For instance, in the $m=5$ setting, approximately the first 25\% of the domain exhibits a change at $\theta_1=0.25$, whereas the final 25\% experiences a change at $\theta_2=0.75$. In the context of our empirical temperature analysis in Section \ref{sec:application}, this reflects a given station undergoing a change at relative time 0.25 during the first quarter (roughly January through March) and at relative time 0.75 during the final quarter (roughly October through December), while the second and third quarters experience no changes over the years.

Figure \ref{fig:  size and power Auedata: data0 2cpts} presents the size and power plots of the different change-point methods. In general, the RP-based methods outperform other methods in terms of size control and power. In particular, under \textit{Setting 2}, the RP-based methods achieve 100\% power the fastest. 
The functional methods (FF and DKV) are designed for a single change point, which explains their lower power in this setting. Conversely, FPC methods can capture multiple change points because different FPC directions are considered; indeed, they sometimes exhibit performance similar to that of the RP-based methods. The superior performance of the RP-based methods in this setting demonstrates that even though they are developed for the AMOC setting, they remain a useful tool for detecting the presence of a change point in functional settings when the true number of changes are unknown.

In terms of location estimation, Figure \ref{fig:  CP locations Auedata standcusum: data 2cpts} in Supplementary Material \ref{supp:sec:AMOC} presents boxplots of the location estimates across 1000 simulations. As expected, if only one set of RP (consisting of 200 projections) is considered, the location estimate in the two-change-point scenario is less accurate. However, when 1000 sets of RP (of 200 projections each) are considered, as shown in the violin plots in Figure \ref{fig:violin_grid:data 2cpts} and histograms in Figures \ref{fig:hist_bonf:data 2cpts} and \ref{fig:hist_bh:data 2cpts} of Supplementary Material \ref{supp:sec:AMOC}, the modes of the RP-based methods tend to align with one of the two true change-point locations. This further validates the strategy introduced in Section \ref{subsec:repeat RP method on one data}, even when multiple change points are present.

\begin{figure} [h!]
            \centering
            \hspace{-1 cm}  
            \includegraphics[width=5.5cm, height=5cm]{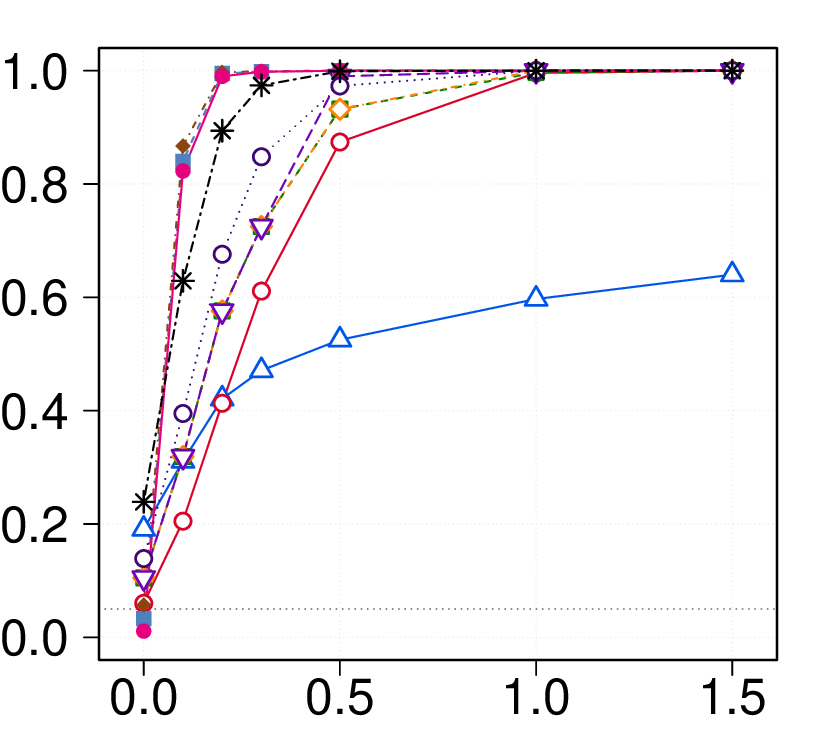}
            \hspace{-1 cm}  
            \includegraphics[width=5.5cm, height=5cm]{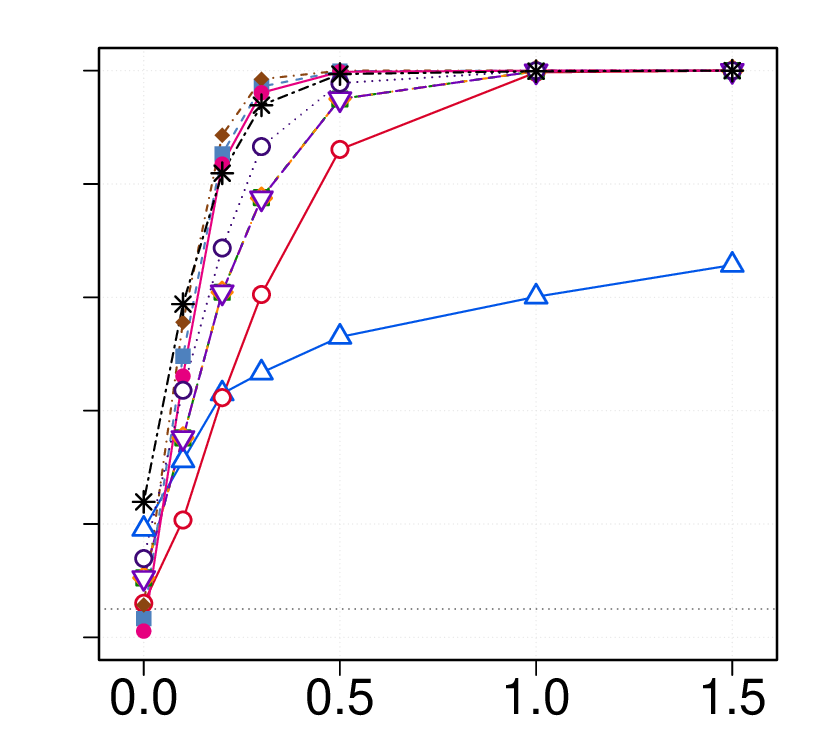}
            \hspace{-1 cm}  
            \includegraphics[width=5.5cm, height=5cm]{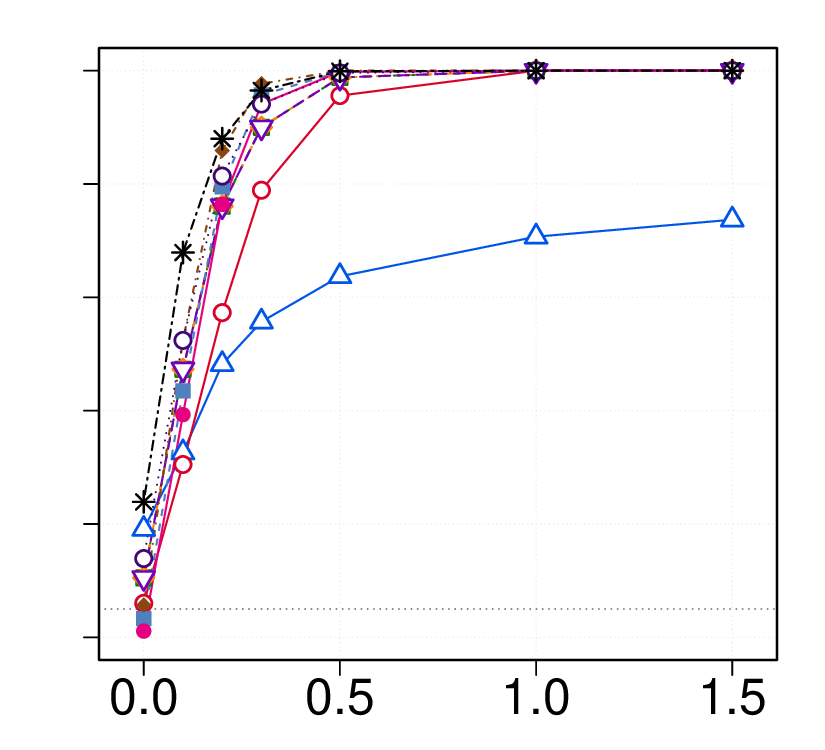}
            \hspace{-1 cm}  
            \vspace{-0.2in}
            \caption*{\small{(a) \textit{Setting 1}, with $m=5,10,15$ } }

            \centering
            \hspace{-1 cm}  
            \includegraphics[width=5.5cm, height=5cm]{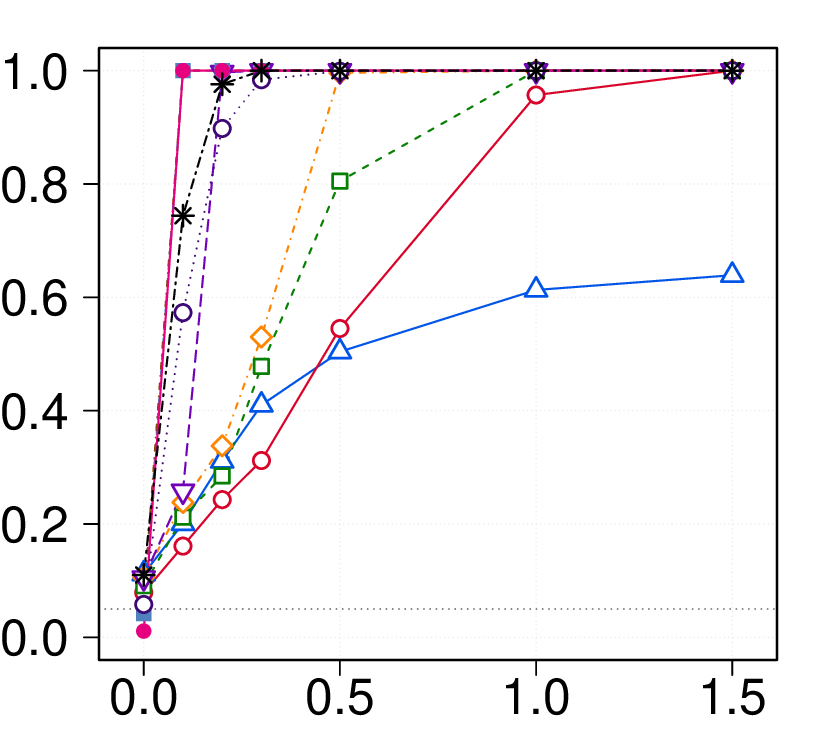}
            \hspace{-1 cm}  
            \includegraphics[width=5.5cm, height=5cm]{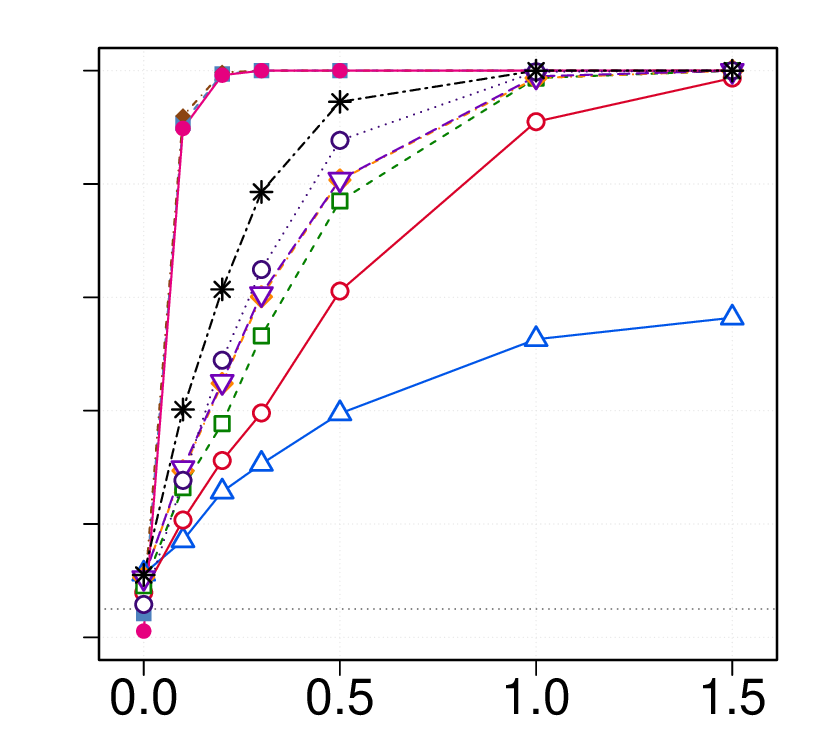}
            \hspace{-1 cm}
            \includegraphics[width=5.5cm, height=5cm]{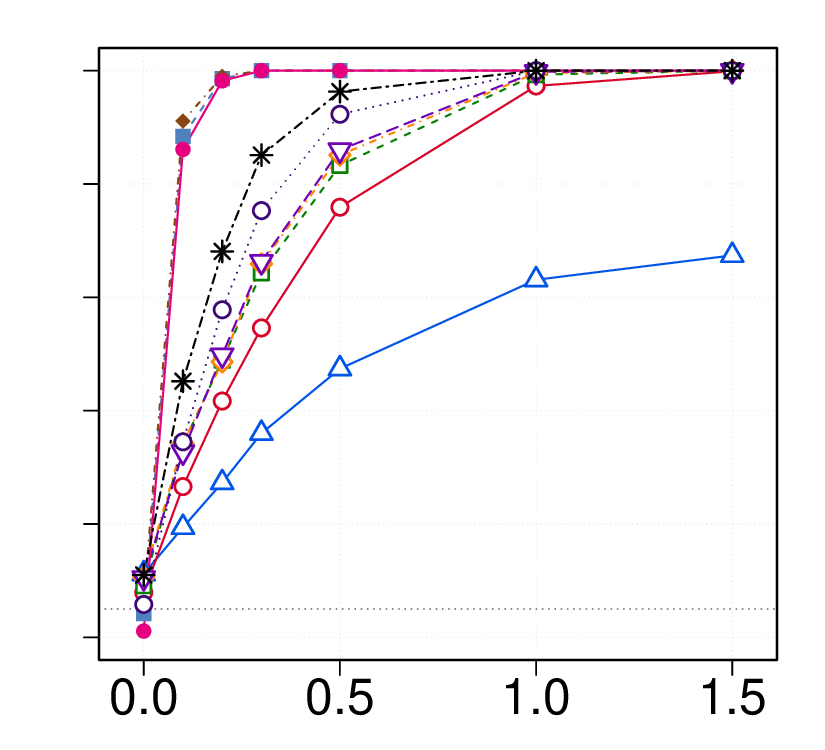}
            \hspace{-1 cm}
            \vspace{-0.2in}
            \caption*{\small{(b) \textit{Setting 2}, with $m=5,10,15$ } }

            \centering
            \hspace{-1 cm}
            \includegraphics[width=5.5cm, height=5cm]{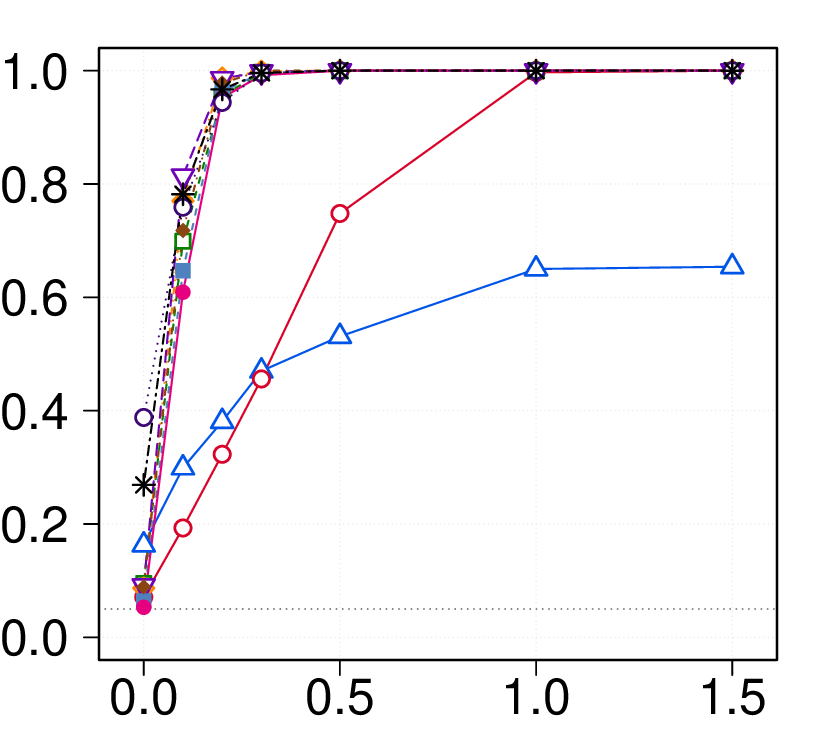}
            \hspace{-1 cm}  
            \includegraphics[width=5.5cm, height=5cm]{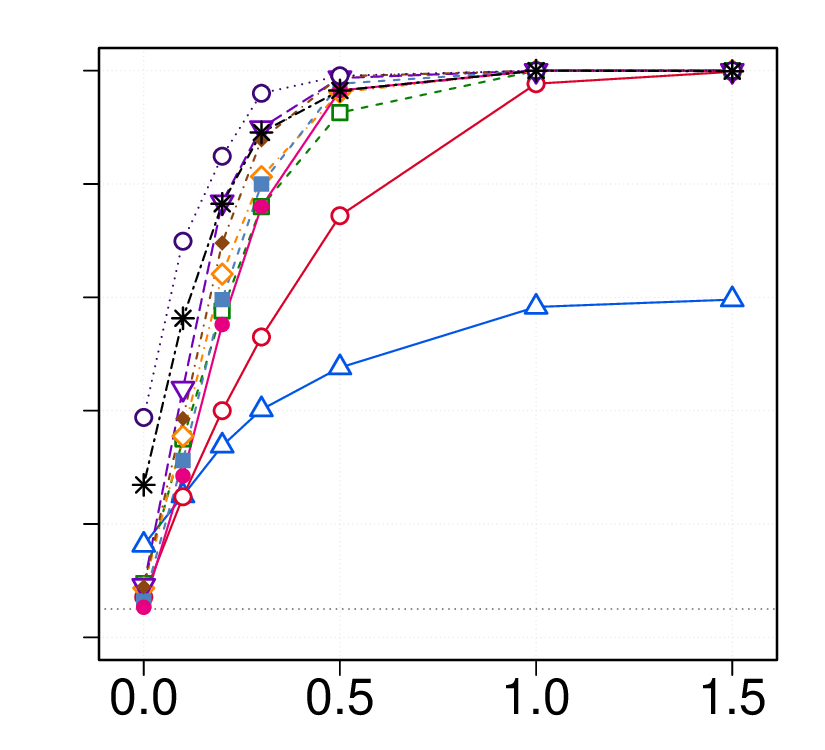}
            \hspace{-1 cm}
            \includegraphics[width=5.5cm, height=5cm]{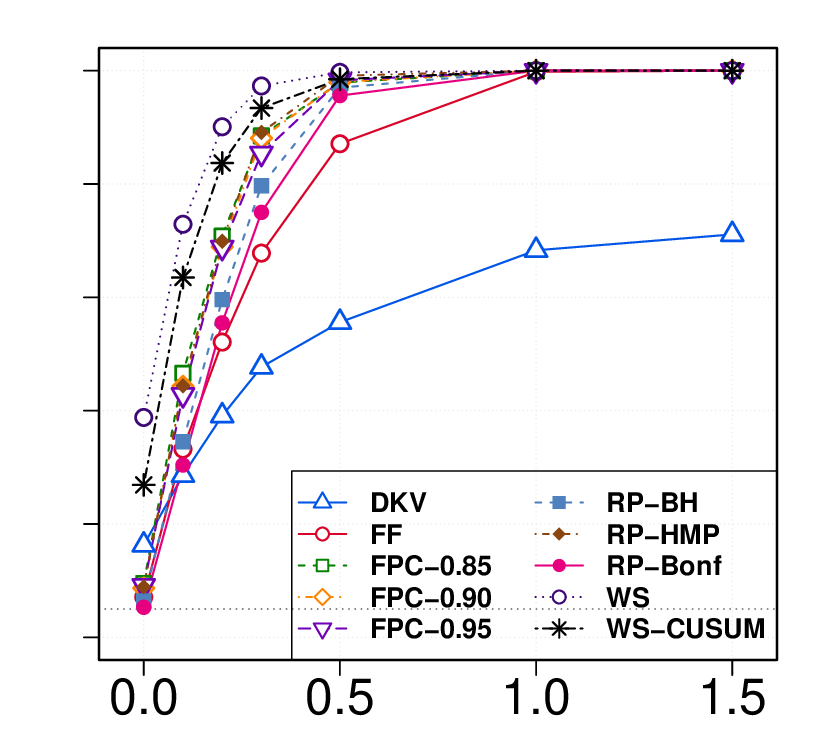}
            \hspace{-1 cm}
            \vspace{-0.2in}
            \caption*{\small{(c) \textit{Setting 3}, with $m=5,10,15$ } }

            \caption{\small{Departures from AMOC setting in Section \ref{subsec: deviations from AMOC}. empirical rejection rates of various change point detection methods, including projection-based and fully functional methods, for various values of $SNR$ in the $x$-axis.
            The data-generating process follows (\ref{eq:data generating process}), (\ref{eq:dgp-2cpts}), and (\ref{bspline break 2cpts}), where the standard deviation $\sigma_{g}$ follows \textit{Settings 1--3}. 
            The change point locations are set at $\theta_1 =0.25$ and $\theta_2=0.75$. The rejection rate is based on 1000 simulations.
            }}
           \label{fig:  size and power Auedata: data0 2cpts}
    \end{figure}

\subsection{High-Dimensional Vector Setting (i.i.d. Data)}\label{subsec: high dimensional vector setting}
Our framework can also be applied in a high-dimensional vector setting. This subsection compares the performance of  high-dimensional or functional change-point tests a high-dimensional vector setting without any temporal or coordinate-wise dependencies. Adapting from the simulation setup in \cite{wang2018high}, the data $\bold{x}_t$ are generated from (\ref{eq: model}) with $\boldsymbol{\mu}=\boldsymbol{0}$ and i.i.d. errors $\varepsilon_{tj}$ from $N(0,1)$ for $t=1,\ldots,n$ and $j=1,\ldots,p$. The change vector is defined as
\begin{equation}\label{wang break}
\boldsymbol{\delta}=\frac{\sqrt{c}}{\Vert{}\boldsymbol{\Delta}\Vert{}}{\boldsymbol{\Delta}},
\end{equation}
where $\boldsymbol{\Delta}=(1,2^{-1/2},\ldots,k^{-1/2},0,\ldots,0)^T\in\mathbb{R}^p$ and $c=\sqrt{k}\vartheta$. Here, $k$ is determined by the signal sparsity $s$ as $k=\lfloor s p\rfloor$. In our simulations, we set $n=50$ and $p=200$. The true change-point location is $\theta=0.25$, corresponding to $z^*=\lfloor \theta n\rfloor = 12$. Sparsity parameters $s \in \{0.25, 0.5, 1\}$ are selected so that $k \in \{50, 100, 200\}$. The signal strength is governed by $\vartheta \in \{0, 0.1, 0.2, 0.3, 0.6, 1.2, 1.8\}$.

Figure \ref{fig:  size and power: data0 Wang} presents the size-power plots in the top row and location estimation boxplots in the bottom row. In this setting, the RP-based approaches lose size control. Instead, the FF method outperforms the other methods in terms of size and power performance, and its location estimates are among the best across all compared methods. The WS-based methods fail to control size; however, as noted in Section \ref{subsec:comparison}, it was developed not for size control but rather as a location estimation procedure. Indeed, its location estimates remain reasonably competitive. Figures \ref{fig:violin_grid:data0 wang}, \ref{fig:hist_bonf:data0 wang}, and \ref{fig:hist_bh:data0 wang} in Supplementary Material \ref{supp:sec:Wang} present the violin plots and histograms of location estimates for specific datasets (Datasets 1 through 5), where the performance of the RP-based methods is noticeably sub-optimal.

Overall, our simulations suggest that RP-based methods boast superior performance for functional or high-dimensional data with dependent coordinates. Even when the AMOC assumption is violated or signals are sparse, RP-based methods remain a powerful tool for detecting the existence of a change. When estimating change-point locations, our method serves as an effective pretest that is robust to both cross-coordinate dependence and an unknown number of change points. Following this pretest, dedicated location estimation tools, such as \cite{wang2018high}, can be applied. 

\begin{figure} [h!]
            \centering
            \hspace{-1 cm}  
            \includegraphics[width=5.5cm, height=5cm]{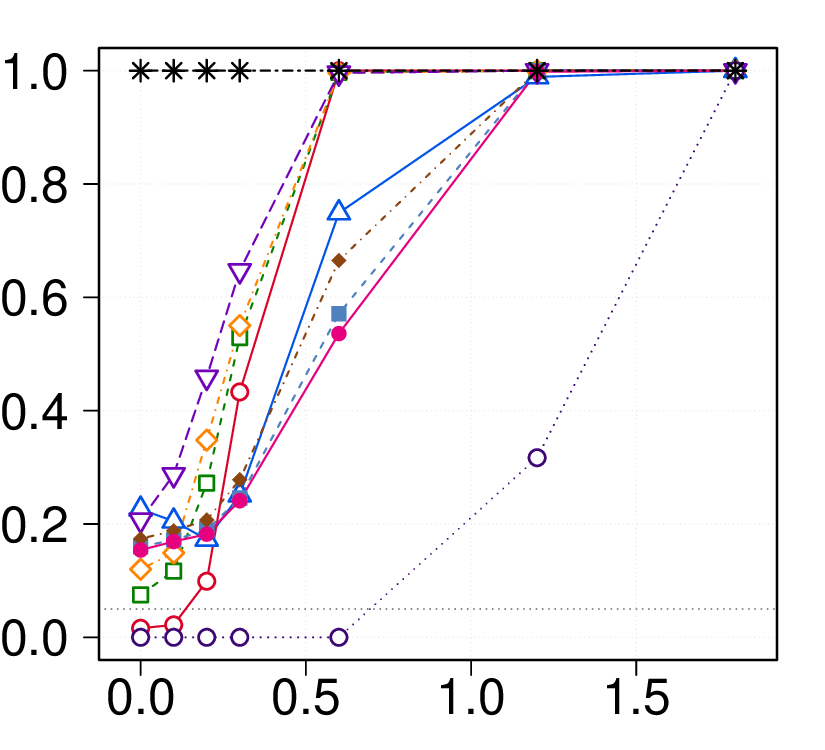}
            \hspace{-1 cm}  
            \includegraphics[width=5.5cm, height=5cm]{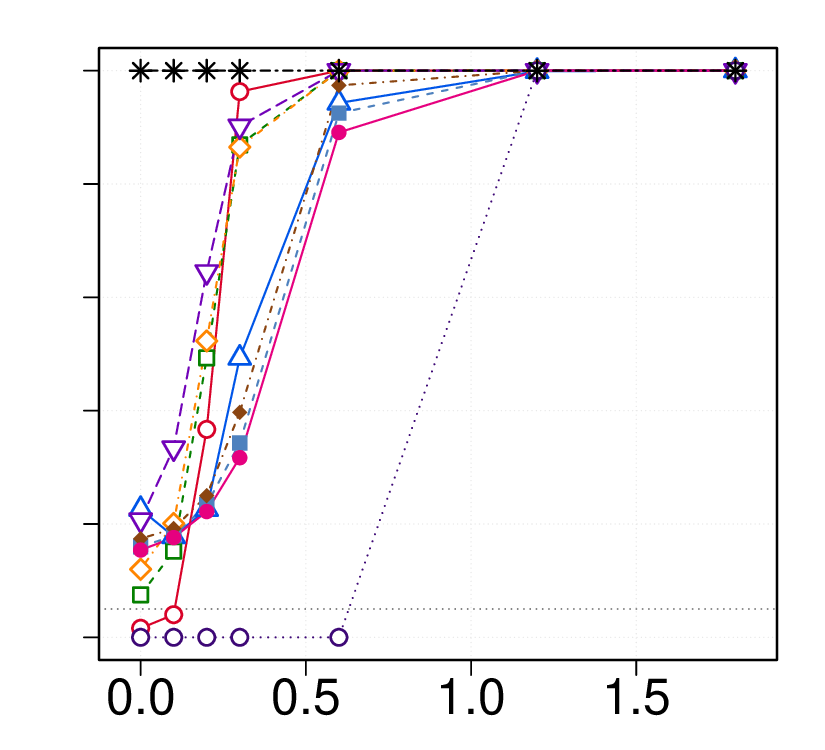}
            \hspace{-1 cm}  
            \includegraphics[width=5.5cm, height=5cm]{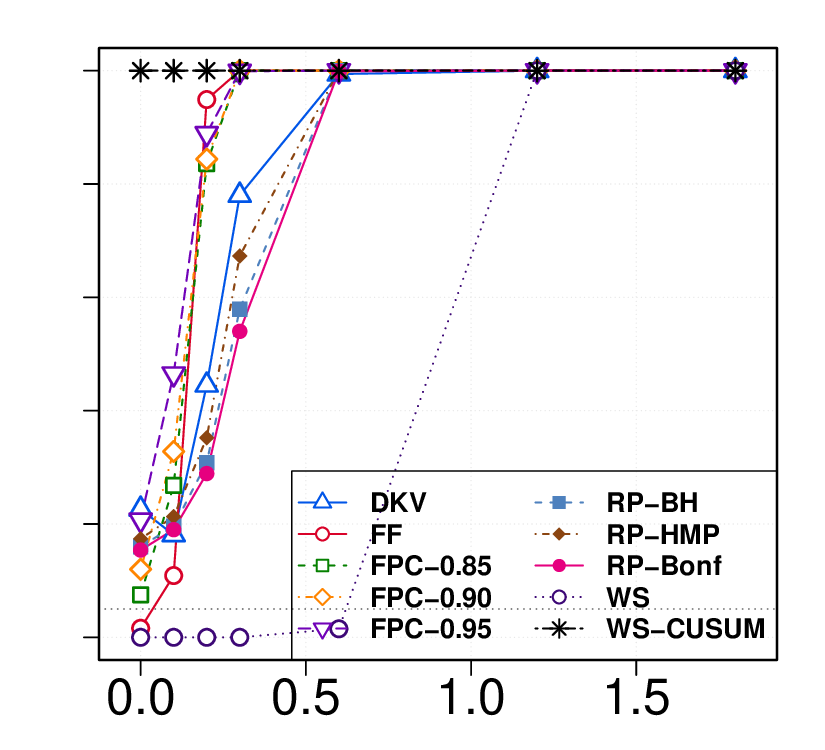}
            \hspace{-1 cm}  
            \vspace{-0.2in}
            \caption*{\small{(a) Size and Power curves with $s=0.25,0.5,1$ } }

           \centering
           \hspace{-1 cm}  
            \includegraphics[width=5.5cm, height=4cm]{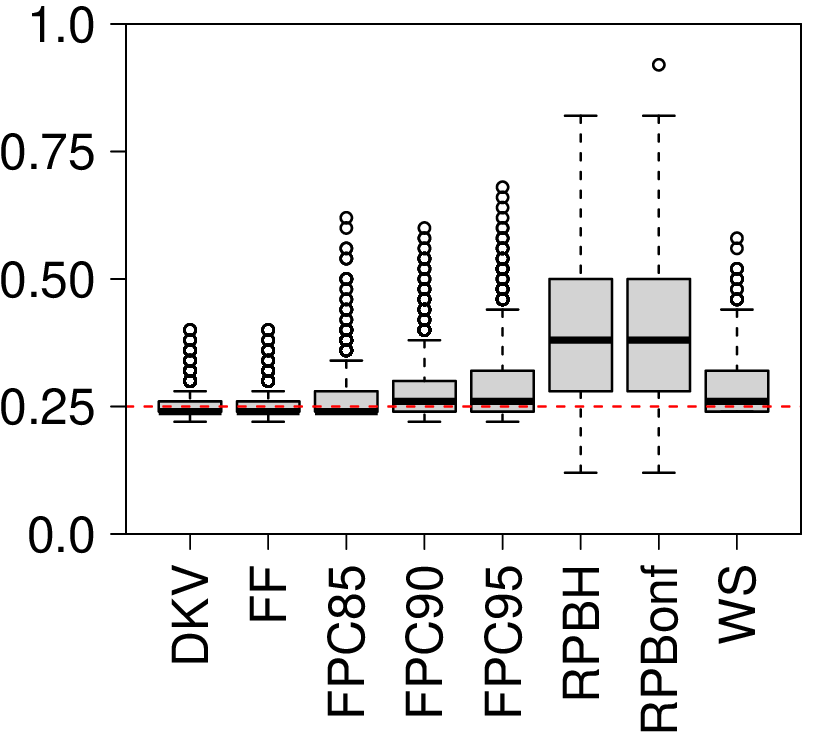}
            \hspace{-1 cm}  
            \includegraphics[width=5.5cm, height=4cm]{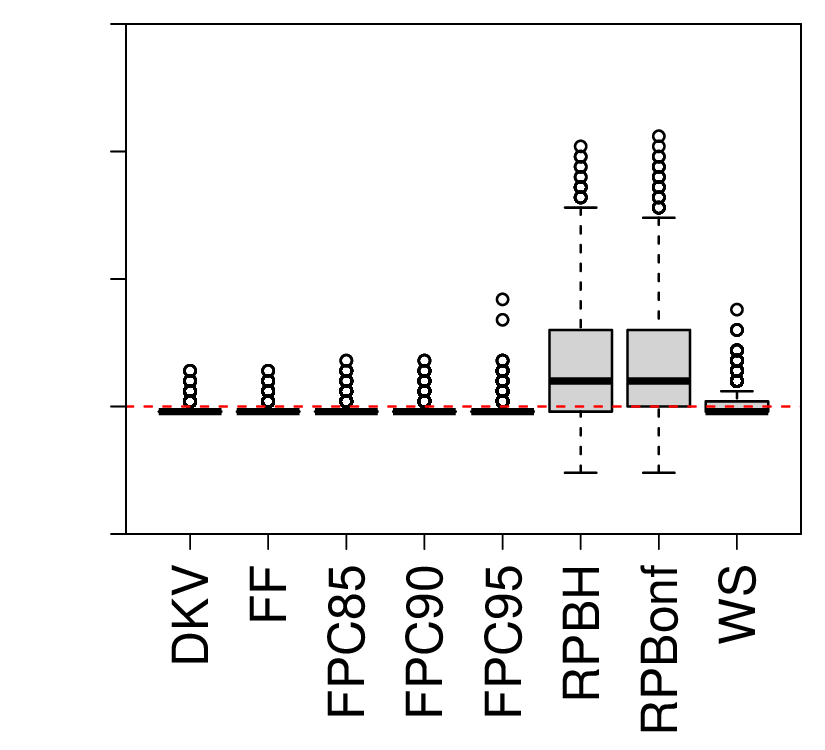}
            \hspace{-1 cm}  
            \includegraphics[width=5.5cm, height=4cm]{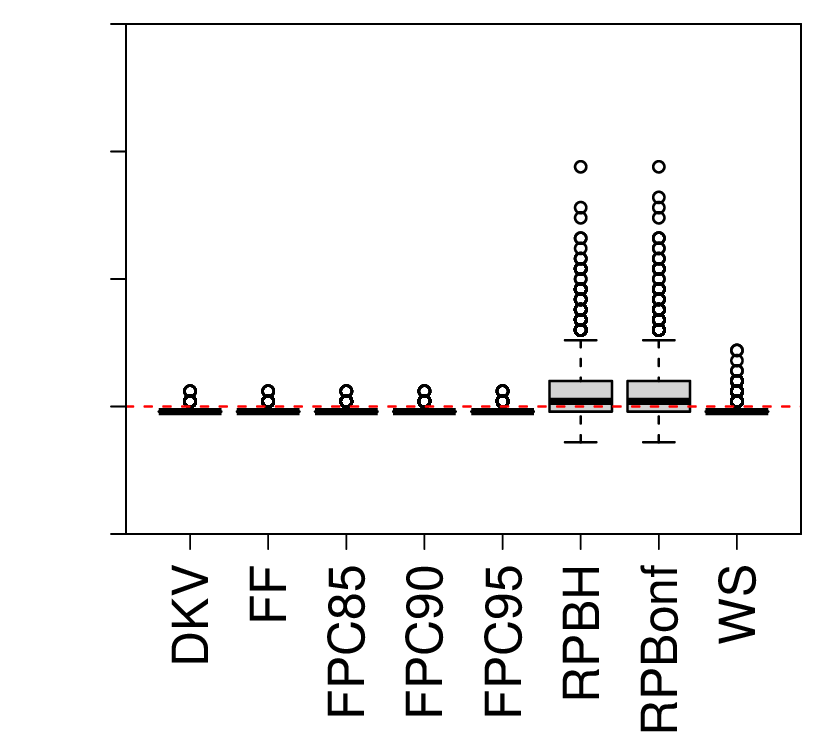}
            \hspace{-1 cm}  
            \vspace{-0.1in}
            \caption*{\small{(b) Location boxplots with $s=0.25, 0.5, 1$  when $\vartheta=0.6$} }

   \caption{\small{High-dimensional vector setting in Section \ref{subsec: high dimensional vector setting}.  Raw empirical rejection rates ($\vartheta$ in the $x$-axis) and estimated change point locations of various change point detection methods, based on 1000 simulations.
            The data-generating process follows 
            (\ref{eq: model}) and (\ref{wang break}) with iid $\varepsilon_{tj}$. The change point location is set at $\theta=0.25$. For plots in the second row, the magnitude of the break function is scaled by $\vartheta=0.6$.
            }}
           \label{fig:  size and power: data0 Wang}
    \end{figure} 

}

\section{Application} \label{sec:application}
We illustrate the performance of the RP method using daily minimum temperatures from eight Australian weather stations, most of which provide records spanning more than 100 years. The station names and numbers are Boulia (038003), Cape Otway (090015), Gayndah (039066), Gunnedah (055024), Hobart (094029), Melbourne (086338),  Robe (026026), and Sydney (066214).
The dataset is available from the Australian Bureau of Meteorology’s on the following website \url{http://www.bom.gov.au/climate/data/acorn-sat/}, which provide long-term climate monitoring. Our objective is to identify the year that the mean shift starts.
For each station, we form a yearly sequence $\{\bold{x}_{t}\}_{t=1}^{n}$, where $n=114$ for most stations.
We remove the extra day in leap years, resulting in $p=365$ observations per year.

\begin{figure} [h!]
        \centering
         \makebox[\textwidth][c]{
            \includegraphics[width=4cm, height=5cm]      {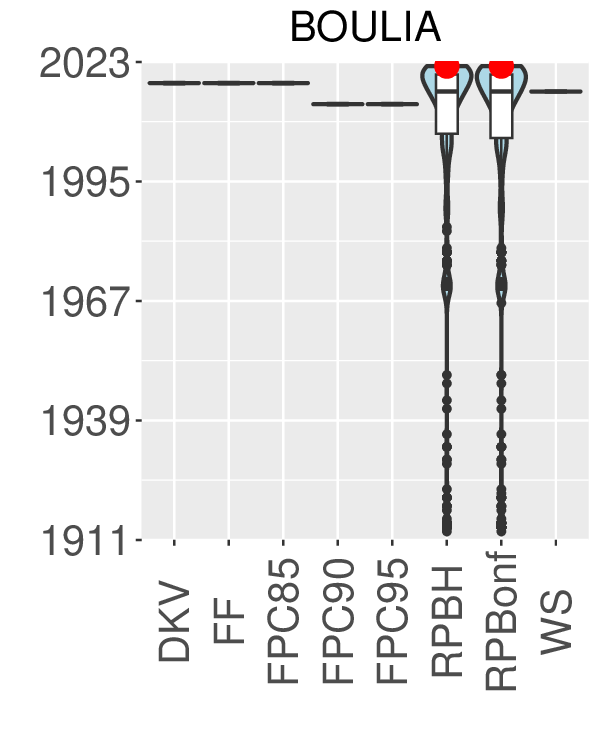}
        
            \includegraphics[width=4cm, height=5cm] {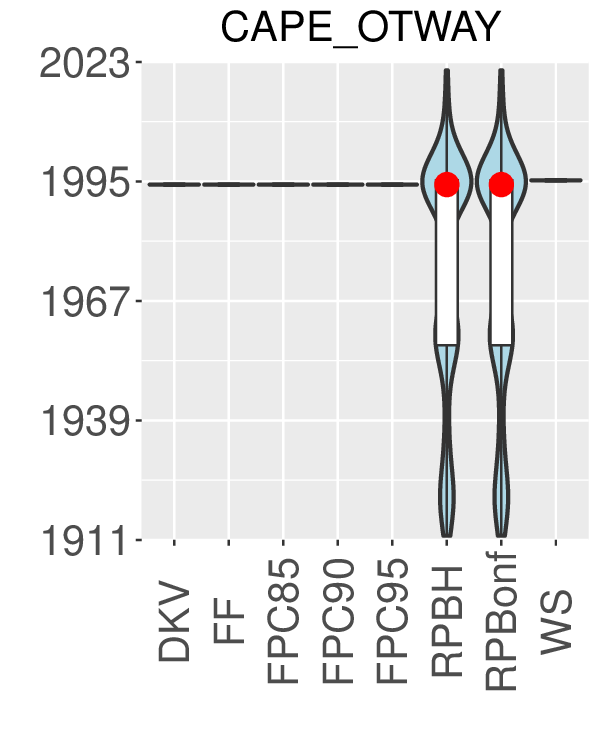}
        
            \includegraphics[width=4cm, height=5cm]   {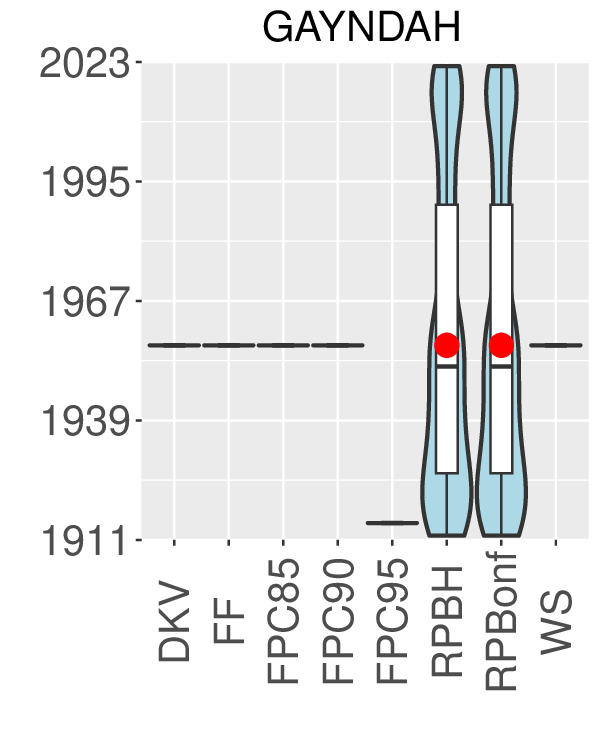}
        
            \includegraphics[width=4cm, height=5cm]      {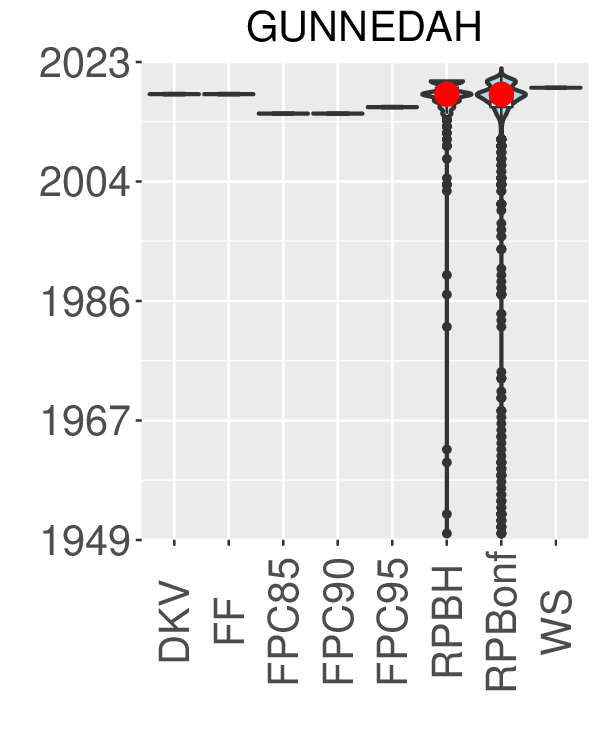}
        }

            \centering              
        \makebox[\textwidth][c]{
           \includegraphics[width=4cm, height=5cm]       {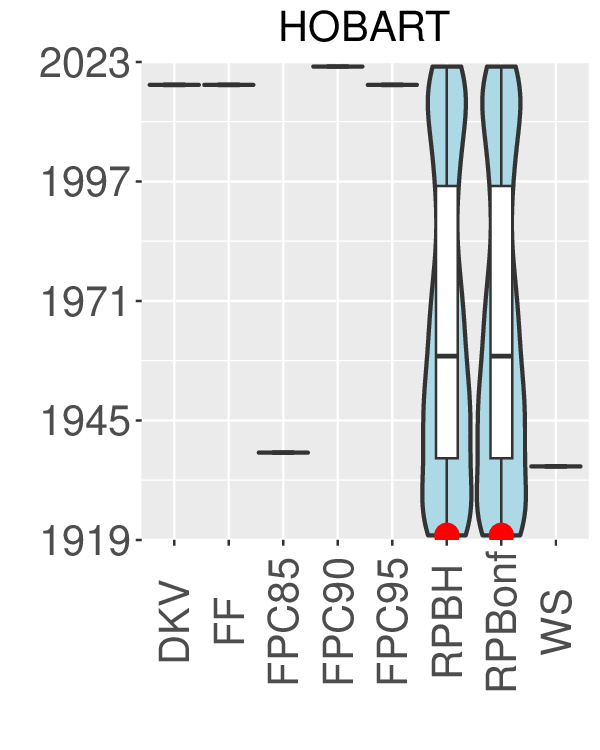}
        
            \includegraphics[width=4cm, height=5cm]     {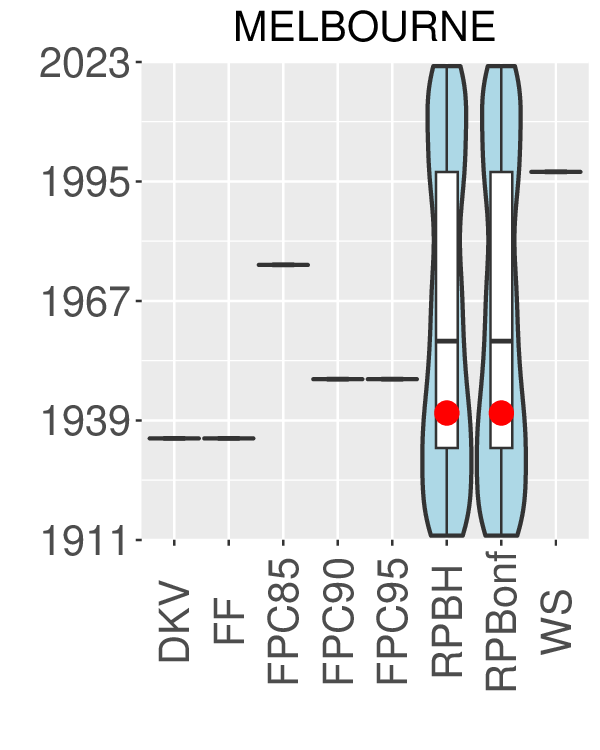}
        
            \includegraphics[width=4cm, height=5cm]      {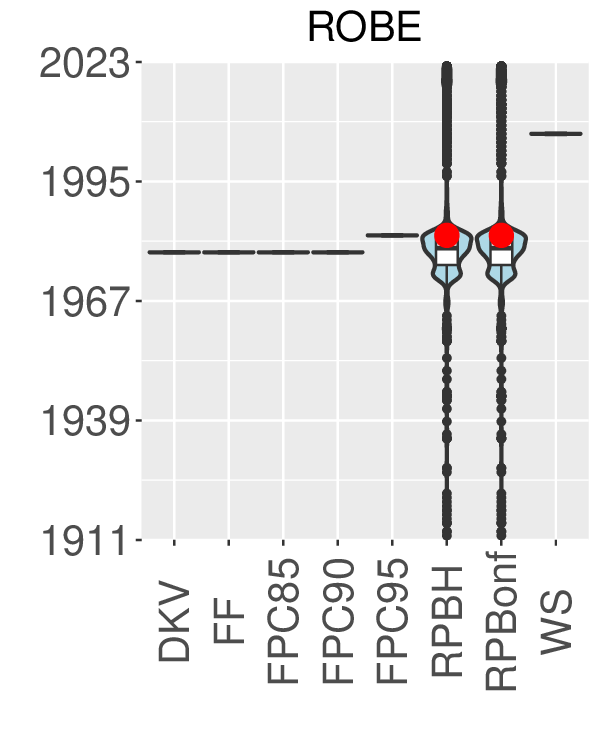}
        
            \includegraphics[width=4cm, height=5cm]      {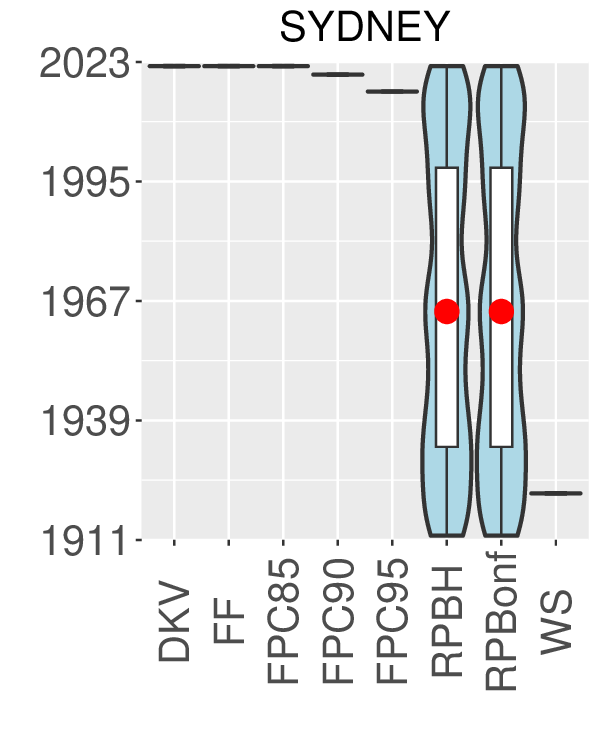}
        }

              \caption{ \small{Estimated change point locations detected by repeating the RP-based methods on the detrended temperature dataset 1000 times and other existing methods. For the RP methods, the mode of the estimated locations across the 1000 repetitions is marked by a red dot.
            }}
            \label{fig:app_temp_diff_comp}

    \end{figure}

 We apply the RP-Bonf {and RP-BH} method{s} with $k=200$ random projections and the standard CUSUM test. Because the estimated change point location can vary across different runs, we repeat the RP-Bonf {and RP-BH} method{s} 1000 times on each station's dataset and report the mode, following the procedure in Subsection \ref{subsec:repeat RP method on one data}. {To account for potential temporal dependence across years, RP-based methods are applied using a HAC variance estimator. While the variance estimator from \cite{horvath2020new} yields results consistent with HAC-based estimation, the latter exhibits greater variability across RP repetitions. For brevity, only the HAC-based results are presented.
 
Since the long-term temperature records may have a deterministic trend, we examine the change point estimation after detrending. To explore the empirical performance of the proposed methods, we compare RP-based methods with the competing change point methods considered in Section \ref{sec:simulation-mean }. Figure \ref{fig:app_temp_diff_comp} presents the estimated locations for the RP-based methods and the competing change point methods using the detrended temperature data, where six stations have common time range 1911--2023. The remaining two stations have different ranges, 1949--2023 and 1919--2023. 
The corresponding histograms of estimated locations for the RP-based methods are presented in Figures \ref{fig:app_diff_temp_hac_hist} and \ref{fig:app_diff_temp_hac_hist_BH} in Supplementary Material \ref{subsec:temp app supp}. These histograms additionally provide the frequency of the mode and reveal other frequently selected candidate years that are not captured by reporting only the mode.

 From Figure \ref{fig:app_temp_diff_comp}, the stations Boulia, Gunnedah and Robe seem to form a sharp peak, indicating that the same calendar year is repeatedly selected across many RP repetitions. In these cases, the mode provides a stable and reproducible estimate of the change point location. The other stations Cape Otway, Gayndah, Hobart, Melbourne and Sydney have more than one peak, suggesting that multiple candidates years or a gradual mean transition. We note that a single change point model may be restrictive for temperature records spanning more than 100 years, particularly for those stations experiencing multiple or smooth breaks. Therefore, the reported change point could be interpreted as the most likely change rather than the only possible change. For most stations, the change point locations chosen by RP-based methods are broadly consistent with those obtained using the existing methods. 
}



\section{Conclusion}
\label{conclusion}
This paper proposes a random projection change point detection method in high-dimensional data. By projecting high-dimensional data onto multiple random directions and applying a standard univariate change point test on each projected series, the proposed method converts a high-dimensional problem into a collection of one-dimensional tasks, which is computationally easy and offers more methods to use. The family-wise error or false discovery rate is controlled by aggregating across projections using a $p$-value combination method. Our simulation results show that the RP method has better size control, higher power, and accurate location detection when the break function is not constant in the considered cases. At the same time, the change location estimate of the RP methods may have high variability. We handle it by repeating RP methods and reporting the mode over repetitions. The real-data analysis illustrates how the proposed RP method can be used in practice.  In summary, the RP method provides a computationally easy and conceptually simple approach for change point analysis in high dimensions.

{The scope of this paper is limited to the single change point case; extending this framework to multiple change points would broaden its applicability and will be addressed in future work using binary segmentation. Additionally, the proposed method targets the detection of a common change point in high-dimensional data rather than identifying specific coordinates affected by the shift. While the estimated change point allows for preliminary exploration of candidate coordinates in the original data, developing formal identification procedures remains an important direction for future research.}

\section*{Acknowledgement}
{The authors would like to thank the anonymous referees and the Editor for their constructive comments that significantly improved the quality of this paper.}
Rho was partially supported by NIH-R15GM135806.

\bibliographystyle{chicago}
\bibliography{References}

\clearpage
\newpage

\setcounter{page}{1}
\setcounter{equation}{0}
\setcounter{table}{0}
\setcounter{figure}{0}
\setcounter{section}{0}

\renewcommand{\thesection}{S.\arabic{section}} 
\setcounter{figure}{0}
\setcounter{table}{0}
\renewcommand{\thetable}{S.\arabic{table}}   
\counterwithin*{figure}{subsubsection}
\counterwithin*{figure}{subsection}
\counterwithin*{figure}{section}

\renewcommand{\thefigure}{%
  \ifnum\value{subsubsection}>0
    \thesubsubsection.\Roman{figure}
  \else
    \ifnum\value{subsection}>0
      \thesubsection.\Roman{figure}
    \else
      \thesection.\Roman{figure}
    \fi
  \fi
}

\begin{center} {\bf\Large
Supplementary Material on ``Change point analysis of high-dimensional data using random projections"}\end{center}
	\centerline{\textsc{Yi Xu and Yeonwoo Rho}\footnote{
	 Author of Correspondence: Y. Rho (yrho@mtu.edu)}}	
		\bigskip
	\centerline {Department of Mathematical Sciences, Michigan Technological University}
    \bigskip
    \centerline{\today}
	\bigskip

The supplementary material presents additional simulation results. Section \ref{subsec: mean-parameter selection} presents additional results for the parameter choice, relating to Section \ref{subsection: tuning}: Subsection \ref{subsec: Result of using HAC estimator} presents the results of the RP method using HAC estimator, and Subsection \ref{subsec: Result of using IID estimator} presents the results of the RP method using \cite{horvath2020new}'s variance estimator $\hat{\sigma}^2_z= \frac{1}{n} [(\sum_{t=1}^{z} (y_t-\Bar{y}_z)^2 + \sum_{t=z+1}^{n} (y_t-\Bar{y}_{n-z})^2 )]$. Section \ref{subsec: weightedCUSUM} presents estimated change locations using weighted CUSUM approaches. {Section \ref{subsec:Repeat methods for data1-5} presents a figure similar to Figure \ref{fig:hist_bonf:data0} in Section \ref{subsec:repeat RP method on one data} but with RP-BH instead of RP-Bonf. Sections \ref{supp:sec:sparse functional}--\ref{supp:sec:Wang} present additional plots for Sections \ref{subsec: sparse functional setting}--\ref{subsec: high dimensional vector setting}. Section \ref{subsec:temp app supp} presents additional plots for the application in Section \ref{sec:application}.}

\section{Additional results for Subsection \ref{subsection: tuning}}
\label{subsec: mean-parameter selection}

\subsection{Results of using the HAC variance estimator}
\label{subsec: Result of using HAC estimator}
\begin{figure} [H]
{
         \centering
            \hspace{-1 cm}  
            \includegraphics[width=5.5cm, height=5cm]{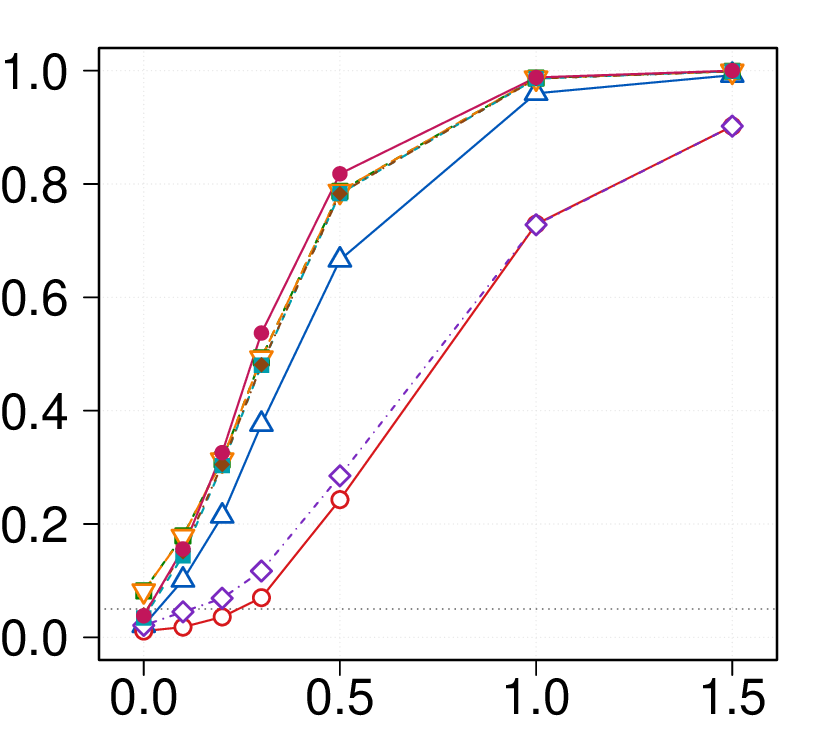}
            \hspace{-1 cm}  
            \includegraphics[width=5.5cm, height=5cm]{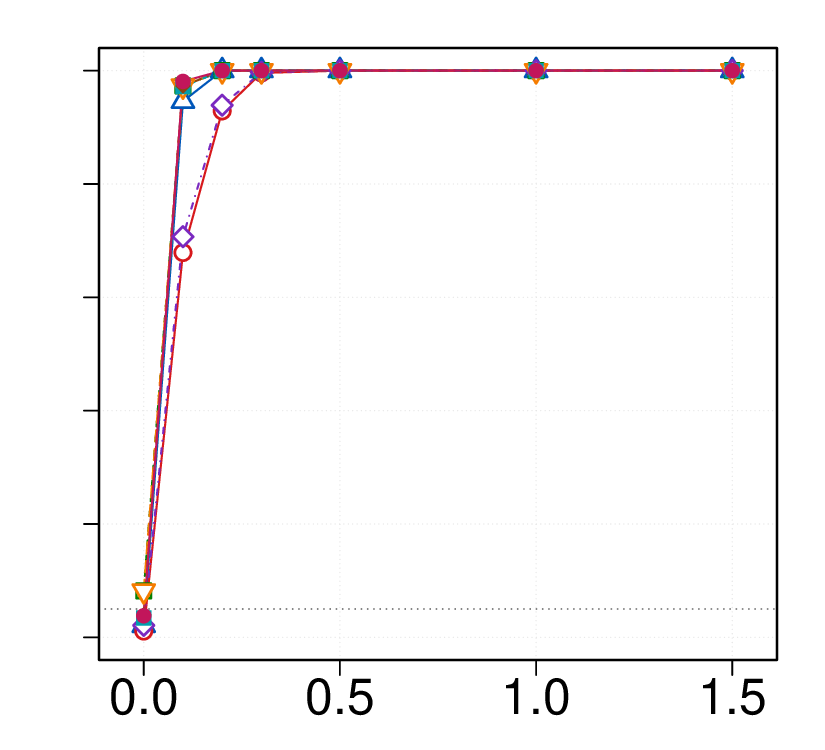}
            \hspace{-1 cm}  
            \includegraphics[width=5.5cm, height=5cm]{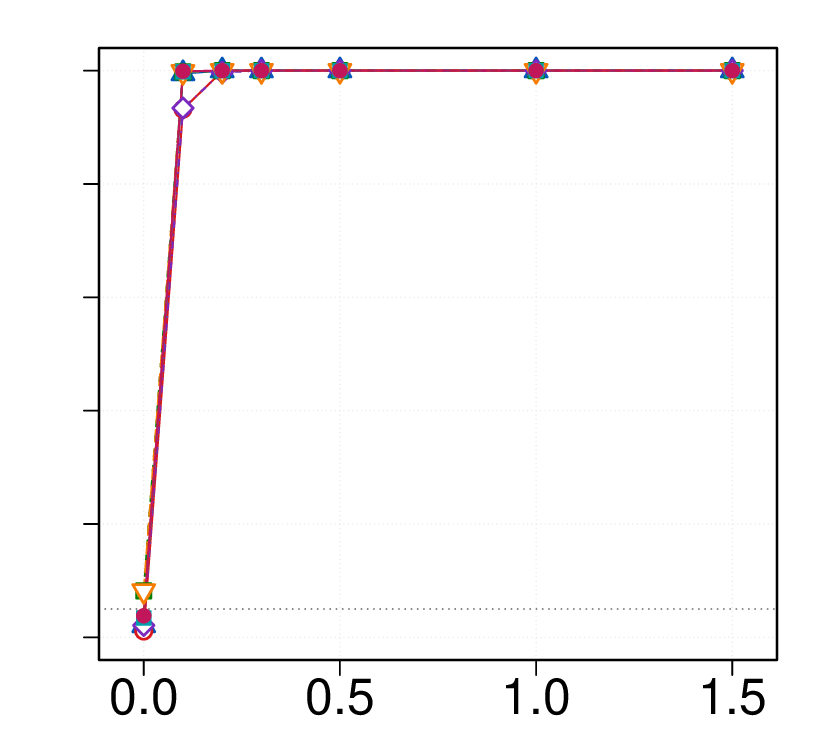}
            \hspace{-1 cm}  
            \vspace{-0.2in}
            \caption*{(a)\small{ \textit{Setting 1}, with $m=1,5,20$ }}

            \centering
            \hspace{-1 cm}  
                \includegraphics[width=5.5cm, height=5cm]{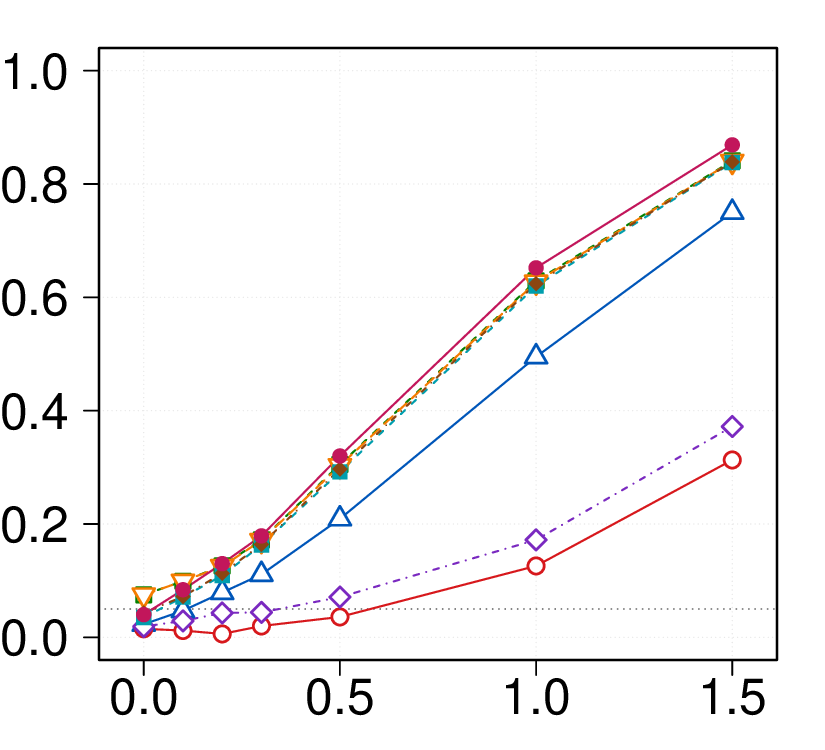}
            \hspace{-1 cm}  
            \includegraphics[width=5.5cm, height=5cm]{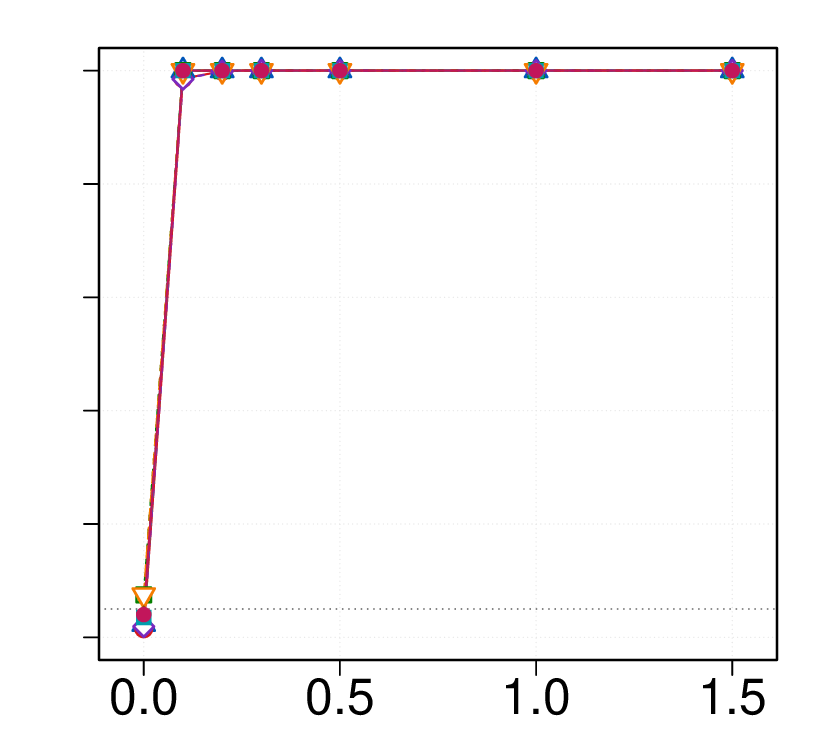}
            \hspace{-1 cm}  
            \includegraphics[width=5.5cm, height=5cm]{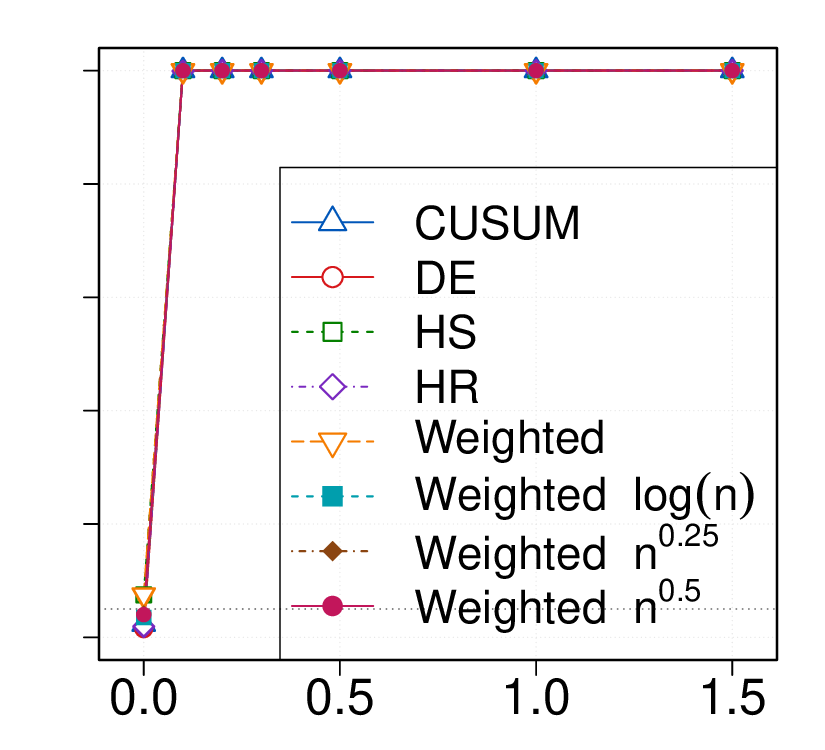}
            \hspace{-1 cm}  
            \vspace{-0.2in}
            \caption*{(b)\small{ \textit{Setting 2}, with $m=1,5,20$}}

            \centering
            \hspace{-1 cm}  
                \includegraphics[width=5.5cm, height=5cm]{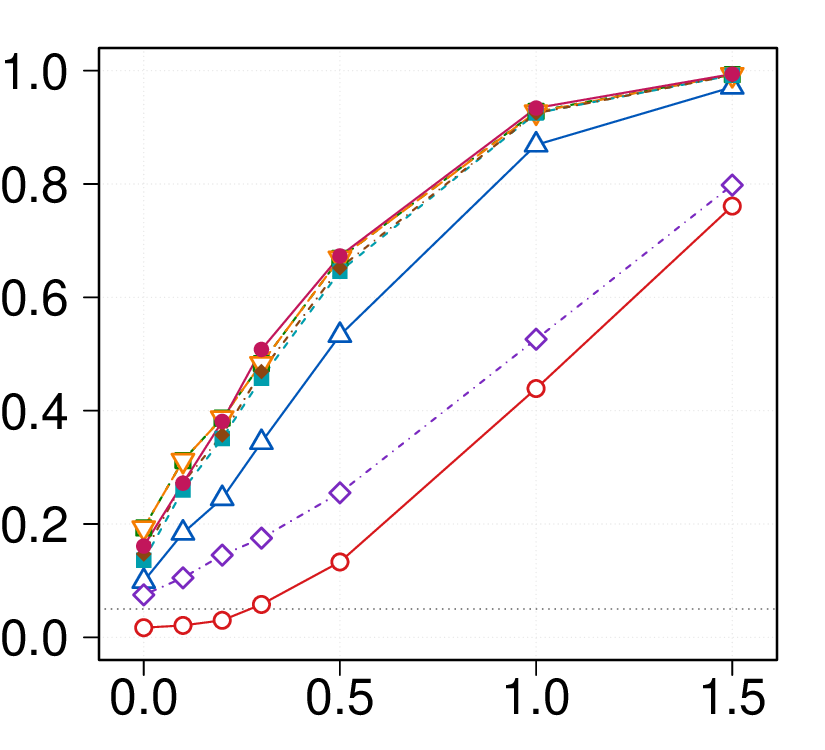}
            \hspace{-1 cm}  
            \includegraphics[width=5.5cm, height=5cm]{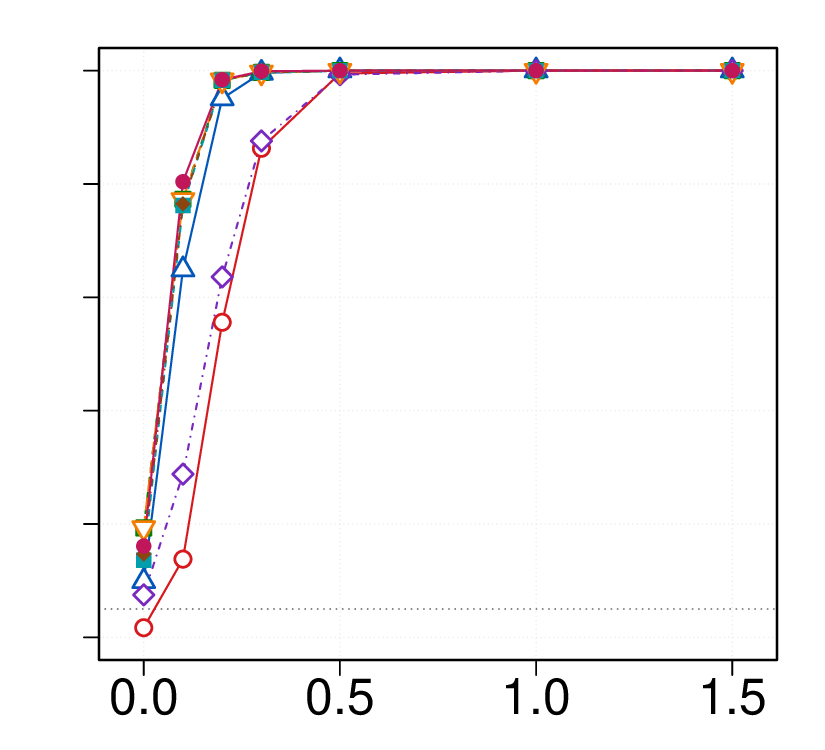}
            \hspace{-1 cm}  
            \includegraphics[width=5.5cm, height=5cm]{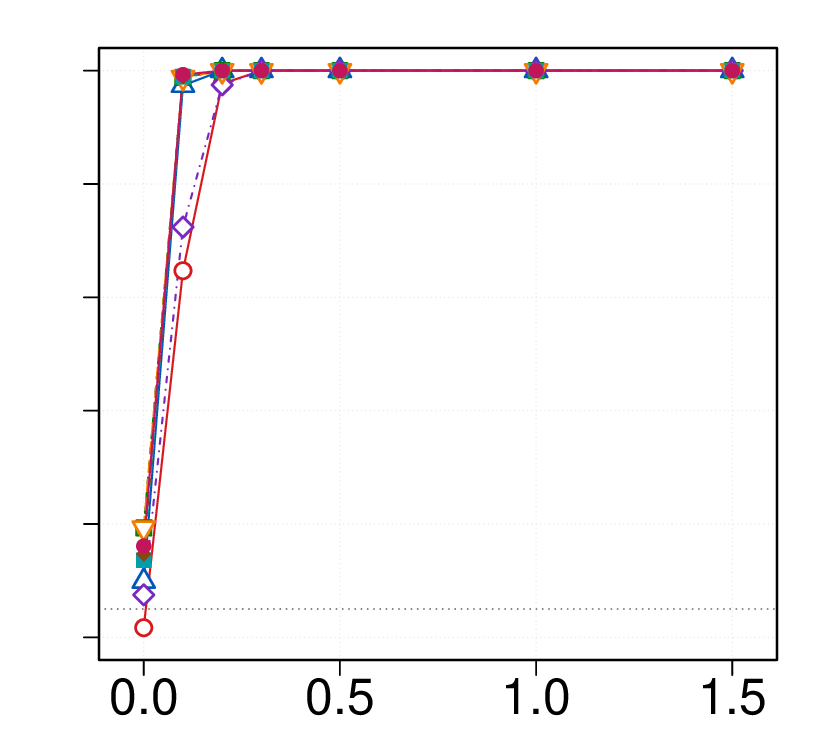}
            \hspace{-1 cm}  
            \vspace{-0.2in}
            \caption*{(c)\small{ \textit{Setting 3}, with $m=1,5,20$ } }

            \caption{\small{Raw empirical rejection rates of the RP methods for various values of $SNR$ in the x-axis. The RP method performs 200 random projections and applies different change point tests (CUSUM, Weighted, DE, HS, HR) and the Bonf combination method. The data-generating process follows (\ref{eq:model for fd}), (\ref{eq:data generating process}), and (\ref{break function}), where the standard deviation $\sigma_{g}$ follows \textit{Settings 1--3}.
            The change point location is set at $\theta=0.25$.
            The empirical rejection rate is based on 1000 simulations.
            }}
}
            \label{fig: tuning cp test (Bonf_HAC)}
    \end{figure} 

\begin{figure} [H]
         \centering
            \hspace{-1 cm}  
            \includegraphics[width=5.5cm, height=5cm]{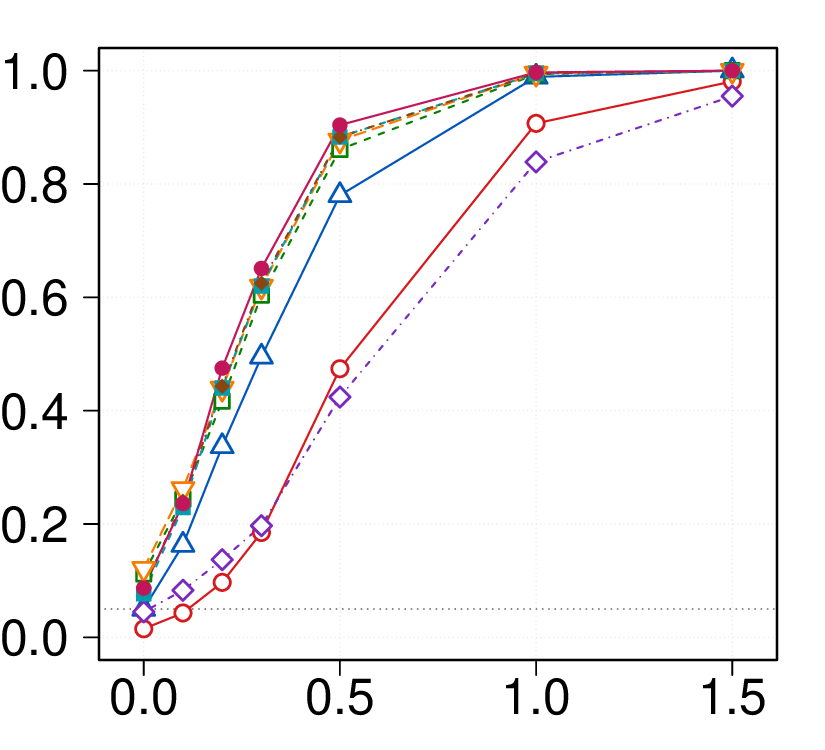}
            \hspace{-1 cm}  
            \includegraphics[width=5.5cm, height=5cm]{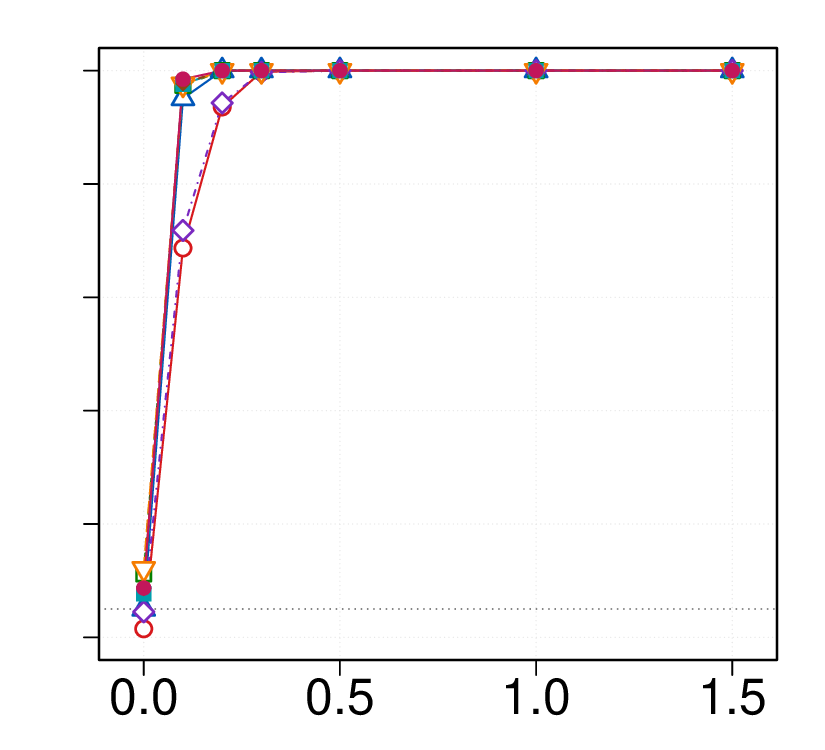}
            \hspace{-1 cm}  
            \includegraphics[width=5.5cm, height=5cm]{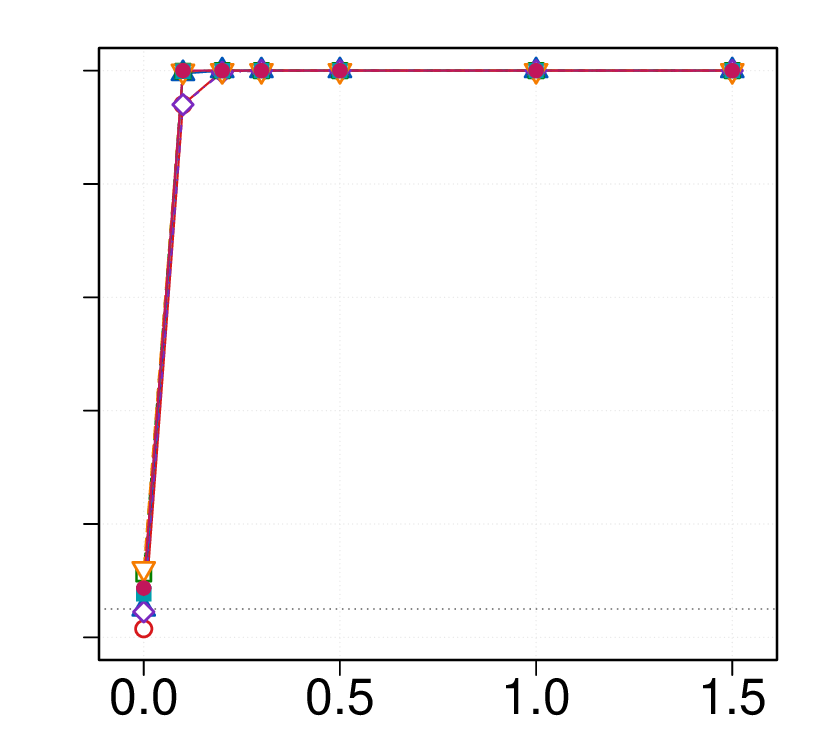}
            \hspace{-1 cm}  
            \vspace{-0.2in}
            \caption*{(a)\small{ \textit{Setting 1}, with $m=1,5,20$ }}

            \centering
            \hspace{-1 cm}  
                \includegraphics[width=5.5cm, height=5cm]{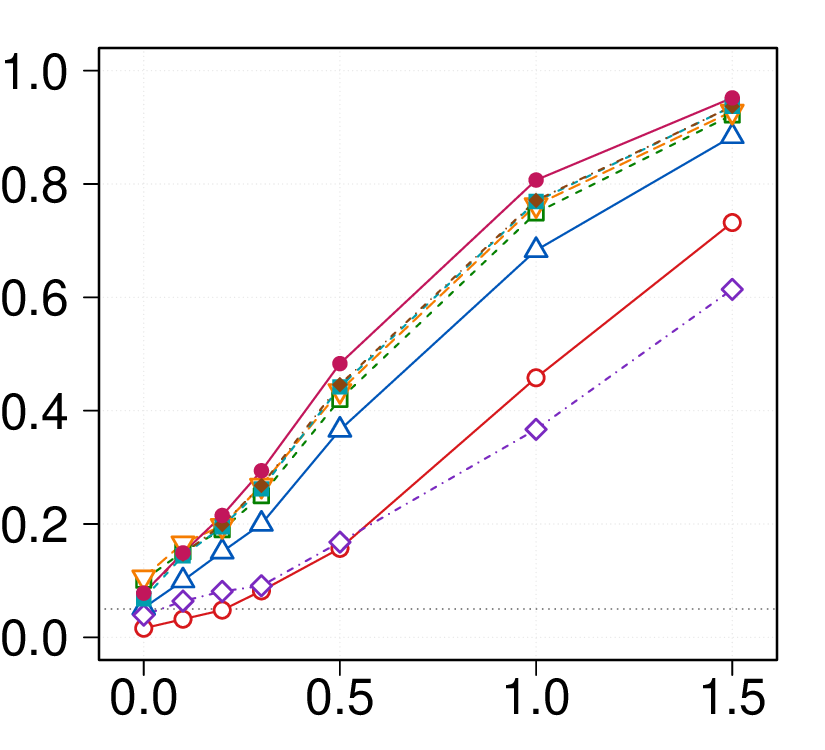}
            \hspace{-1 cm}  
            \includegraphics[width=5.5cm, height=5cm]{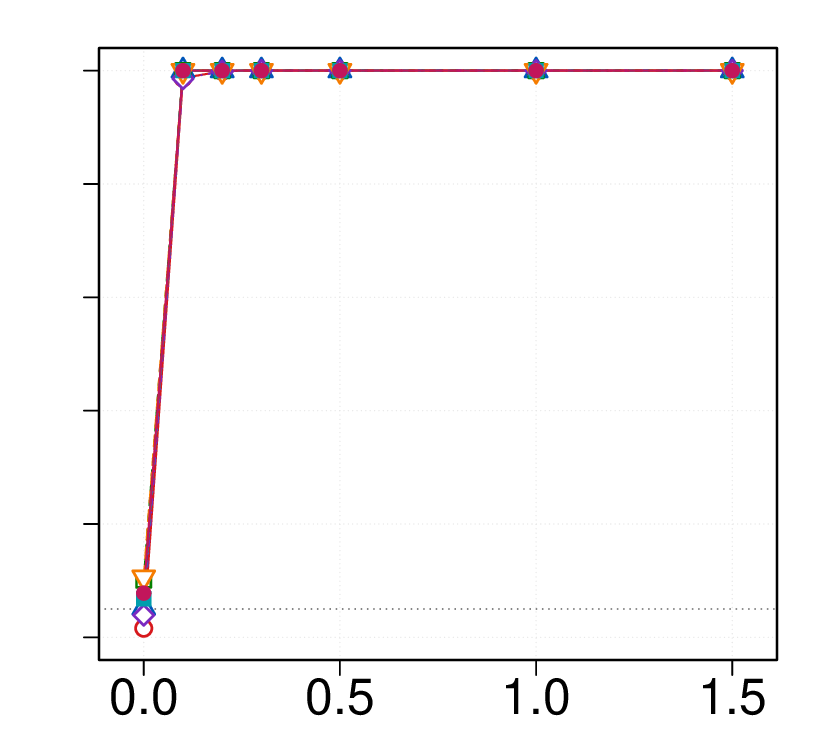}
            \hspace{-1 cm}  
            \includegraphics[width=5.5cm, height=5cm]{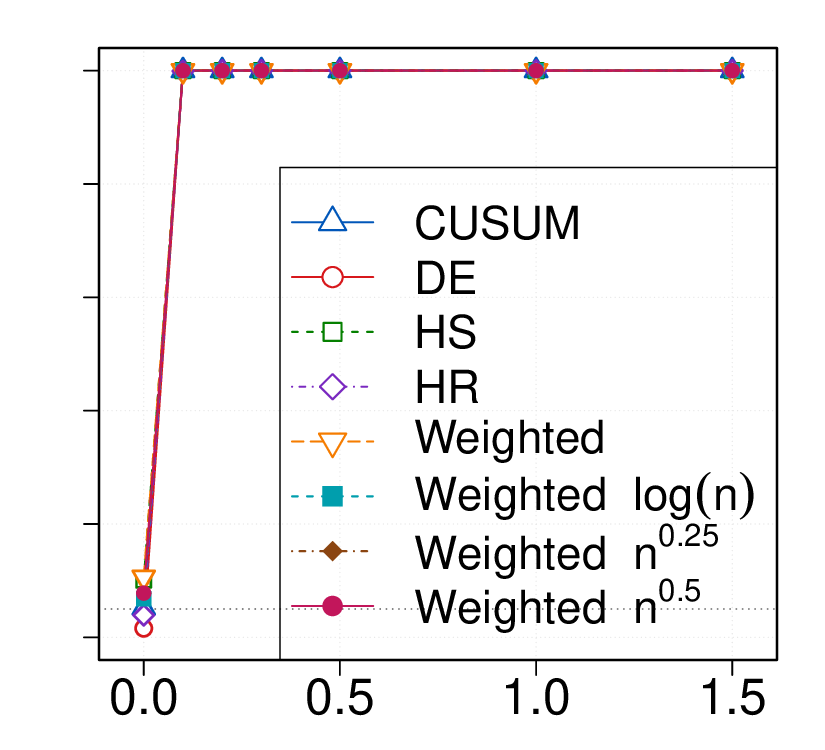}
            \hspace{-1 cm}  
            \vspace{-0.2in}
            \caption*{ (b)\small{ \textit{Setting 2}, with $m=1,5,20$} }

              \centering
            \hspace{-1 cm}  
                \includegraphics[width=5.5cm, height=5cm]{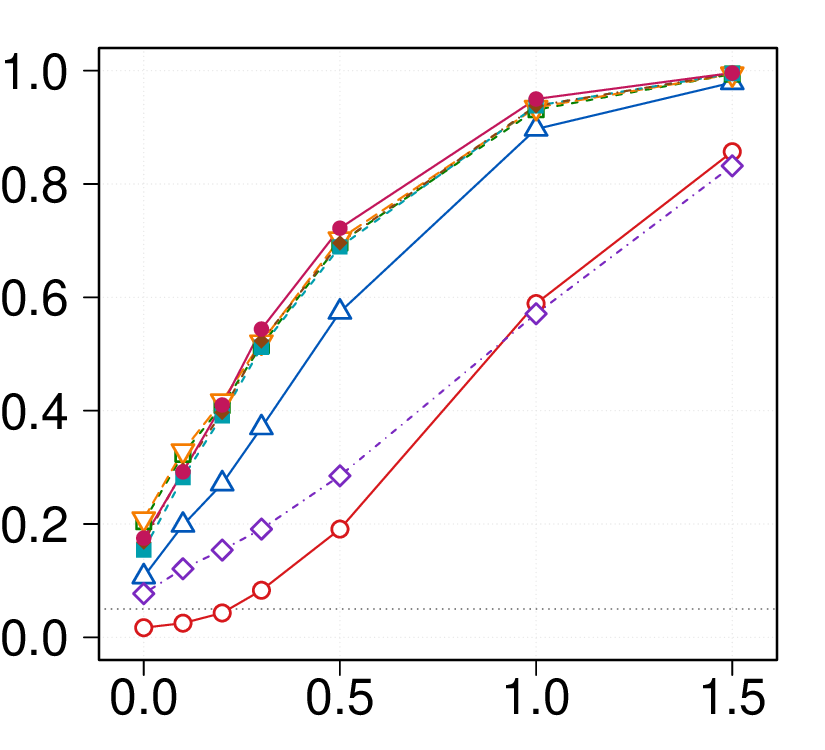}
            \hspace{-1 cm}  
            \includegraphics[width=5.5cm, height=5cm]{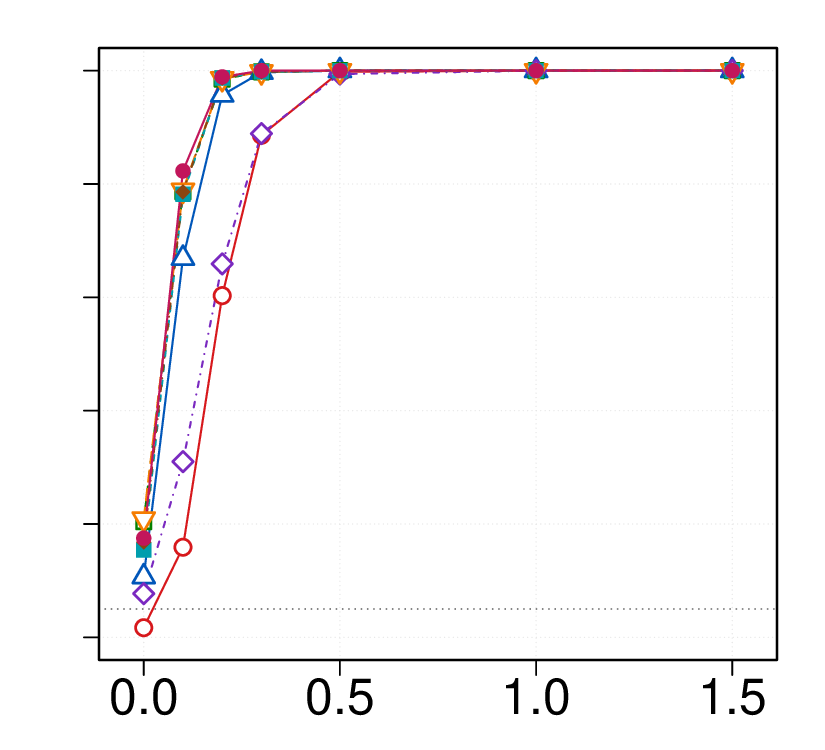}
            \hspace{-1 cm}  
            \includegraphics[width=5.5cm, height=5cm]{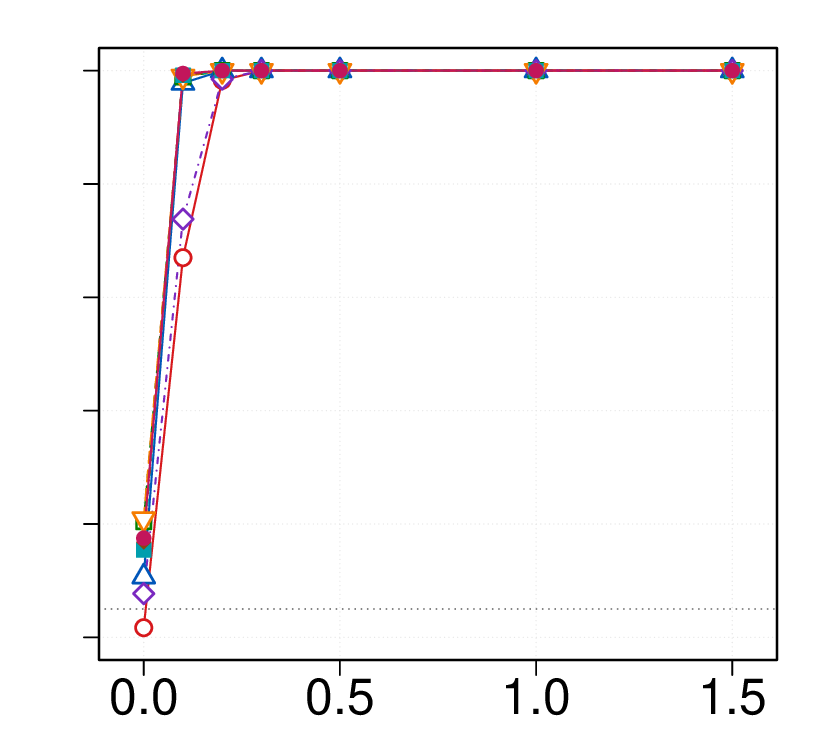}
            \hspace{-1 cm}  
            \vspace{-0.2in}
            \caption*{(c)\small{ \textit{Setting 3}, with $m=1,5,20$} }

            \caption{\small{Raw empirical rejection rates of the RP methods for various values of $SNR$ in the x-axis. The RP method performs 200 random projections and applies different change point tests (CUSUM, Weighted, DE, HS, HR) and the BH combination method. The data-generating process follows (\ref{eq:model for fd}), (\ref{eq:data generating process}), and (\ref{break function}), where the standard deviation $\sigma_{g}$ follows \textit{Settings 1--3}.
            The change point location is set at $\theta=0.25$.
            The empirical rejection rate is based on 1000 simulations.
            }}
            \label{fig: tuning cp test (BH_HAC)}
    \end{figure} 

\begin{figure} [H]
        \centering
            \hspace{-1 cm}  
            \includegraphics[width=5.5cm, height=5cm]{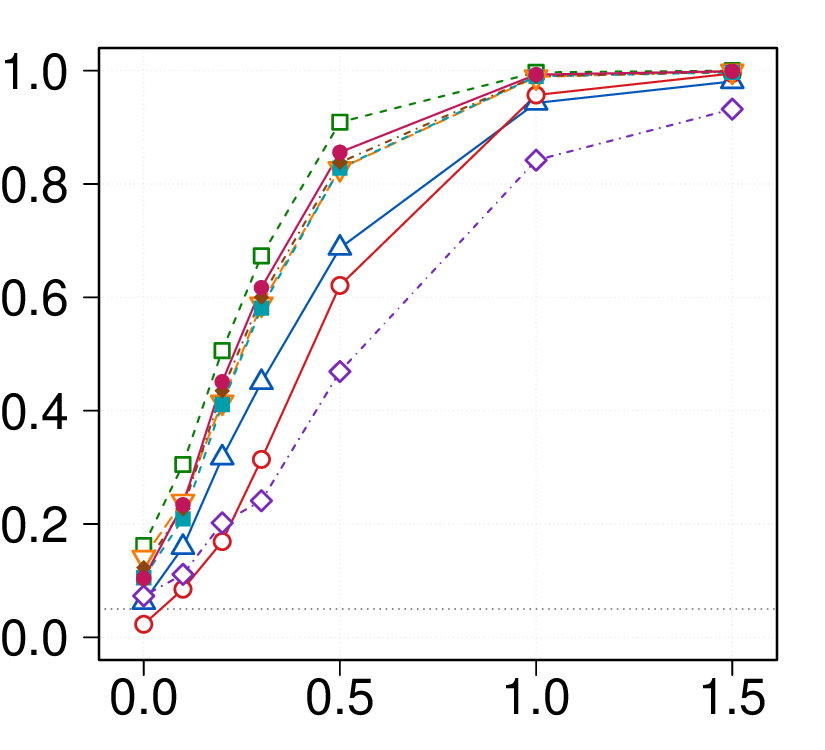}
            \hspace{-1 cm}  
            \includegraphics[width=5.5cm, height=5cm]{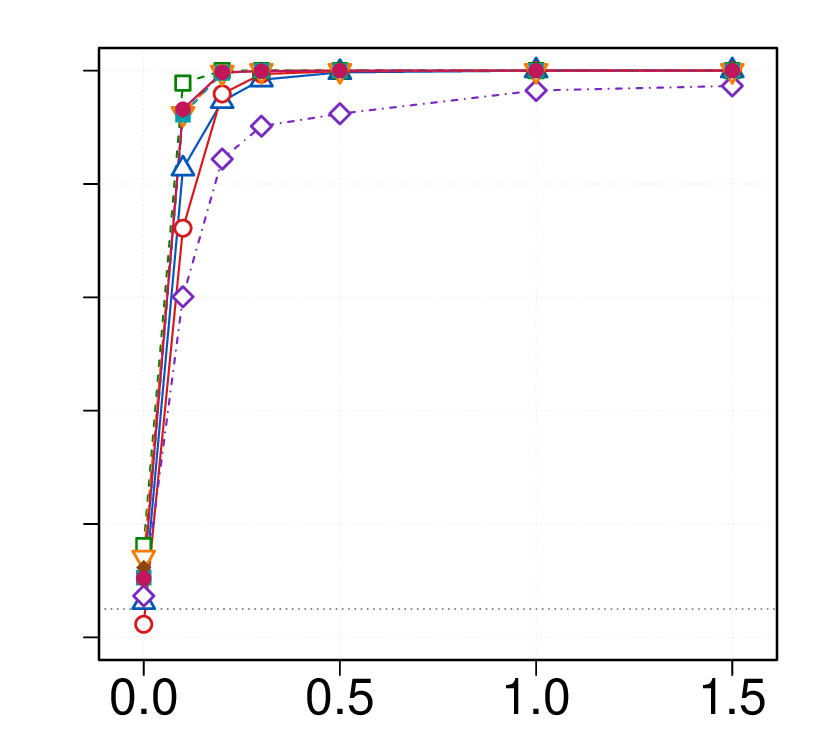}
            \hspace{-1 cm}  
            \includegraphics[width=5.5cm, height=5cm]{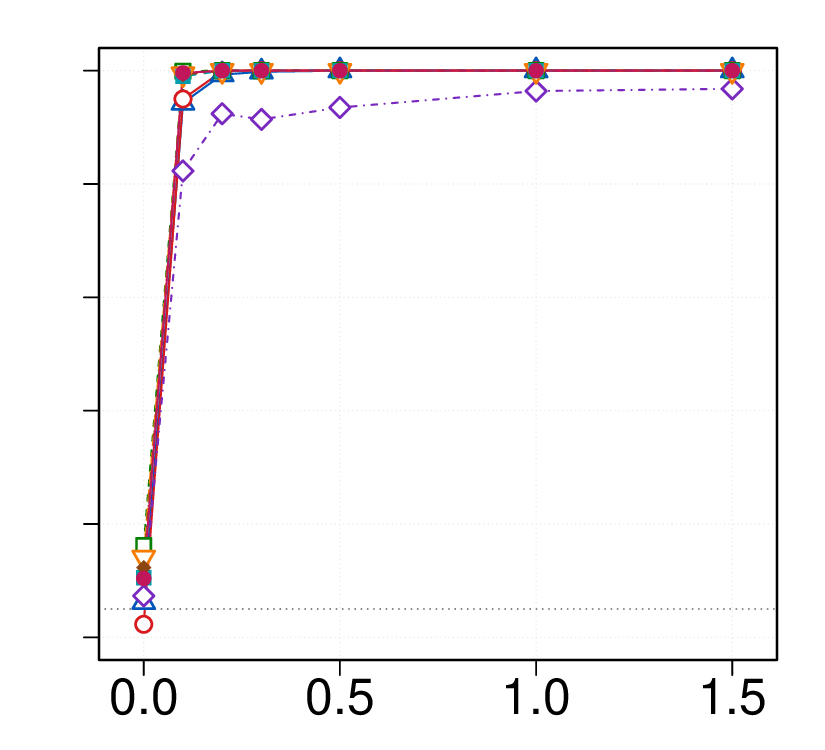}
            \hspace{-1 cm}  
            \vspace{-0.2in}
            \caption*{(a) \small{\textit{Setting 1}, with $m=1,5,20$}}

            \centering
            \hspace{-1 cm}  
                \includegraphics[width=5.5cm, height=5cm]{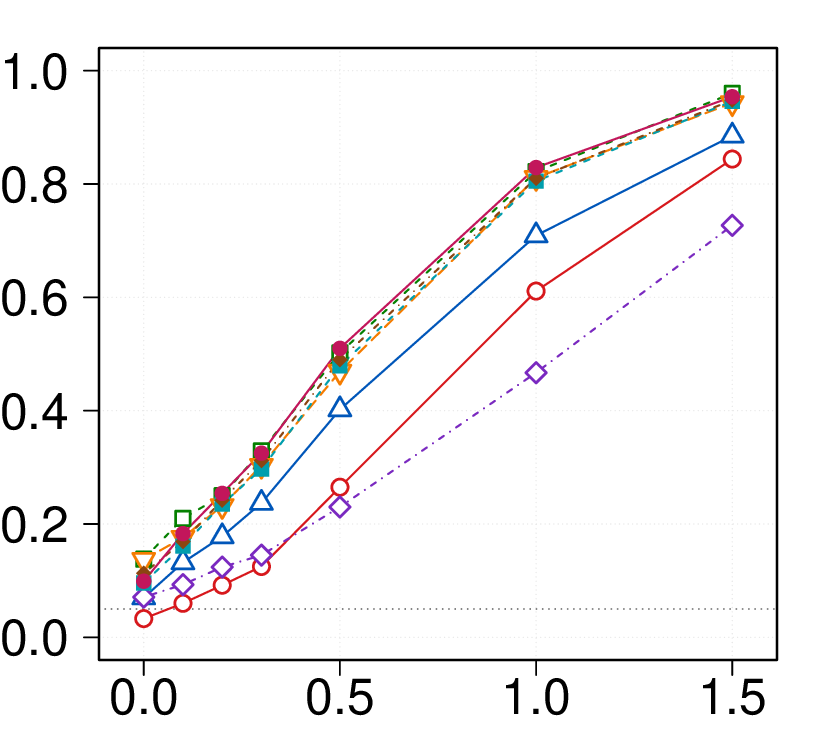}
            \hspace{-1 cm}  
            \includegraphics[width=5.5cm, height=5cm]{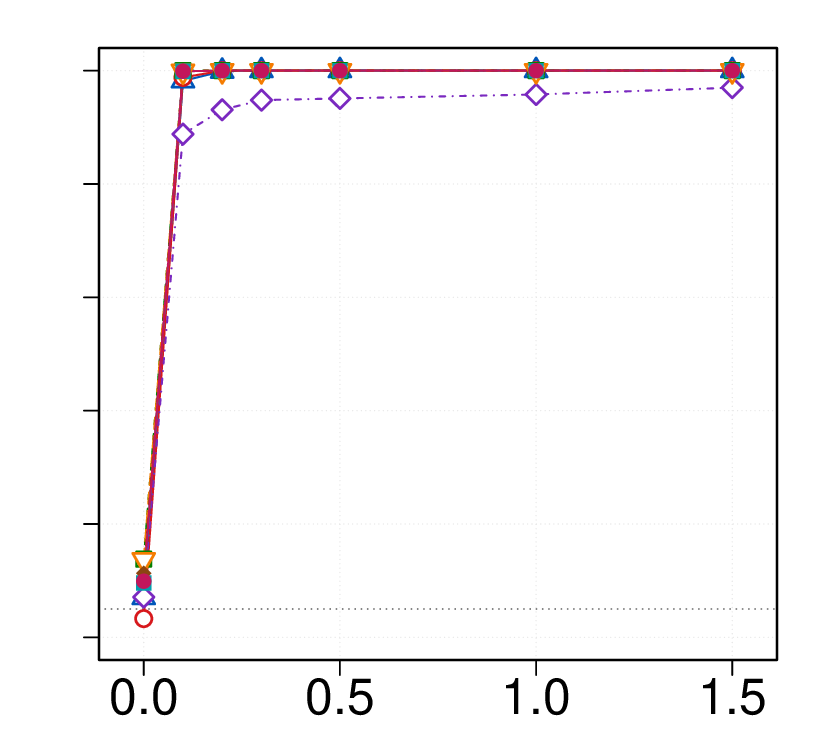}
            \hspace{-1 cm}  
            \includegraphics[width=5.5cm, height=5cm]{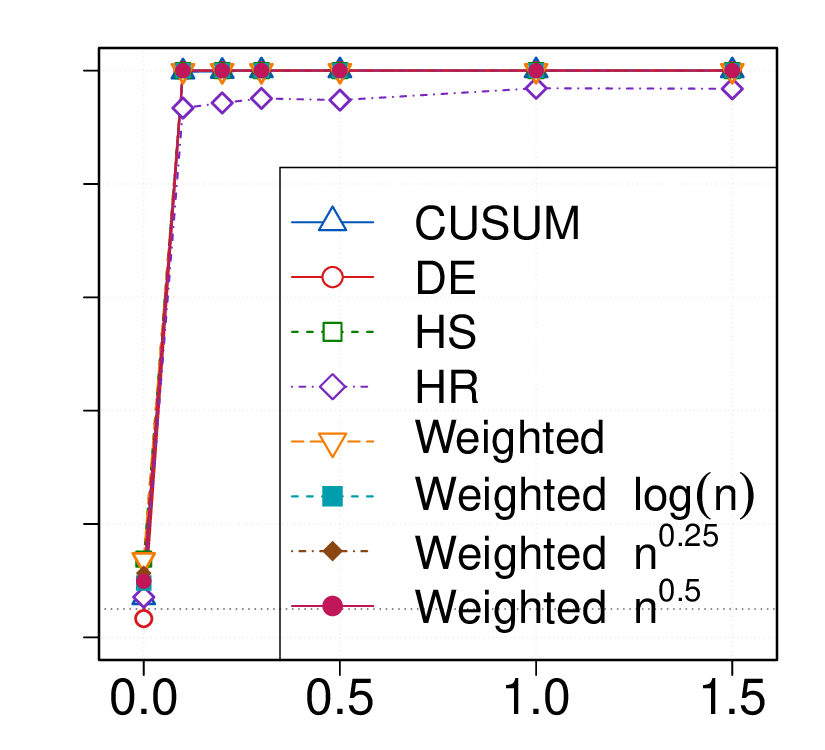}
            \hspace{-1 cm}  
            \vspace{-0.2in}
            \caption*{ (b)\small{ \textit{Setting 2}, with $m=1,5,20$} }

              \centering
            \hspace{-1 cm}  
                \includegraphics[width=5.5cm, height=5cm]{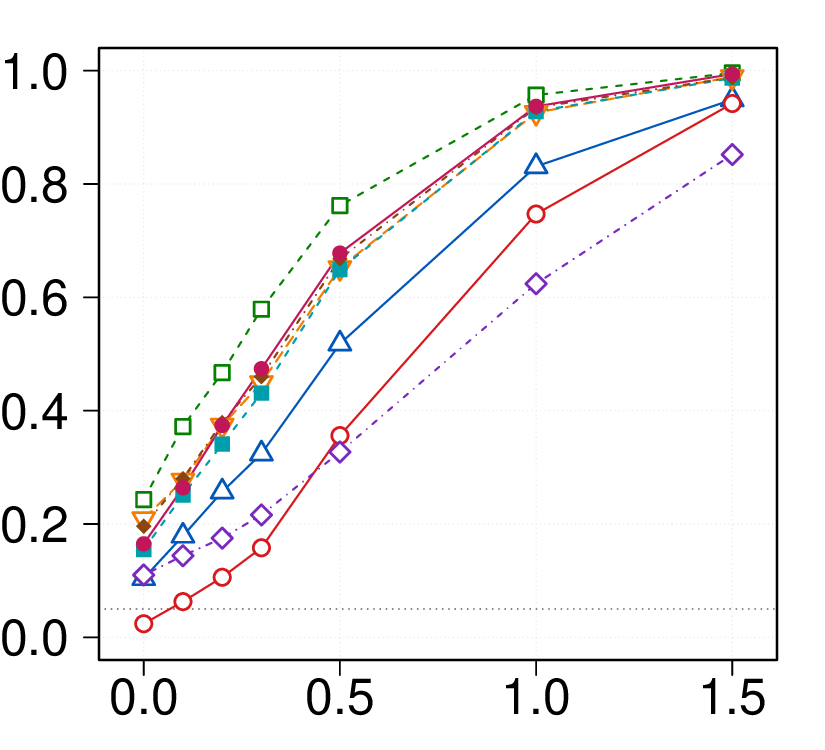}
            \hspace{-1 cm}  
            \includegraphics[width=5.5cm, height=5cm]{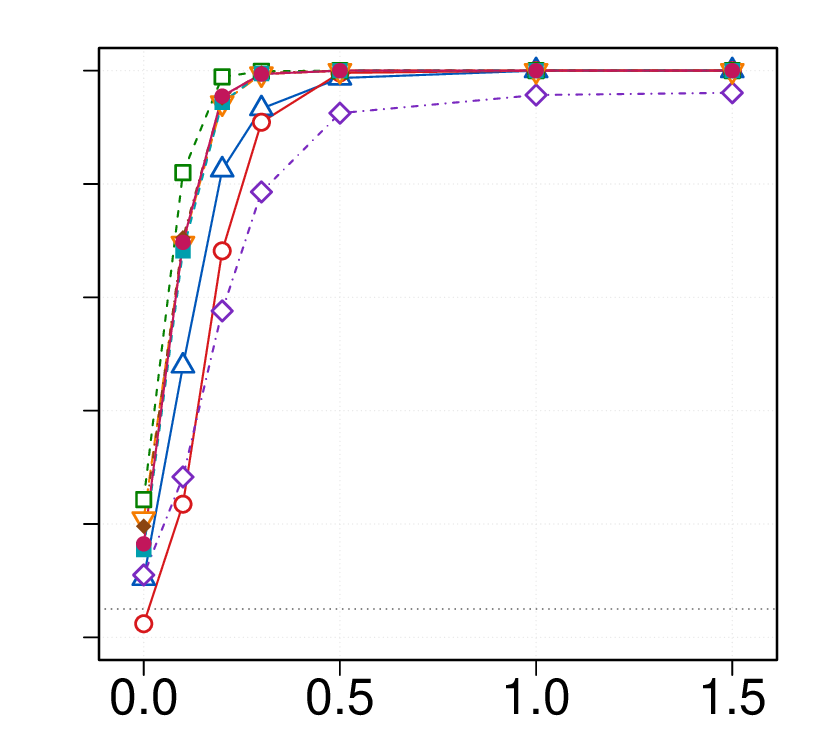}
            \hspace{-1 cm}  
            \includegraphics[width=5.5cm, height=5cm]{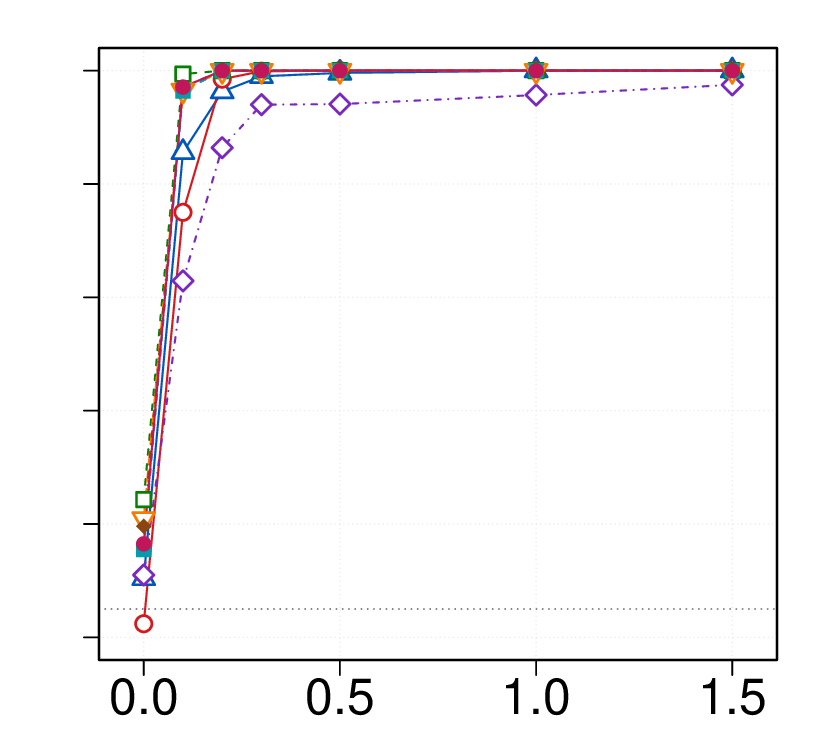}
            \hspace{-1 cm}  
            \vspace{-0.2in}
            \caption*{(c)\small{ \textit{Setting 3}, with $m=1,5,20$} }

            \caption{\small{Raw empirical rejection rates of the RP methods for various values of $SNR$ in the x-axis. The RP method performs 200 random projections and applies different change point tests (CUSUM, Weighted, DE, HS, HR) and the CCT combination method. The data-generating process follows (\ref{eq:model for fd}), (\ref{eq:data generating process}), and (\ref{break function}), where the standard deviation $\sigma_{g}$ follows \textit{Settings 1--3}.
            The change point location is set at $\theta=0.25$.
            The empirical rejection rate is based on 1000 simulations.
            }}
            \label{fig: tuning cp test (CCT_HAC) }
    \end{figure} 

\begin{figure} [H]
        \centering
            \hspace{-1 cm}  
            \includegraphics[width=5.5cm, height=5cm]{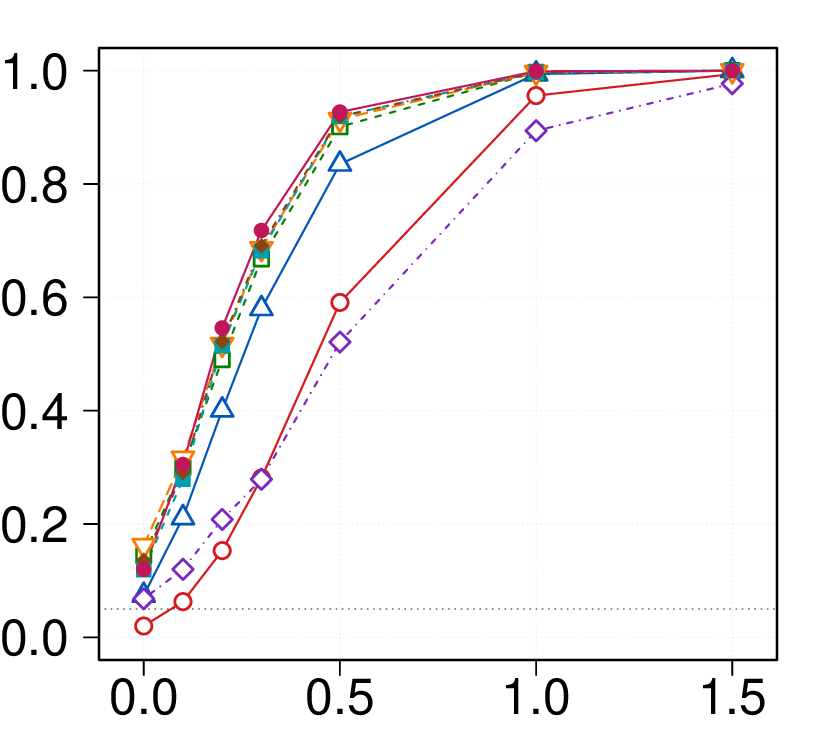}
            \hspace{-1 cm}  
            \includegraphics[width=5.5cm, height=5cm]{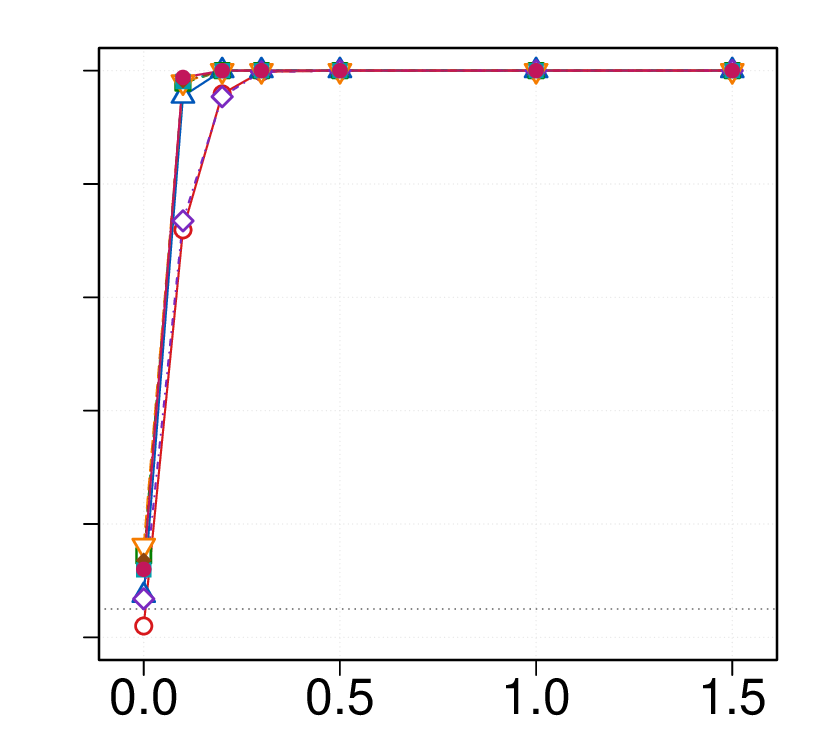}
            \hspace{-1 cm}  
            \includegraphics[width=5.5cm, height=5cm]{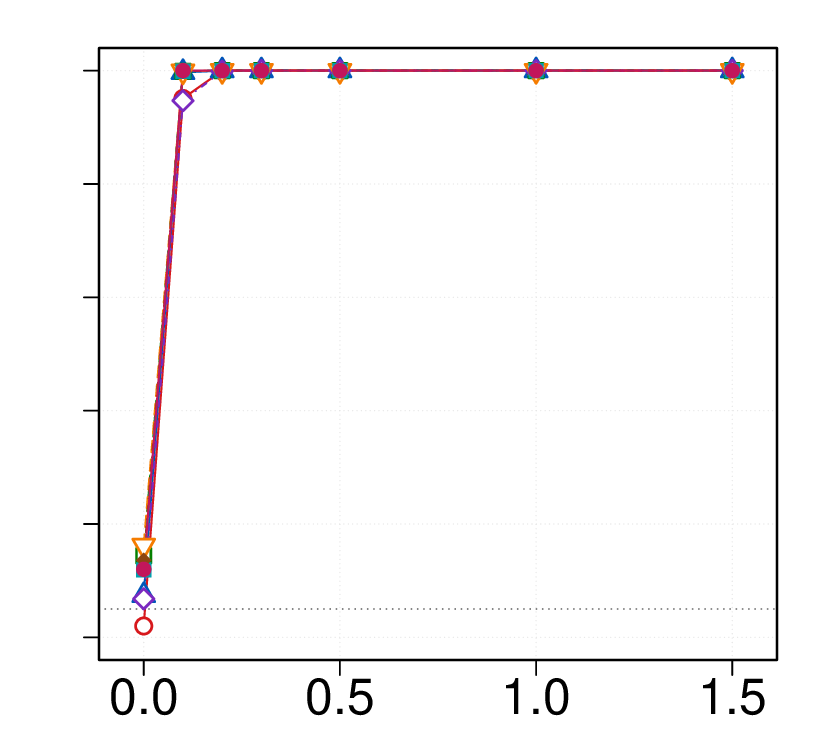}
            \hspace{-1 cm}  
            \vspace{-0.2in}
            \caption*{(a)\small{  \textit{Setting 1}, with $m=1,5,20$  } }

            \centering
            \hspace{-1 cm}  
                \includegraphics[width=5.5cm, height=5cm]{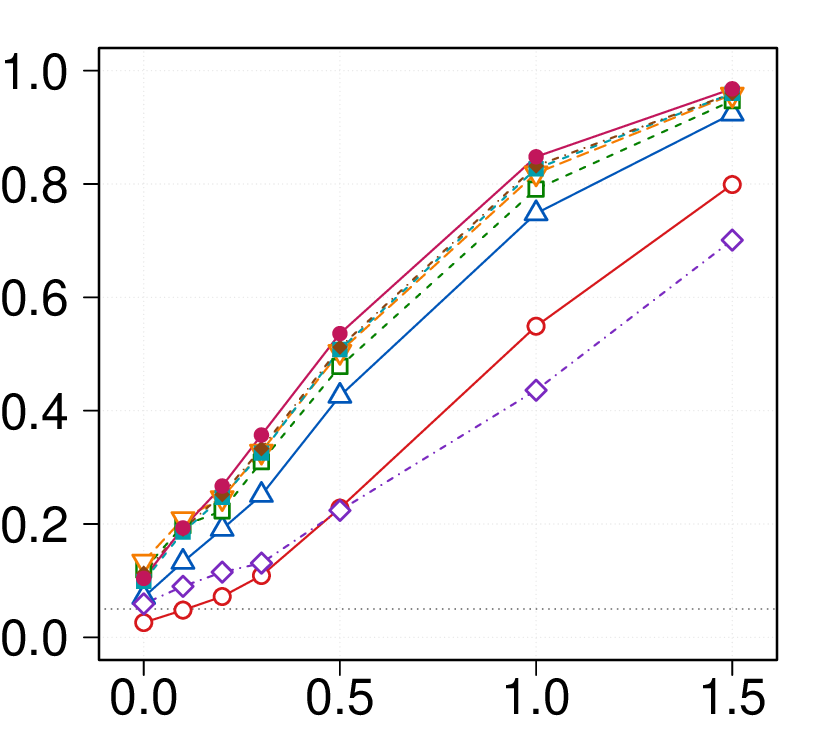}
            \hspace{-1 cm}  
            \includegraphics[width=5.5cm, height=5cm]{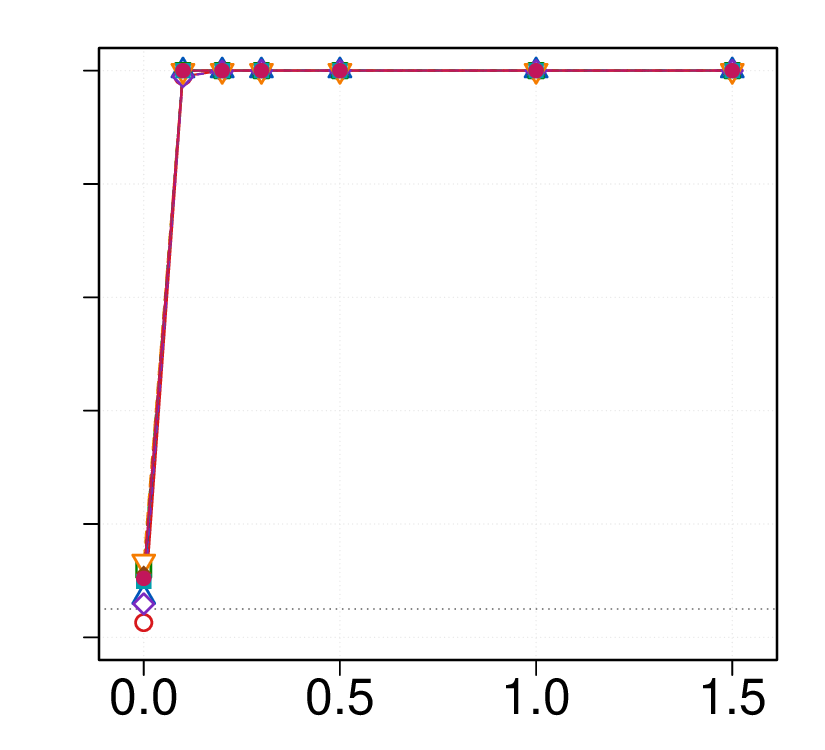}
            \hspace{-1 cm}  
            \includegraphics[width=5.5cm, height=5cm]{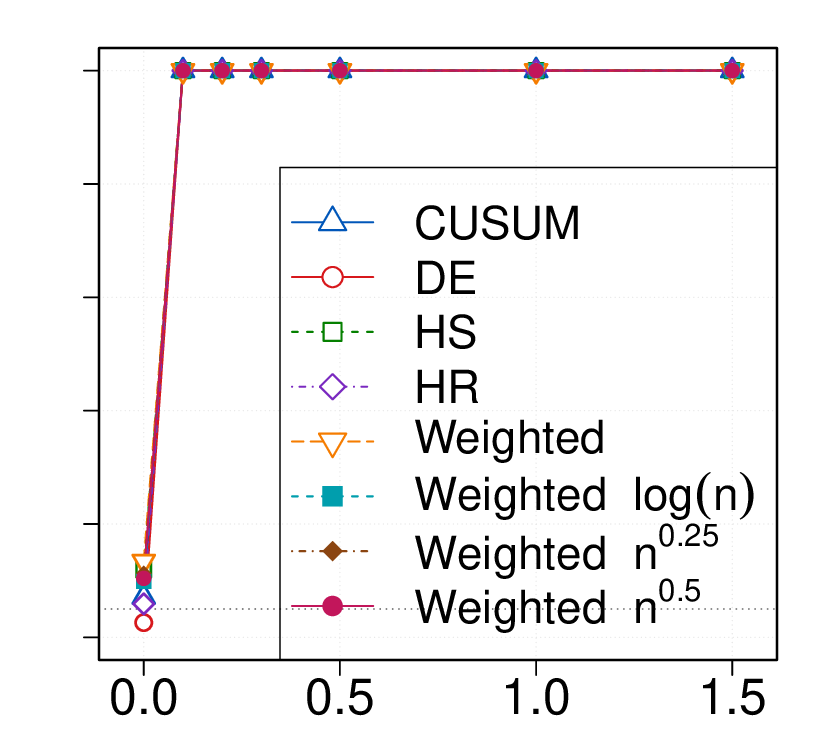}
            \hspace{-1 cm}  
            \vspace{-0.2in}
            \caption*{ (b) \small{ \textit{Setting 2}, with $m=1,5,20$ } }

              \centering
            \hspace{-1 cm}  
                \includegraphics[width=5.5cm, height=5cm]{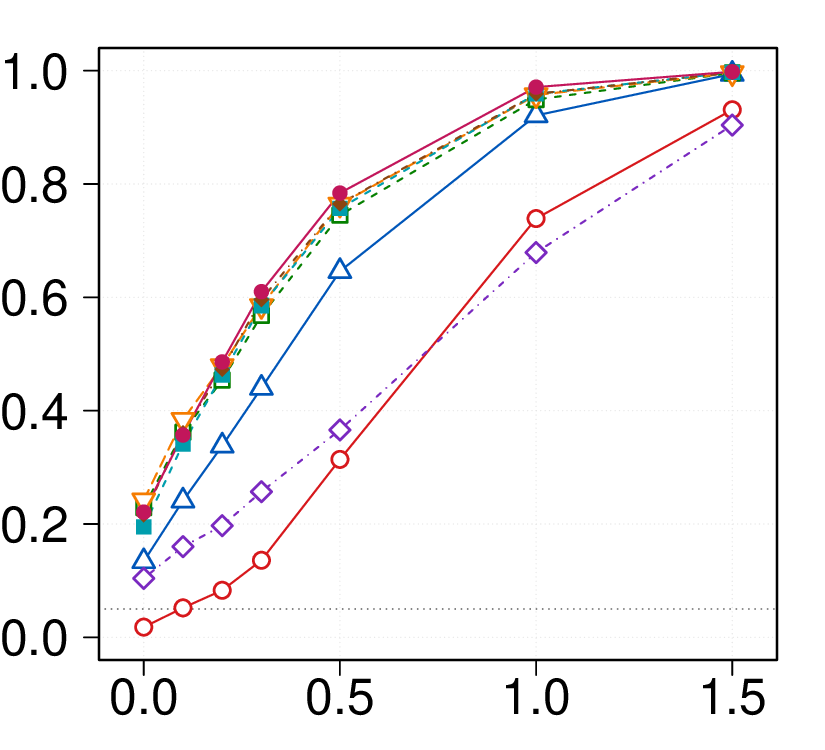}
            \hspace{-1 cm}  
            \includegraphics[width=5.5cm, height=5cm]{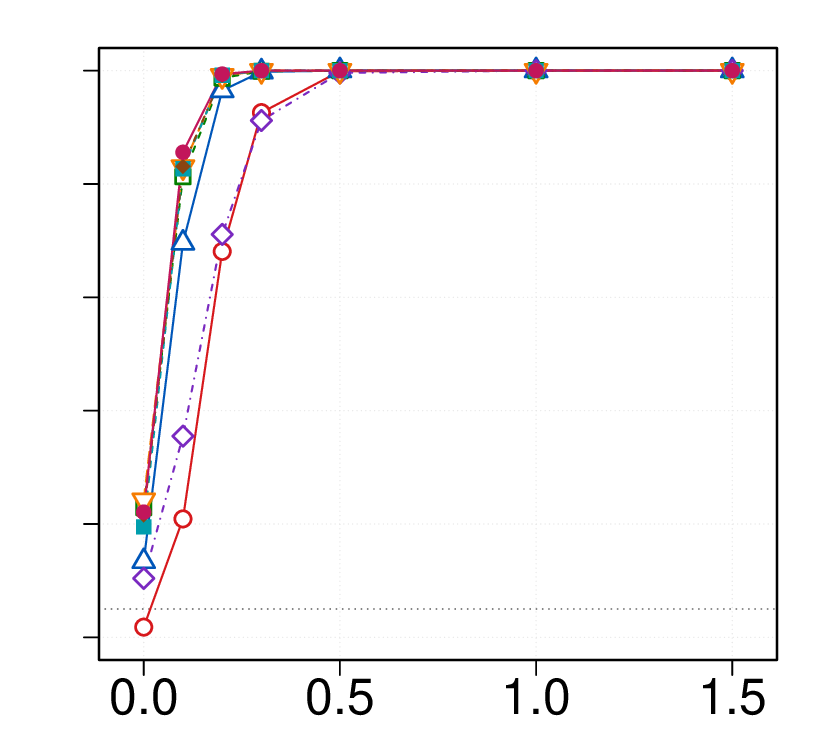}
            \hspace{-1 cm}  
            \includegraphics[width=5.5cm, height=5cm]{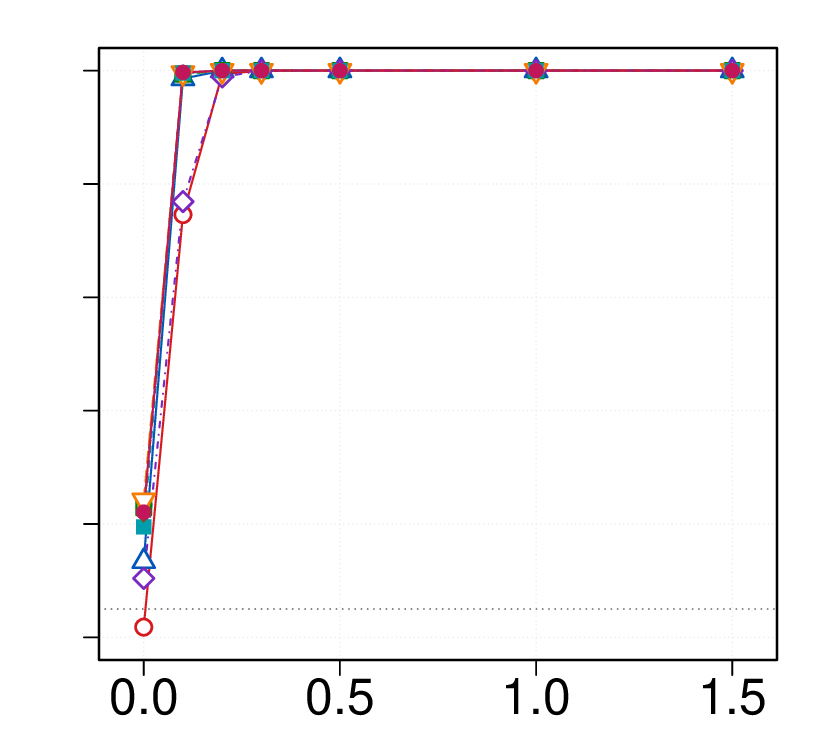}
            \hspace{-1 cm}  
            \vspace{-0.2in}
            \caption*{(c) \small{ \textit{Setting 3} , with $m=1,5,20$} }

            \caption{\small{Raw empirical rejection rates of the RP methods for various values of $SNR$ in the x-axis. The RP method performs 200 random projections and applies different change point tests (CUSUM, Weighted, DE, HS, HR) and the HMP combination method. The data-generating process follows (\ref{eq:model for fd}), (\ref{eq:data generating process}), and (\ref{break function}), where the standard deviation $\sigma_{g}$ follows \textit{Settings 1--3}.
            The change point location is set at $\theta=0.25$.
            The empirical rejection rate is based on 1000 simulations.
            }}
            \label{fig: tuning cp test (HMP_HAC)}
    \end{figure} 

\begin{figure} [H]
        \centering
        \hspace{-1 cm}
            \includegraphics[width=5.5cm, height=5cm]{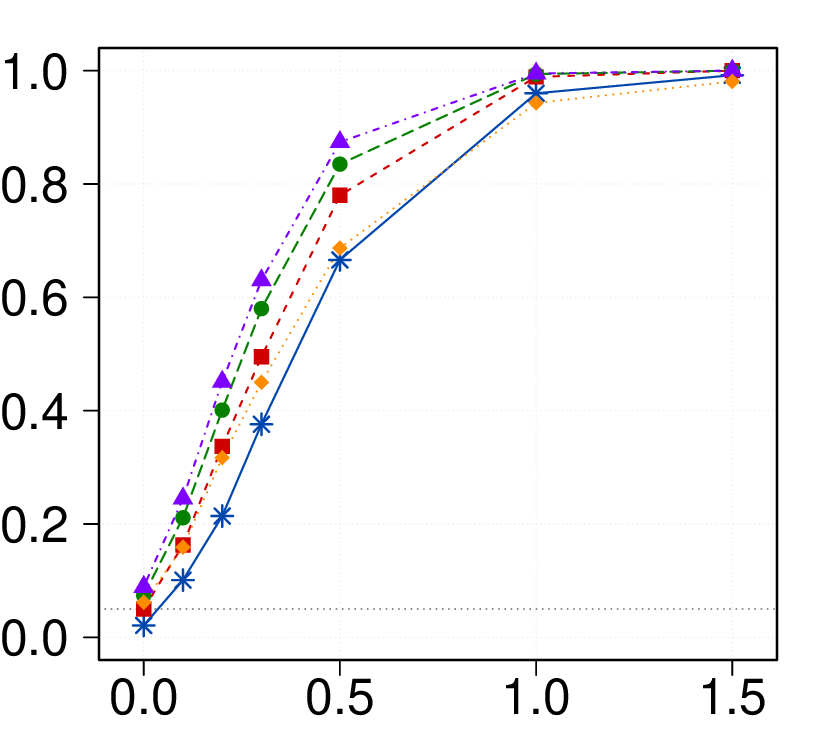}
         \hspace{-1 cm}
            \includegraphics[width=5.5cm, height=5cm]{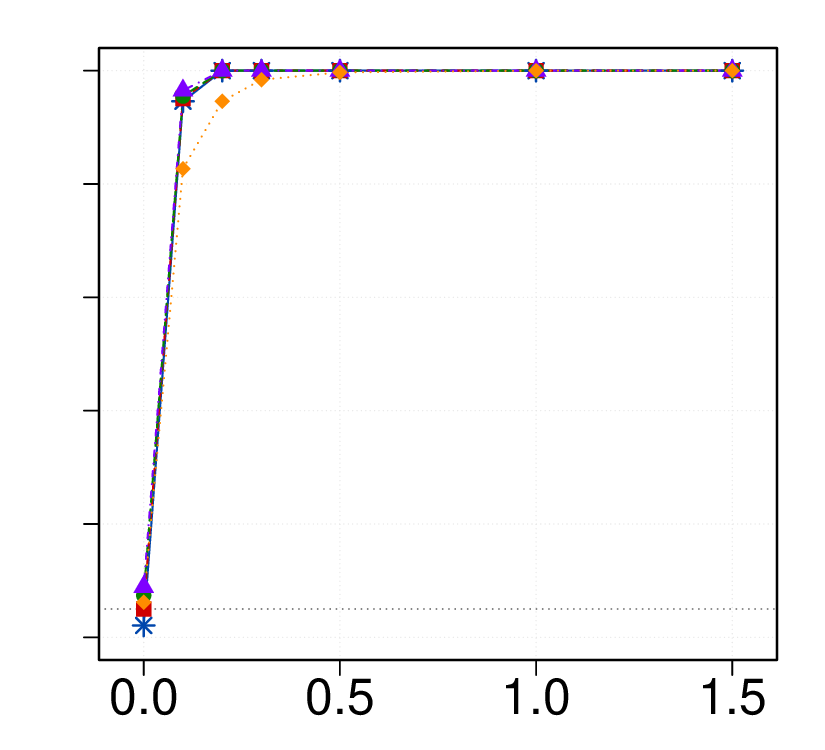}
         \hspace{-1 cm}
            \includegraphics[width=5.5cm, height=5cm]{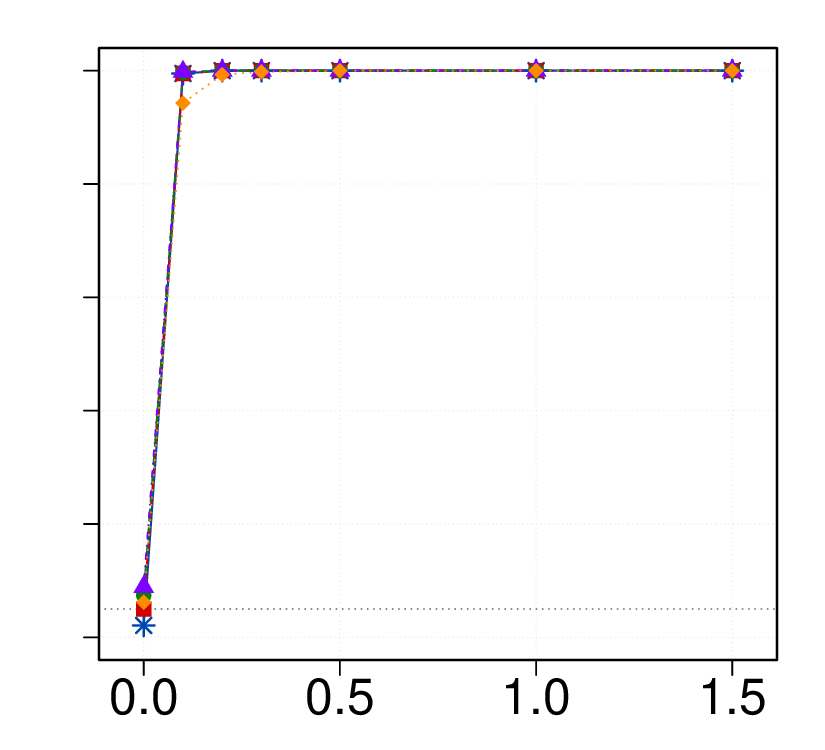}
         \hspace{-1 cm} 
         \vspace{-0.2in}
            \caption*{(a) \small{\textit{Setting 1}, with $m=1,5,20$ } }

            \centering
            \hspace{-1 cm}
            \includegraphics[width=5.5cm, height=5cm]{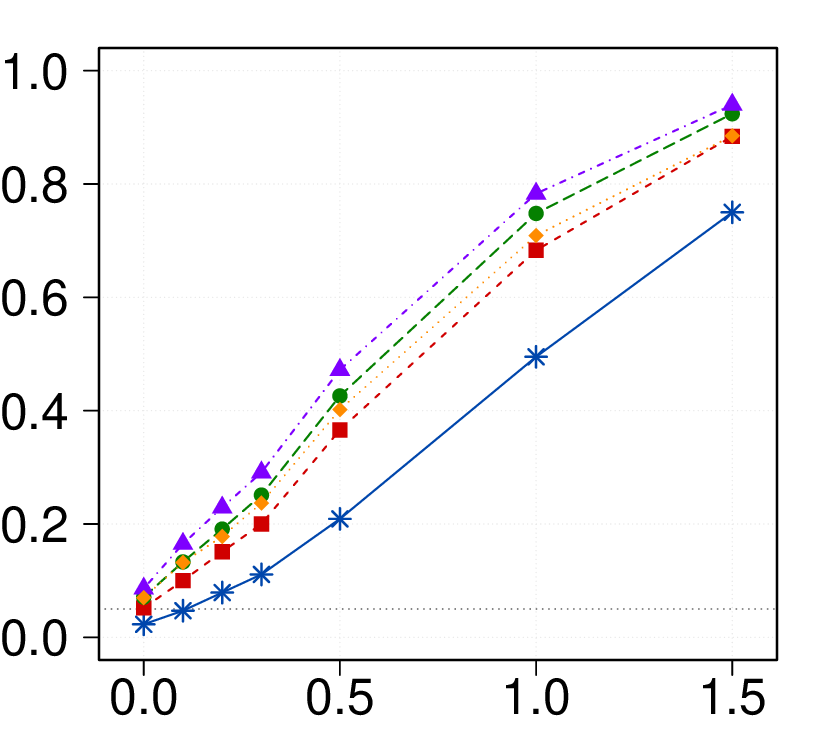}  
         \hspace{-1 cm}
            \includegraphics[width=5.5cm, height=5cm]{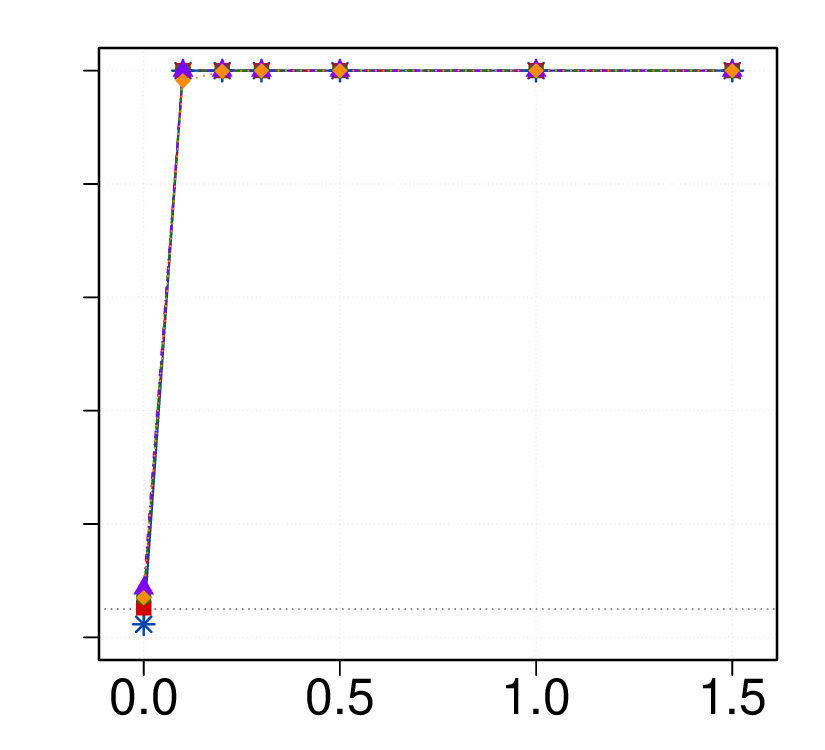}   \hspace{-1 cm}
            \includegraphics[width=5.5cm, height=5cm]{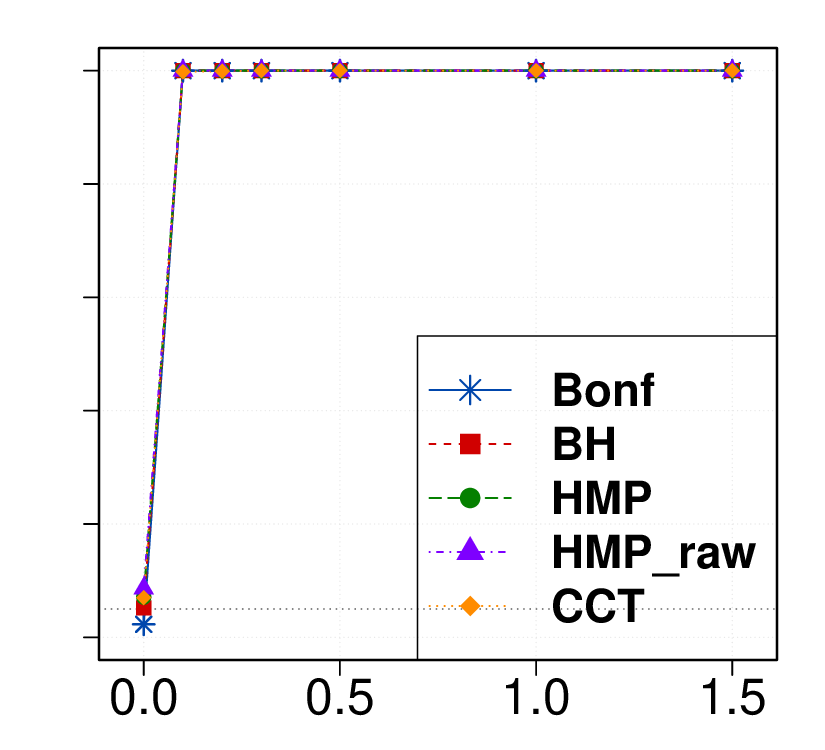}  \hspace{-1 cm} 
            \vspace{-0.2in}
            \caption*{(b) \small{\textit{Setting 2}, with $m=1,5,20$} }

            \centering
       \hspace{-1 cm}
            \includegraphics[width=5.5cm, height=5cm]{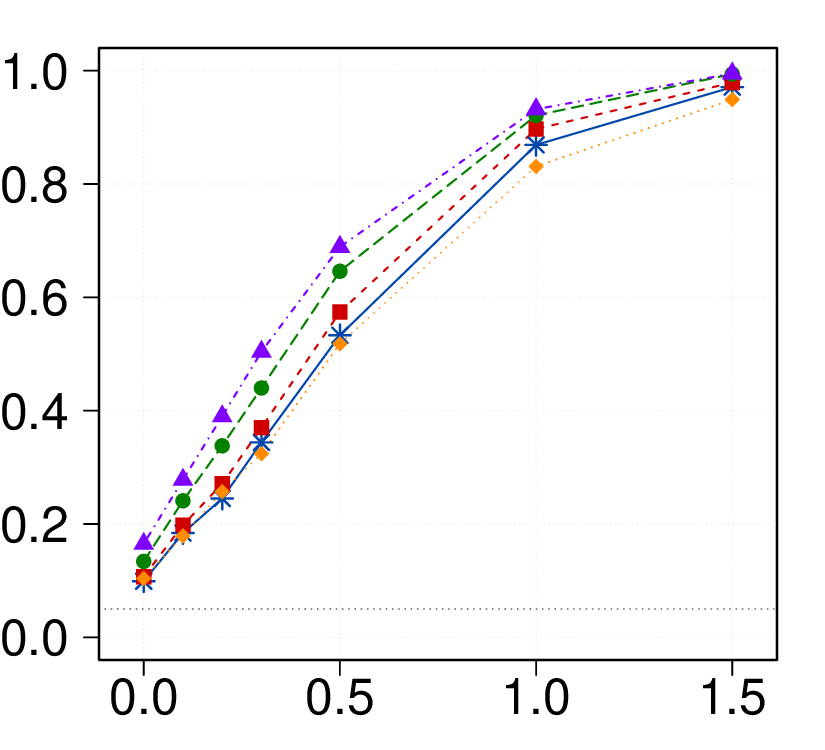} 
         \hspace{-1 cm}
            \includegraphics[width=5.5cm, height=5cm]{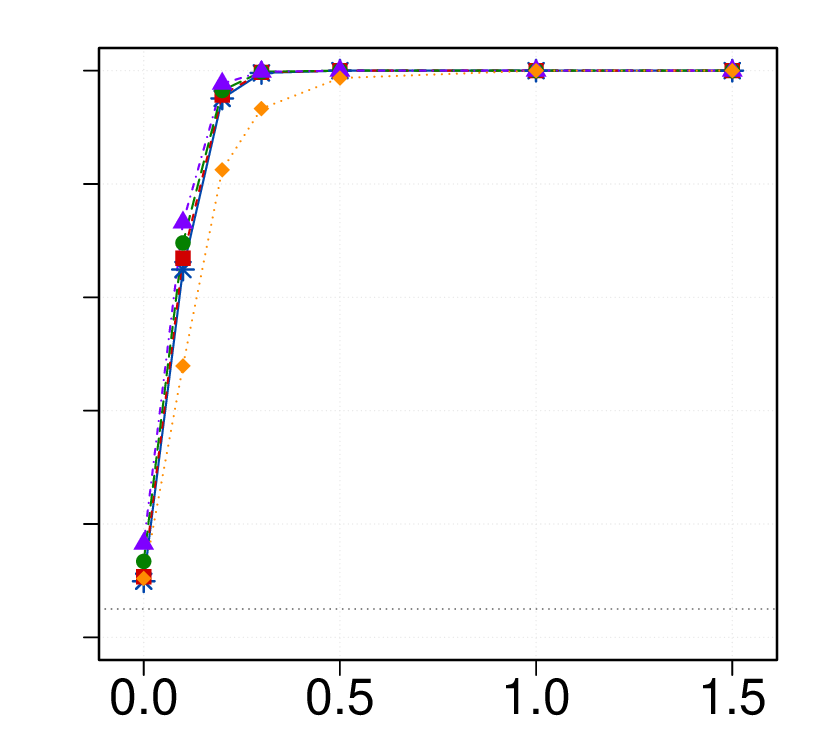}  
         \hspace{-1 cm}
            \includegraphics[width=5.5cm, height=5cm]{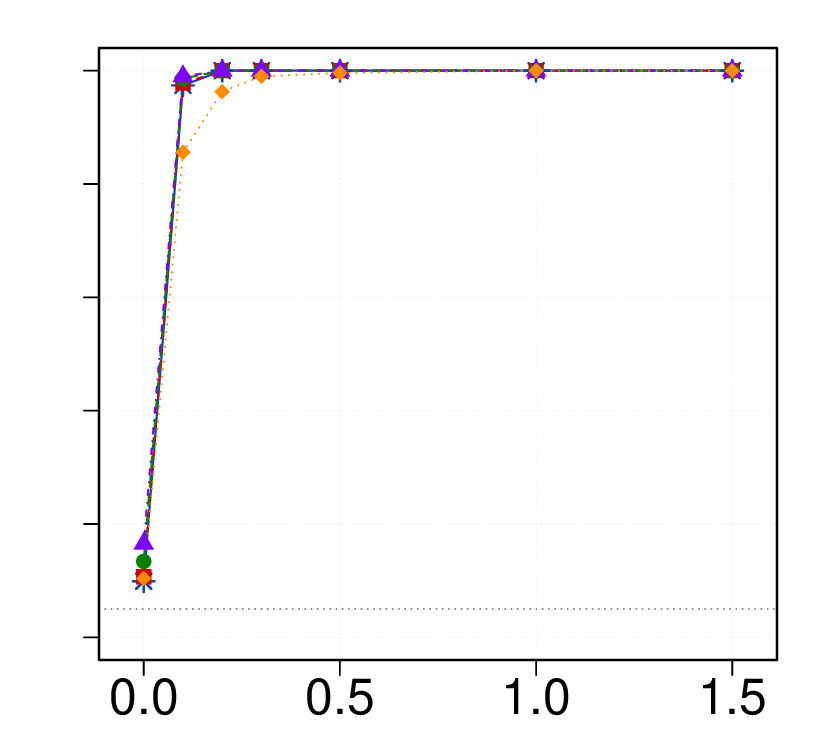} 
         \hspace{-1 cm} 
            \vspace{-0.2in}
            \caption*{(c) \small{\textit{Setting 3}, with $m=1,5,20$} }

            \caption{\small{Raw empirical rejection rates of the RP-Bonf, RP-BH, RP-HMP, RP-HMP-raw, and RP-CCT methods with the standard CUSUM test for various values of $SNR$ in the x-axis. The RP method performs 200 random projections. The data-generating process follows (\ref{eq:model for fd}), (\ref{eq:data generating process}), and (\ref{break function}), where the standard deviation $\sigma_{g}$ follows \textit{Settings 1--3}.
            The change point location is set at $\theta=0.25$.
            The empirical rejection rate is based on 1000 simulations.
            }}
            \label{fig: tuning Pvalue-comb (HAC)}
    \end{figure} 

\newpage
\subsection{Results using the variance estimator in \cite{horvath2020new}}
\label{subsec: Result of using IID estimator}
\begin{figure} [H]
             \centering
            \hspace{-1 cm}  
            \includegraphics[width=5.5cm, height=5cm]{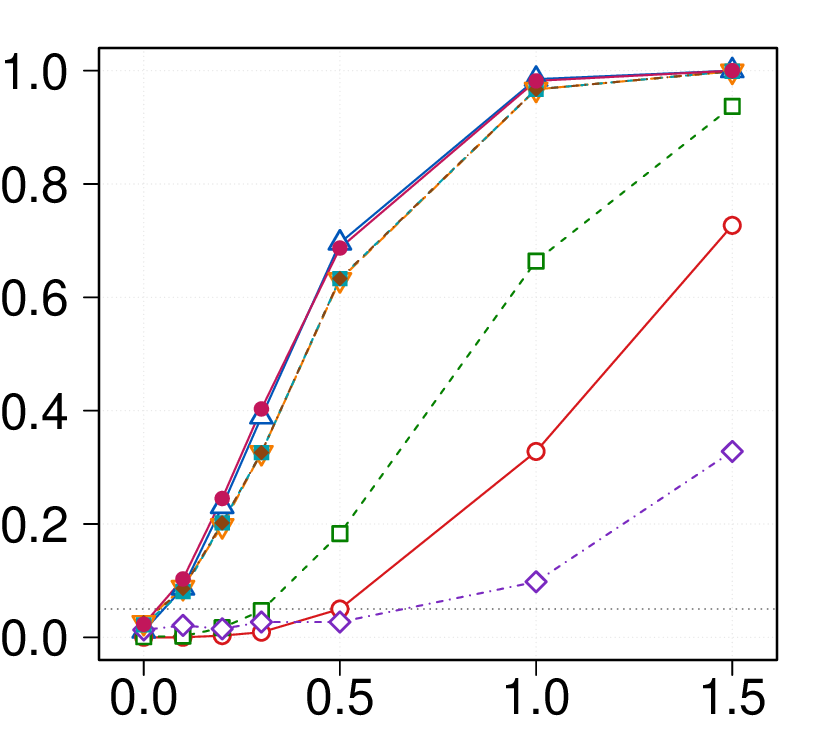}
            \hspace{-1 cm}  
            \includegraphics[width=5.5cm, height=5cm]{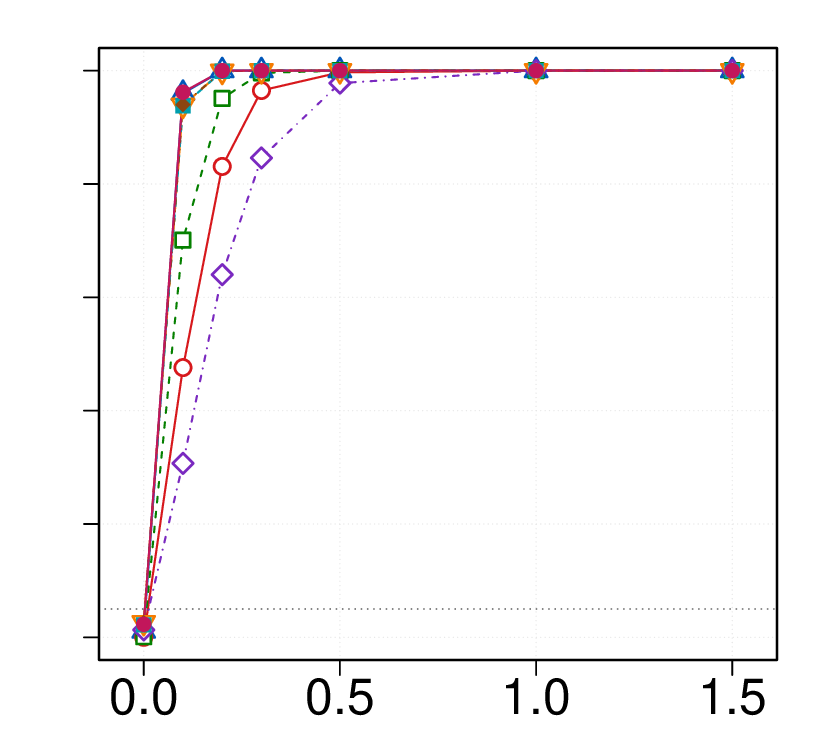}
            \hspace{-1 cm}  
            \includegraphics[width=5.5cm, height=5cm]{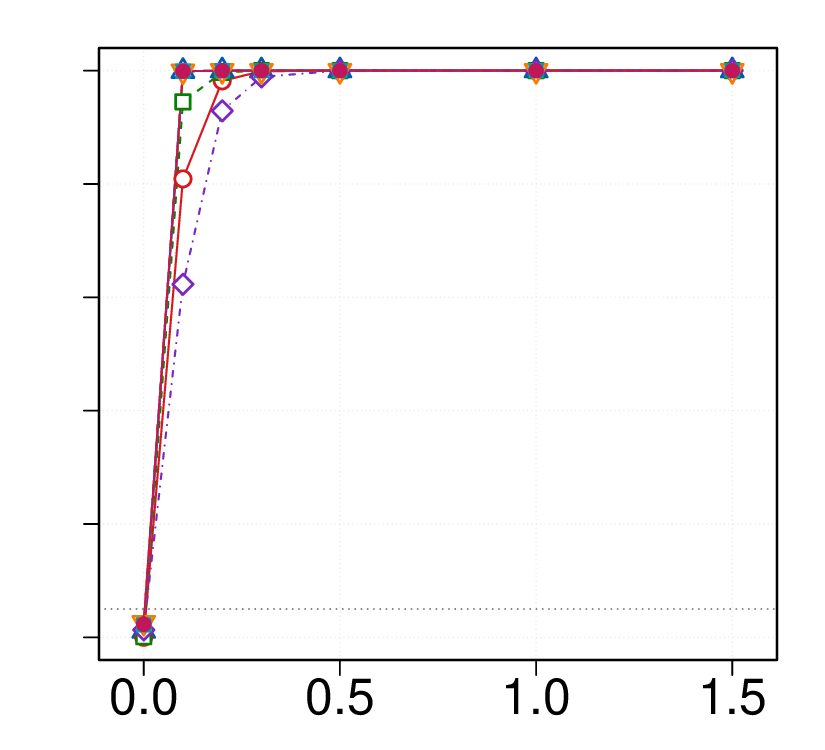}
            \hspace{-1 cm}  
            \vspace{-0.2in}
            \caption*{(a) \small{ \textit{Setting 1}, with $m=1,5,20$  } }

            \centering
            \hspace{-1 cm}  
                \includegraphics[width=5.5cm, height=5cm]{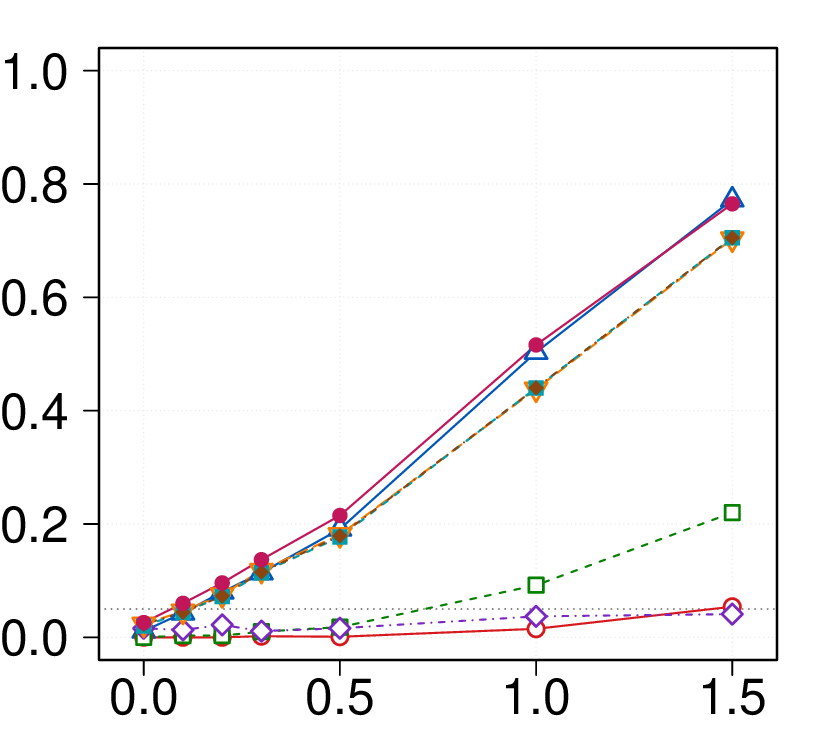}
            \hspace{-1 cm}  
            \includegraphics[width=5.5cm, height=5cm]{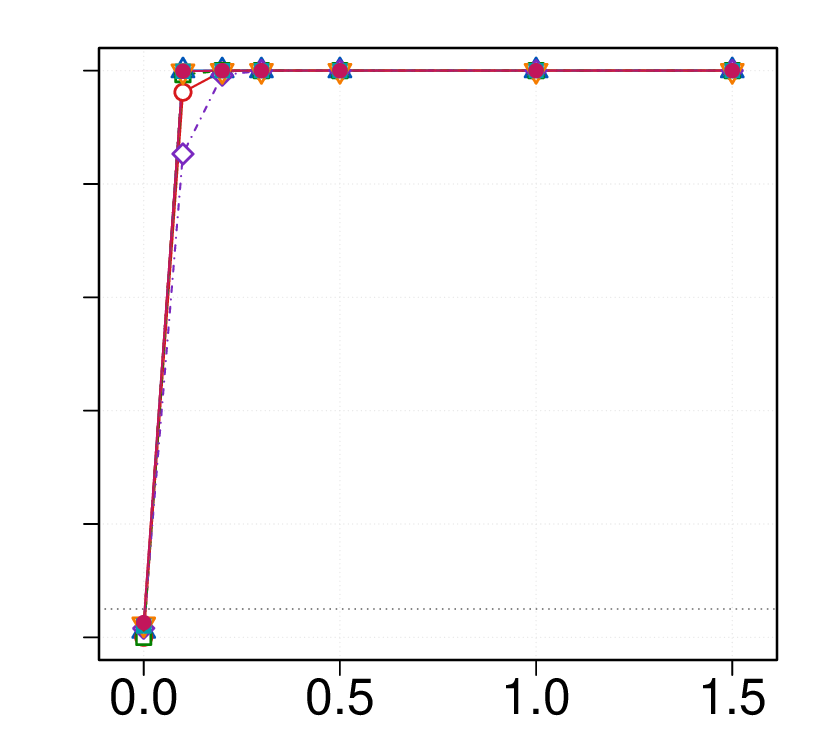}
            \hspace{-1 cm}  
            \includegraphics[width=5.5cm, height=5cm]{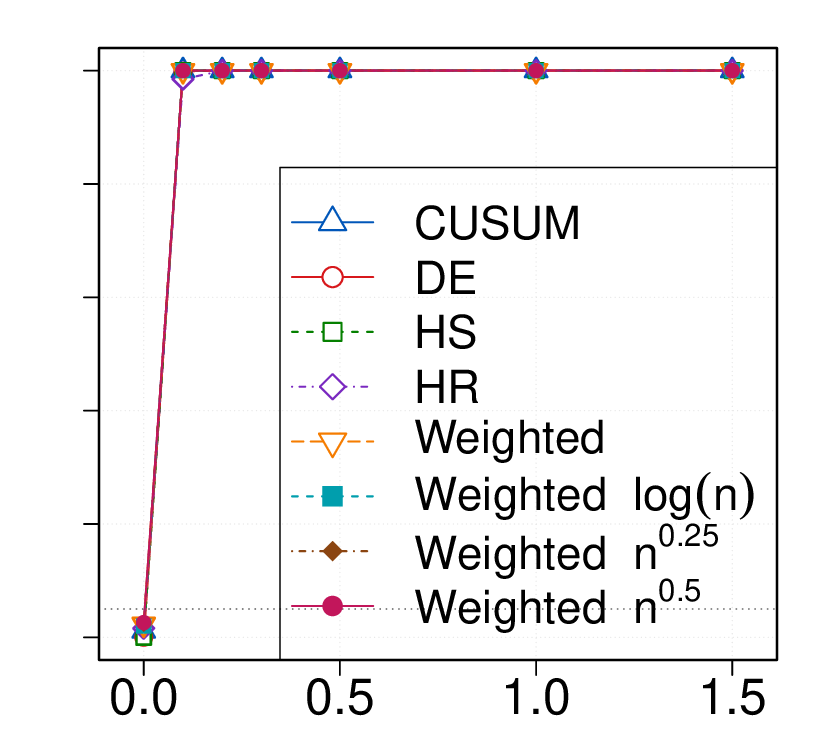}
            \hspace{-1 cm}  
            \vspace{-0.2in}
            \caption*{(b)\small{ \textit{Setting 2}, with $m=1,5,20$  } }

              \centering
            \hspace{-1 cm}  
                \includegraphics[width=5.5cm, height=5cm]{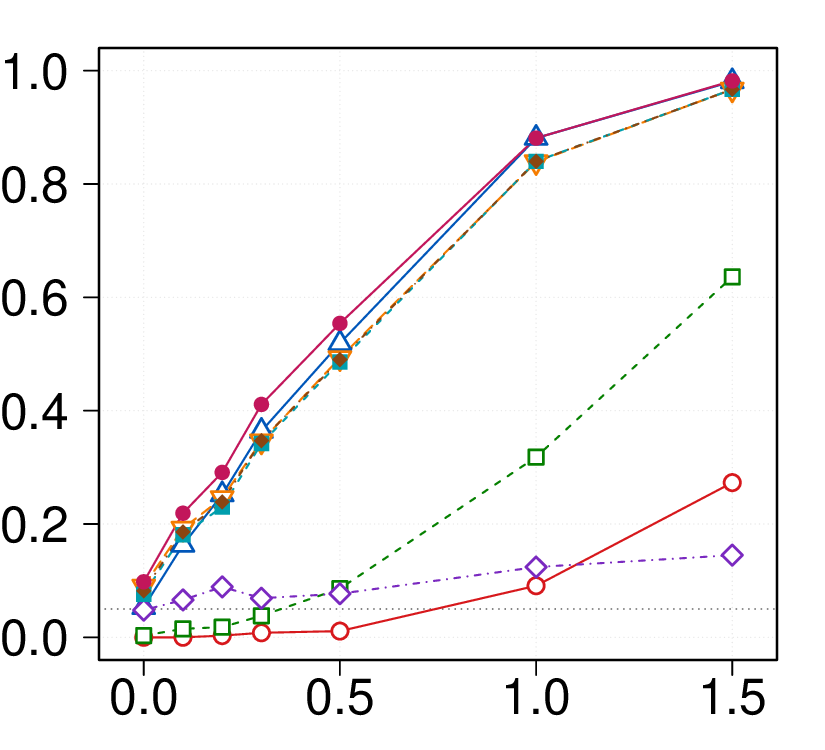}
            \hspace{-1 cm}  
            \includegraphics[width=5.5cm, height=5cm]{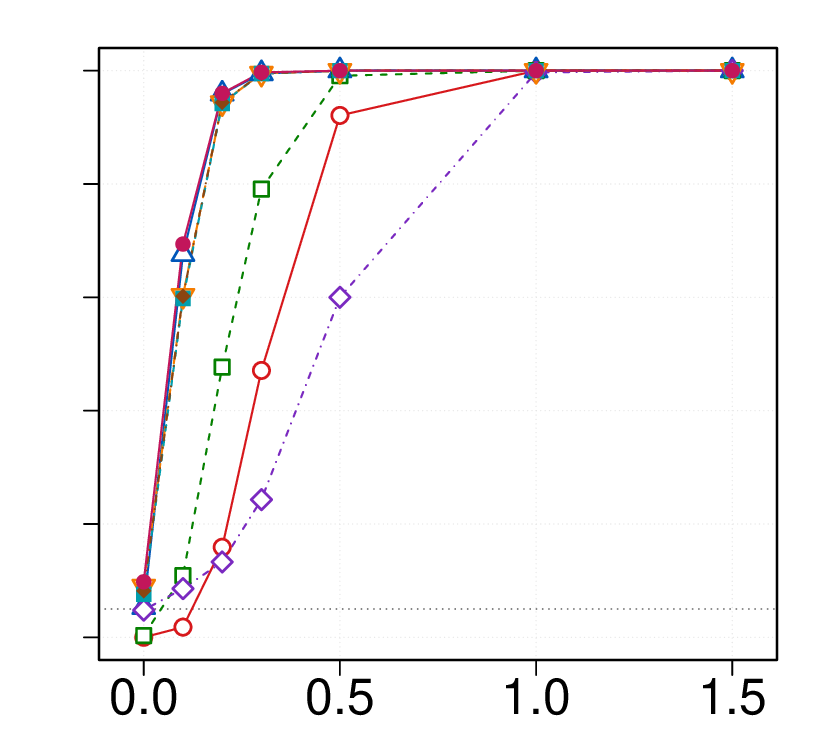}
            \hspace{-1 cm}  
            \includegraphics[width=5.5cm, height=5cm]{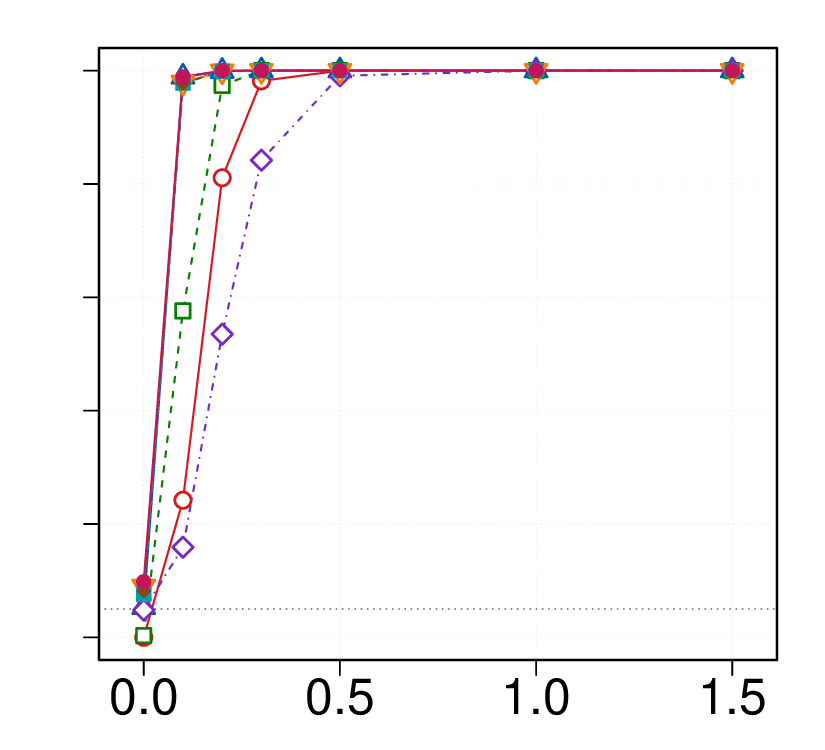}
            \hspace{-1 cm}  
            \vspace{-0.2in}
            \caption*{(c)\small{\textit{Setting 3}, with $m=1,5,20$  } }
            
             \caption{\small{
             Raw empirical rejection rates of the RP methods for various values of $SNR$ in the x-axis. The RP method performs 200 random projections and applies different change point tests (CUSUM, Weighted, DE, HS, HR) and the Bonf combination method. The data-generating process follows (\ref{eq:model for fd}), (\ref{eq:data generating process}), and (\ref{break function}), where the standard deviation $\sigma_{g}$ follows \textit{Settings 1--3}.
            The change point location is set at $\theta=0.5$.
            The empirical rejection rate is based on 1000 simulations.
             }}
             \label{fig: tuning cp test Bonf-theta=0.5}
        
    \end{figure} 
\begin{figure} [H]
            \centering
            \hspace{-1 cm}  
            \includegraphics[width=5.5cm, height=5cm]{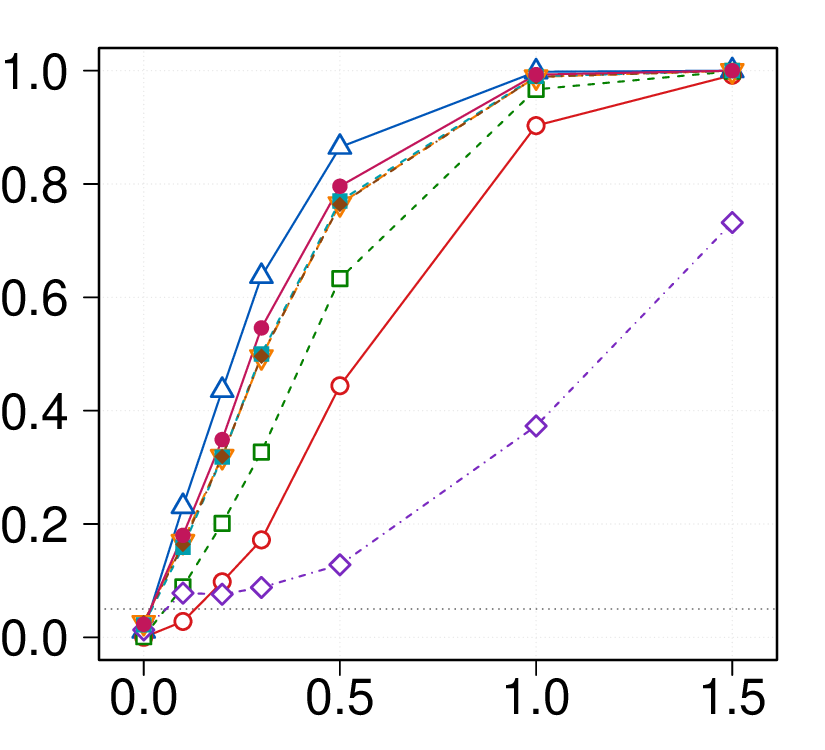}
            \hspace{-1 cm}  
            \includegraphics[width=5.5cm, height=5cm]{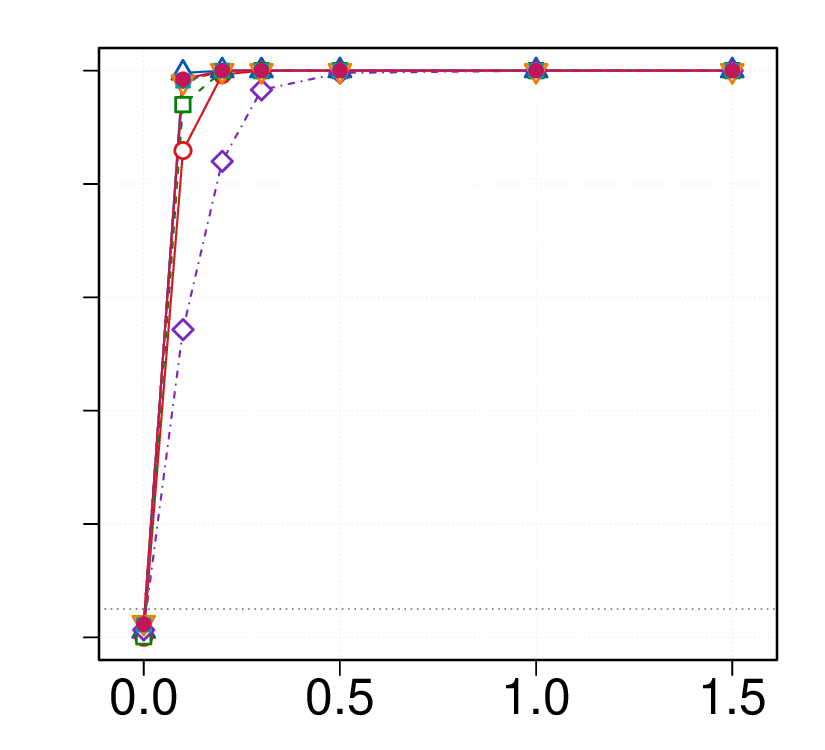}
            \hspace{-1 cm}  
            \includegraphics[width=5.5cm, height=5cm]{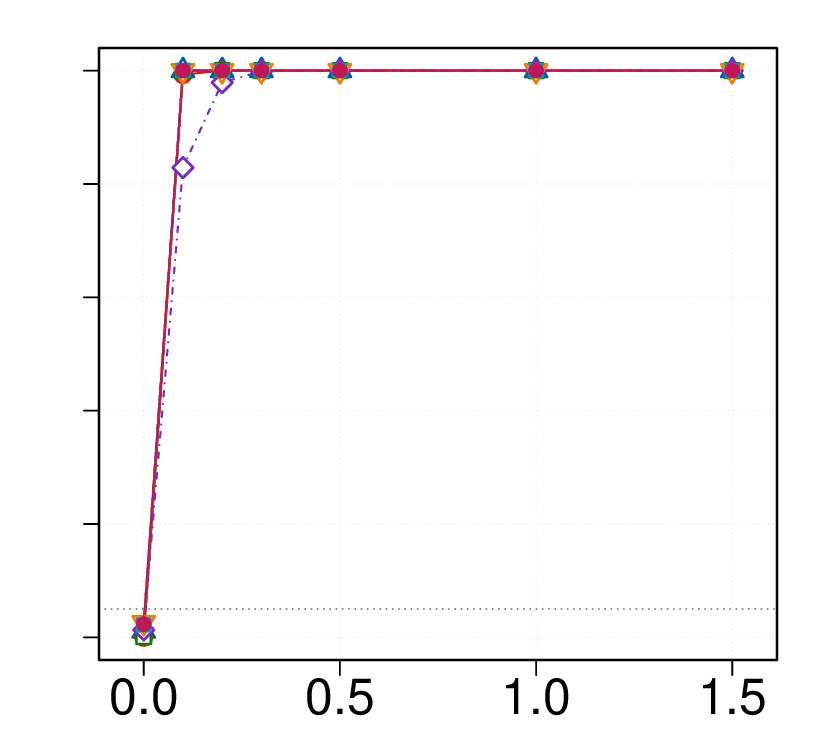}
            \hspace{-1 cm}  
            \vspace{-0.2in}
            \caption*{(a)\small{ \textit{Setting 1}, with $m=1,5,20$  } }

            \centering
            \hspace{-1 cm}  
                \includegraphics[width=5.5cm, height=5cm]{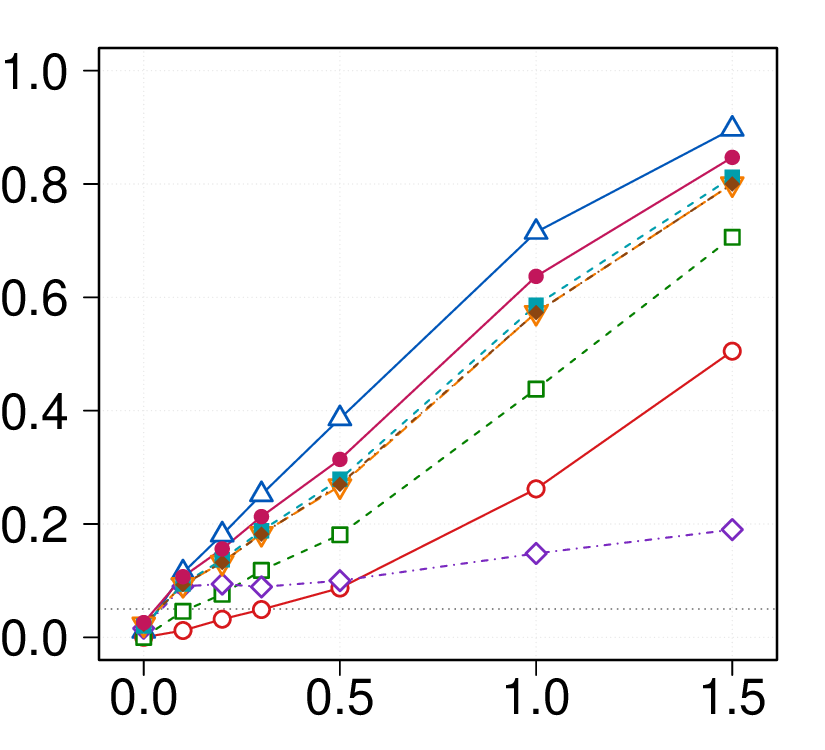}
            \hspace{-1 cm}  
            \includegraphics[width=5.5cm, height=5cm]{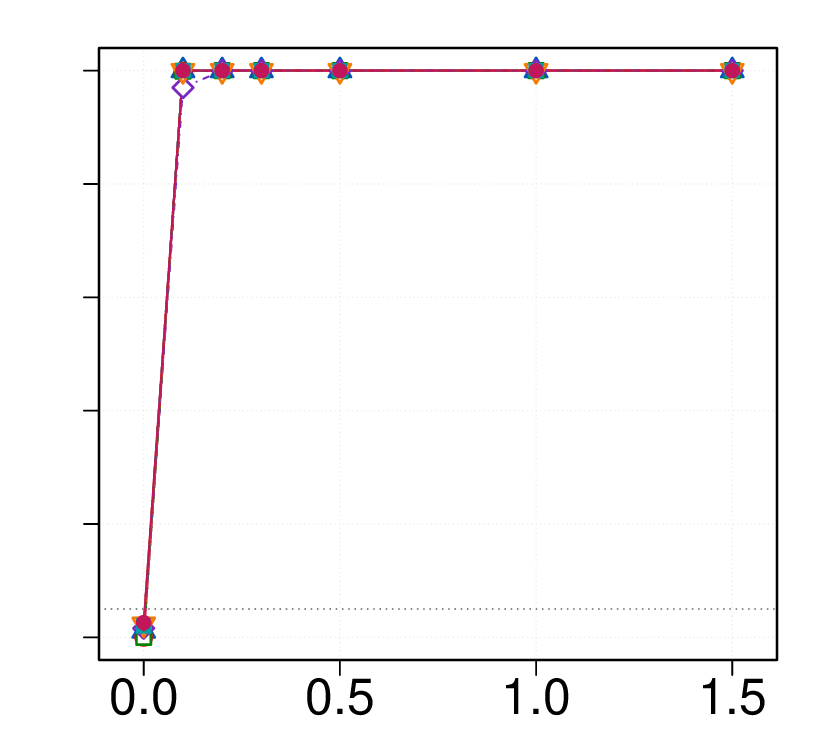}
            \hspace{-1 cm}  
            \includegraphics[width=5.5cm, height=5cm]{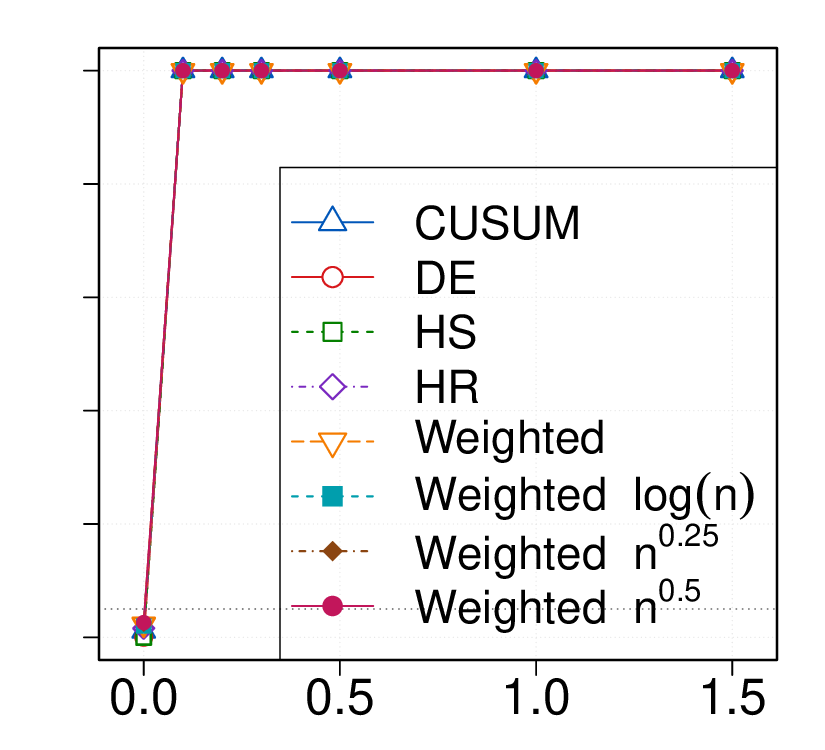}
            \hspace{-1 cm}  
            \vspace{-0.2in}
            \caption*{(b)\small{\textit{Setting 2}, with $m=1,5,20$  } }

              \centering
            \hspace{-1 cm}  
                \includegraphics[width=5.5cm, height=5cm]{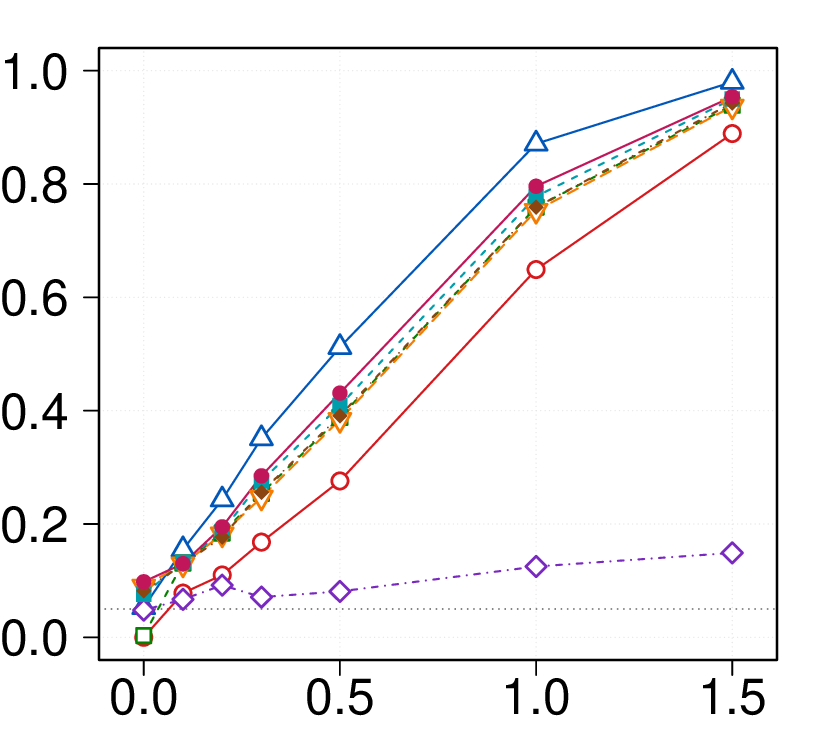}
            \hspace{-1 cm}  
            \includegraphics[width=5.5cm, height=5cm]{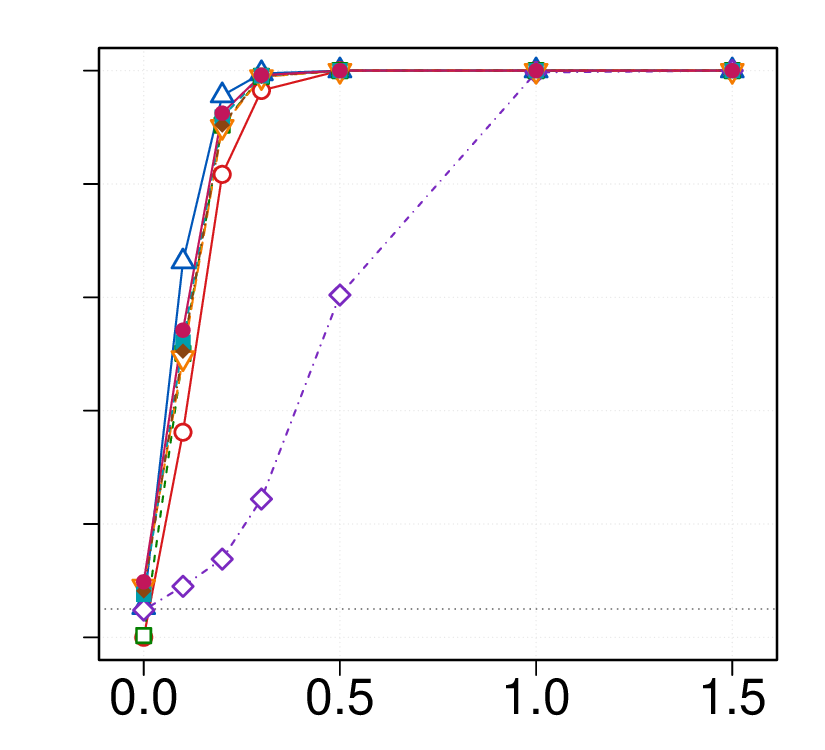}
            \hspace{-1 cm}  
            \includegraphics[width=5.5cm, height=5cm]{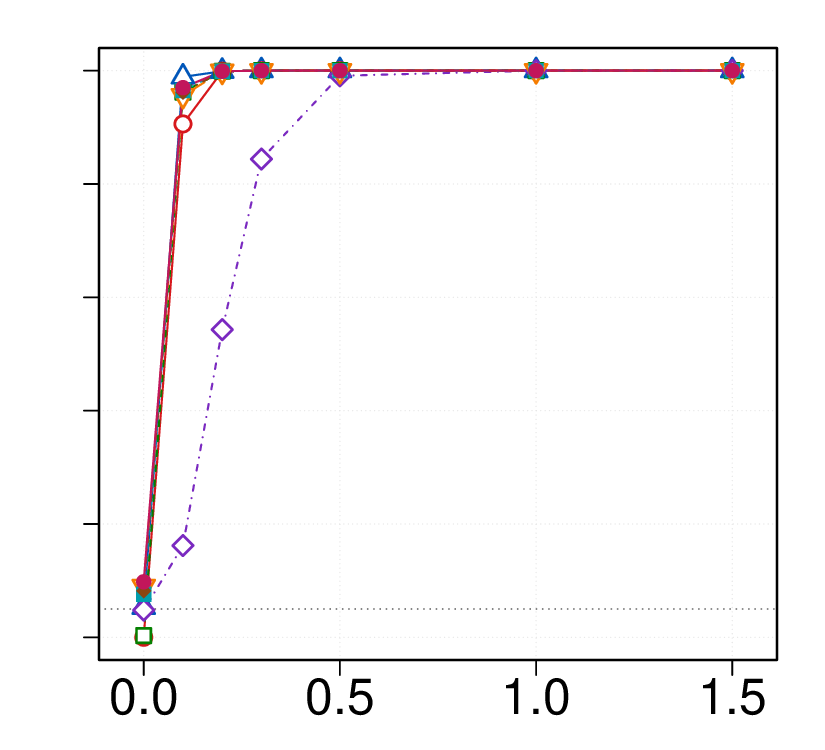}
            \hspace{-1 cm}  
            \vspace{-0.2in}
            \caption*{(c)\small{ \textit{Setting 3}, with $m=1,5,20$  } }

             \caption{\small{
             Adjusted empirical rejection rates of the RP methods for various values of $SNR$ in the x-axis. The RP method performs 200 random projections and applies different change point tests (CUSUM, Weighted, DE, HS, HR) and the Bonf combination method. The data-generating process follows (\ref{eq:model for fd}), (\ref{eq:data generating process}), and (\ref{break function}), where the standard deviation $\sigma_{g}$ follows \textit{Settings 1--3}.
            The change point location is set at $\theta=0.5$.
            The empirical rejection rate is based on 1000 simulations.
             }}
             \label{fig: tuning cp test Bonf.adj-theta=0.5}
        
    \end{figure}


\newpage
\begin{figure} [h!]
        \centering
            \hspace{-1 cm}  
            \includegraphics[width=5.5cm, height=5cm]{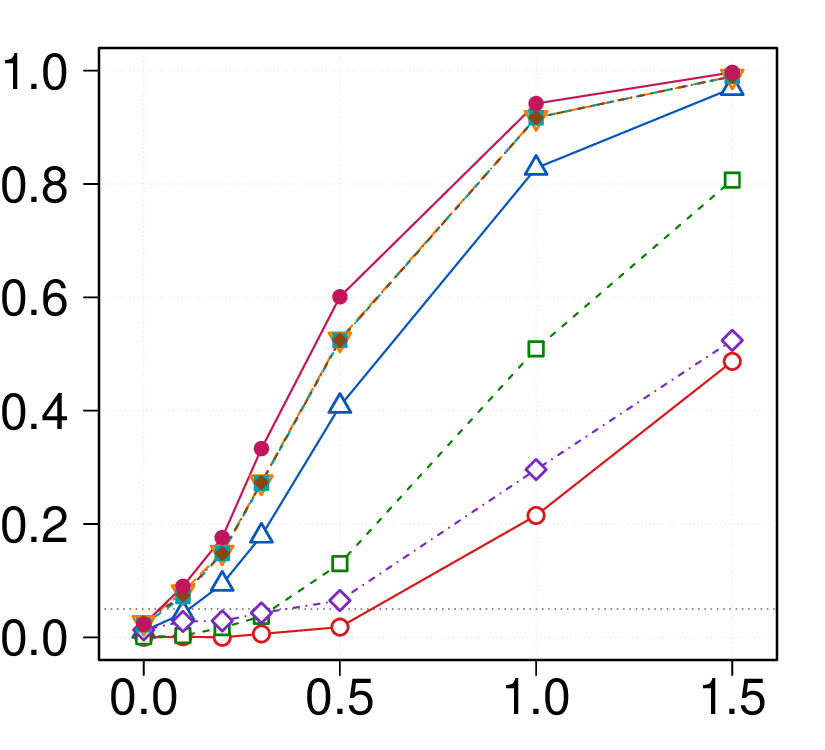}
            \hspace{-1 cm}  
            \includegraphics[width=5.5cm, height=5cm]{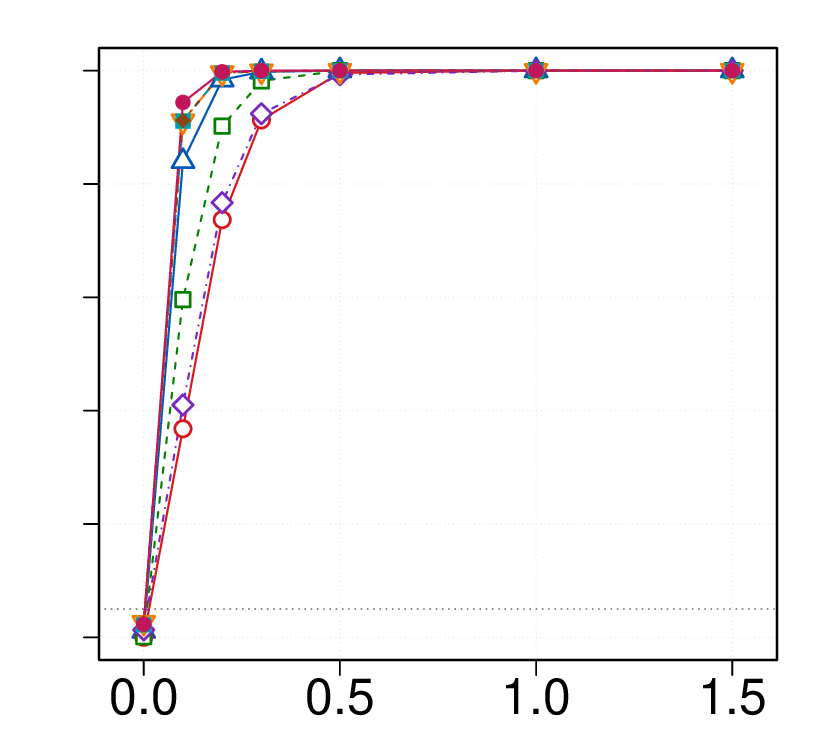}
            \hspace{-1 cm}  
            \includegraphics[width=5.5cm, height=5cm]{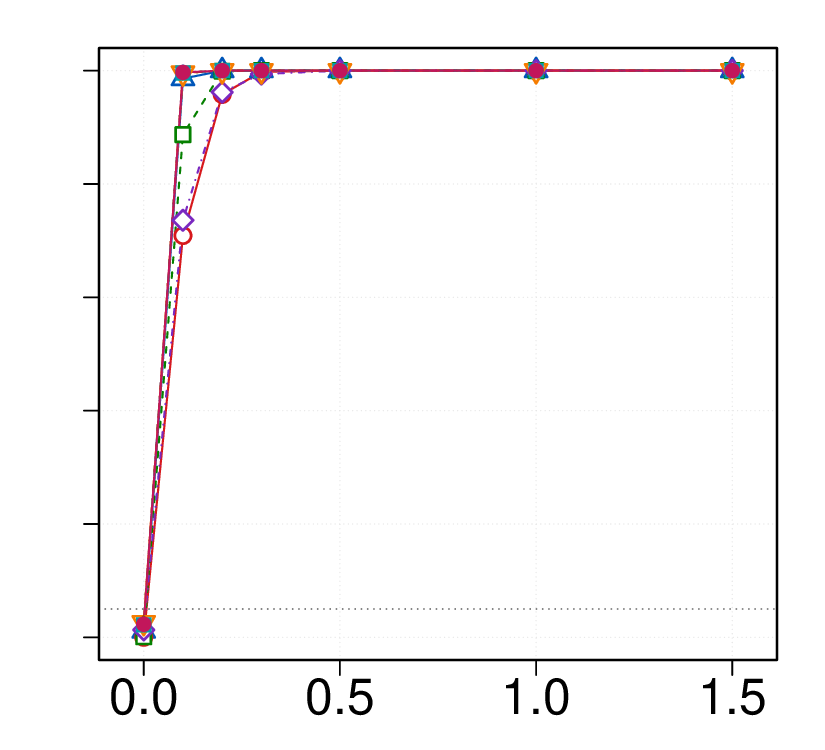}
            \hspace{-1 cm}  
            \vspace{-0.2in}
            \caption*{\small{(a) \textit{Setting 1}, with $m=1,5,20$  } }

            \centering
            \hspace{-1 cm}                 \includegraphics[width=5.5cm, height=5cm]{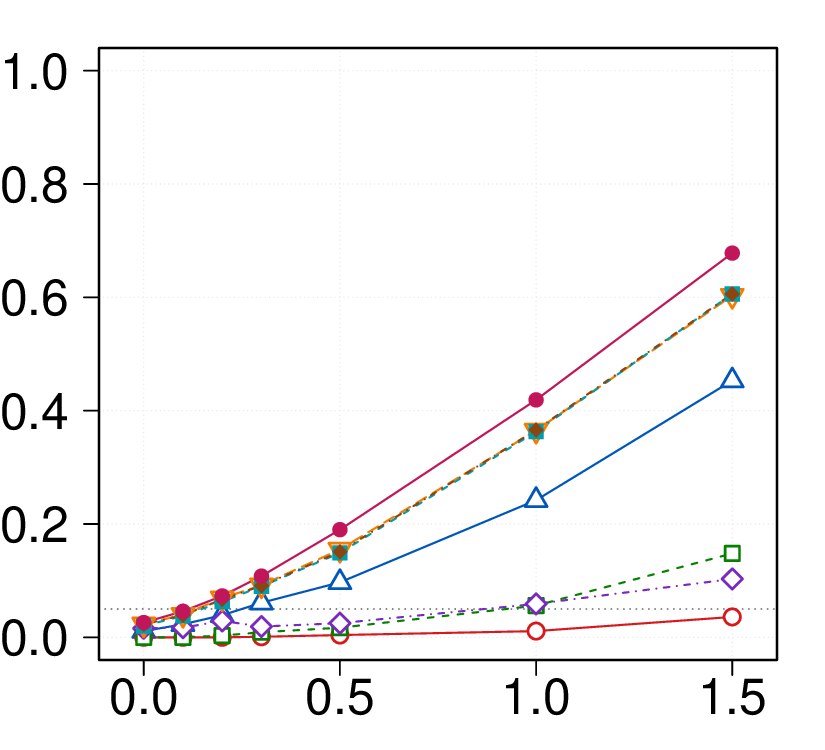}
            \hspace{-1 cm}  
            \includegraphics[width=5.5cm, height=5cm]{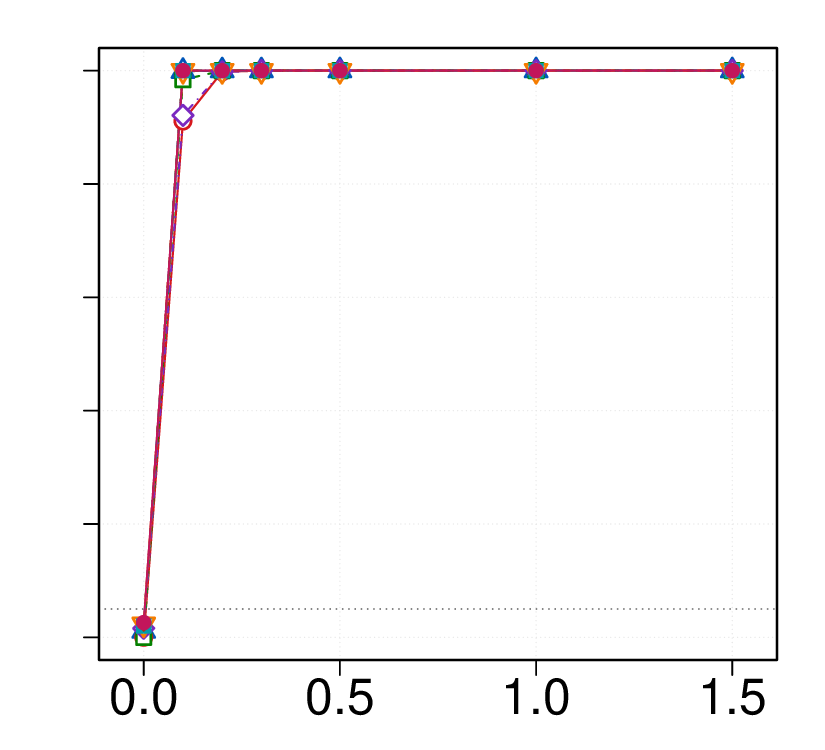}
            \hspace{-1 cm}  
            \includegraphics[width=5.5cm, height=5cm]{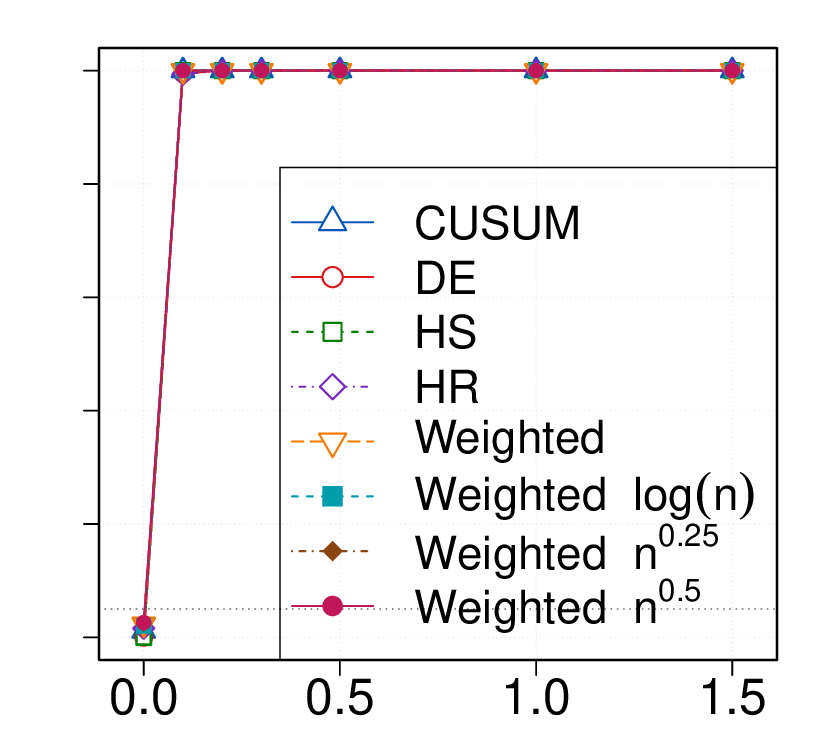}
            \hspace{-1 cm}  
            \vspace{-0.2in}
            \caption*{\small{(b) \textit{Setting 2}, with $m=1,5,20$  } }

              \centering
            \hspace{-1 cm}  \includegraphics[width=5.5cm, height=5cm]{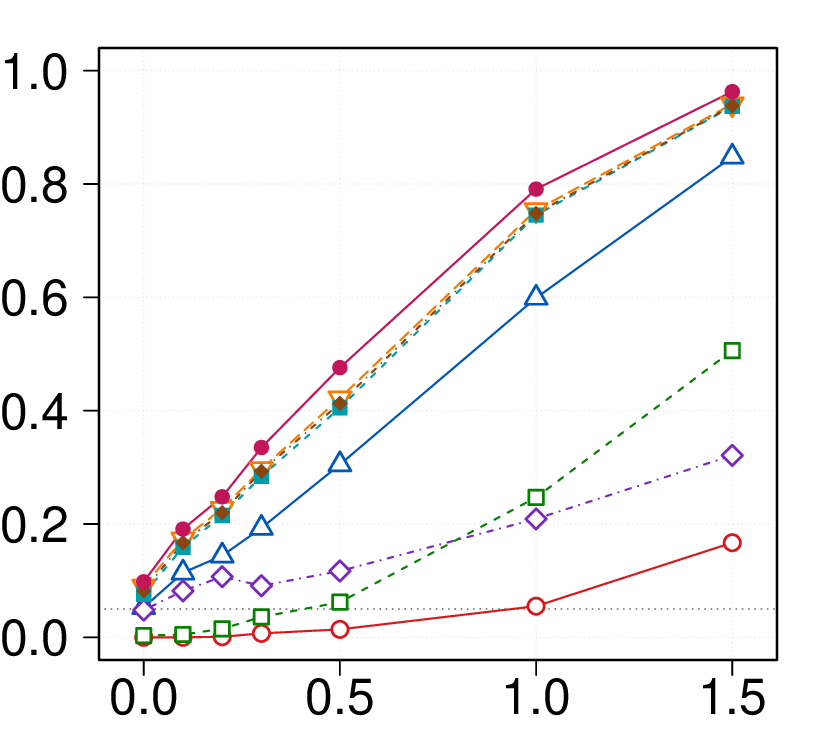}
            \hspace{-1 cm}  
            \includegraphics[width=5.5cm, height=5cm]{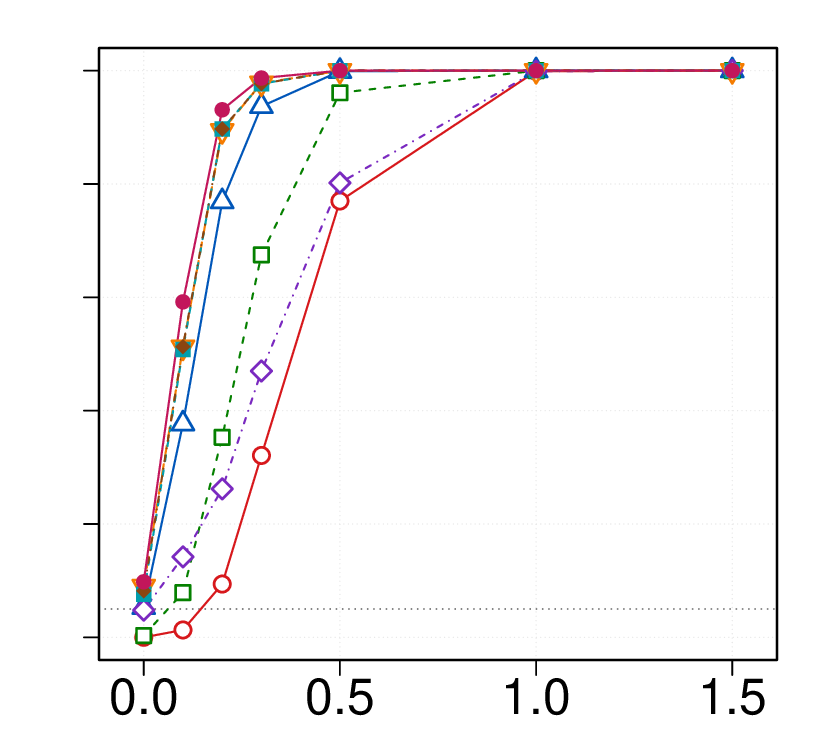}
            \hspace{-1 cm}  
            \includegraphics[width=5.5cm, height=5cm]{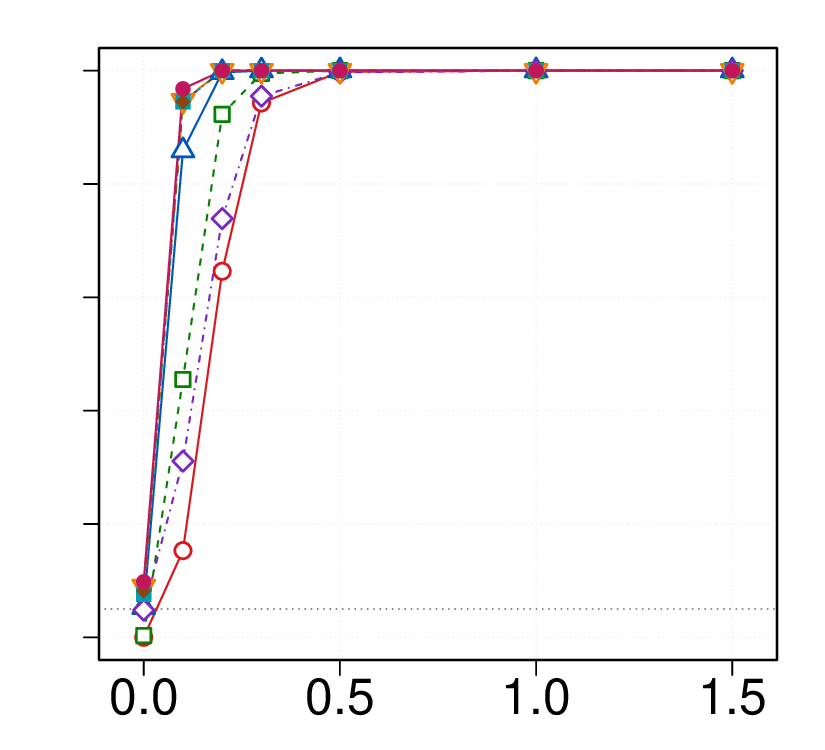}
            \hspace{-1 cm}  
            \vspace{-0.2in}
            \caption*{\small{(c) \textit{Setting 3}, with $m=1,5,20$  } }

            \caption{\small{ Raw empirical rejection rates of the RP methods for various values of $SNR$ in the x-axis. The RP method performs 200 random projections and applies different change point tests (CUSUM, Weighted, DE, HS, HR) and the Bonf combination method. The data-generating process follows (\ref{eq:model for fd}), (\ref{eq:data generating process}), and (\ref{break function}), where the standard deviation $\sigma_{g}$ follows \textit{Settings 1--3}.
            The change point location is set at $\theta=0.25$.
            The empirical rejection rate is based on 1000 simulations.
            }}
            \label{fig: tuning cp test Bonf}
    \end{figure} 

\begin{figure} [H]
        \centering
            \hspace{-1 cm}  
            \includegraphics[width=5.5cm, height=5cm]{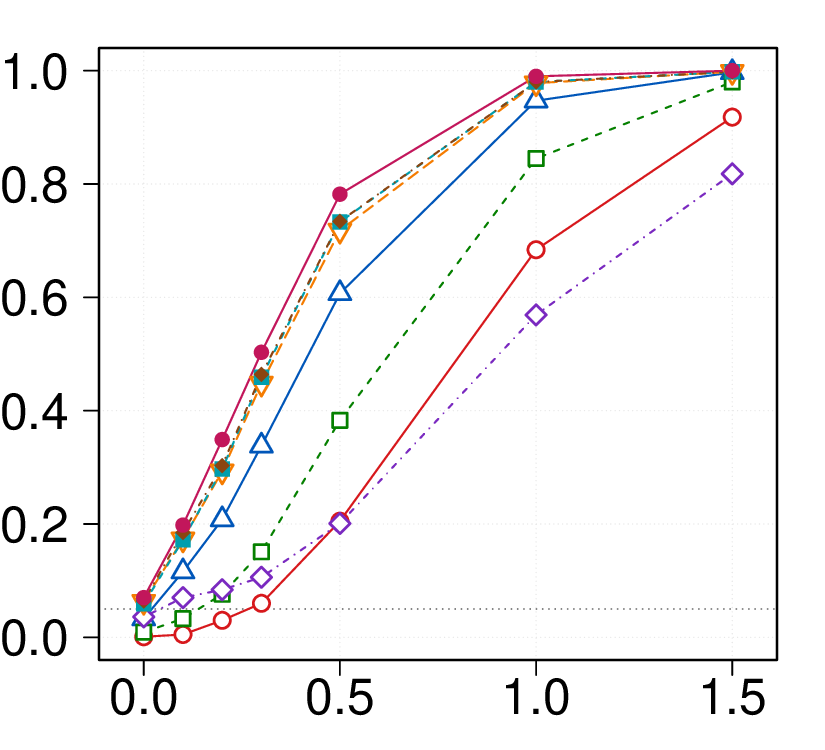}
            \hspace{-1 cm}  
            \includegraphics[width=5.5cm, height=5cm]{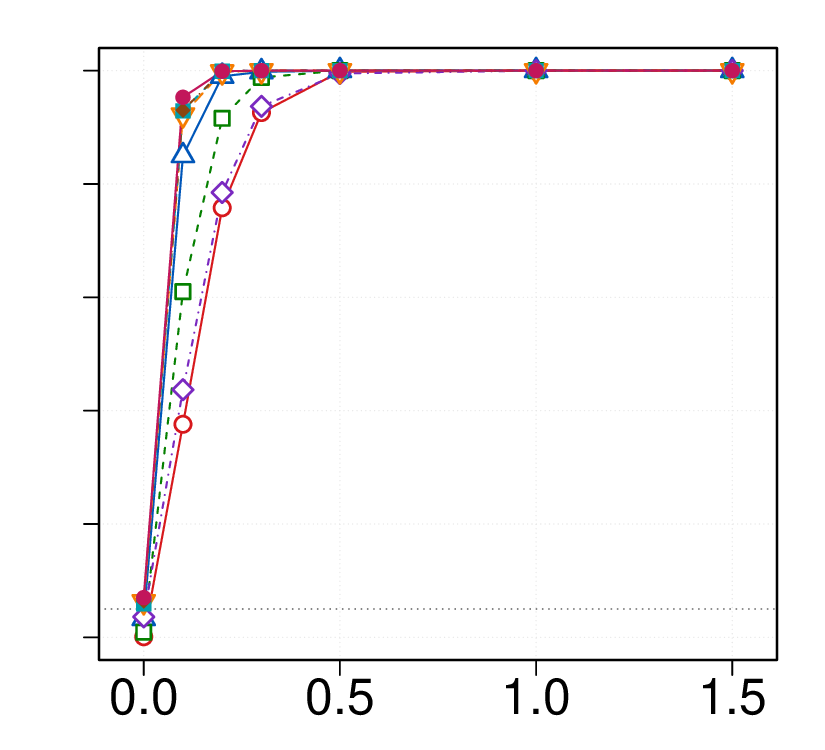}
            \hspace{-1 cm}  
            \includegraphics[width=5.5cm, height=5cm]{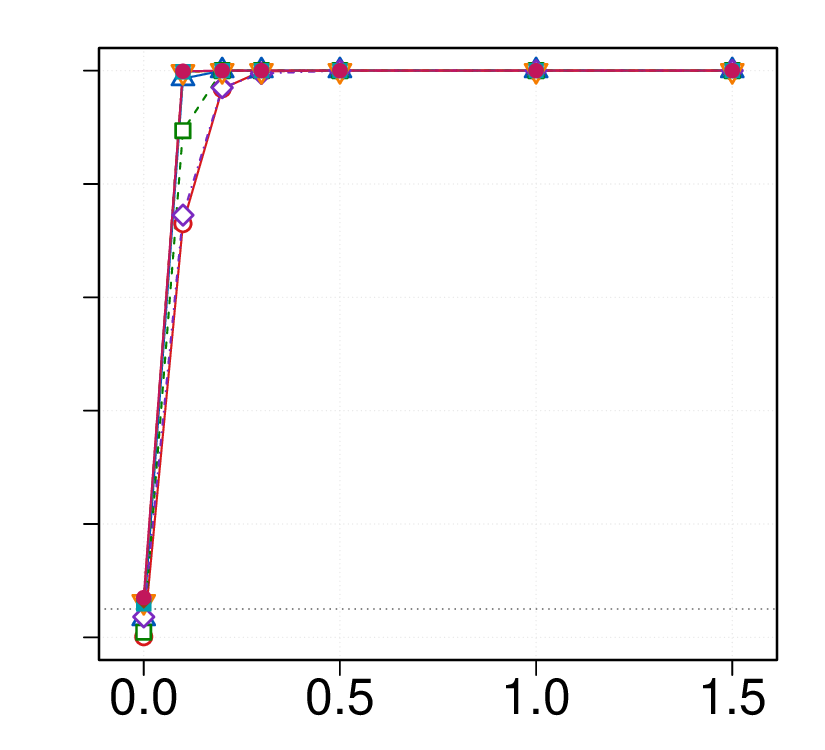}
            \hspace{-1 cm}  
            \vspace{-0.2in}
            \caption*{\small{(a) \textit{Setting 1}, with $m=1,5,20$  } }

            \centering
            \hspace{-1 cm}  
                \includegraphics[width=5.5cm, height=5cm]{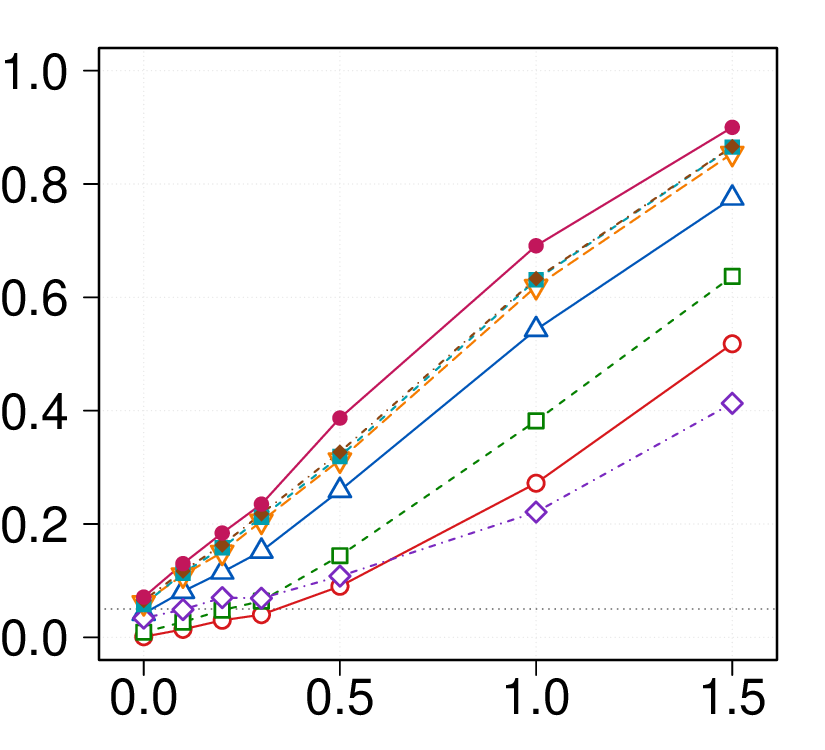}
            \hspace{-1 cm}  
            \includegraphics[width=5.5cm, height=5cm]{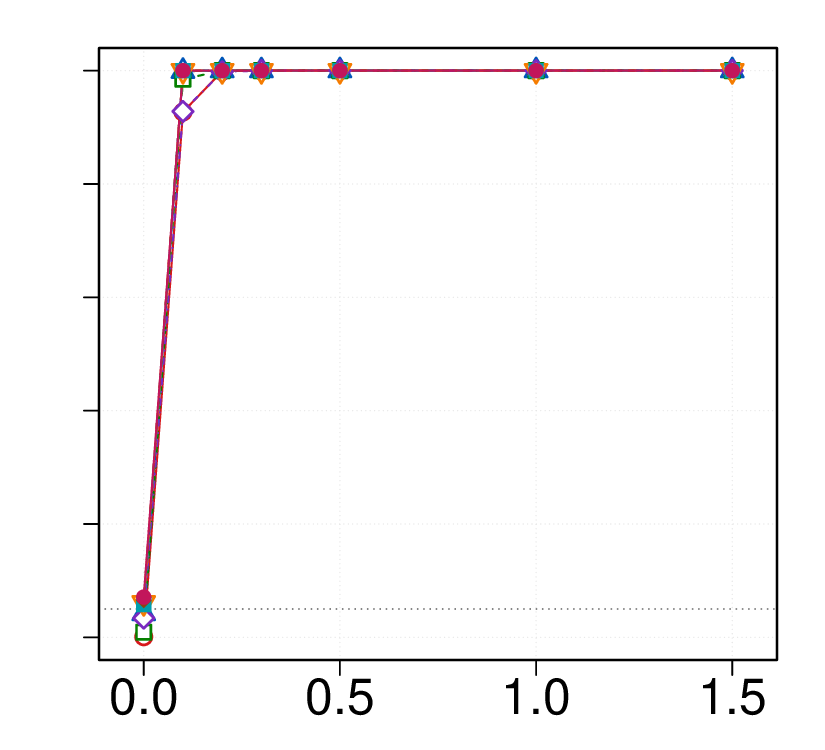}
            \hspace{-1 cm}  
            \includegraphics[width=5.5cm, height=5cm]{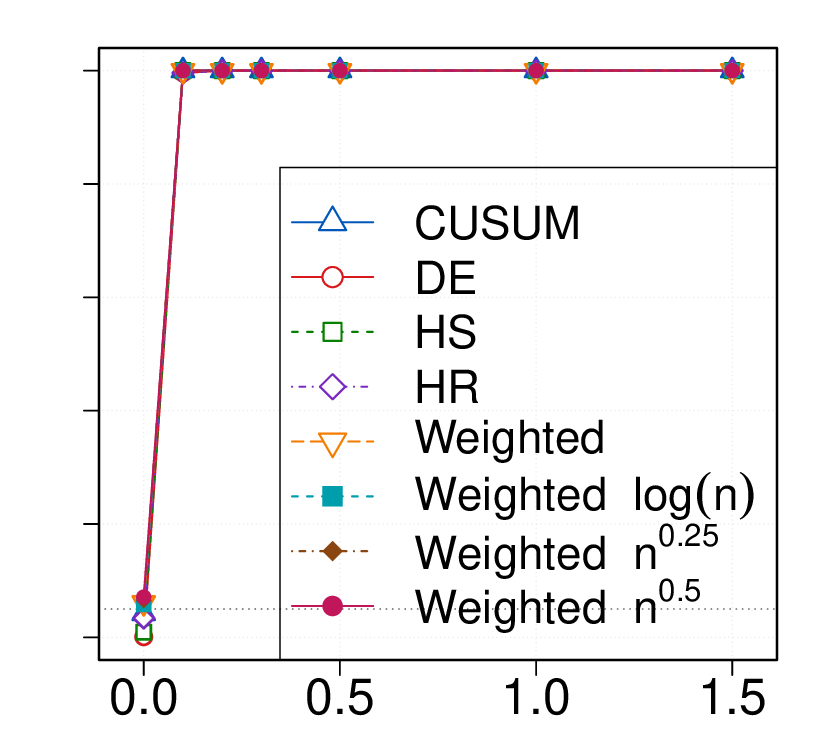}
            \hspace{-1 cm}  
            \vspace{-0.2in}
            \caption*{\small{(b) \textit{Setting 2}, with $m=1,5,20$  } }

              \centering
            \hspace{-1 cm}  
                \includegraphics[width=5.5cm, height=5cm]{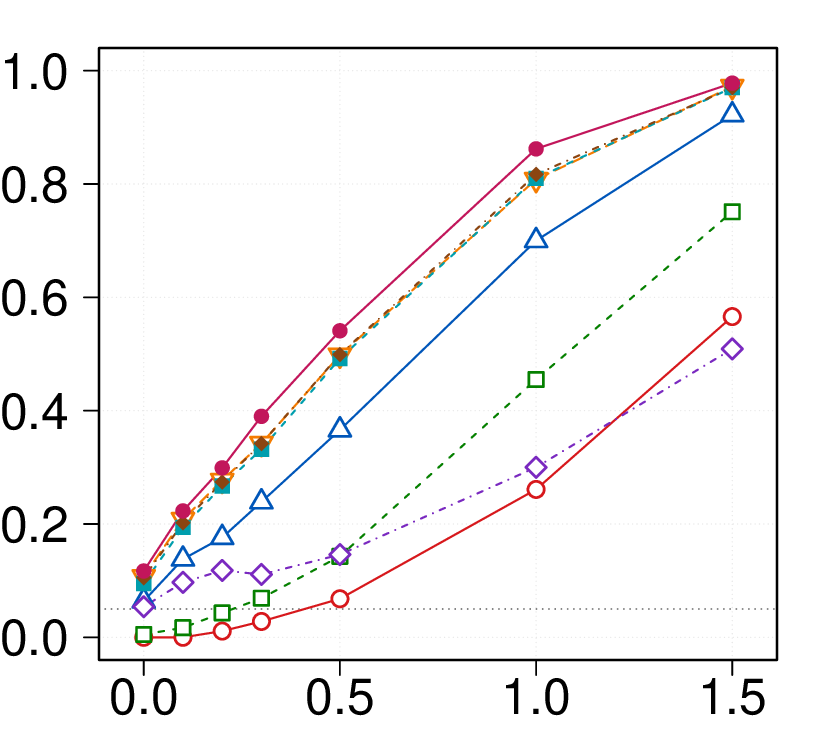}
            \hspace{-1 cm}  
            \includegraphics[width=5.5cm, height=5cm]{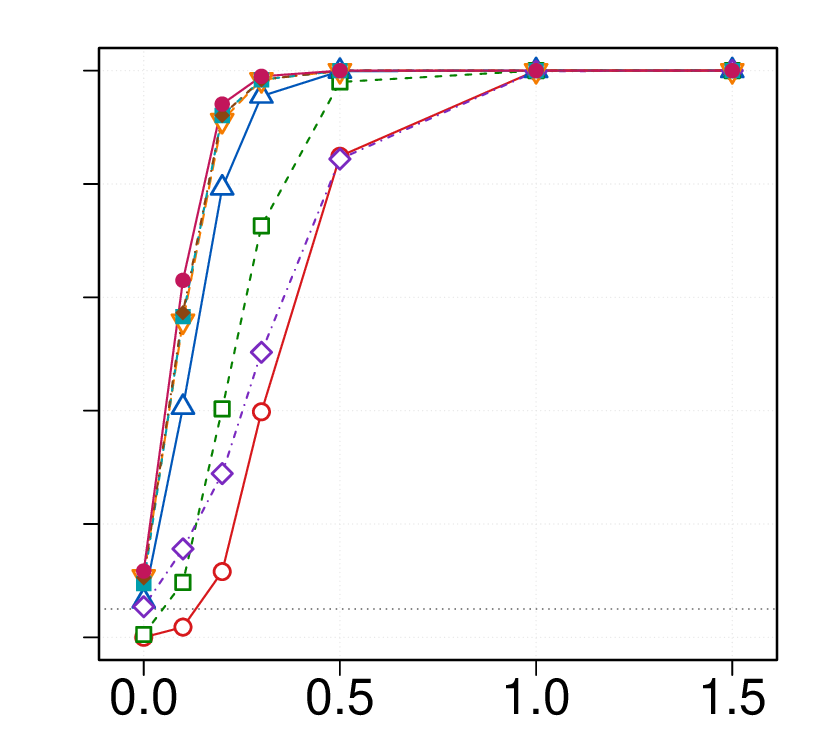}
            \hspace{-1 cm}  
            \includegraphics[width=5.5cm, height=5cm]{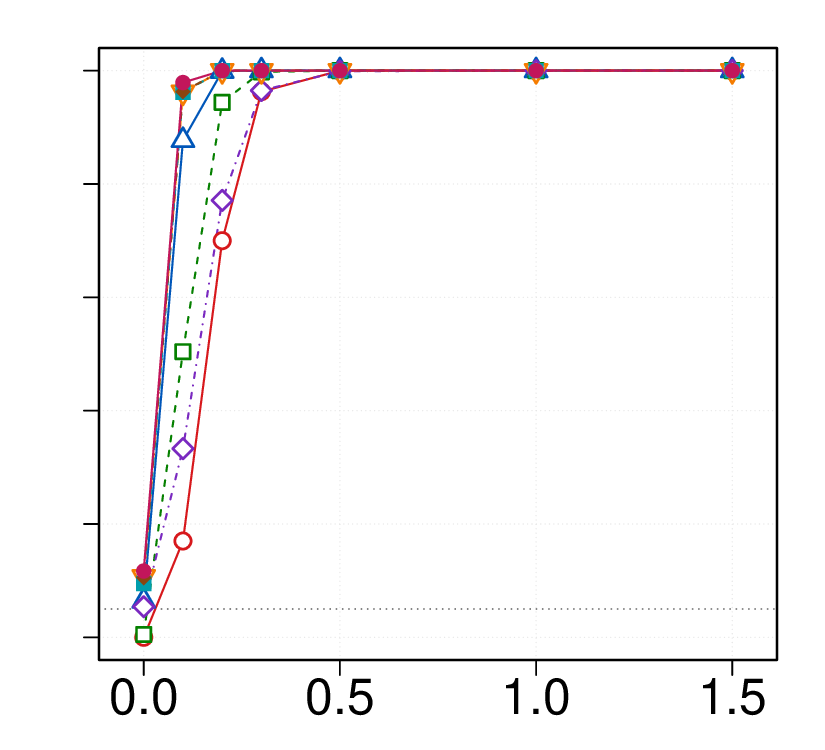}
            \hspace{-1 cm}  
            \vspace{-0.2in}
            \caption*{\small{(c) \textit{Setting 3}, with $m=1,5,20$  } }

            \caption{\small{ Raw empirical rejection rates of the RP methods for various values of $SNR$ in the x-axis. The RP method performs 200 random projections and applies different change point tests (CUSUM, Weighted, DE, HS, HR) and the BH combination method. The data-generating process follows (\ref{eq:model for fd}), (\ref{eq:data generating process}), and (\ref{break function}), where the standard deviation $\sigma_{g}$ follows \textit{Settings 1--3}.
            The change point location is set at $\theta=0.25$.
            The empirical rejection rate is based on 1000 simulations.
            }}
            \label{fig: tuning cp test BH}
    \end{figure} 

\begin{figure} [H]
        \centering
            \hspace{-1 cm}  
            \includegraphics[width=5.5cm, height=5cm]{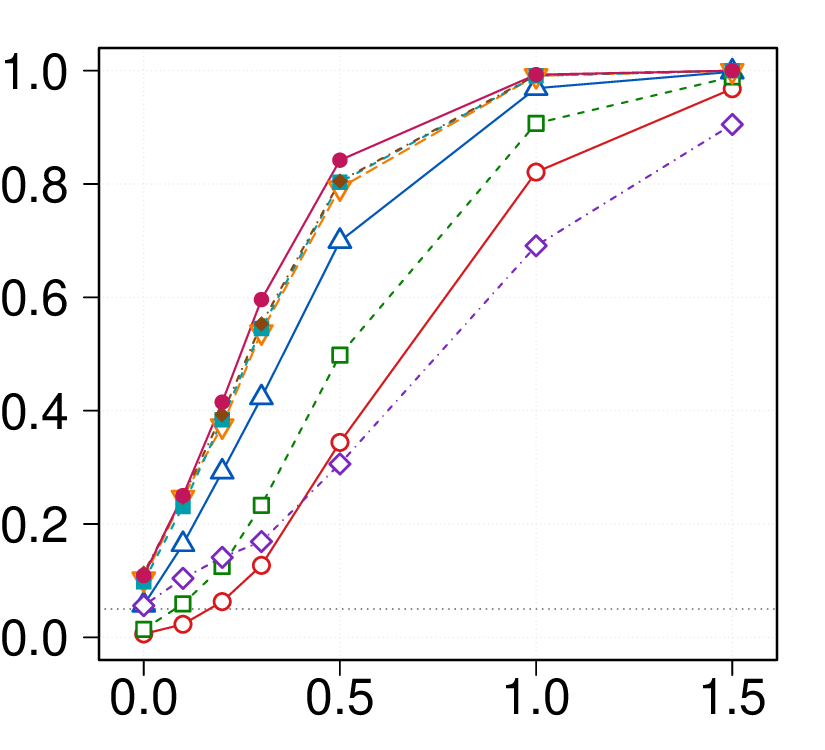}
            \hspace{-1 cm}  
            \includegraphics[width=5.5cm, height=5cm]{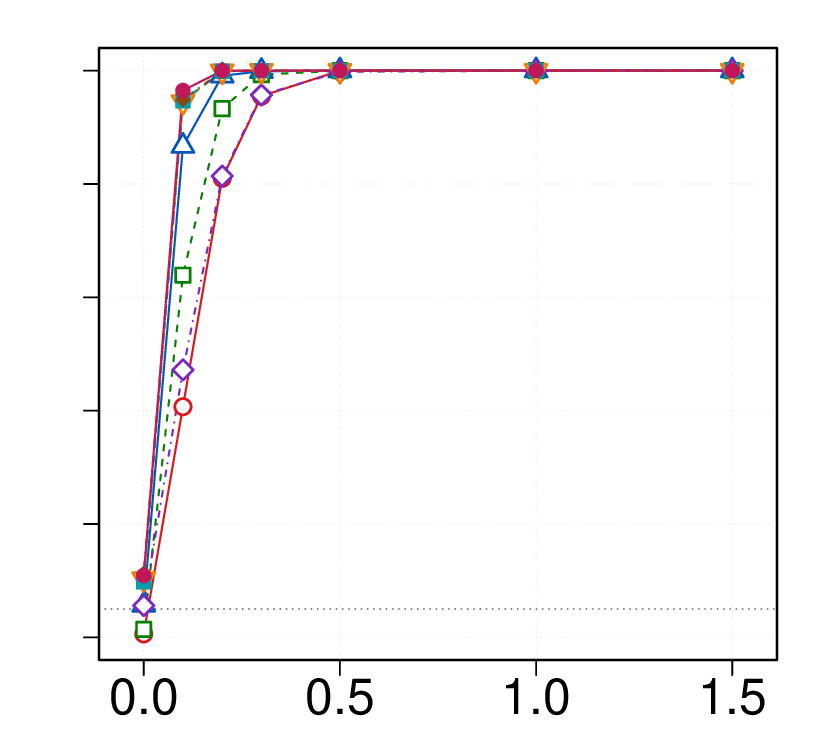}
            \hspace{-1 cm}  
            \includegraphics[width=5.5cm, height=5cm]{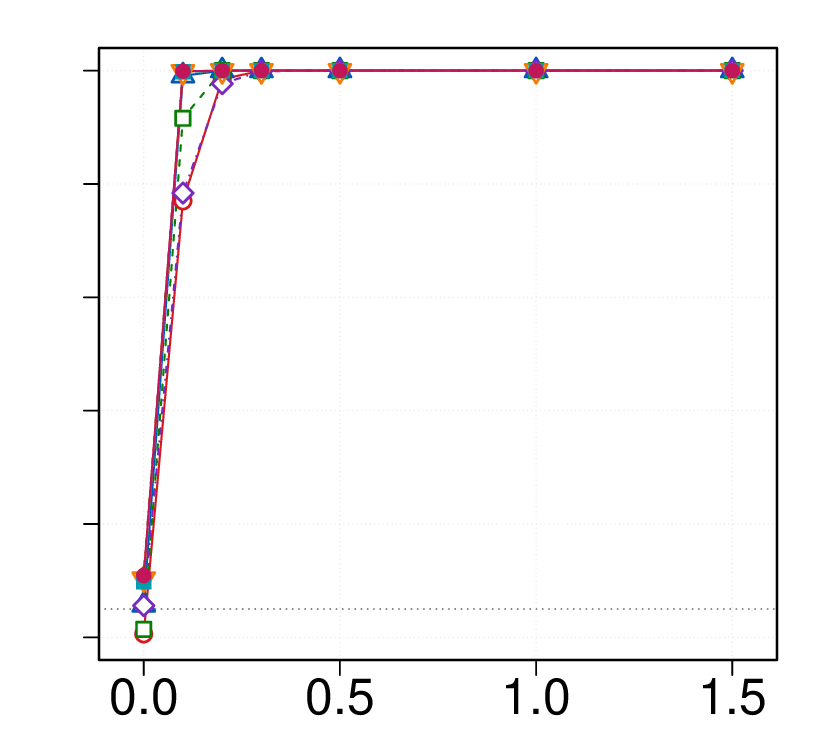}
            \hspace{-1 cm}  
            \vspace{-0.2in}
            \caption*{\small{(a) \textit{Setting 1}, with $m=1,5,20$  } }

            \centering
            \hspace{-1 cm}  
                \includegraphics[width=5.5cm, height=5cm]{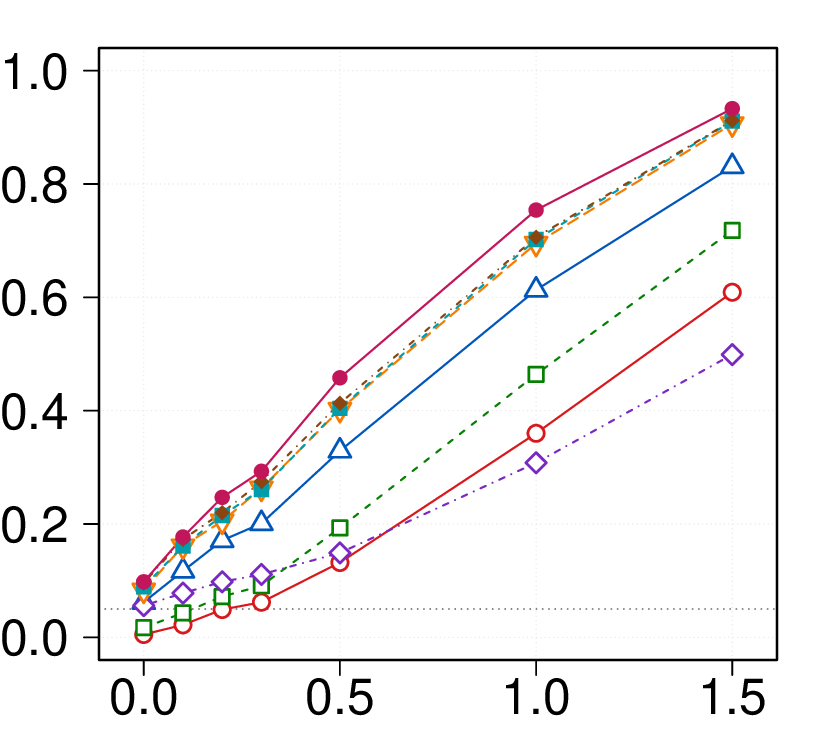}
            \hspace{-1 cm}  
            \includegraphics[width=5.5cm, height=5cm]{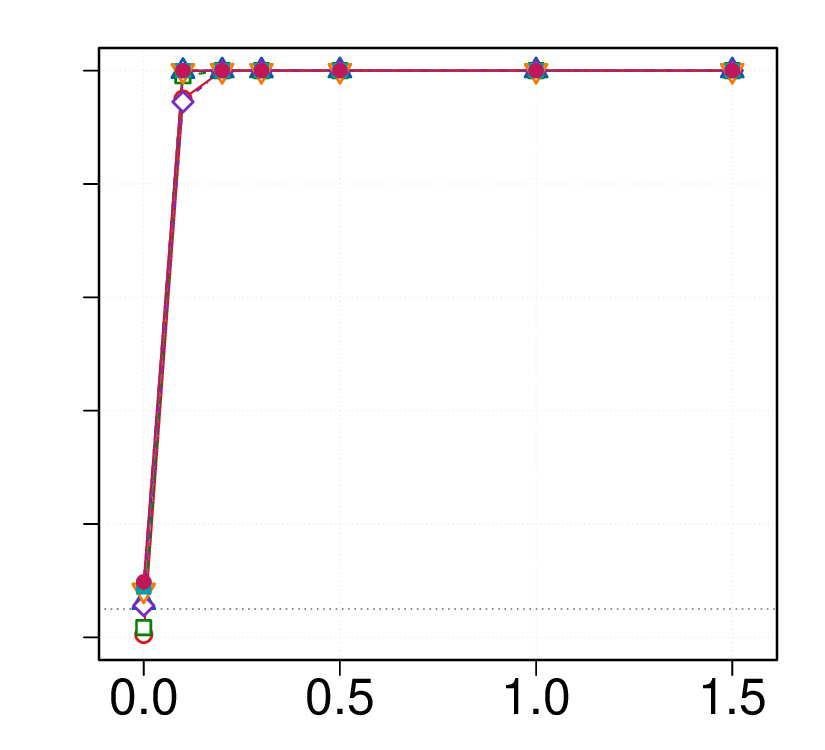}
            \hspace{-1 cm}  
            \includegraphics[width=5.5cm, height=5cm]{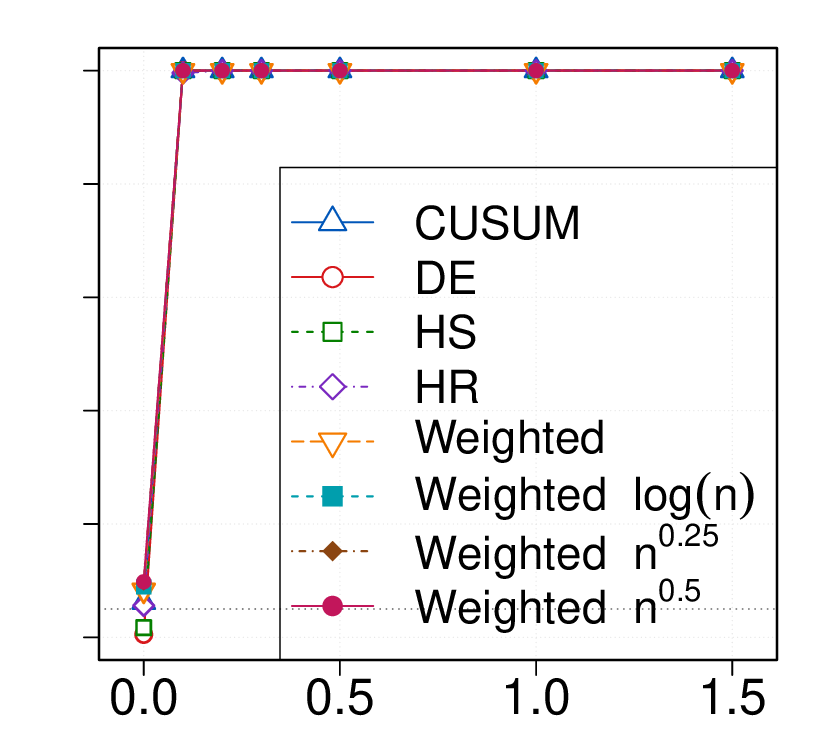}
            \hspace{-1 cm}  
            \vspace{-0.2in}
            \caption*{\small{(b) \textit{Setting 2}, with $m=1,5,20$  } }

              \centering
            \hspace{-1 cm}  
                \includegraphics[width=5.5cm, height=5cm]{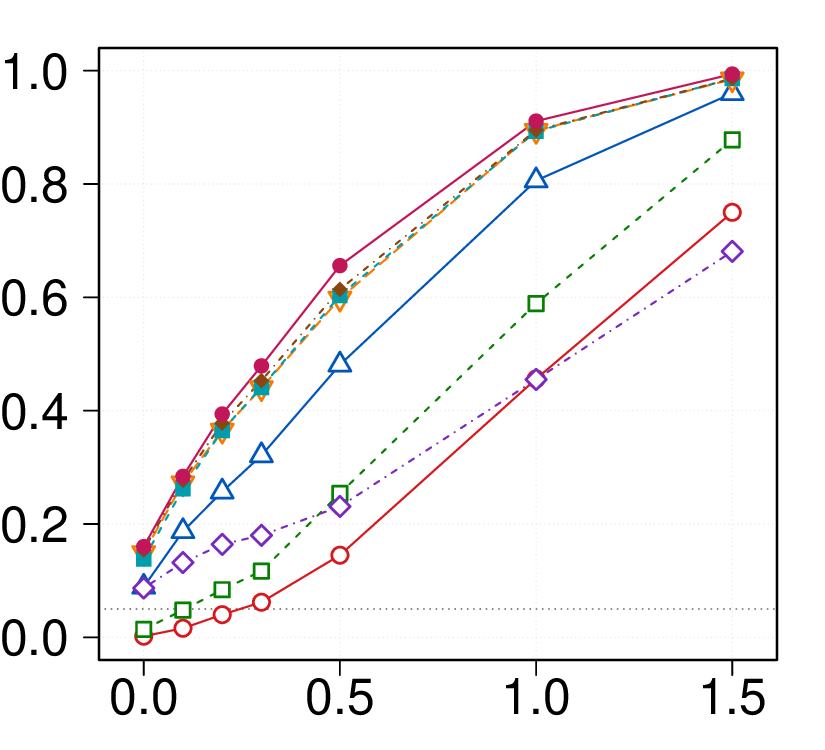}
            \hspace{-1 cm}  
            \includegraphics[width=5.5cm, height=5cm]{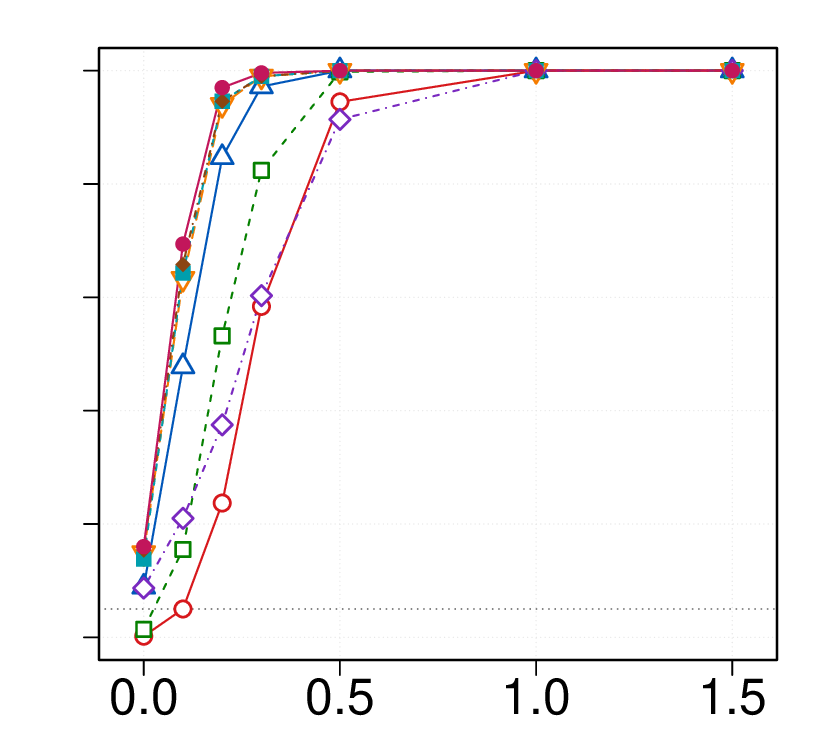}
            \hspace{-1 cm}  
            \includegraphics[width=5.5cm, height=5cm]{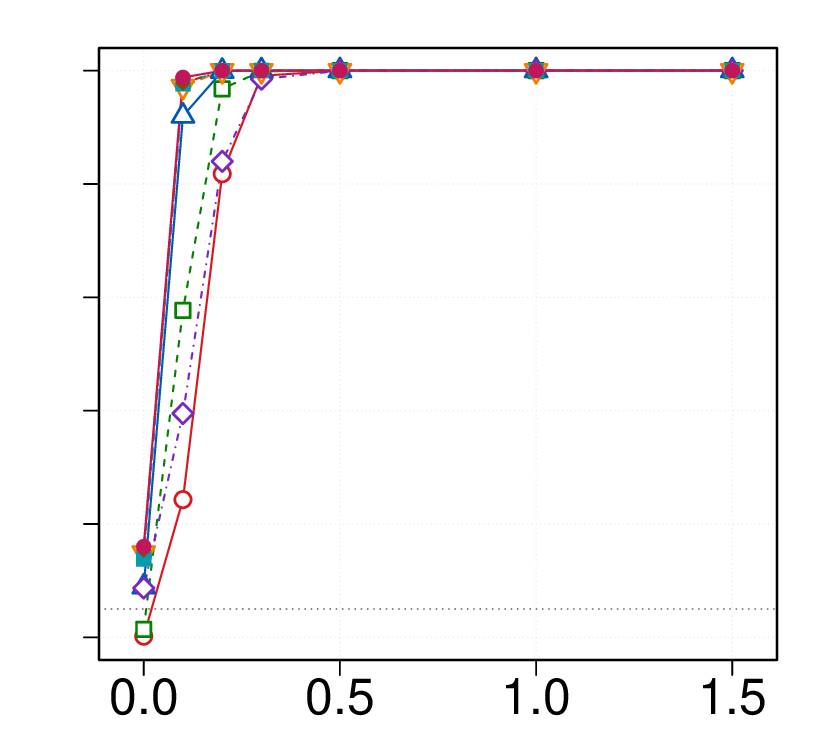}
            \hspace{-1 cm}  
            \vspace{-0.2in}
            \caption*{\small{(c) \textit{Setting 3}, with $m=1,5,20$  } }

            \caption{\small{ Raw empirical rejection rates of the RP methods for various values of $SNR$ in the x-axis. The RP method performs 200 random projections and applies different change point tests (CUSUM, Weighted, DE, HS, HR) and the HMP combination method. The data-generating process follows (\ref{eq:model for fd}), (\ref{eq:data generating process}), and (\ref{break function}), where the standard deviation $\sigma_{g}$ follows \textit{Settings 1--3}.
            The change point location is set at $\theta=0.25$.
            The empirical rejection rate is based on 1000 simulations.
            }}
            \label{fig: tuning cp test HMP}
    \end{figure} 

\begin{figure} [H]
        \centering
            \hspace{-1 cm}  
            \includegraphics[width=5.5cm, height=5cm]{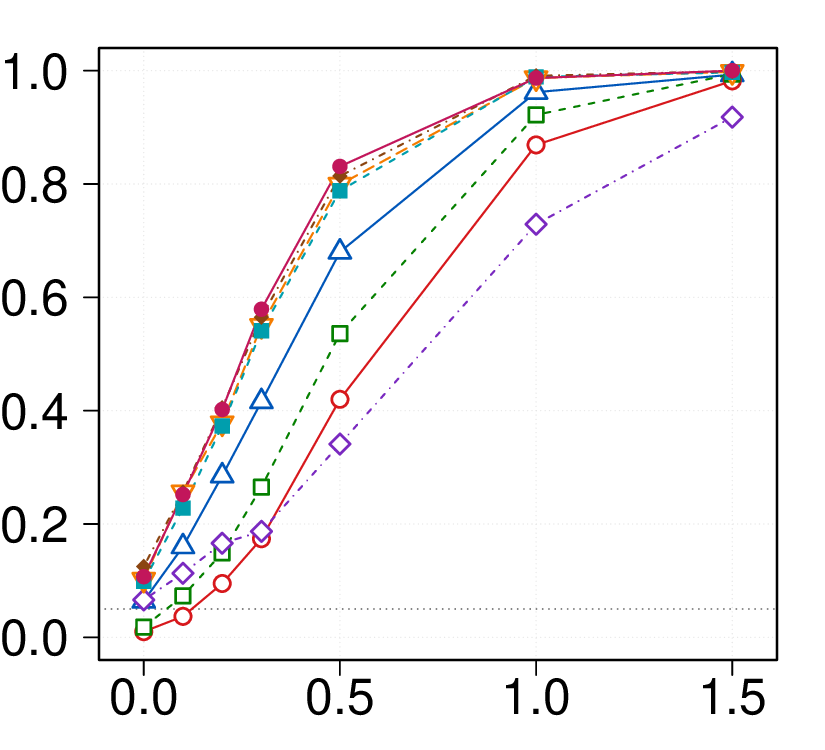}
            \hspace{-1 cm}  
            \includegraphics[width=5.5cm, height=5cm]{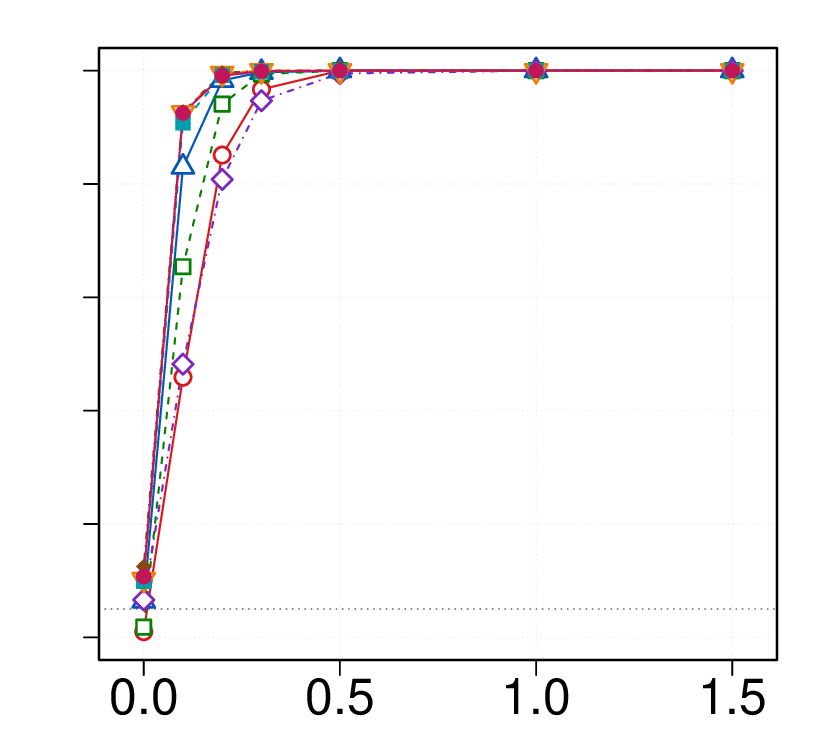}
            \hspace{-1 cm}  
            \includegraphics[width=5.5cm, height=5cm]{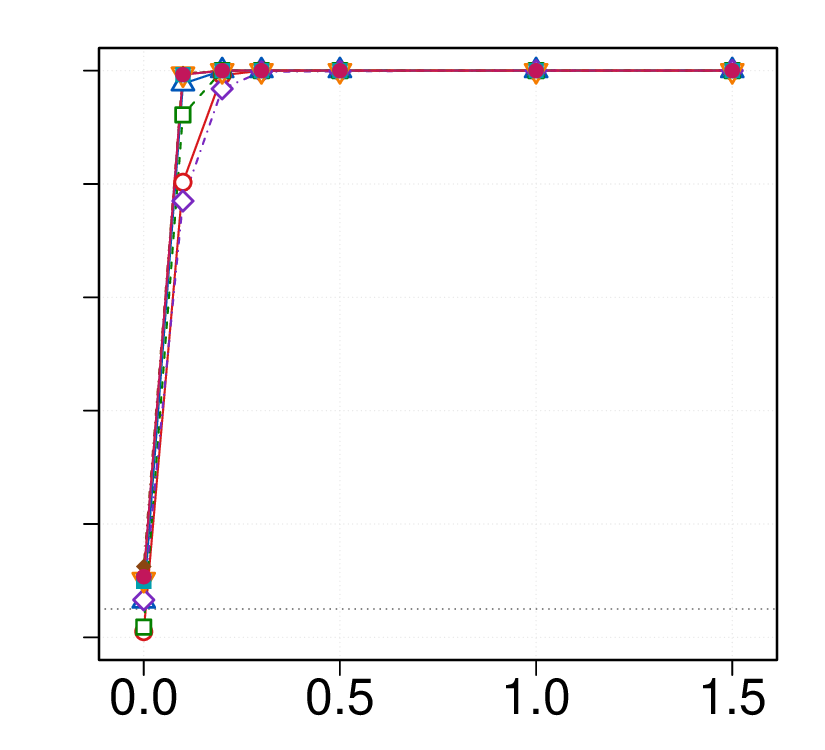}
            \hspace{-1 cm}  
            \vspace{-0.2in}
            \caption*{\small{(a) \textit{Setting 1}, with $m=1,5,20$  } }

            \centering
            \hspace{-1 cm}  
                \includegraphics[width=5.5cm, height=5cm]{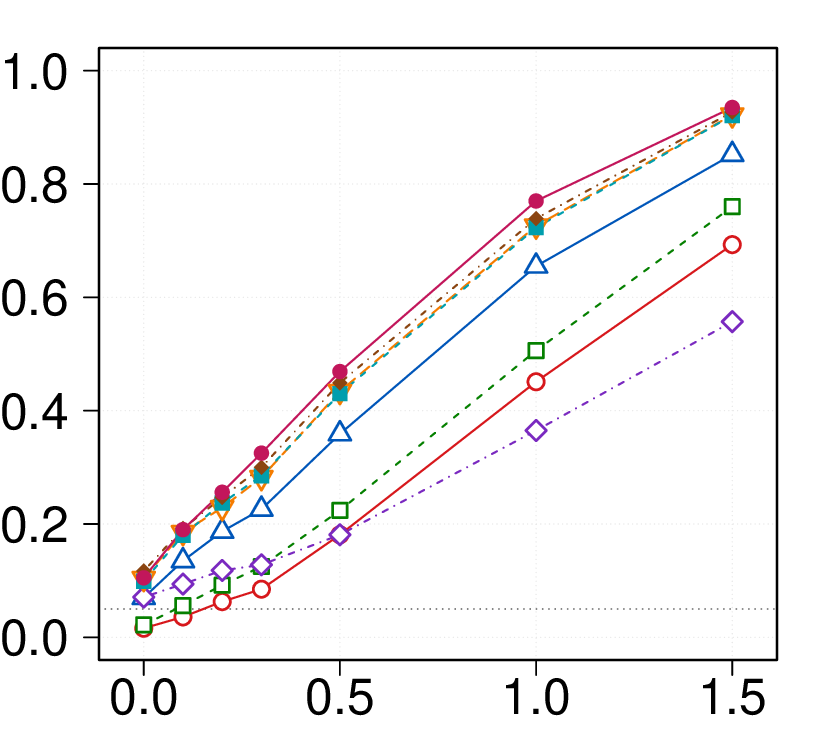}
            \hspace{-1 cm}  
            \includegraphics[width=5.5cm, height=5cm]{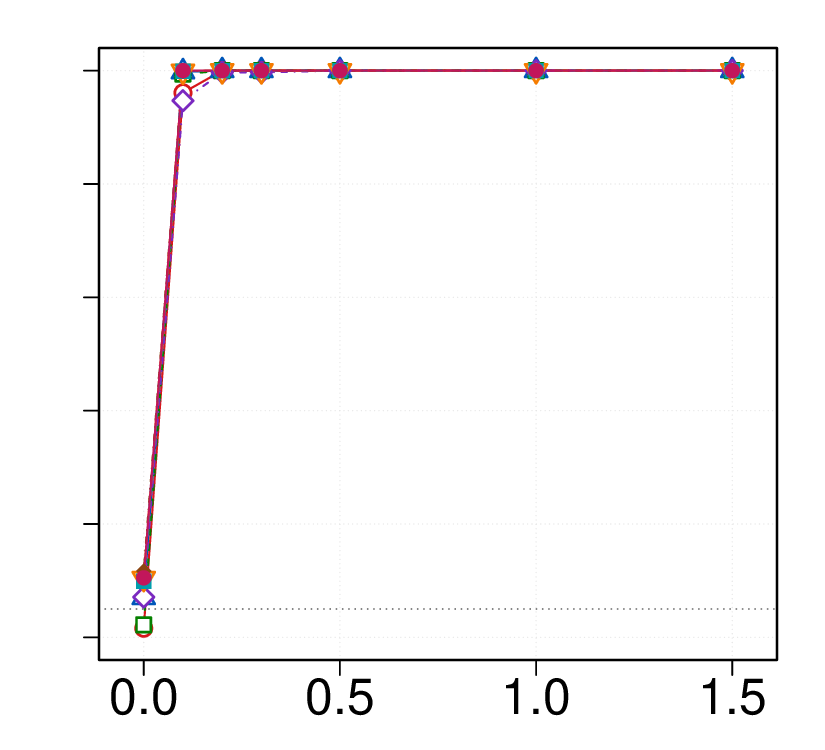}
            \hspace{-1 cm}  
            \includegraphics[width=5.5cm, height=5cm]{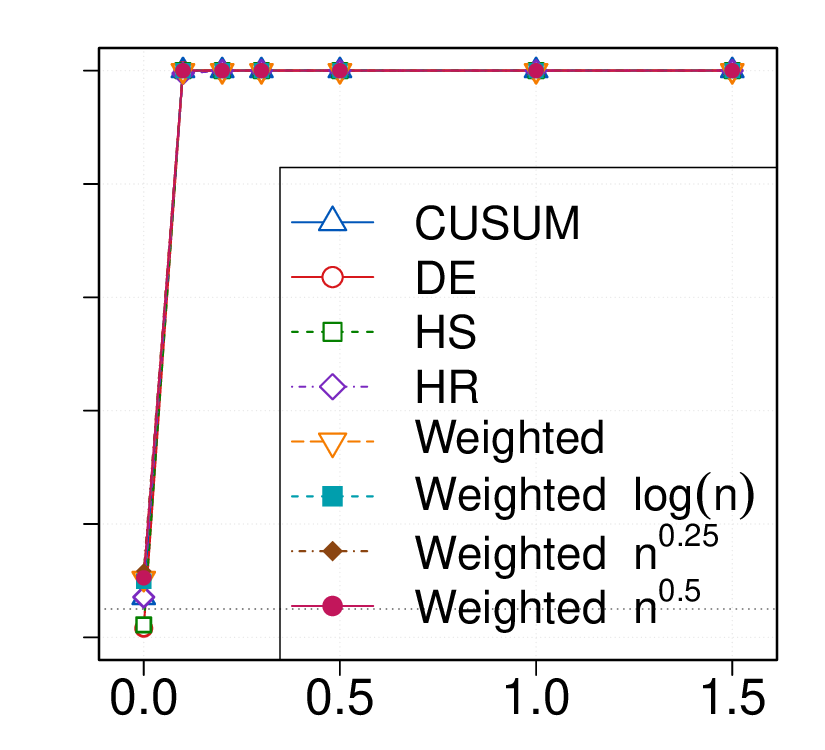}
            \hspace{-1 cm}  
            \vspace{-0.2in}
            \caption*{\small{(b) \textit{Setting 2}, with $m=1,5,20$  } }

              \centering
            \hspace{-1 cm}  
                \includegraphics[width=5.5cm, height=5cm]{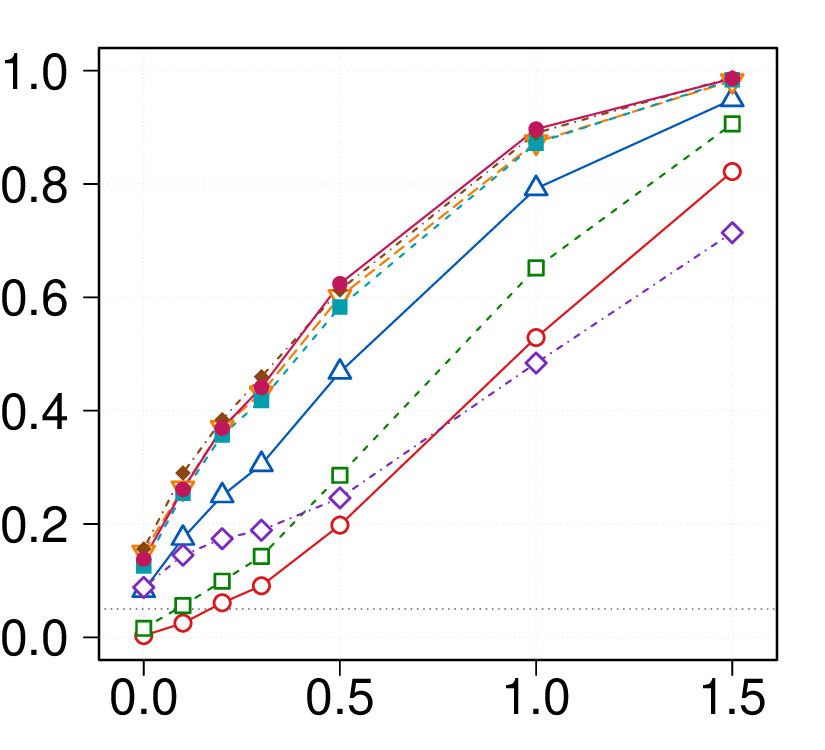}
            \hspace{-1 cm}  
            \includegraphics[width=5.5cm, height=5cm]{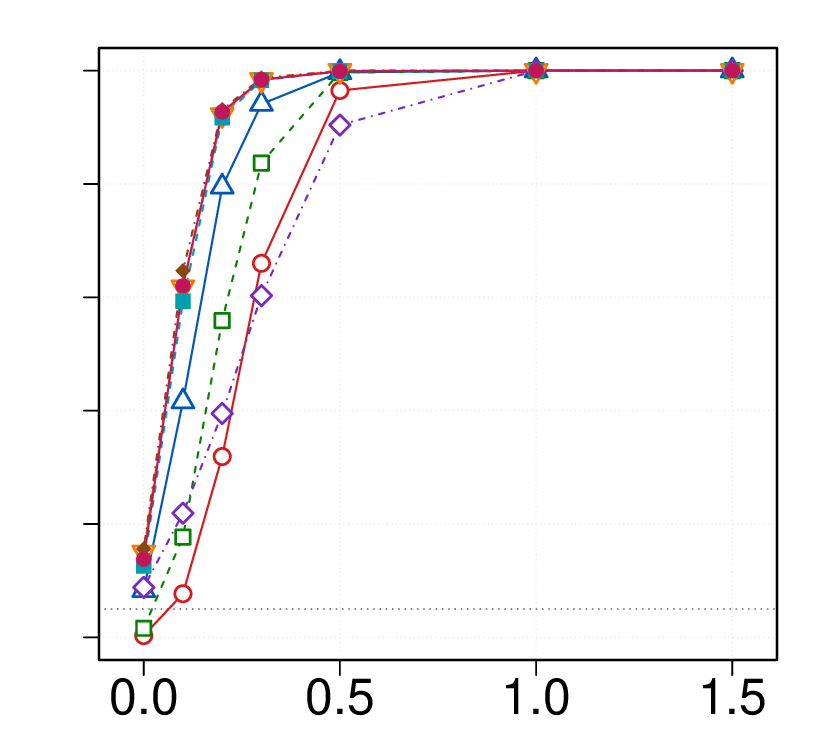}
            \hspace{-1 cm}  
            \includegraphics[width=5.5cm, height=5cm]{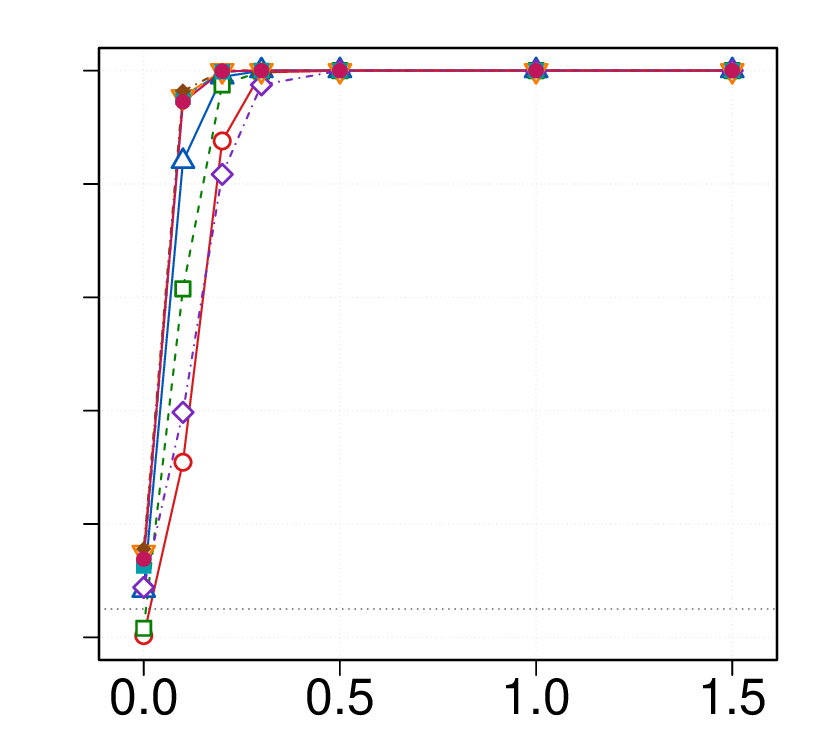}
            \hspace{-1 cm}  
            \vspace{-0.2in}
            \caption*{\small{(c) \textit{Setting 3}, with $m=1,5,20$  } }

            \caption{\small{ Raw empirical rejection rates of the RP methods for various values of $SNR$ in the x-axis. The RP method performs 200 random projections and applies different change point tests (CUSUM, Weighted, DE, HS, HR) and the CCT combination method. The data-generating process follows (\ref{eq:model for fd}), (\ref{eq:data generating process}), and (\ref{break function}), where the standard deviation $\sigma_{g}$ follows \textit{Settings 1--3}.
            The change point location is set at $\theta=0.25$.
            The empirical rejection rate is based on 1000 simulations.
            }}
            \label{fig: tuning cp test CCT}
    \end{figure} 

\newpage
\begin{figure} [H]
         \centering
            \hspace{-1 cm}  
            \includegraphics[width=5.5cm, height=5cm]{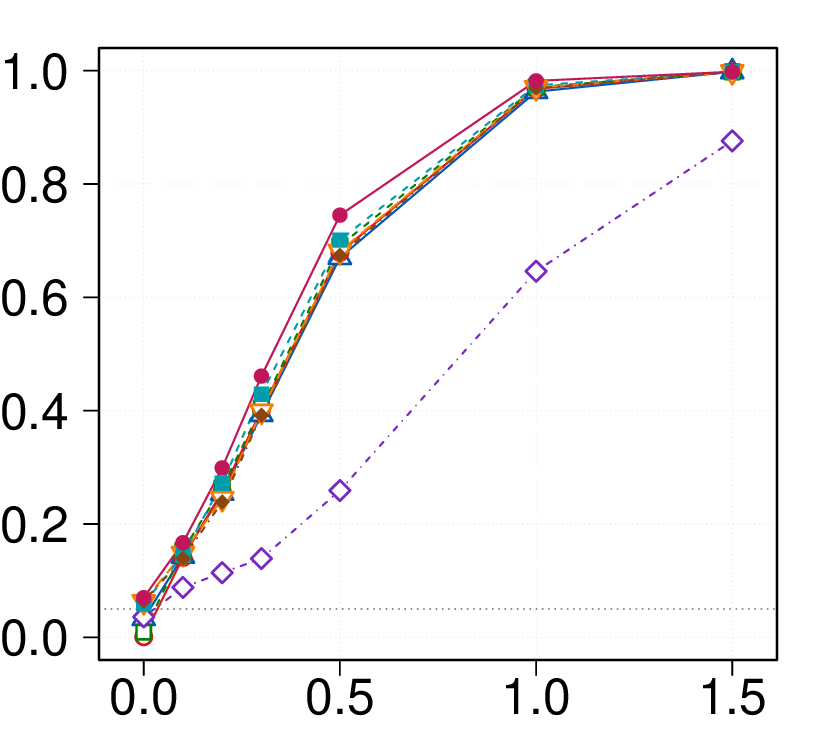}
            \hspace{-1 cm}  
            \includegraphics[width=5.5cm, height=5cm]{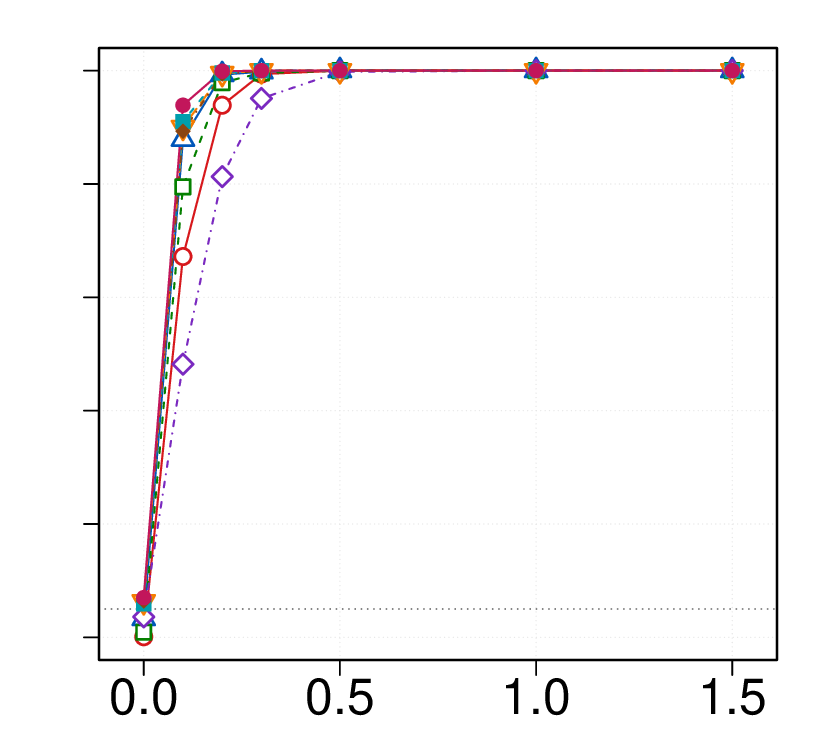}
            \hspace{-1 cm}  
            \includegraphics[width=5.5cm, height=5cm]{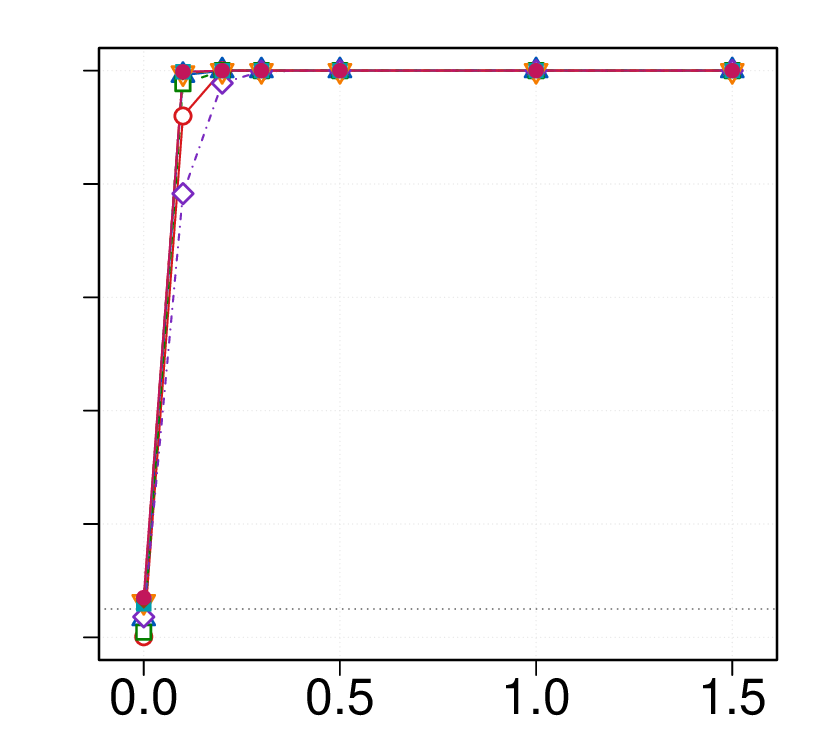}
            \hspace{-1 cm}  
            \vspace{-0.2in}
            \caption*{\small{(a) \textit{Setting 1}, with $m=1,5,20$  } }

            \centering
            \hspace{-1 cm}  
                \includegraphics[width=5.5cm, height=5cm]{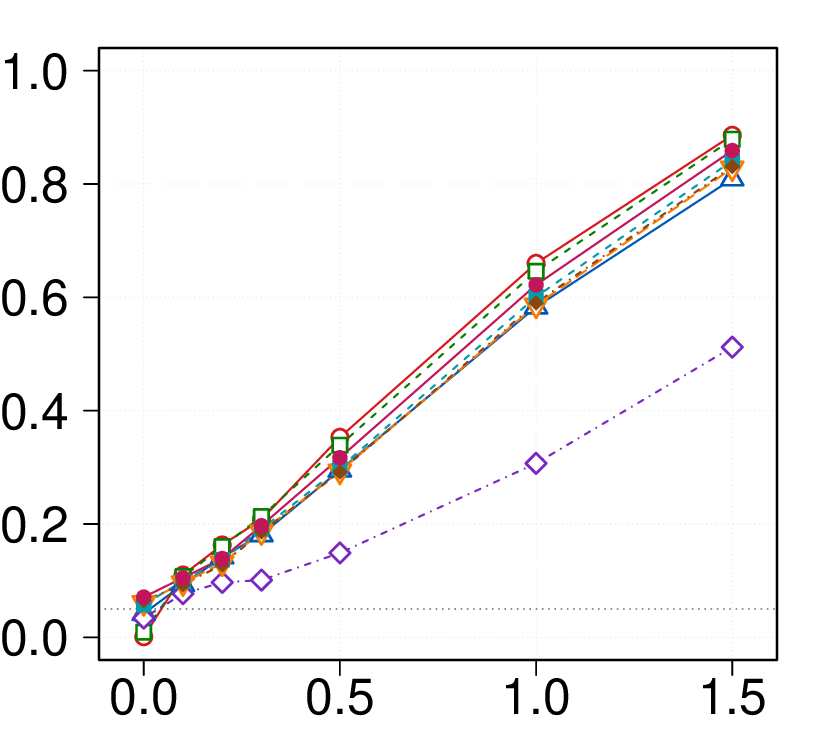}
            \hspace{-1 cm}  
            \includegraphics[width=5.5cm, height=5cm]{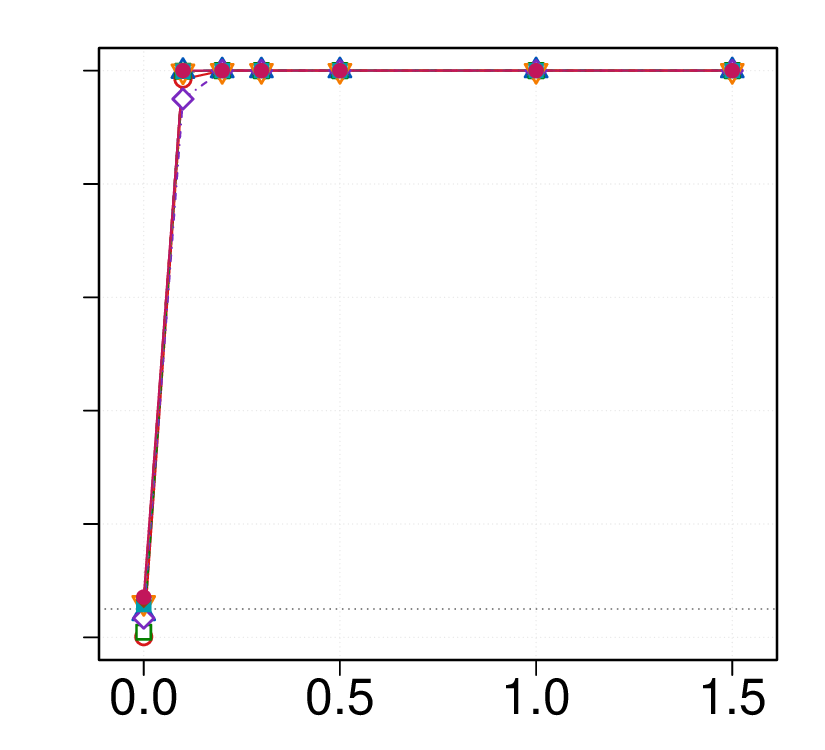}
            \hspace{-1 cm}  
            \includegraphics[width=5.5cm, height=5cm]{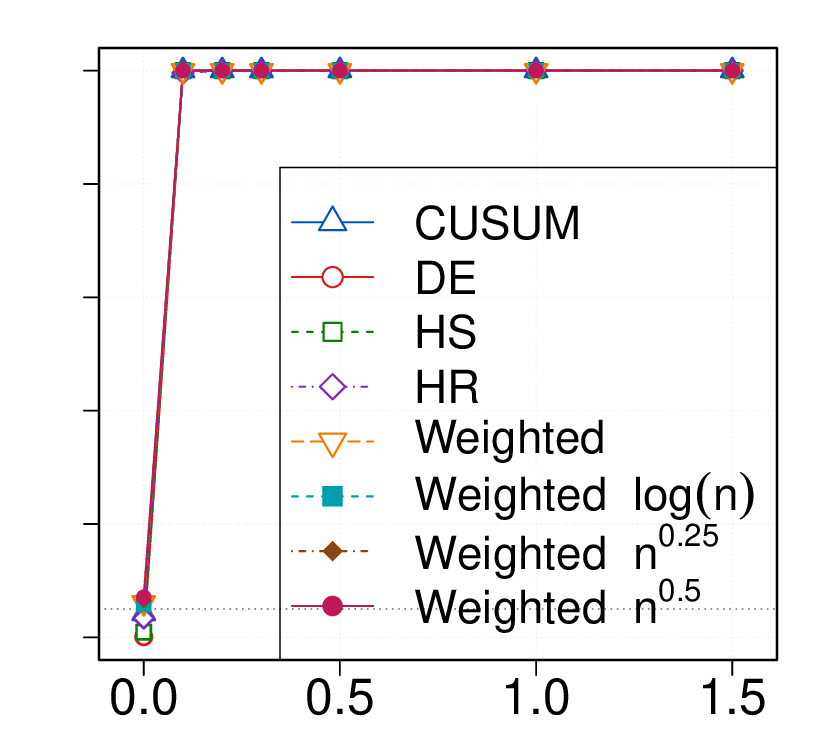}
            \hspace{-1 cm}  
            \vspace{-0.2in}
            \caption*{\small{(b) \textit{Setting 2}, with $m=1,5,20$  } }

              \centering
            \hspace{-1 cm}  
                \includegraphics[width=5.5cm, height=5cm]{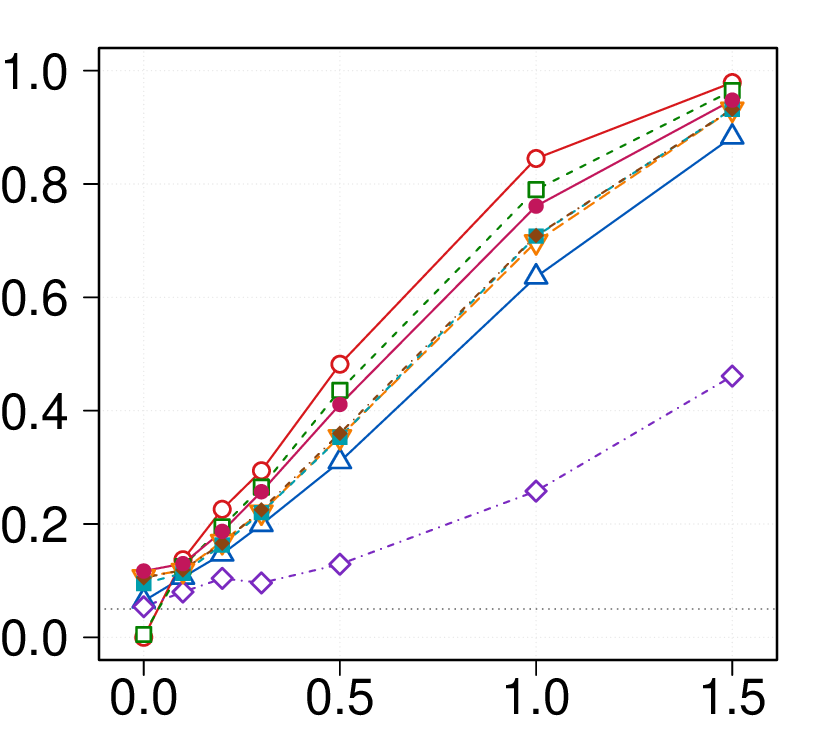}
            \hspace{-1 cm}  
            \includegraphics[width=5.5cm, height=5cm]{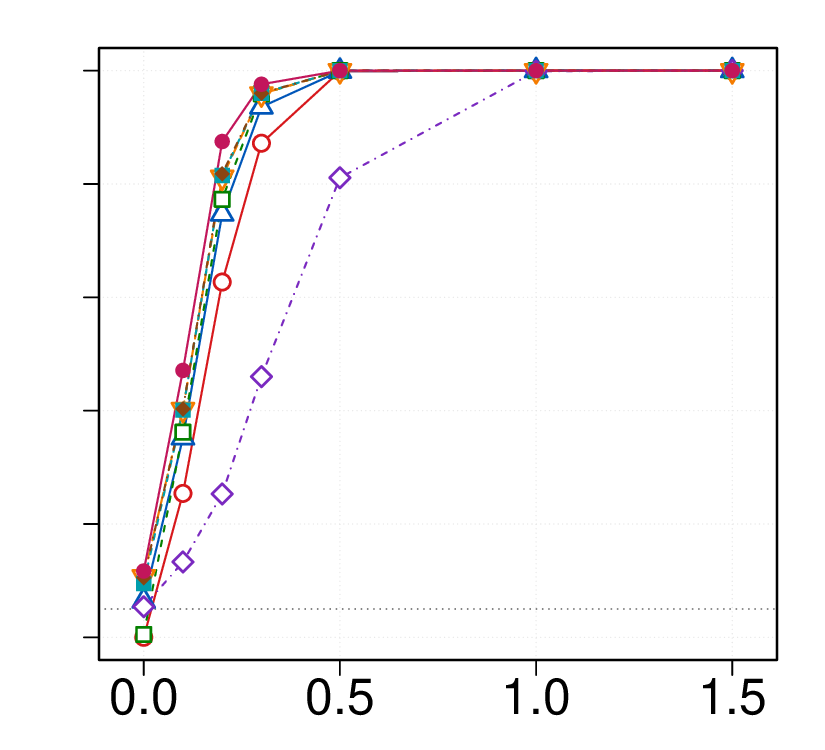}
            \hspace{-1 cm}  
            \includegraphics[width=5.5cm, height=5cm]{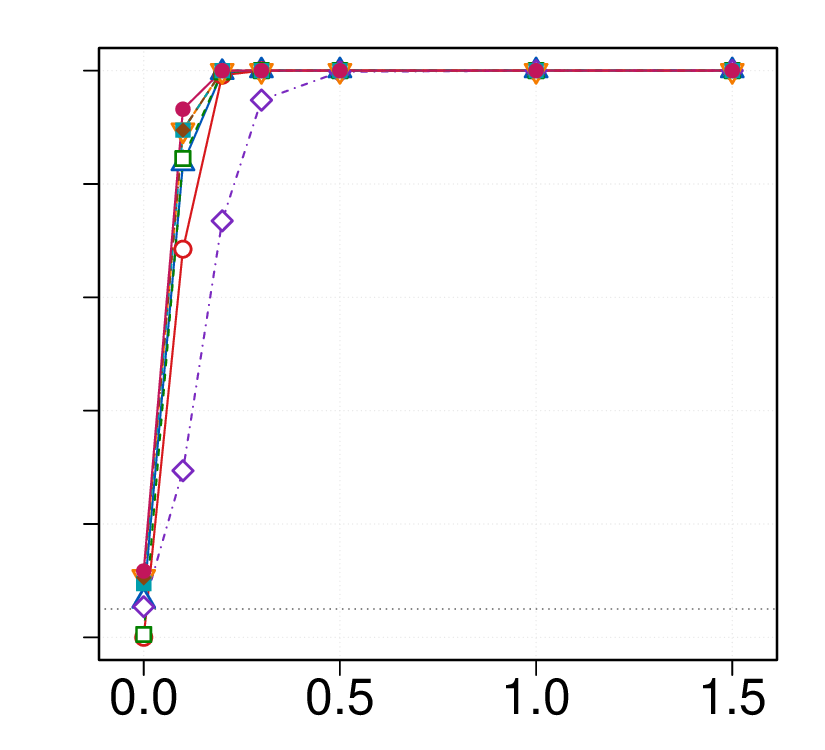}
            \hspace{-1 cm}  
            \vspace{-0.2in}
            \caption*{\small{(c) \textit{Setting 3}, with $m=1,5,20$  } }

            \caption{\small{Adjusted empirical rejection rates of the RP methods for various values of $SNR$ in the x-axis. The RP method performs 200 random projections and applies different change point tests (CUSUM, Weighted, DE, HS, HR) and the BH combination method. The data-generating process follows (\ref{eq:model for fd}), (\ref{eq:data generating process}), and (\ref{break function}), where the standard deviation $\sigma_{g}$ follows \textit{Settings 1--3}.
            The change point location is set at $\theta=0.25$.
            The empirical rejection rate is based on 1000 simulations.
            }}
            \label{fig: tuning cp test BH.adj}
    \end{figure} 

\begin{figure} [H]
         \centering
            \hspace{-1 cm}  
            \includegraphics[width=5.5cm, height=5cm]{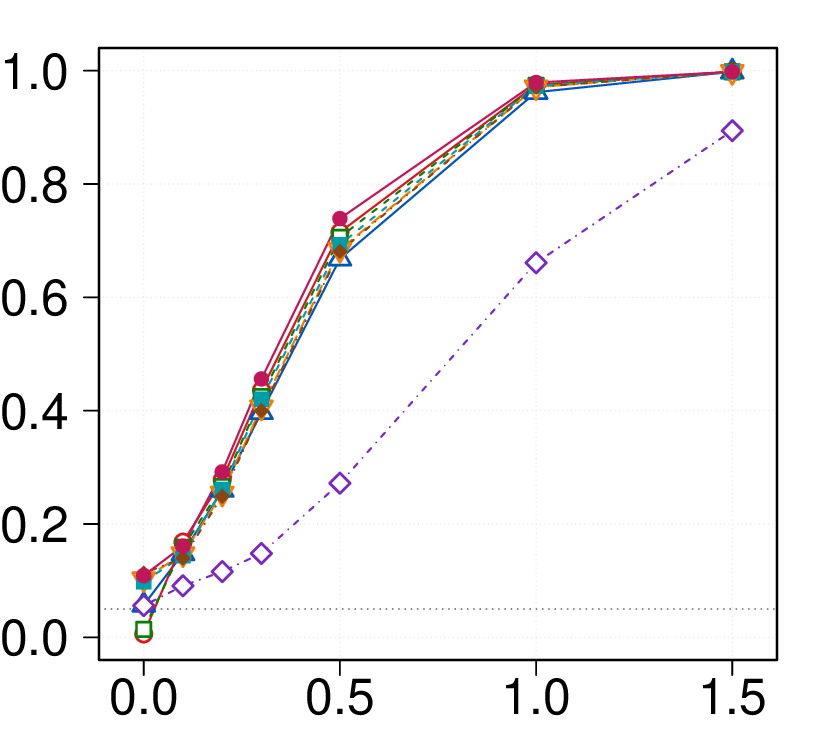}
            \hspace{-1 cm}  
            \includegraphics[width=5.5cm, height=5cm]{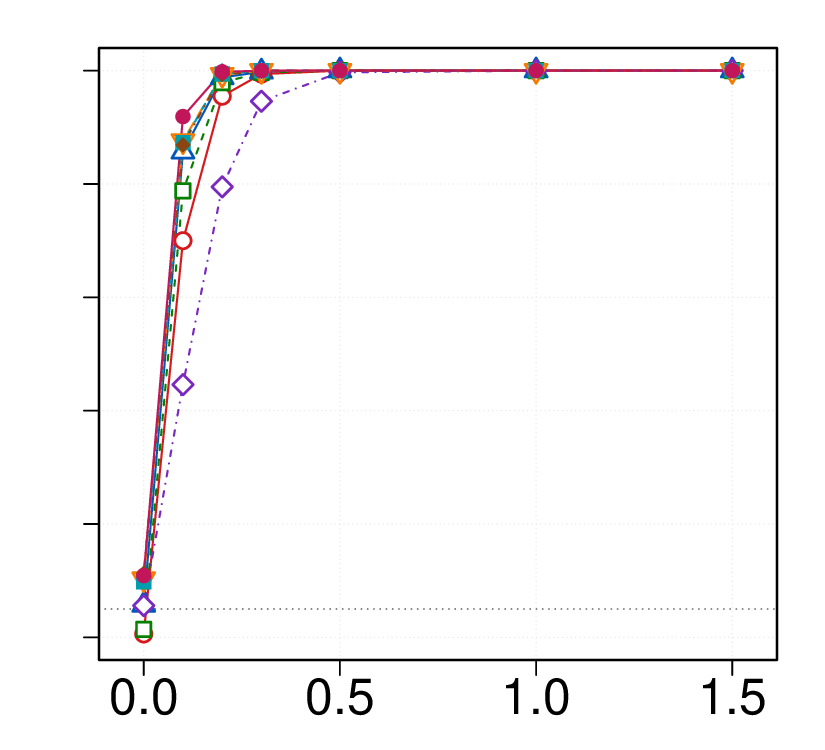}
            \hspace{-1 cm}  
            \includegraphics[width=5.5cm, height=5cm]{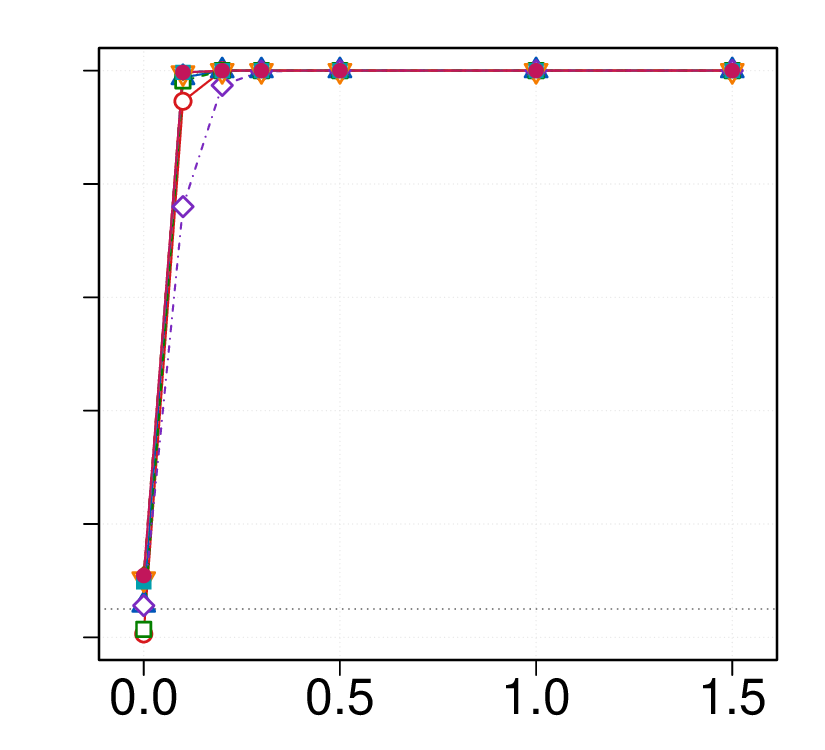}
            \hspace{-1 cm}  
            \vspace{-0.2in}
            \caption*{\small{(a) \textit{Setting 1}, with $m=1,5,20$  } }

            \centering
            \hspace{-1 cm}  
                \includegraphics[width=5.5cm, height=5cm]{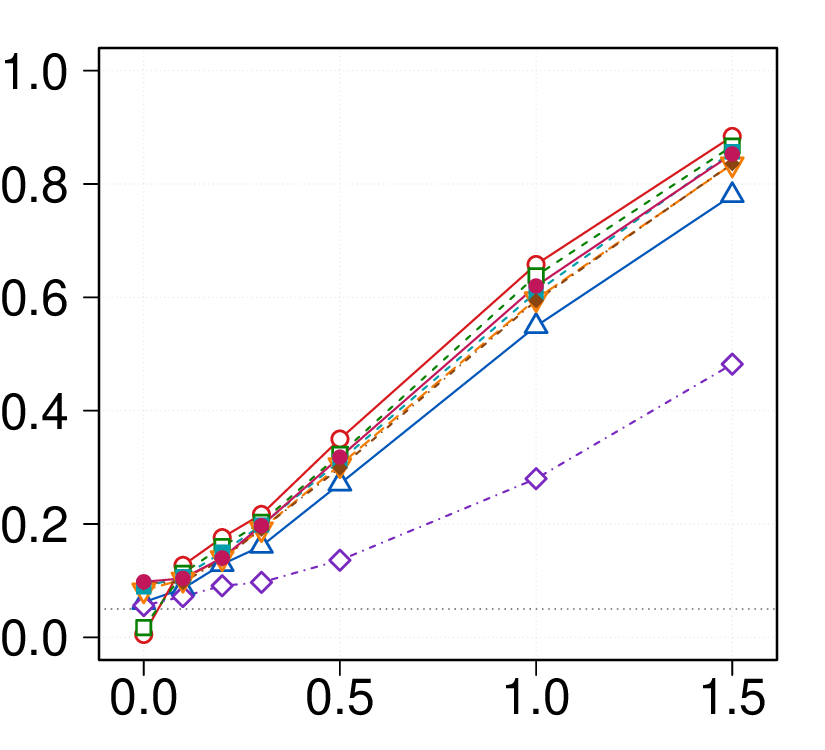}
            \hspace{-1 cm}  
            \includegraphics[width=5.5cm, height=5cm]{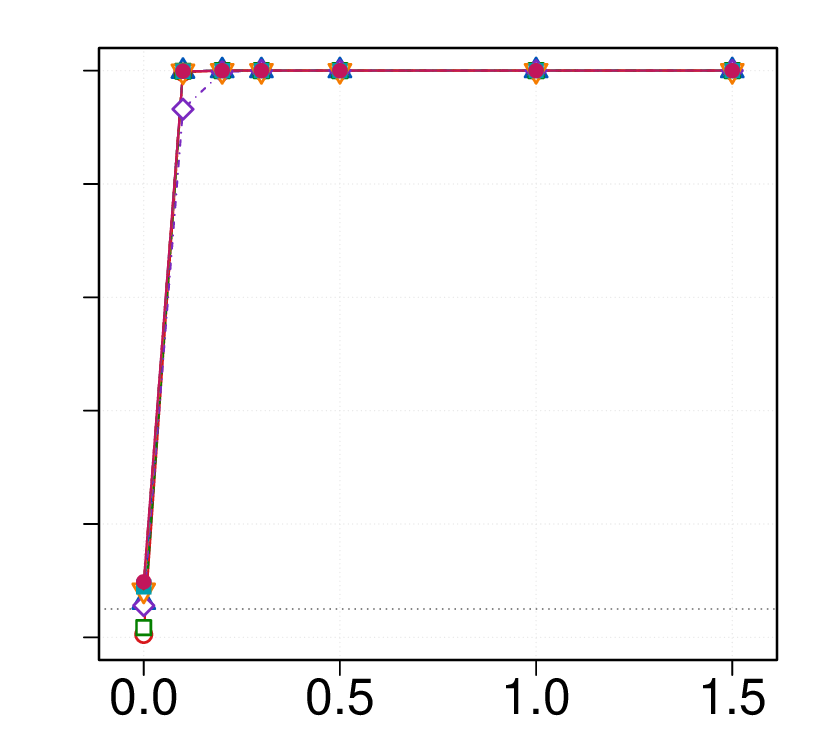}
            \hspace{-1 cm}  
            \includegraphics[width=5.5cm, height=5cm]{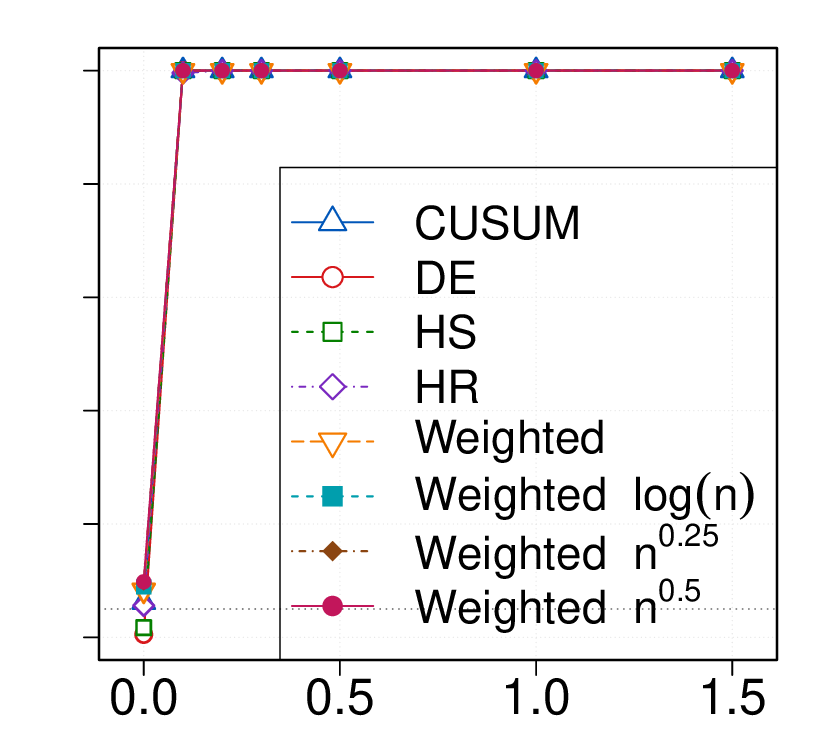}
            \hspace{-1 cm}  
            \vspace{-0.2in}
            \caption*{\small{(b) \textit{Setting 2}, with $m=1,5,20$  } }

              \centering
            \hspace{-1 cm}  
                \includegraphics[width=5.5cm, height=5cm]{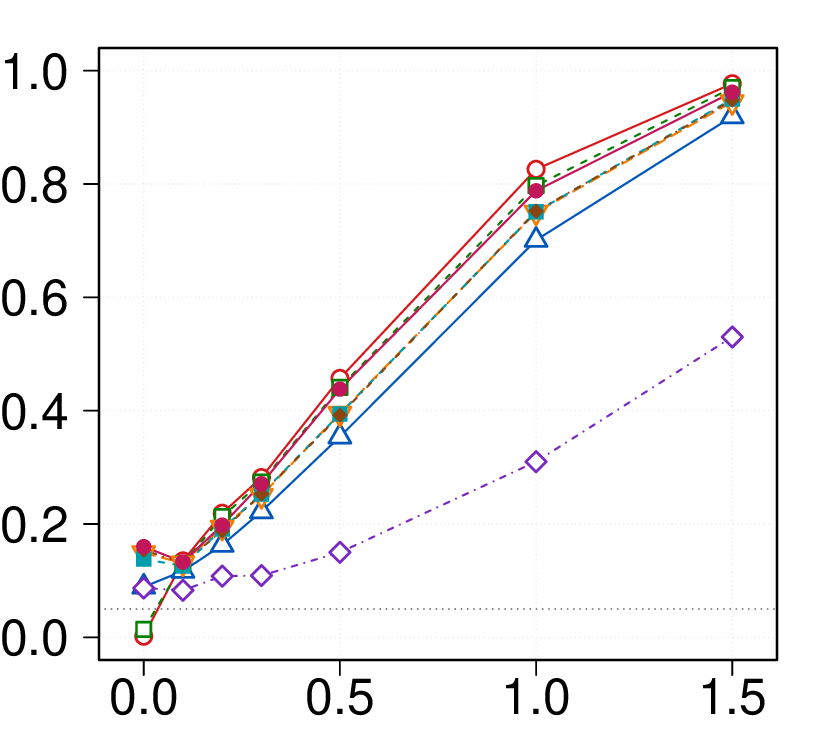}
            \hspace{-1 cm}  
            \includegraphics[width=5.5cm, height=5cm]{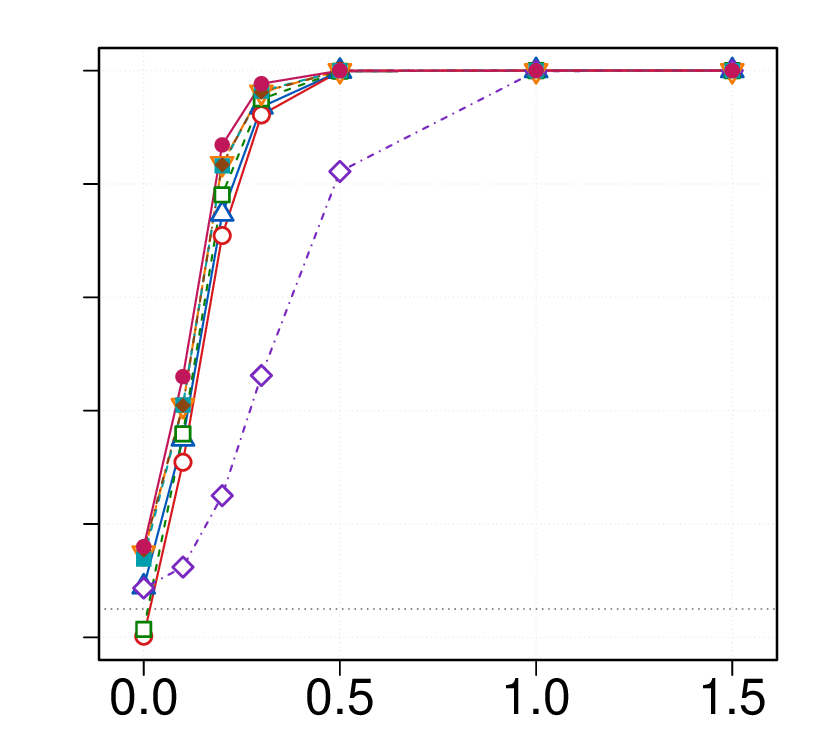}
            \hspace{-1 cm}  
            \includegraphics[width=5.5cm, height=5cm]{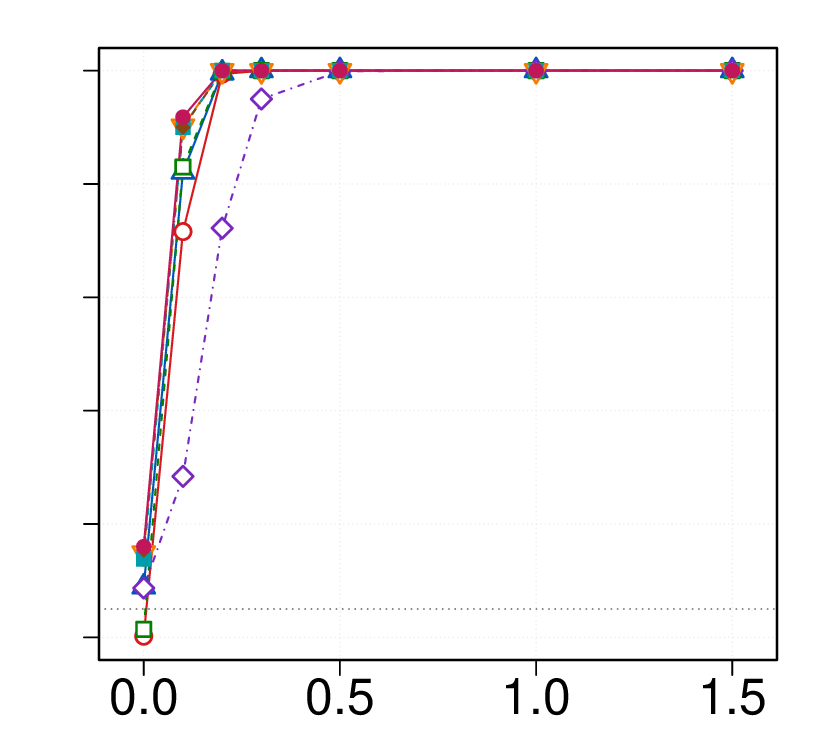}
            \hspace{-1 cm}  
            \vspace{-0.2in}
            \caption*{\small{(c) \textit{Setting 3}, with $m=1,5,20$  } }

            \caption{\small{Adjusted empirical rejection rates of the RP methods for various values of $SNR$ in the x-axis. The RP method performs 200 random projections and applies different change point tests (CUSUM, Weighted, DE, HS, HR) and the HMP combination method. The data-generating process follows (\ref{eq:model for fd}), (\ref{eq:data generating process}), and (\ref{break function}), where the standard deviation $\sigma_{g}$ follows \textit{Settings 1--3}.
            The change point location is set at $\theta=0.25$.
            The empirical rejection rate is based on 1000 simulations.
            }}
            \label{fig: tuning cp test HMP.adj}
    \end{figure} 
    
\begin{figure} [H]
         \centering
            \hspace{-1 cm}  
            \includegraphics[width=5.5cm, height=5cm]{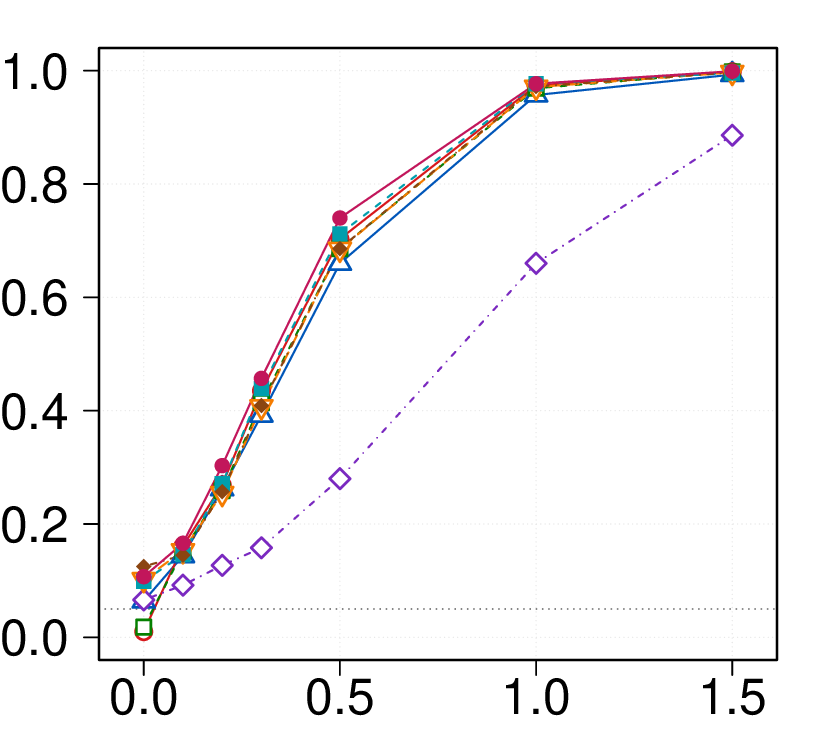}
            \hspace{-1 cm}  
            \includegraphics[width=5.5cm, height=5cm]{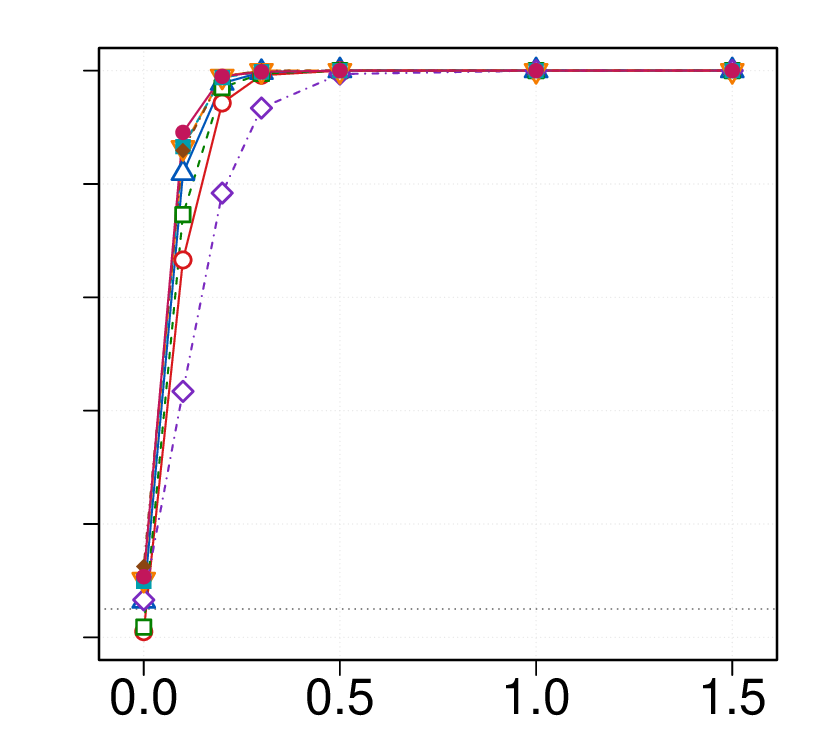}
            \hspace{-1 cm}  
            \includegraphics[width=5.5cm, height=5cm]{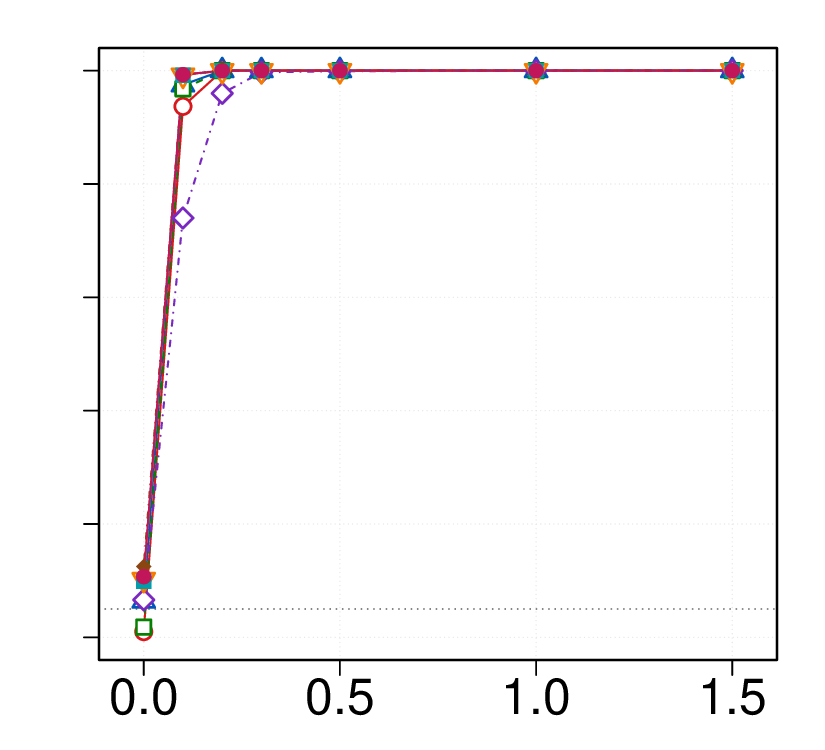}
            \hspace{-1 cm}  
            \vspace{-0.2in}
            \caption*{\small{(a) \textit{Setting 1}, with $m=1,5,20$  } }

            \centering
            \hspace{-1 cm}  
                \includegraphics[width=5.5cm, height=5cm]{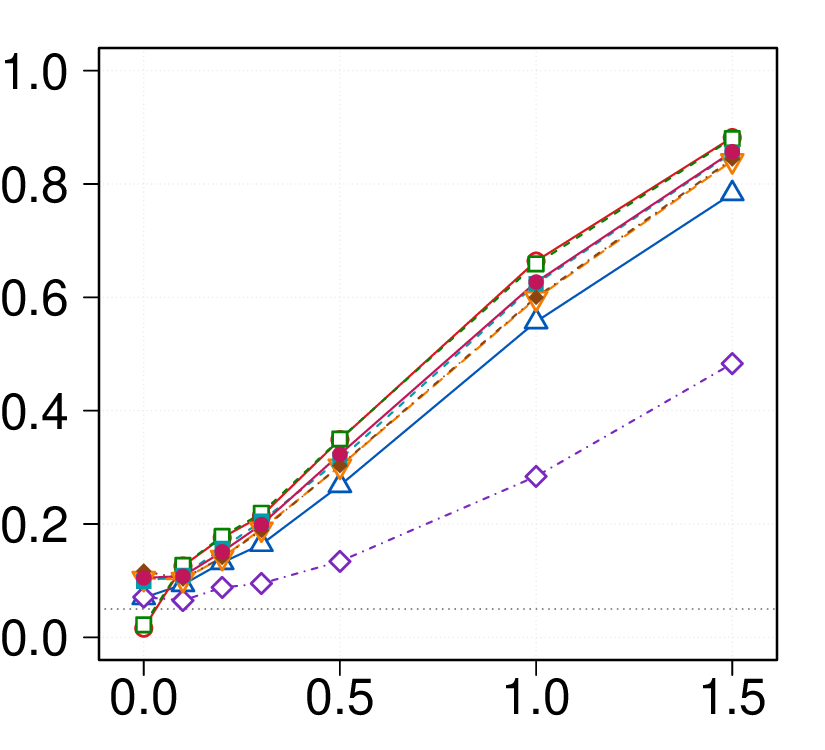}
            \hspace{-1 cm}  
            \includegraphics[width=5.5cm, height=5cm]{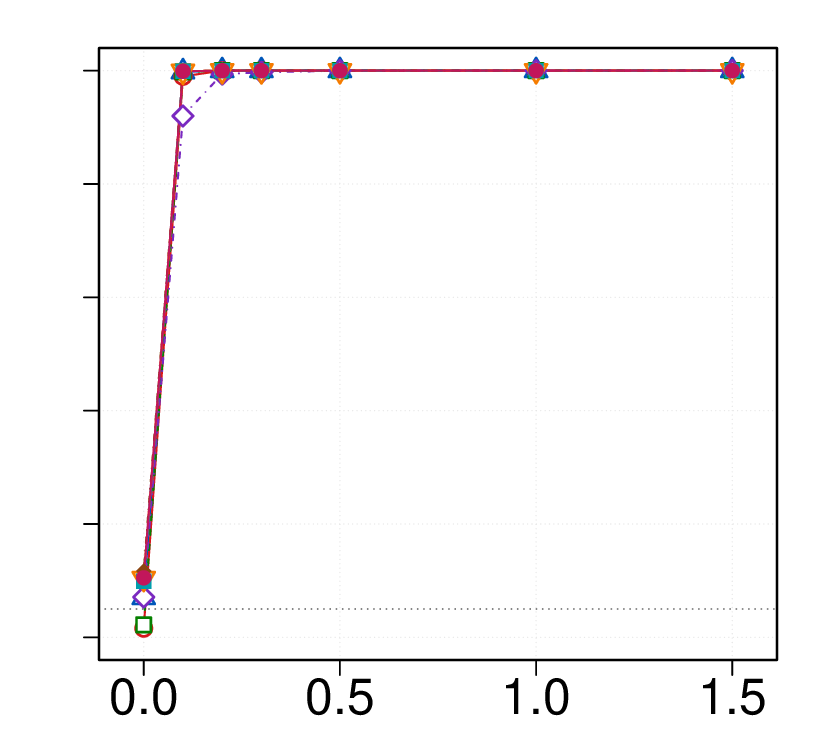}
            \hspace{-1 cm}  
            \includegraphics[width=5.5cm, height=5cm]{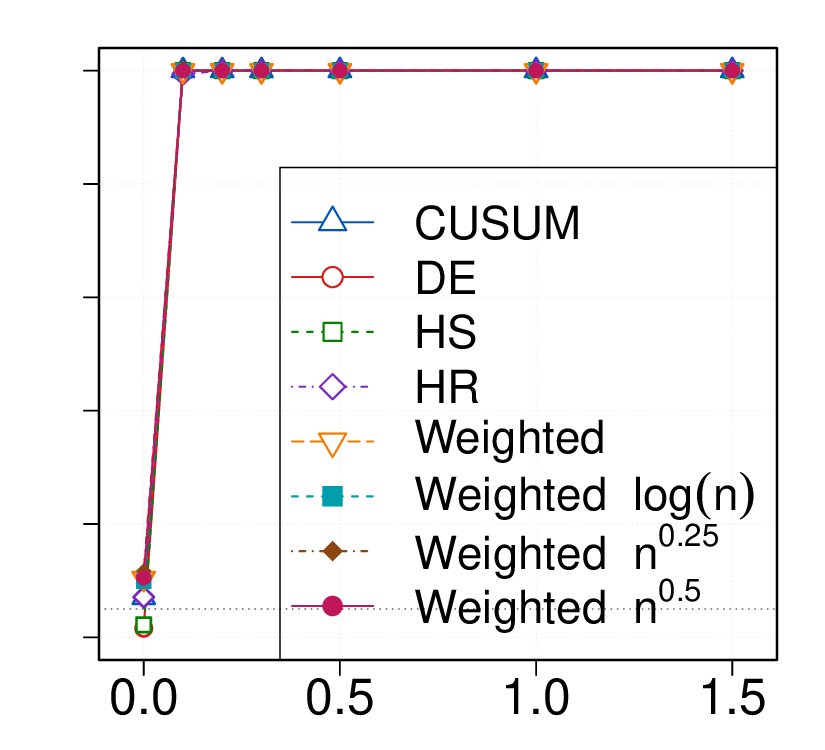}
            \hspace{-1 cm}  
            \vspace{-0.2in}
            \caption*{\small{(b) \textit{Setting 2}, with $m=1,5,20$  } }

              \centering
            \hspace{-1 cm}  
                \includegraphics[width=5.5cm, height=5cm]{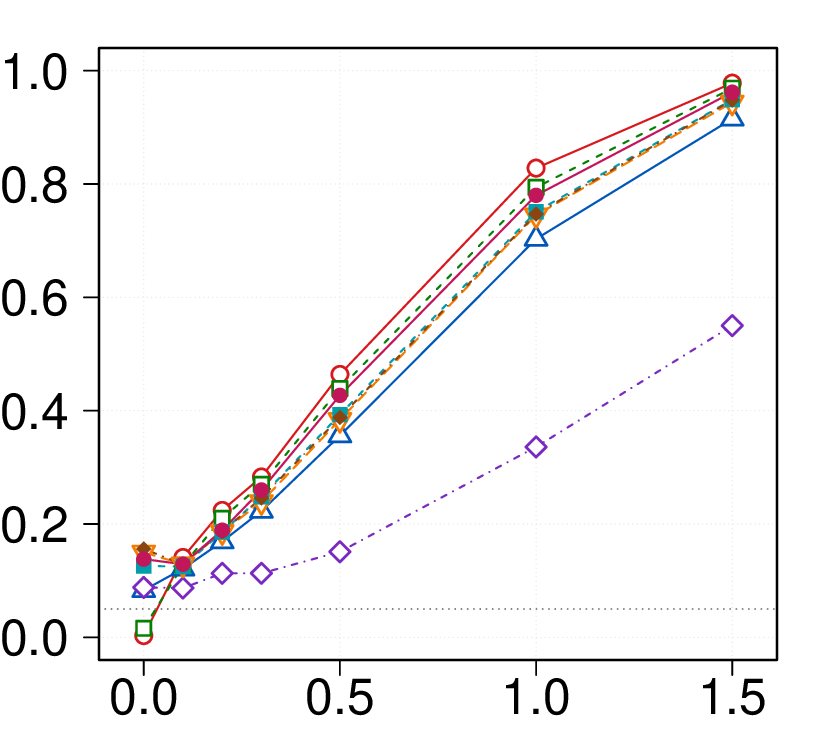}
            \hspace{-1 cm}  
            \includegraphics[width=5.5cm, height=5cm]{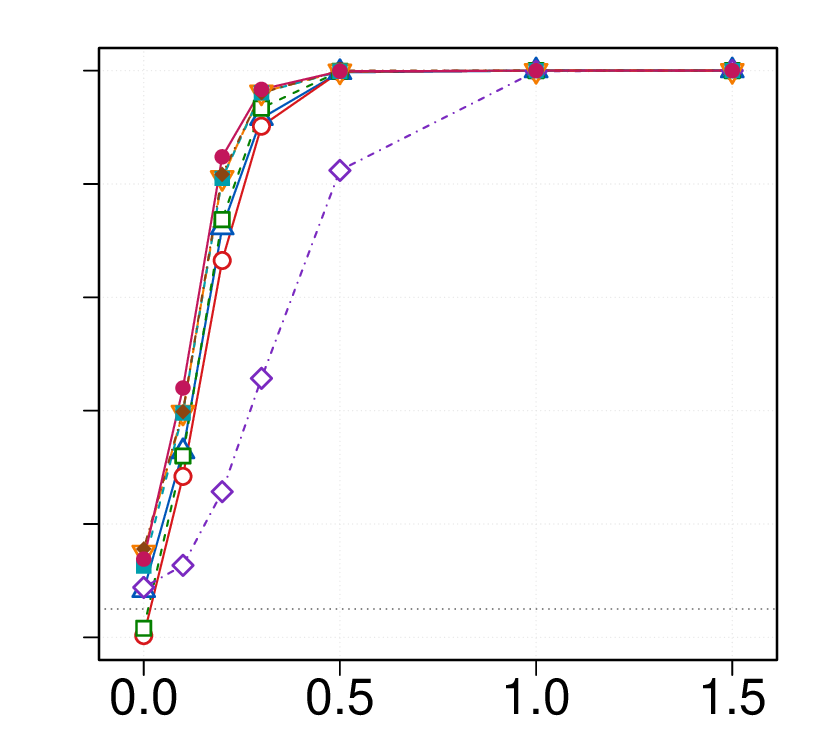}
            \hspace{-1 cm}  
            \includegraphics[width=5.5cm, height=5cm]{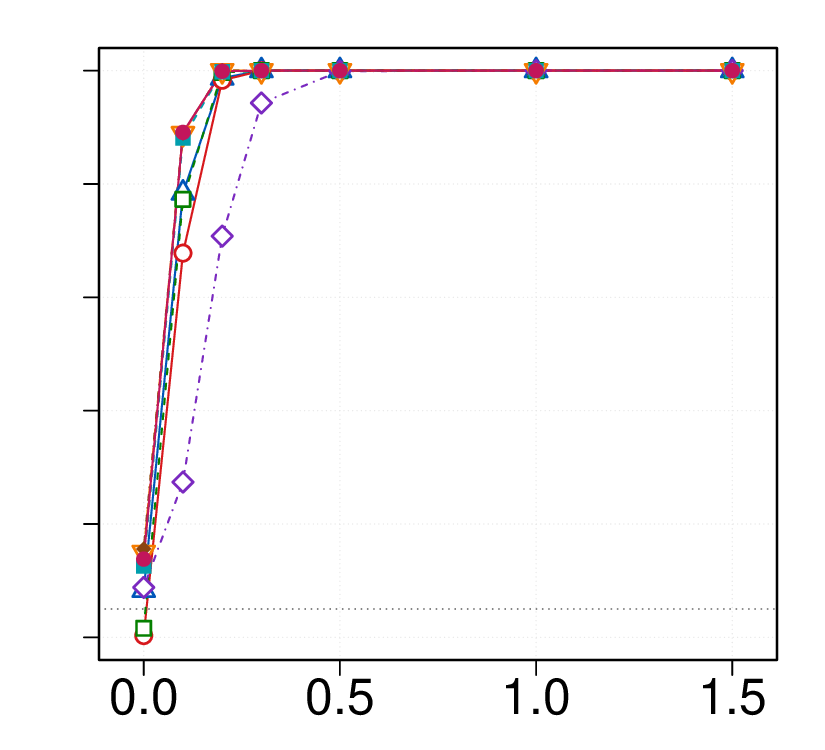}
            \hspace{-1 cm}  
            \vspace{-0.2in}
            \caption*{\small{(c) \textit{Setting 3}, with $m=1,5,20$ } }
            
            \caption{\small{Adjusted empirical rejection rates of the RP methods for various values of $SNR$ in the x-axis. The RP method performs 200 random projections and applies different change point tests (CUSUM, Weighted, DE, HS, HR) and the CCT combination method. The data-generating process follows (\ref{eq:model for fd}), (\ref{eq:data generating process}), and (\ref{break function}), where the standard deviation $\sigma_{g}$ follows \textit{Settings 1--3}.
            The change point location is set at $\theta=0.25$.
            The empirical rejection rate is based on 1000 simulations.
            }}
            \label{fig: tuning cp test CCT.adj}
    \end{figure}

\newpage
\begin{figure} [h!]
         \centering
        \hspace{-1 cm}
            \includegraphics[width=5.5cm, height=5cm]{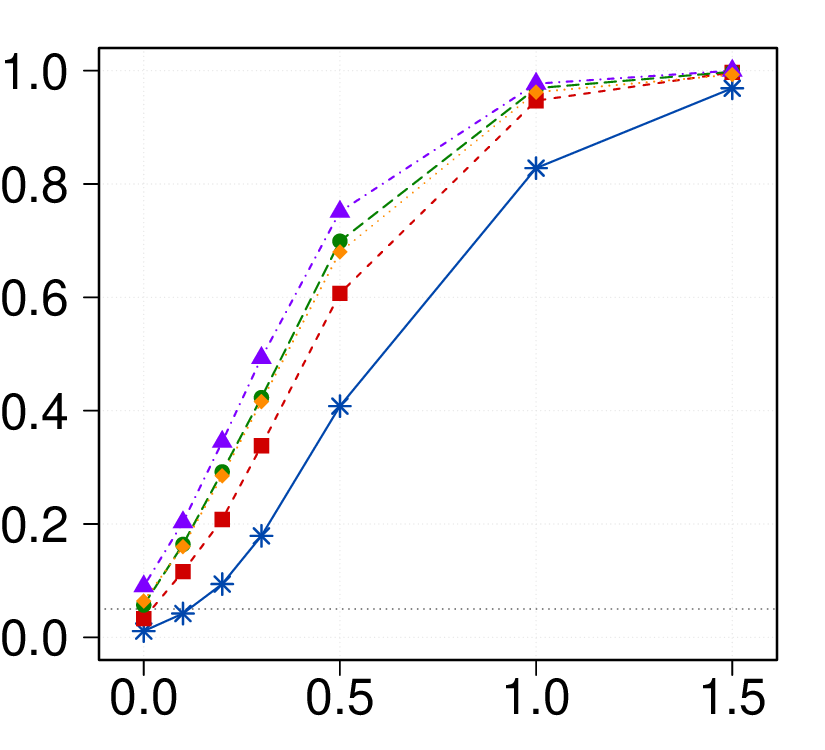}
         \hspace{-1 cm}
            \includegraphics[width=5.5cm, height=5cm]{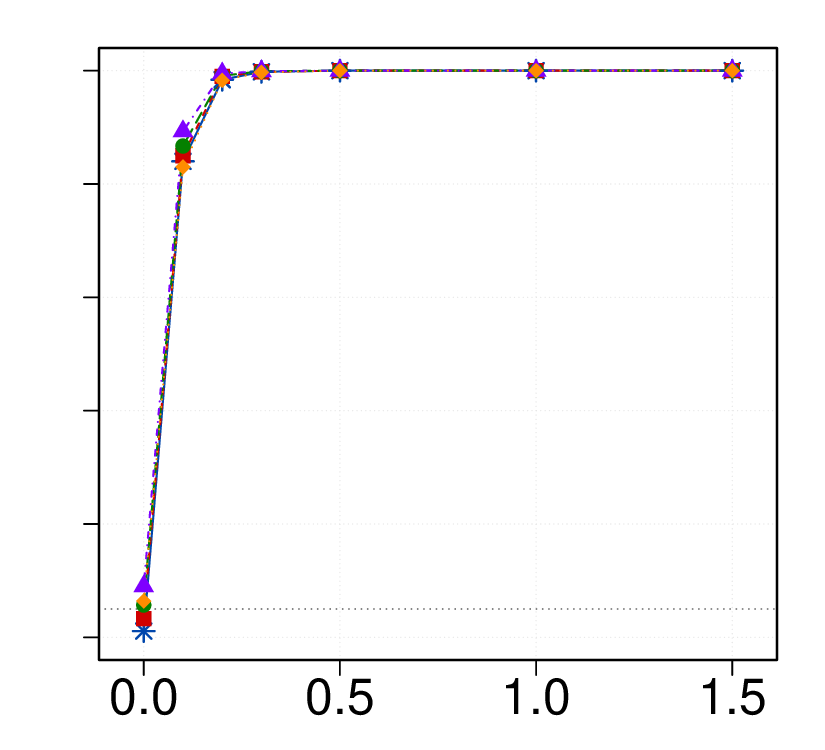}
         \hspace{-1 cm}
            \includegraphics[width=5.5cm, height=5cm]{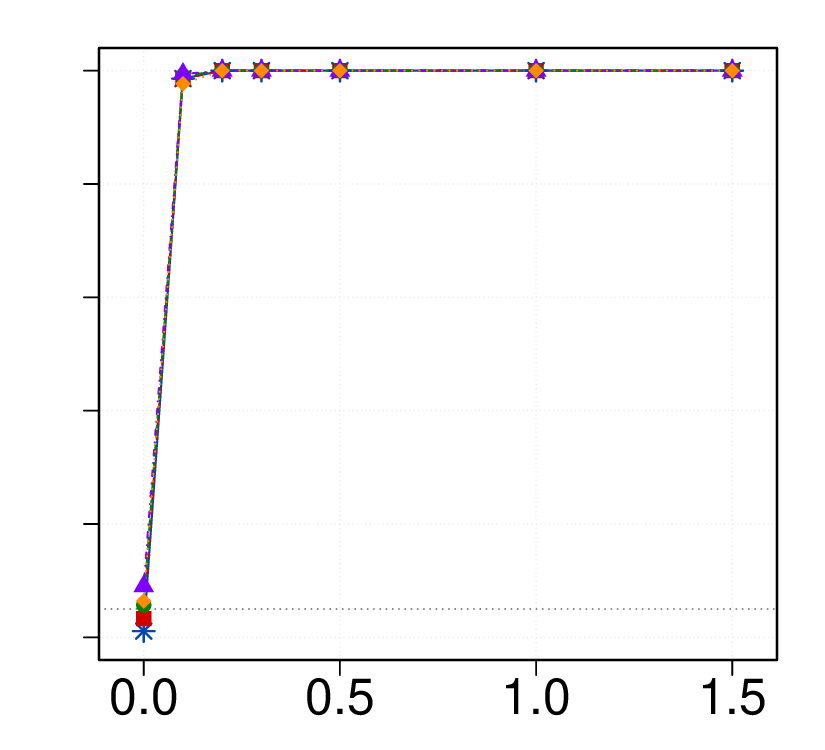}
         \hspace{-1 cm} 
         \vspace{-0.2in}
            \caption*{(a) \small{\textit{Setting 1}, with $m=1,5,20$  } }

            \centering
            \hspace{-1 cm}
            \includegraphics[width=5.5cm, height=5cm]{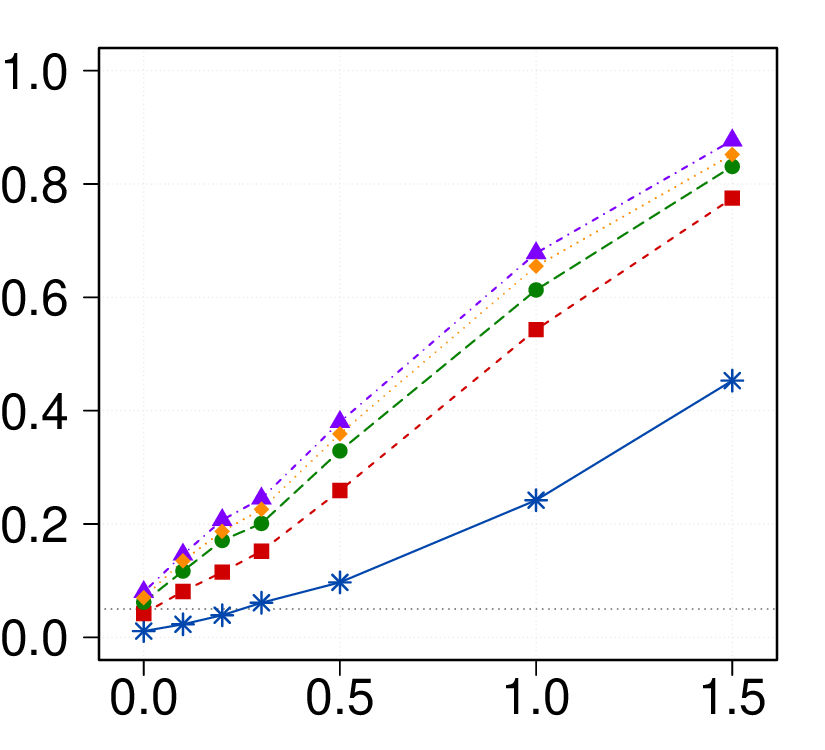}  
         \hspace{-1 cm}
            \includegraphics[width=5.5cm, height=5cm]{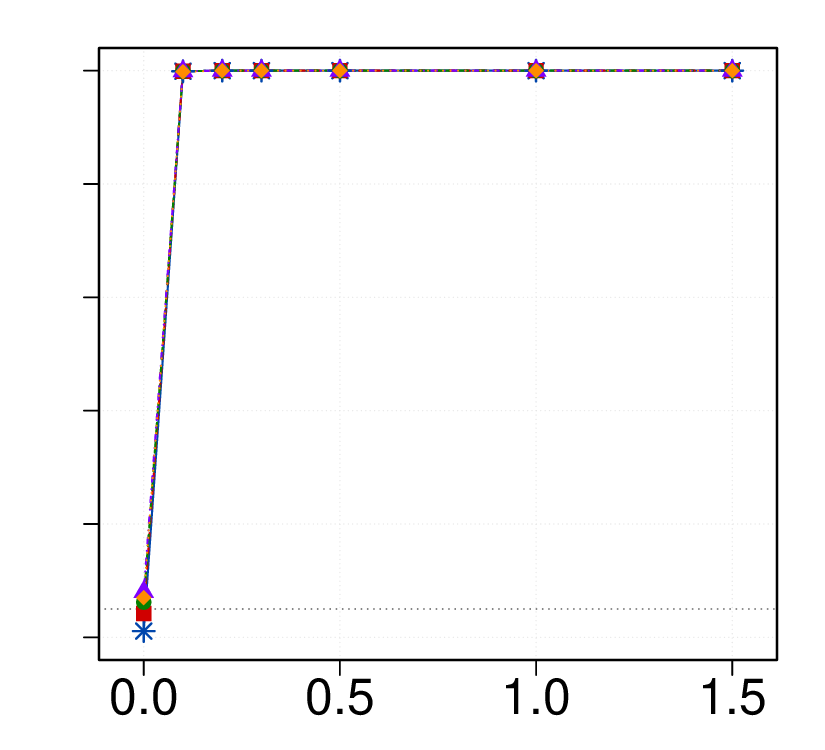}   \hspace{-1 cm}
            \includegraphics[width=5.5cm, height=5cm]{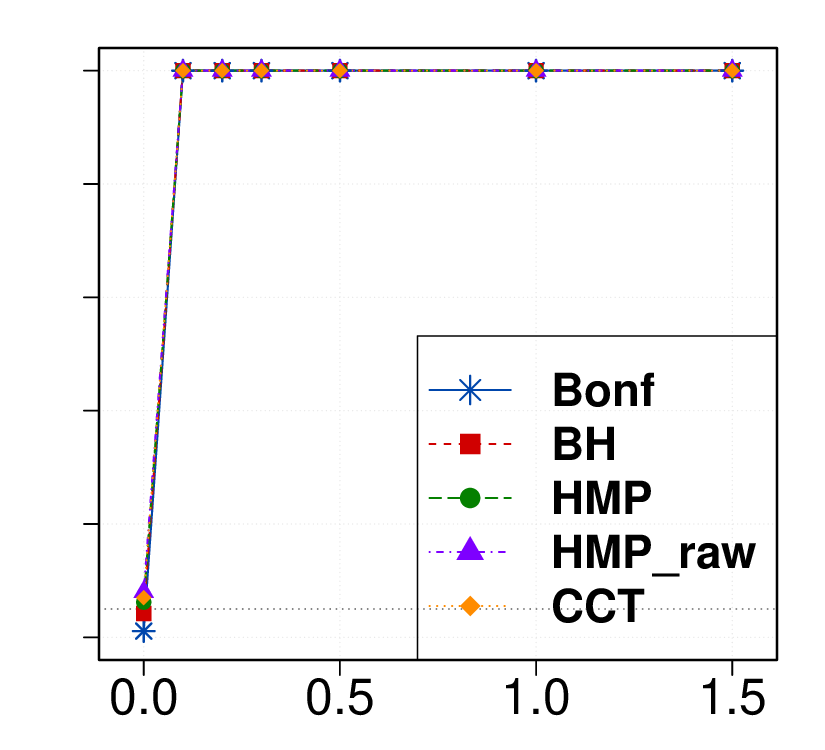}  \hspace{-1 cm} 
            \vspace{-0.2in}
            \caption*{\small{\textit{Setting 2}, with $m=1,5,20$ } }

            \centering
       \hspace{-1 cm}
            \includegraphics[width=5.5cm, height=5cm]{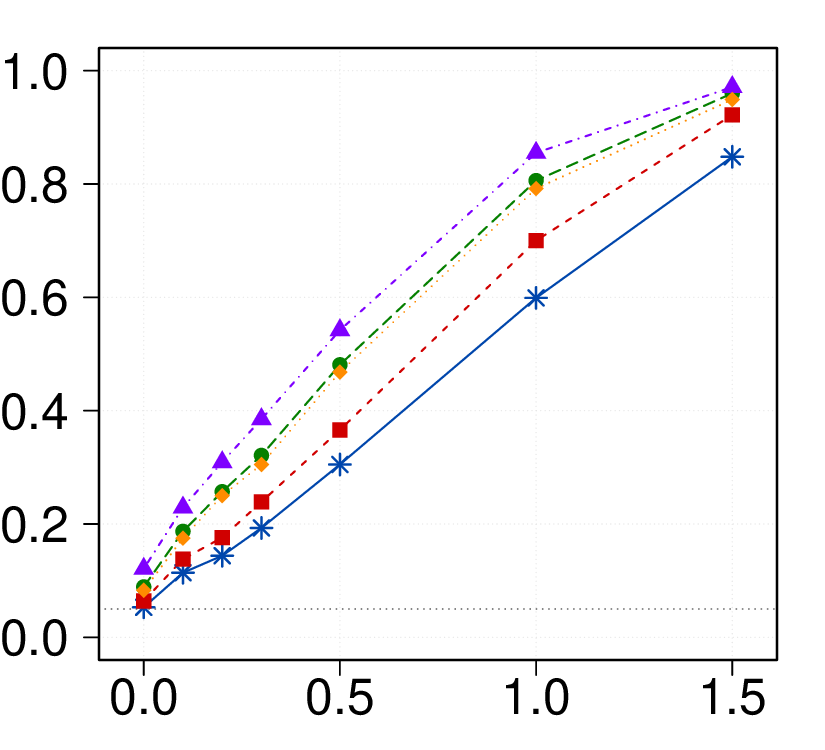} 
         \hspace{-1 cm}
            \includegraphics[width=5.5cm, height=5cm]{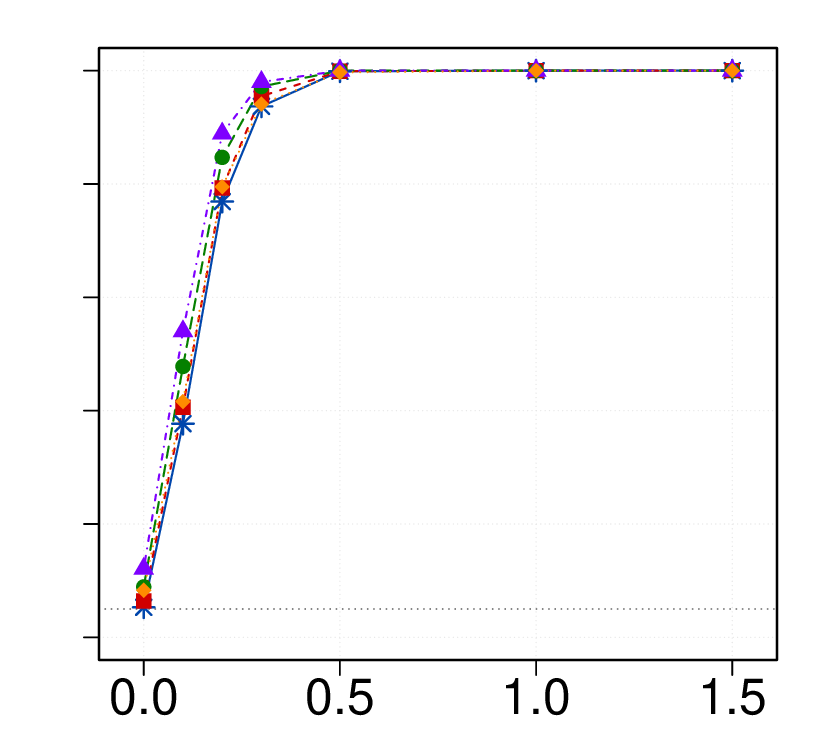}  
         \hspace{-1 cm}
            \includegraphics[width=5.5cm, height=5cm]{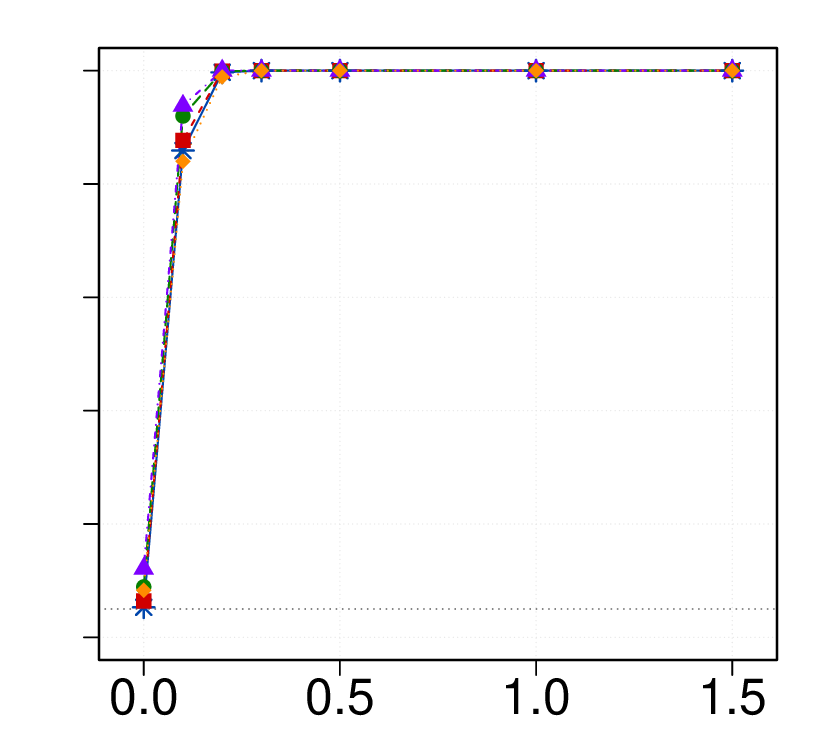} 
         \hspace{-1 cm} 
            \vspace{-0.2in}
            \caption*{(c) \small{\textit{Setting 3}, with $m=1,5,20$ } }

            \caption{\small{
             Raw empirical rejection rates of the RP-Bonf, RP-BH, RP-HMP, RP-HMP-raw, and RP-CCT methods with the standard CUSUM test for various values of $SNR$ in the x-axis. The RP method performs 200 random projections. The data-generating process follows (\ref{eq:model for fd}), (\ref{eq:data generating process}), and (\ref{break function}), where the standard deviation $\sigma_{g}$ follows \textit{Settings 1--3}.
            The change point location is set at $\theta=0.25$.
            The empirical rejection rate is based on 1000 simulations.
            }}
            \label{fig: tuning Pvalue-comb}
    \end{figure}

\newpage
\begin{table}[H]
\centering
\begin{tabular}{ 
|p{1.5cm}||p{2.5cm}|p{1.5cm}|p{1.5cm}|p{1.5cm}|p{1.5cm}|p{1.5cm}| }
 \hline
 & & Bonf & BH & HMP & HMP-raw & CCT \\
 \hline
 \multirow{8}{*}{\textit{Setting 1}} 

 & CUSUM & 0.011 & 0.033 & 0.057 & 0.090 & 0.064 \\ 
 & DE & 0.000 & 0.001 & 0.006 & 0.010 & 0.010 \\ 
 & HS & 0.001 & 0.009 & 0.014 & 0.024 & 0.018 \\ 
 & HR & 0.013 & 0.036 & 0.056 & 0.078 & 0.066 \\ 
  & Weighted & 0.026 & 0.063 & 0.103 & 0.132 & 0.102 \\ 
  & Weighted $n^{0.25}$  & 0.026 & 0.064 & 0.113 & 0.141 & 0.125 \\ 
  & Weighted $n^{0.5}$ & 0.023 & 0.070 & 0.109 & 0.132 & 0.107 \\ 
  & Weighted $log(n)$ & 0.022 & 0.059 & 0.098 & 0.127 & 0.099 \\ 

 \hline
 \multirow{8}{*}{\textit{Setting 2}} 

   & CUSUM & 0.011 & 0.042 & 0.062 & 0.080 & 0.070 \\ 
  & DE  & 0.000 & 0.001 & 0.005 & 0.016 & 0.016 \\ 
  & HS & 0.000 & 0.009 & 0.017 & 0.023 & 0.022 \\ 
  & HR & 0.016 & 0.034 & 0.056 & 0.075 & 0.071 \\ 
  & Weighted & 0.023 & 0.061 & 0.083 & 0.121 & 0.104 \\ 
  & Weighted $n^{0.25}$ & 0.023 & 0.065 & 0.096 & 0.126 & 0.117 \\ 
  & Weighted $n^{0.5}$  & 0.026 & 0.071 & 0.098 & 0.119 & 0.105 \\ 
  & Weighted $log(n)$  & 0.020 & 0.058 & 0.089 & 0.115 & 0.099 \\ 

 \hline
 \multirow{8}{*}{\textit{Setting 3}} 

  & CUSUM & 0.053 & 0.064 & 0.089 & 0.121 & 0.083 \\ 
  & DE  & 0.000 & 0.000 & 0.002 & 0.006 & 0.003 \\ 
  & HS & 0.003 & 0.005 & 0.014 & 0.023 & 0.016 \\ 
  & HR  & 0.048 & 0.054 & 0.087 & 0.115 & 0.088 \\ 
  & Weighted & 0.090 & 0.107 & 0.149 & 0.199 & 0.150 \\ 
  & Weighted $n^{0.25}$ & 0.082 & 0.105 & 0.154 & 0.204 & 0.156 \\ 
  & Weighted $n^{0.5}$ & 0.098 & 0.117 & 0.160 & 0.203 & 0.138 \\ 
  & Weighted $log(n)$  & 0.076 & 0.095 & 0.138 & 0.181 & 0.126 \\ 

 \hline
\end{tabular}

\caption{\small{Empirical rejection rates under the null of RP method using different univariate change point tests and $p$-value combination methods with $k=200$, at significance level $0.05$.
}}
\label{tab:IID_sizes}
\end{table}

\newpage
  \begin{figure} [H]
        \centering
            \includegraphics[width=5.5cm, height=5cm]{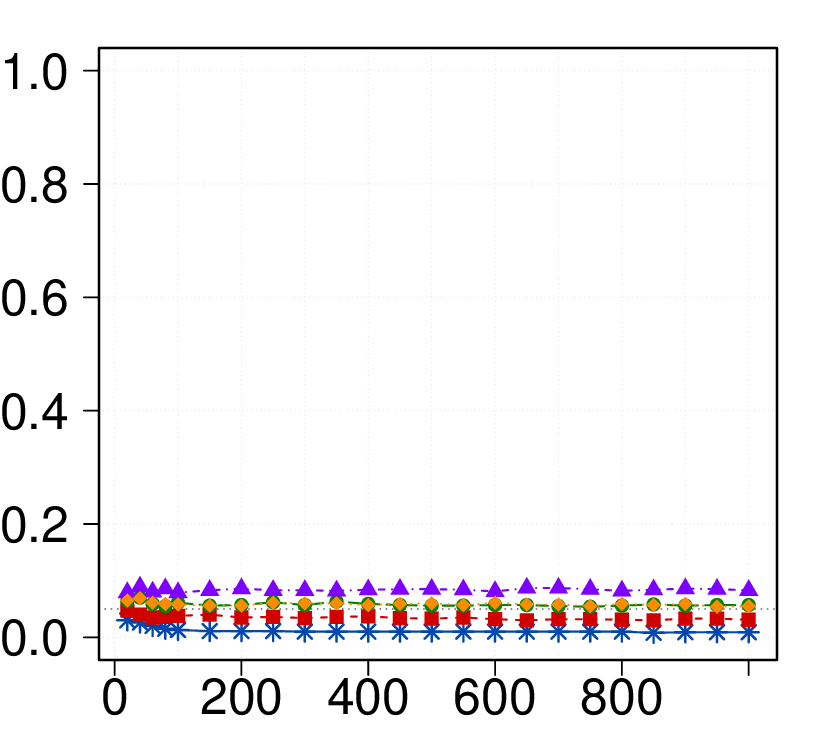}
        \vspace{-0.2in}
            \caption*{\small{(a) $SNR=0$} }

        \hspace{-1 cm}
            \includegraphics[width=5.5cm, height=5cm]{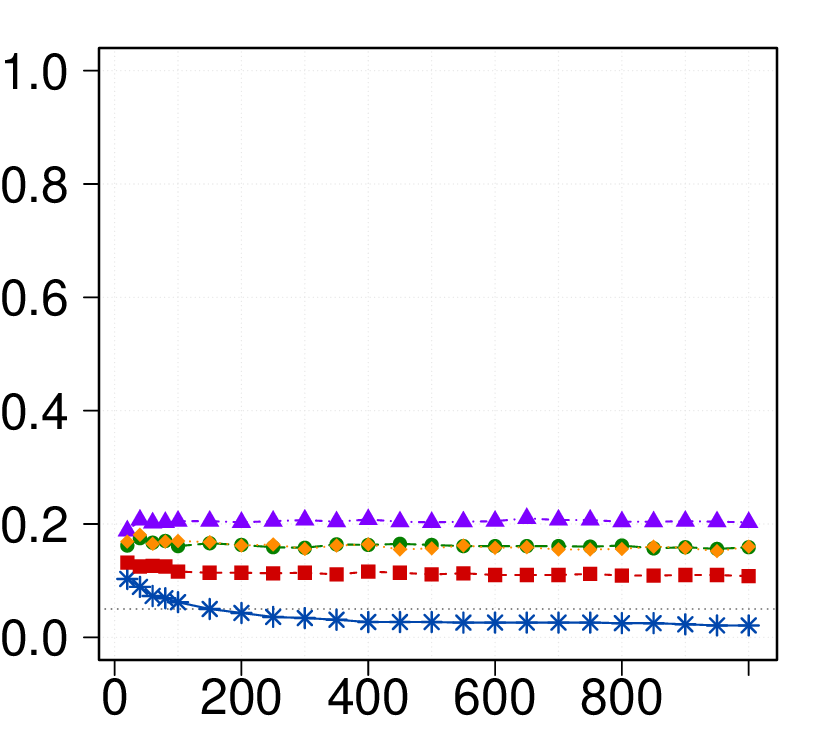}
        \hspace{-1 cm}
            \includegraphics[width=5.5cm, height=5cm]{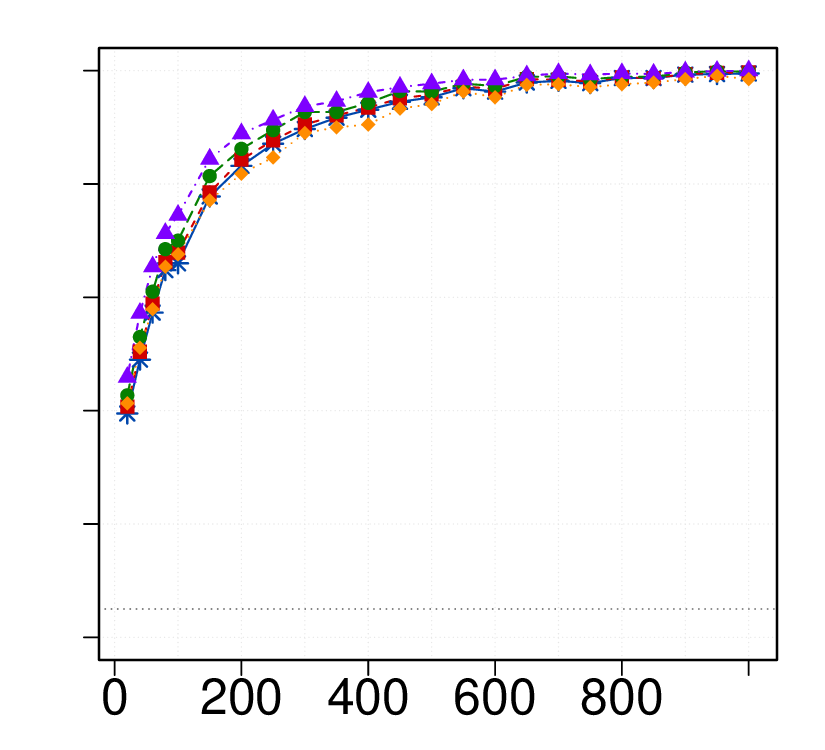}
        \hspace{-1 cm}
            \includegraphics[width=5.5cm, height=5cm]    {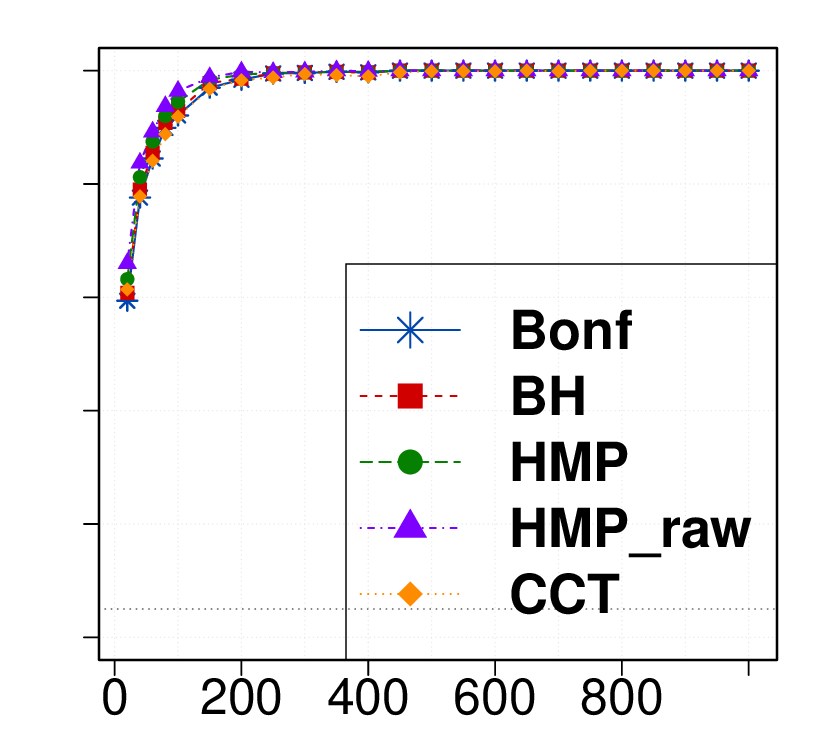}
        \hspace{-1 cm}
        \vspace{-0.2in}
            \caption*{\small{(b) $SNR=0.5$, with $m=1,5,20$ } }
            
        \hspace{-1 cm}    
             \includegraphics[width=5.5cm, height=5cm]{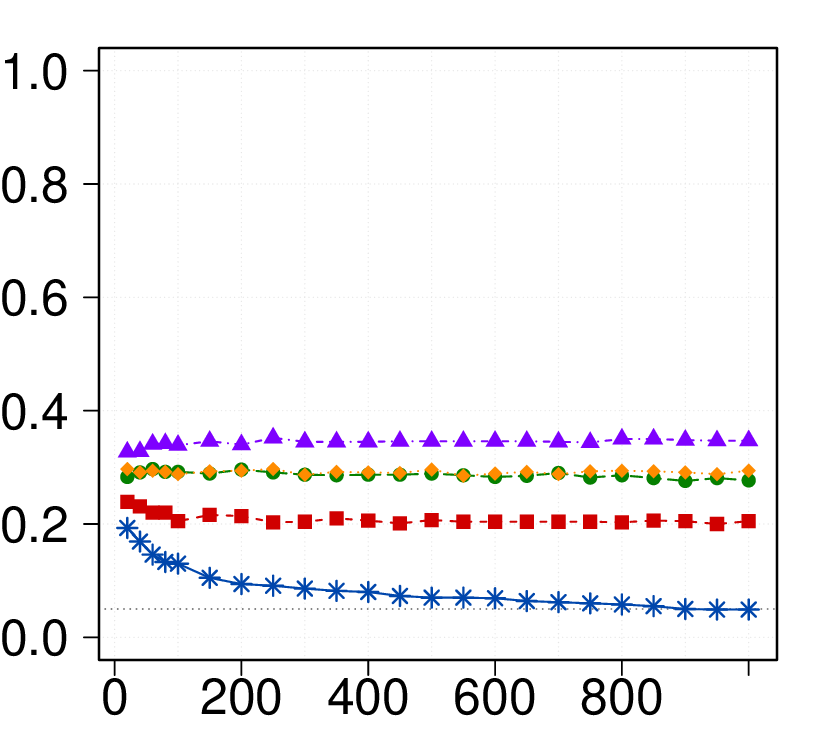}
        \hspace{-1 cm}
            \includegraphics[width=5.5cm, height=5cm]{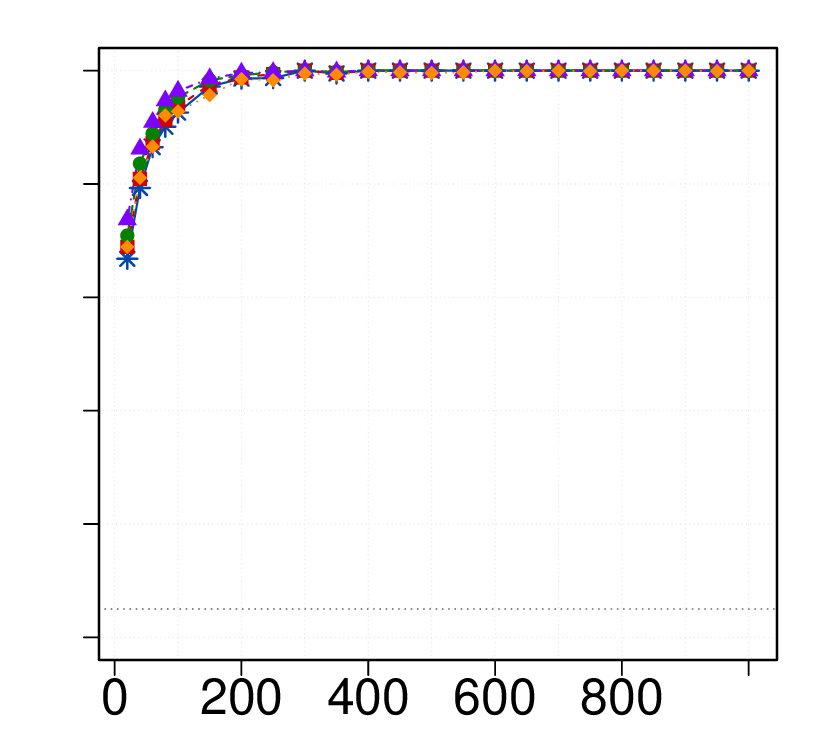}
        \hspace{-1 cm}
            \includegraphics[width=5.5cm, height=5cm]    {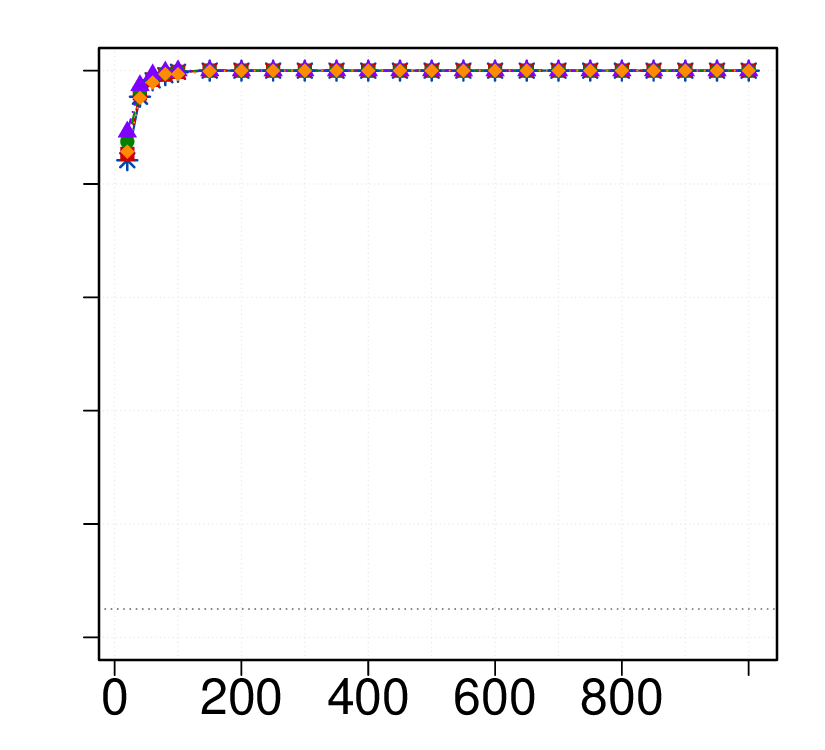}
        \hspace{-1 cm}
            \vspace{-0.2in}
            \caption*{\small{(c) $SNR=1.5$, with $m=1,5,20$ } }
            
            \caption{\small{
             Raw empirical rejection rates of the RP-Bonf, RP-BH, RP-HMP, RP-HMP-raw, and RP-CCT methods with the standard CUSUM test for various choices of number $k$ of random projections in the x-axis. 
            The data-generating process follows (\ref{eq:model for fd}), (\ref{eq:data generating process}), and (\ref{break function}), where the standard deviation $\sigma_{g}$ follows \textit{Setting 1}.
            The change point location is set at $\theta=0.25$.
            The empirical rejection rate is based on 1000 simulations.
            }}
            \label{fig:  tuning num_rps_sigma1}
    \end{figure} 
\newpage
  \begin{figure} [H]
        \centering
            \includegraphics[width=5.5cm, height=5cm]{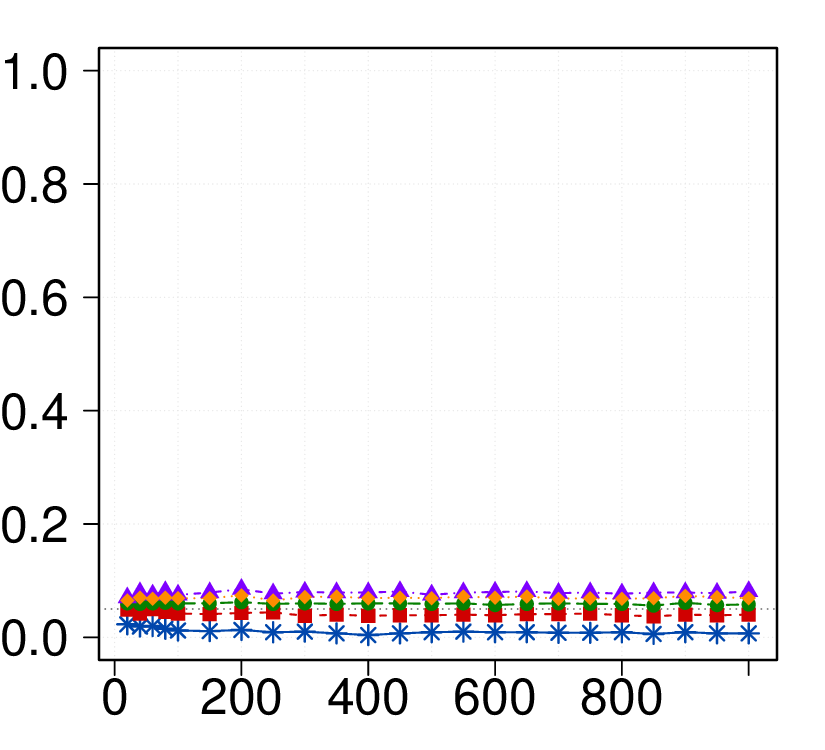}
        \vspace{-0.2in}
            \caption*{\small{(a) $SNR=0$} }

        \hspace{-1 cm}
            \includegraphics[width=5.5cm, height=5cm]{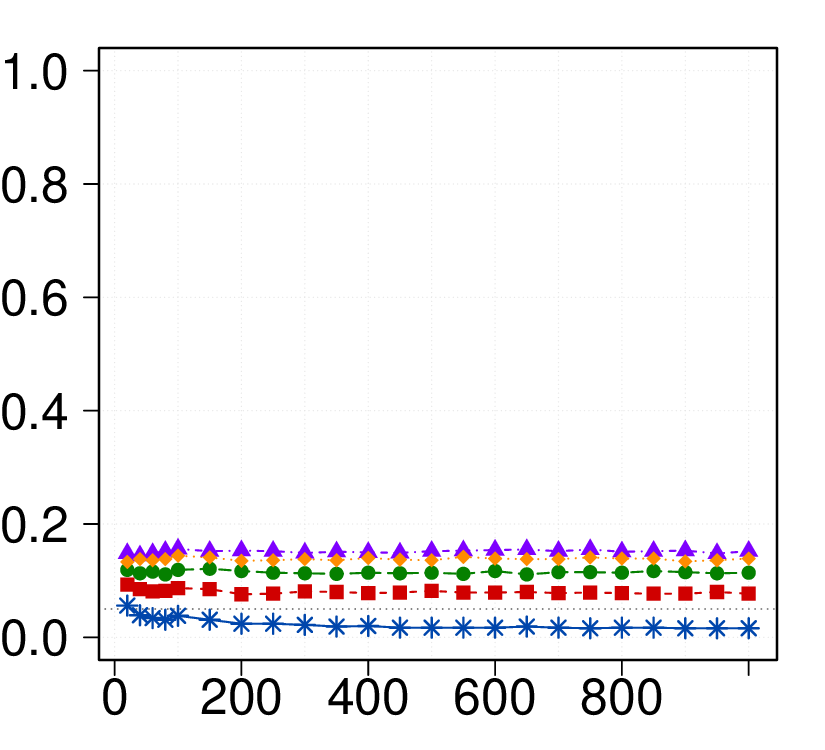}
        \hspace{-1 cm}
            \includegraphics[width=5.5cm, height=5cm]{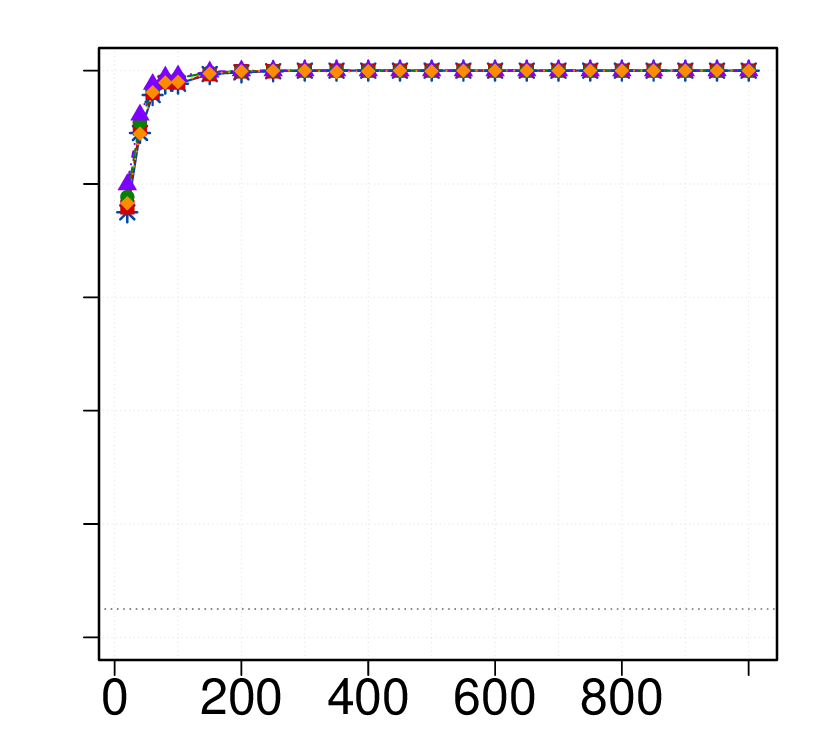}
        \hspace{-1 cm}
            \includegraphics[width=5.5cm, height=5cm]{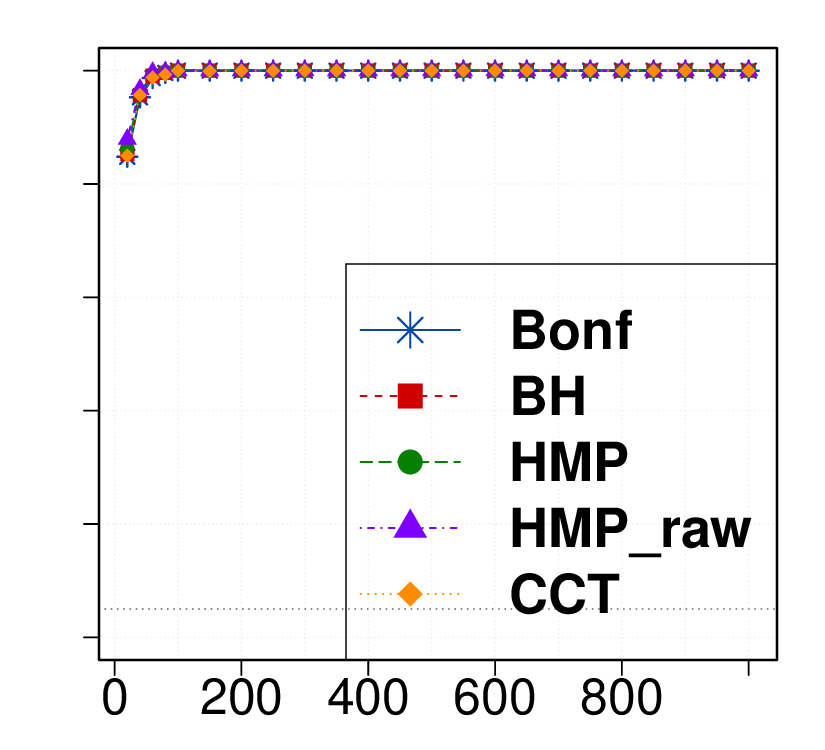}
        \hspace{-1 cm}
        \vspace{-0.2in}
            \caption*{\small{(b) $SNR=0.5$, with $m=1,5,20$ } }
            
        \hspace{-1 cm}    
             \includegraphics[width=5.5cm, height=5cm]{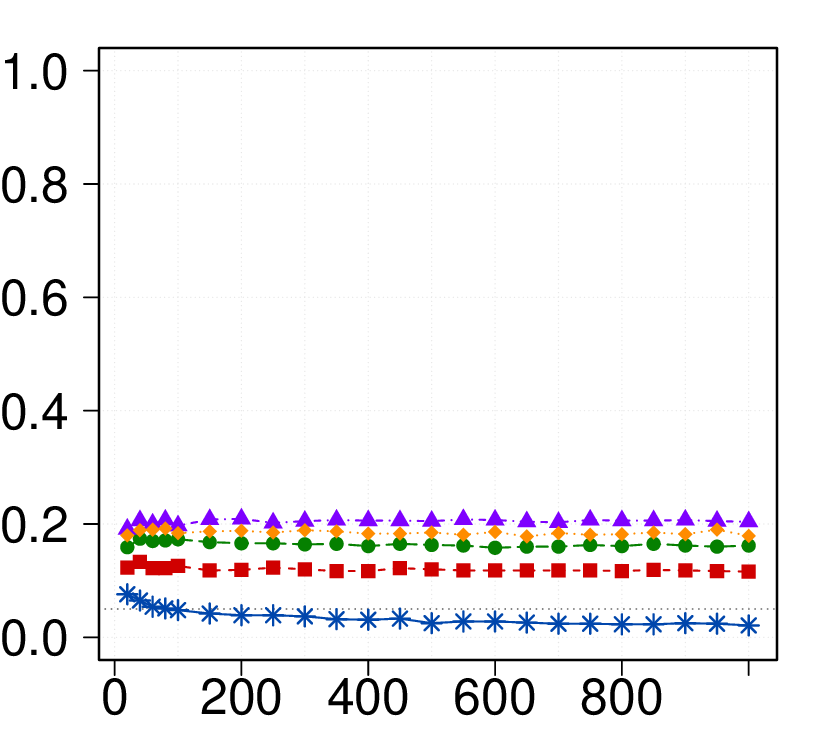}
        \hspace{-1 cm}
            \includegraphics[width=5.5cm, height=5cm]{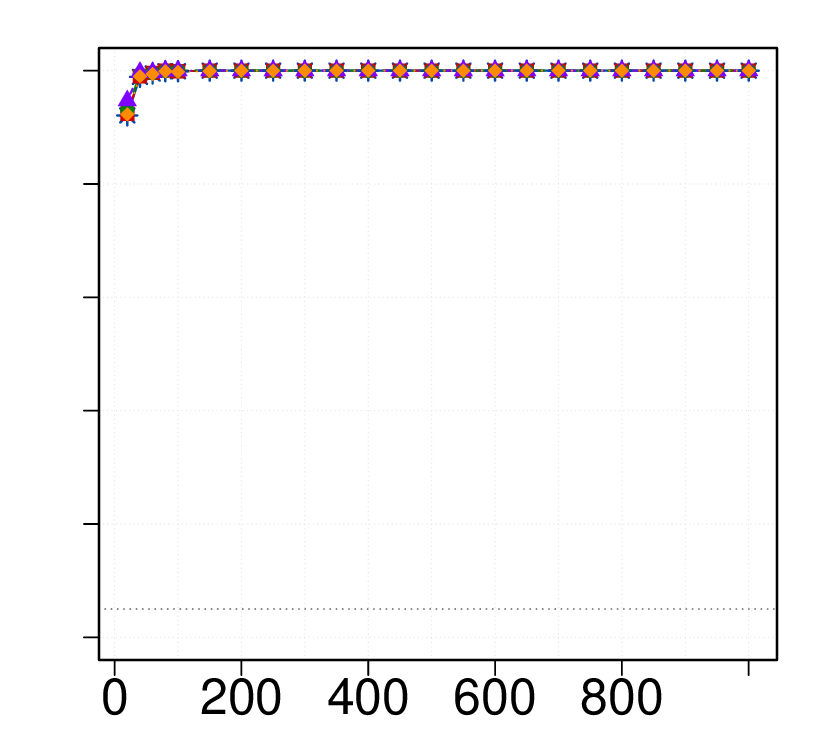}
        \hspace{-1 cm}
            \includegraphics[width=5.5cm, height=5cm]{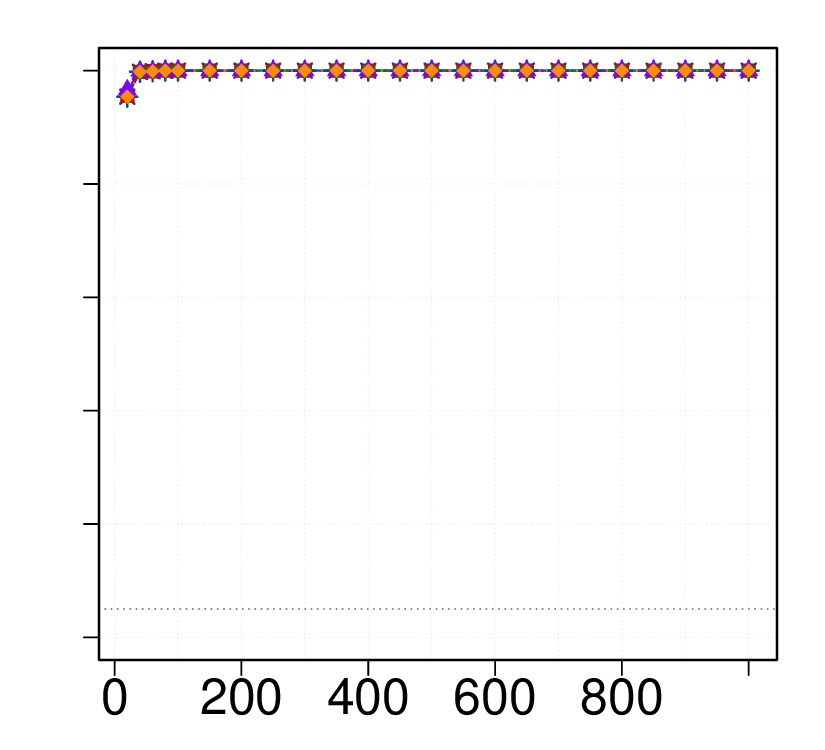}
        \hspace{-1 cm}
            \vspace{-0.2in}
            \caption*{\small{(c) $SNR=1.5$, with $m=1,5,20$ } }
            
            \caption{\small{
             Raw empirical rejection rates of the RP-Bonf, RP-BH, RP-HMP, RP-HMP-raw, and RP-CCT methods with the standard CUSUM test for various choices of number $k$ of random projections in the x-axis. 
            The data-generating process follows (\ref{eq:model for fd}), (\ref{eq:data generating process}), and (\ref{break function}), where the standard deviation $\sigma_{g}$ follows \textit{Setting 2}.
            The change point location is set at $\theta=0.25$.
            The empirical rejection rate is based on 1000 simulations.
            }}
            \label{fig:  tuning num_rps_sigma2}
    \end{figure} 
\newpage
  \begin{figure} [H]
        \centering
            \includegraphics[width=5.5cm, height=5cm]{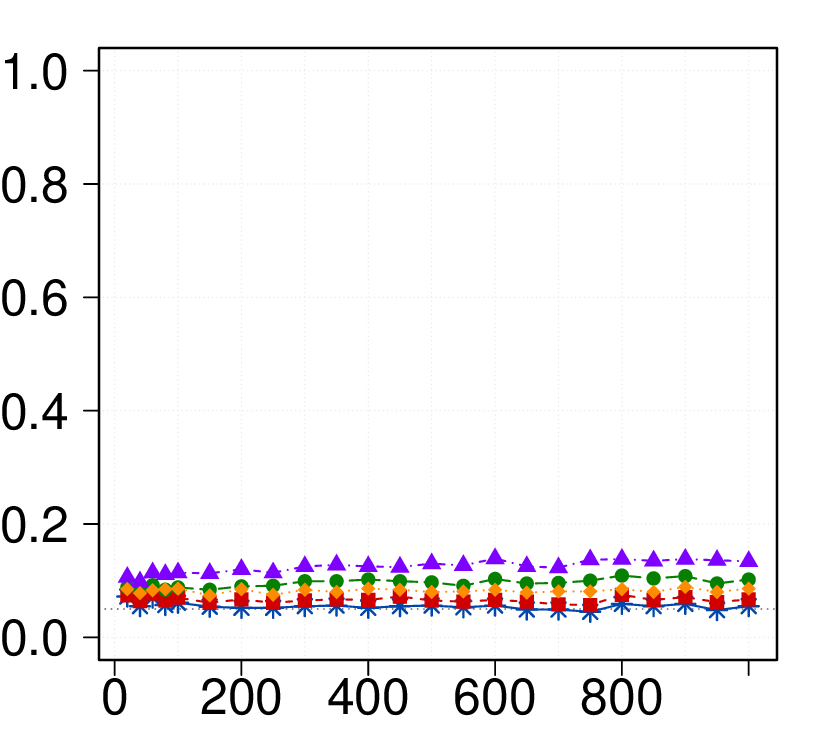}
        \vspace{-0.2in}
            \caption*{\small{(a) $SNR=0$} }

        \hspace{-1 cm}
            \includegraphics[width=5.5cm, height=5cm]{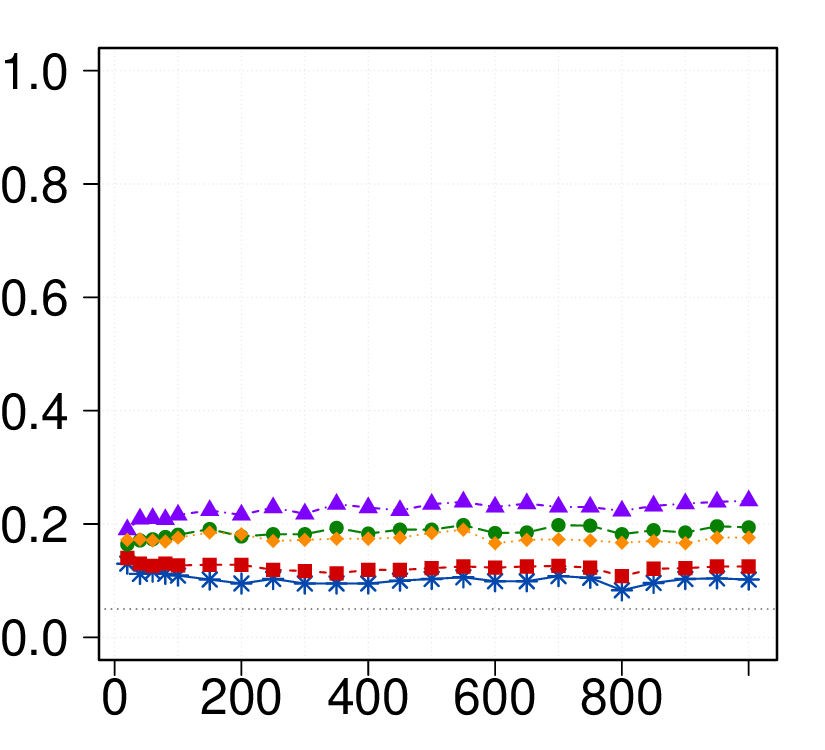}
        \hspace{-1 cm}
            \includegraphics[width=5.5cm, height=5cm]{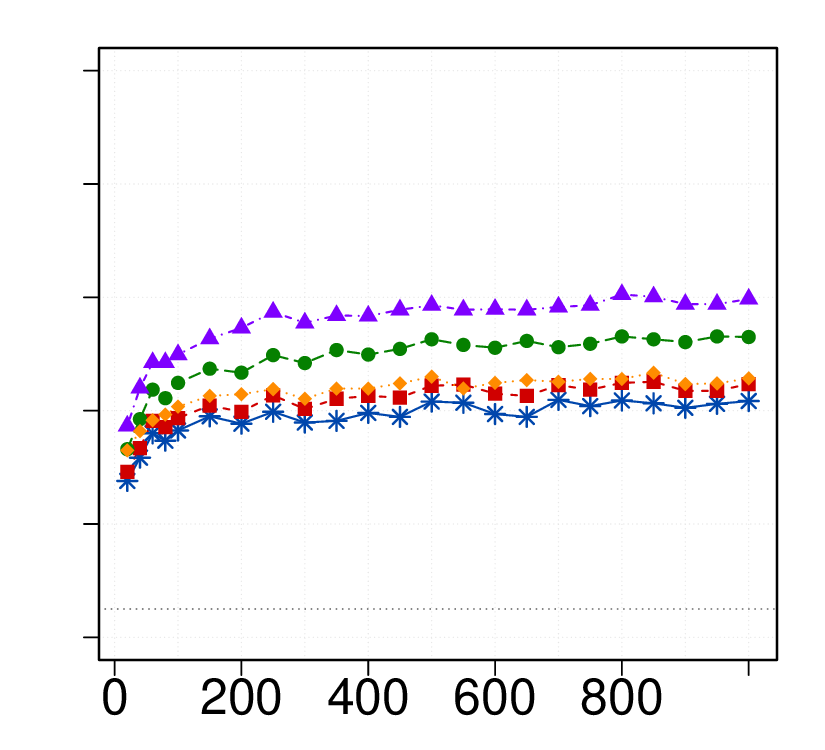}
        \hspace{-1 cm}
            \includegraphics[width=5.5cm, height=5cm]    {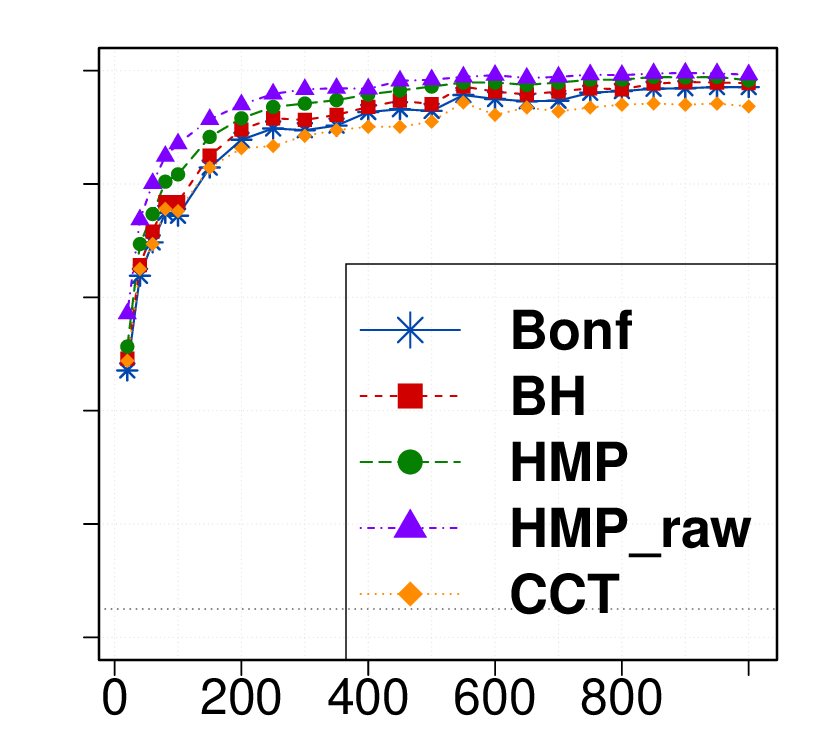}
        \hspace{-1 cm}
        \vspace{-0.2in}
            \caption*{\small{(b) $SNR=0.5$, with $m=1,5,20$ } }
            
        \hspace{-1 cm}    
             \includegraphics[width=5.5cm, height=5cm]{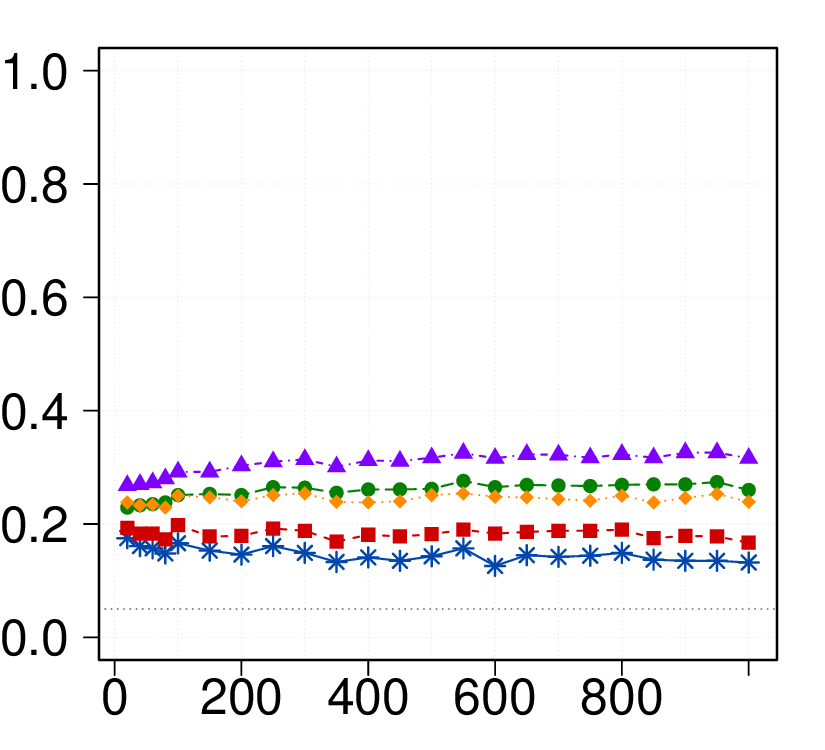}
        \hspace{-1 cm}
            \includegraphics[width=5.5cm, height=5cm]{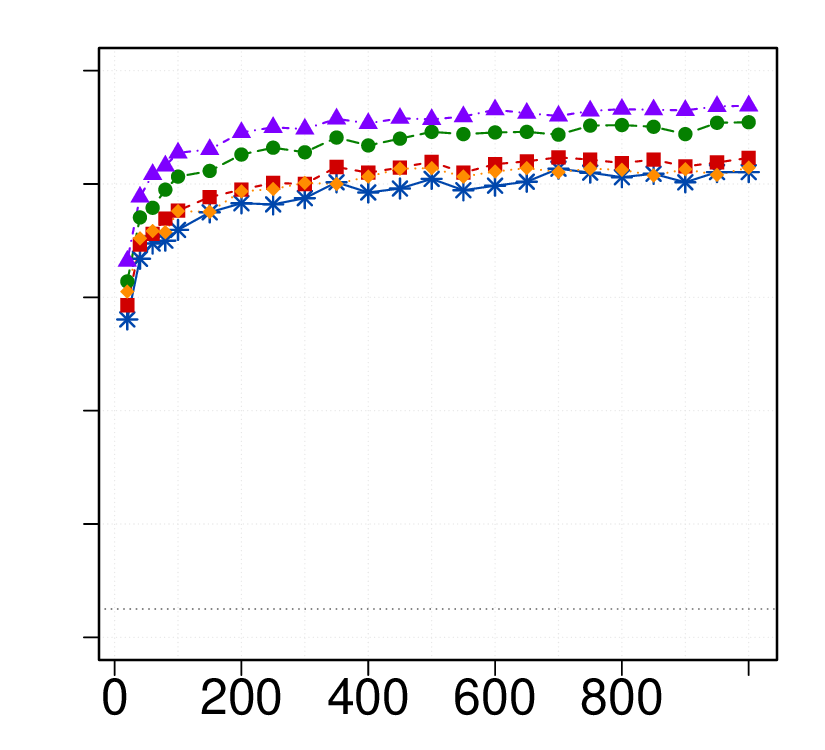}
        \hspace{-1 cm}
            \includegraphics[width=5.5cm, height=5cm]    {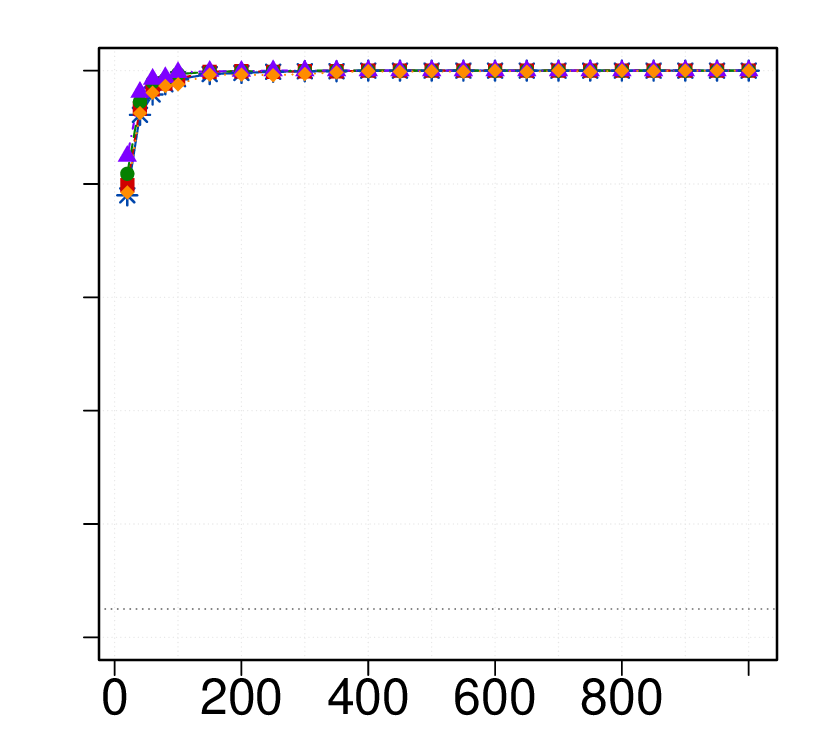}
        \hspace{-1 cm}
            \vspace{-0.2in}
            \caption*{\small{(c) $SNR=1.5$, with $m=1,5,20$ } }
            
            \caption{\small{
             Raw empirical rejection rates of the RP-Bonf, RP-BH, RP-HMP, RP-HMP-raw, and RP-CCT methods with the standard CUSUM test for various choices of number $k$ of random projections in the x-axis. 
            The data-generating process follows (\ref{eq:model for fd}), (\ref{eq:data generating process}), and (\ref{break function}), where the standard deviation $\sigma_{g}$ follows \textit{Setting 3}.
            The change point location is set at $\theta=0.25$.
            The empirical rejection rate is based on 1000 simulations.
            }}
            \label{fig:  tuning num_rps_sigma3}
    \end{figure} 
\newpage
  \begin{figure} [H]
        \centering
            \includegraphics[width=5.5cm, height=5cm]{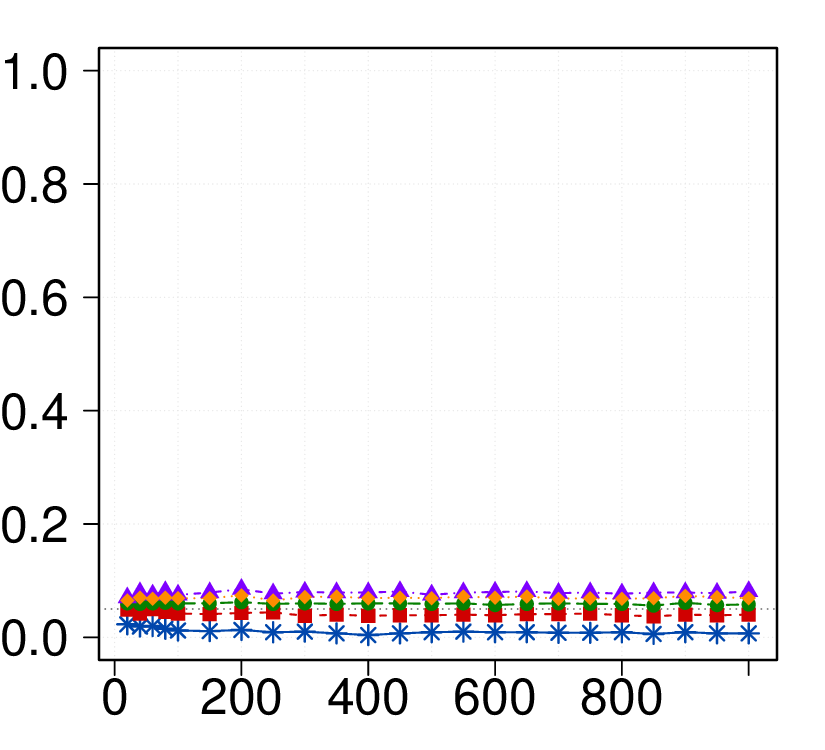}
        \vspace{-0.2in}
        \caption*{\small{(a) $SNR=0$} }

        \hspace{-1 cm}
            \includegraphics[width=5.5cm, height=5cm]{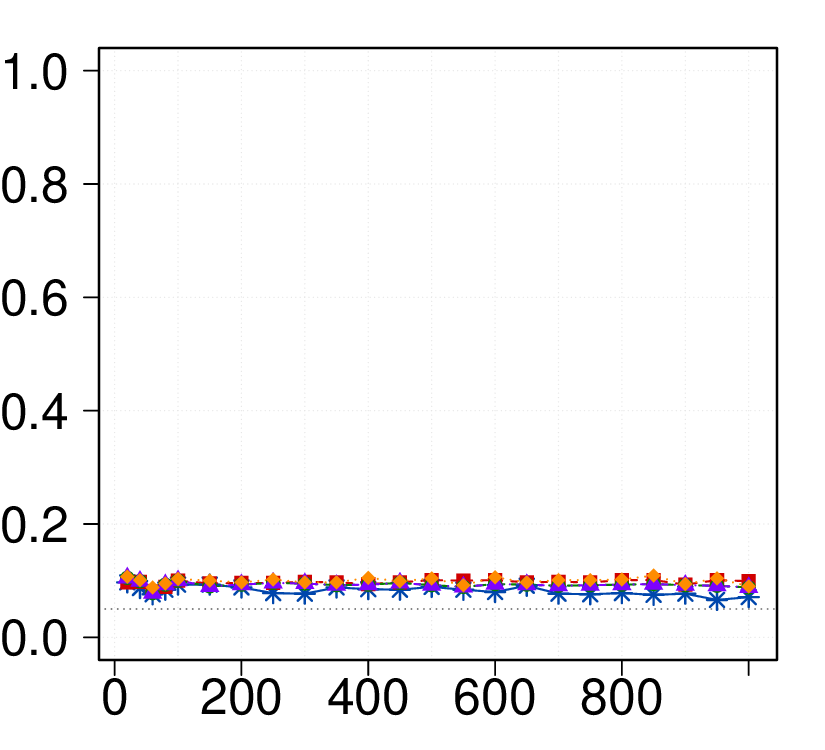}
        \hspace{-1 cm}
            \includegraphics[width=5.5cm, height=5cm]{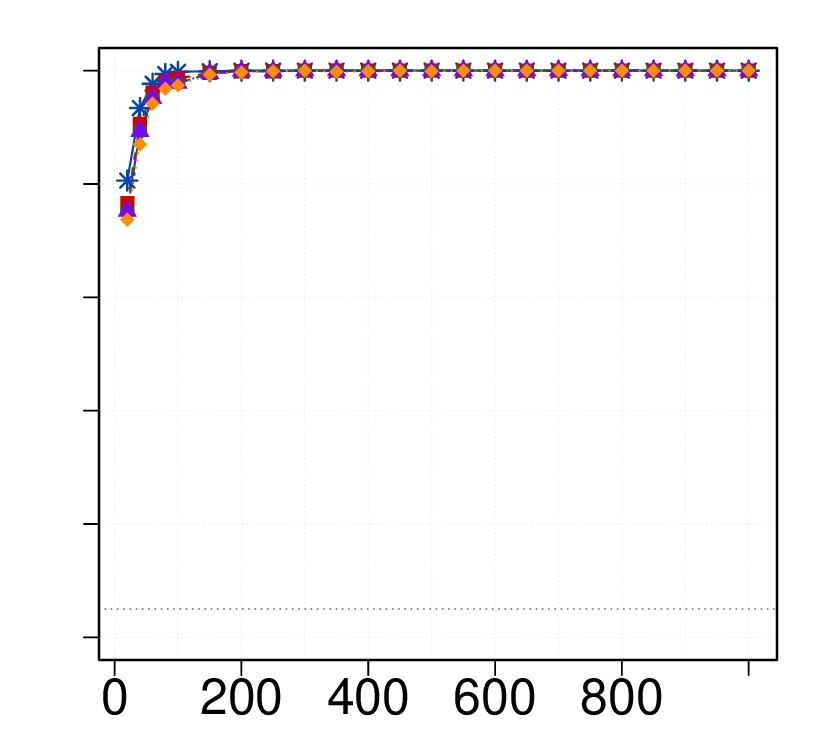}
        \hspace{-1 cm}
            \includegraphics[width=5.5cm, height=5cm]    {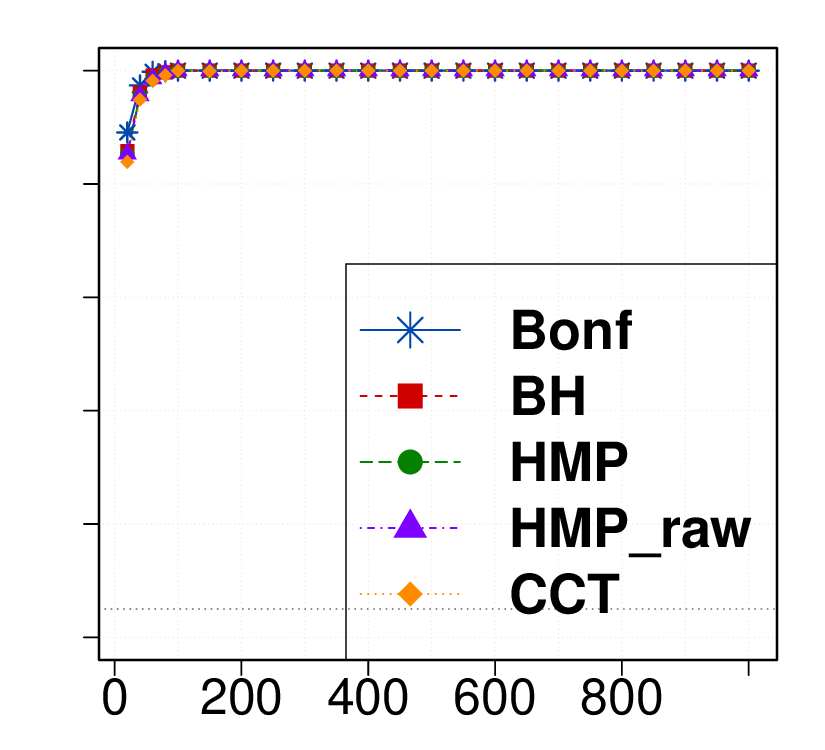}
        \hspace{-1 cm}
        \vspace{-0.2in}
            \caption*{\small{(b) $SNR=0.5$, with $m=1,5,20$  } }
            
        \hspace{-1 cm}    
             \includegraphics[width=5.5cm, height=5cm]{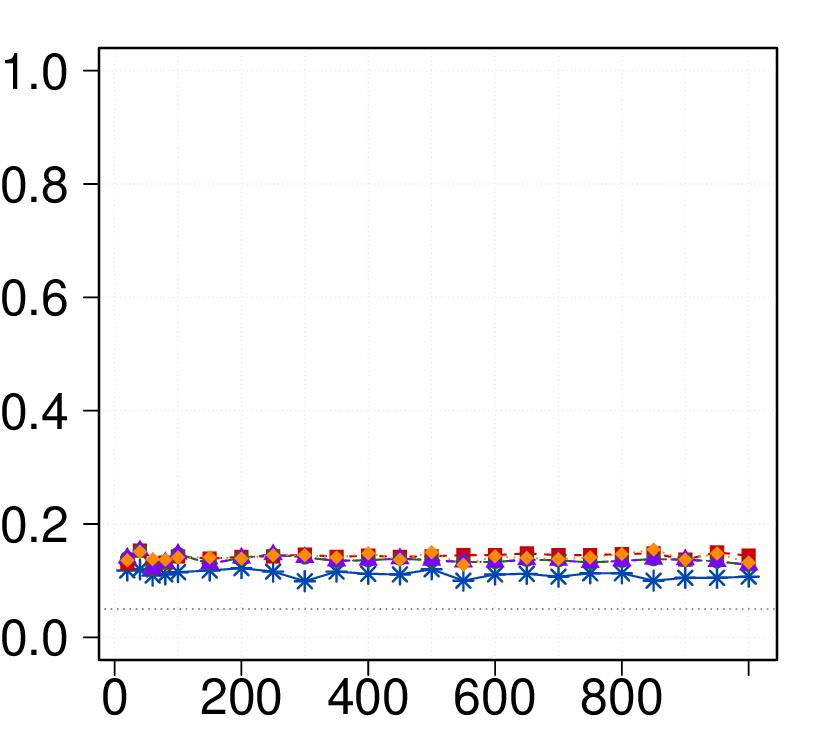}
        \hspace{-1 cm}
            \includegraphics[width=5.5cm, height=5cm]{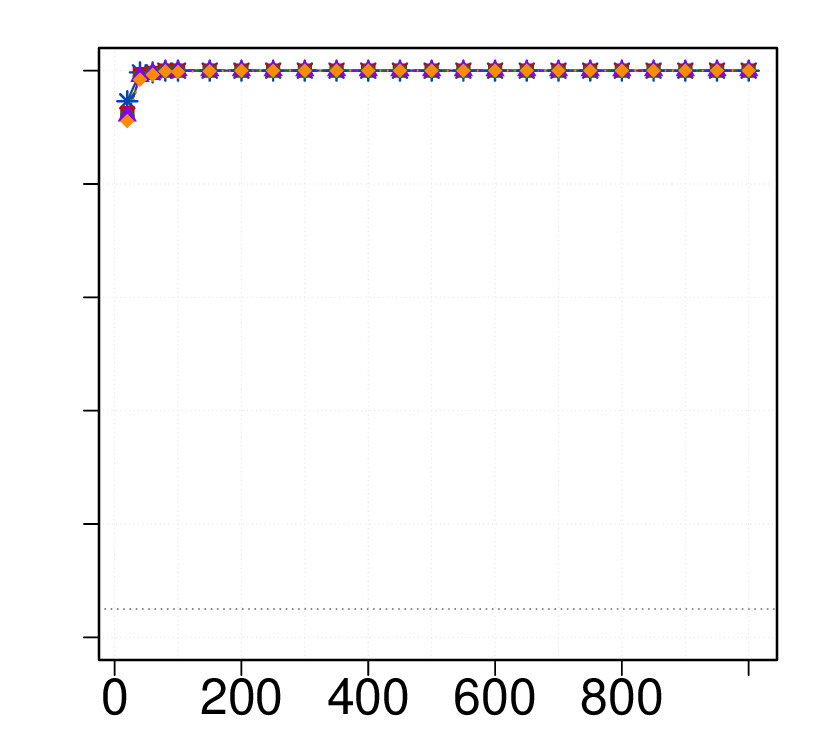}
        \hspace{-1 cm}
            \includegraphics[width=5.5cm, height=5cm]    {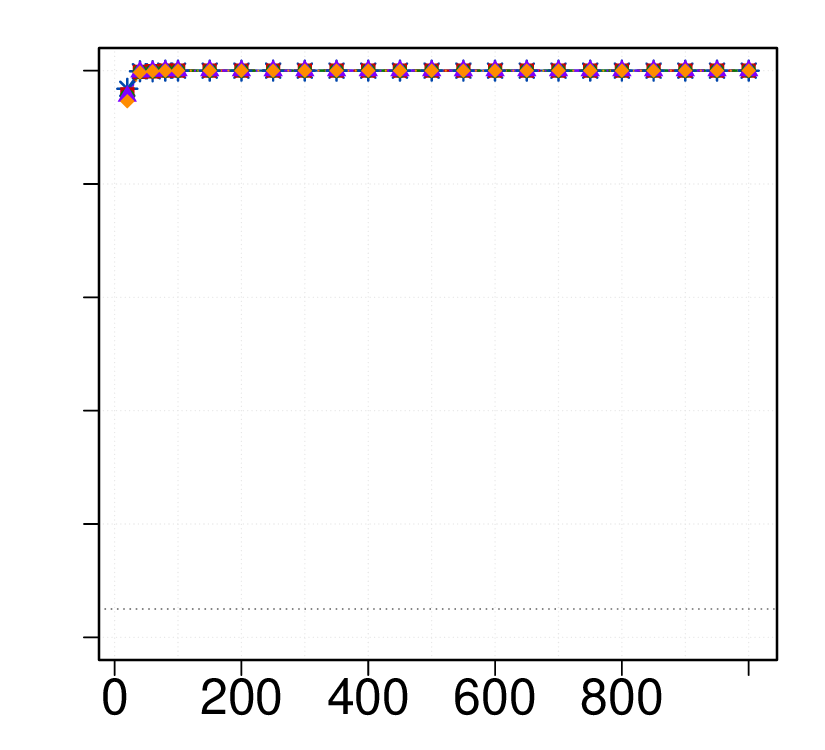}
        \hspace{-1 cm}
            \vspace{-0.2in}
            \caption*{\small{(c) $SNR=1.5$, with $m=1,5,20$  } }
            
            \caption{\small{
             Adjusted empirical rejection rates of the RP-Bonf, RP-BH, RP-HMP, RP-HMP-raw, and RP-CCT methods with the standard CUSUM test for various choices of number $k$ of random projections in the x-axis. 
            The data-generating process follows (\ref{eq:model for fd}), (\ref{eq:data generating process}), and (\ref{break function}), where the standard deviation $\sigma_{g}$ follows \textit{Setting 2}.
            The change point location is set at $\theta=0.25$.
            The empirical rejection rate is based on 1000 simulations.
            }}
            \label{fig:  tuning num_rps_sigma2(adj)}
    \end{figure} 
\newpage
  \begin{figure} [H]
        \centering
            \includegraphics[width=5.5cm, height=5cm]{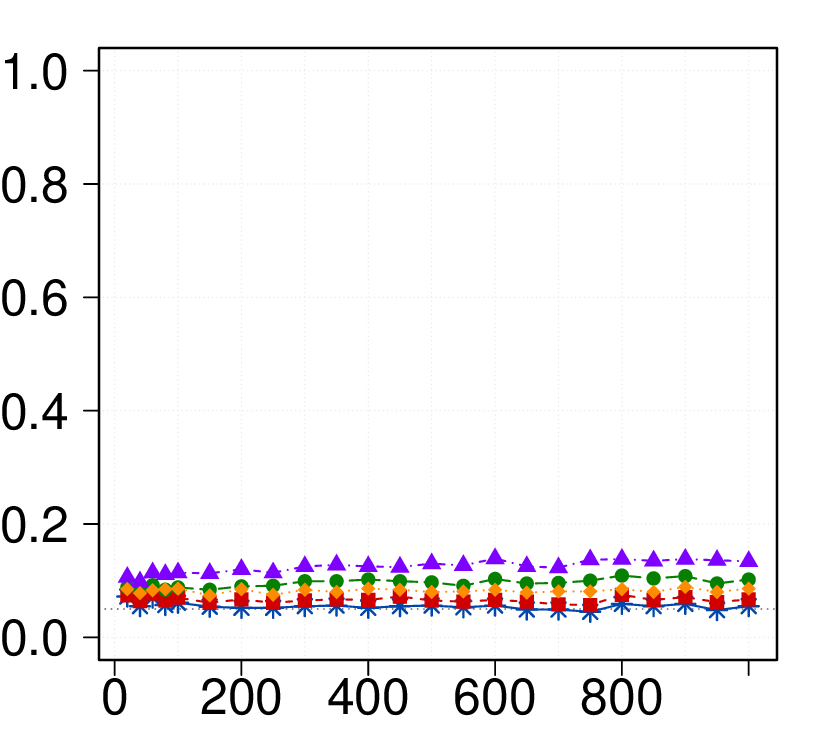}
        \vspace{-0.2in}
            \caption*{\small{(a) $SNR=0$} }

        \hspace{-1 cm}
            \includegraphics[width=5.5cm, height=5cm]{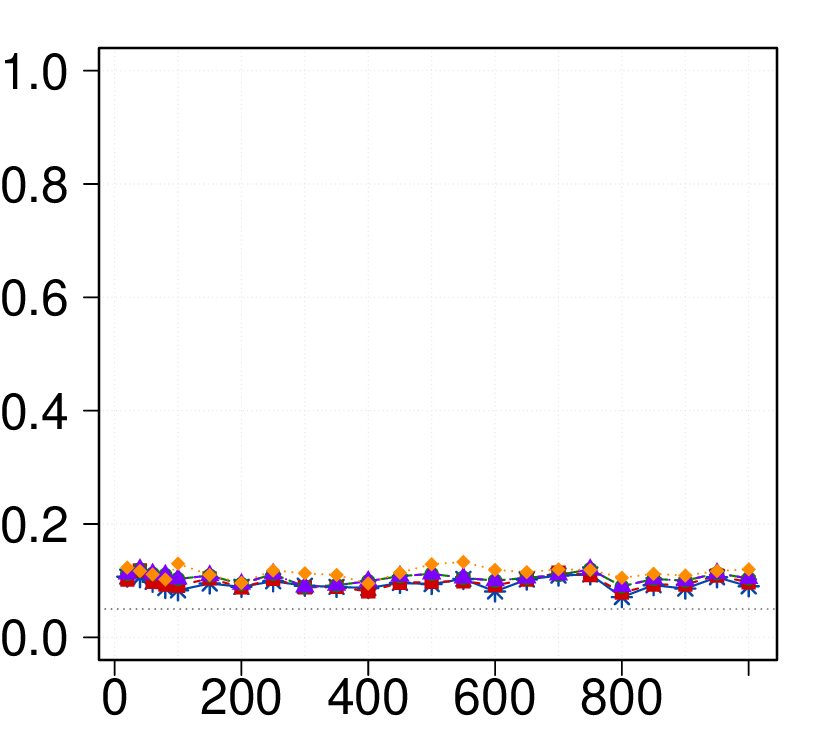}
        \hspace{-1 cm}
            \includegraphics[width=5.5cm, height=5cm]{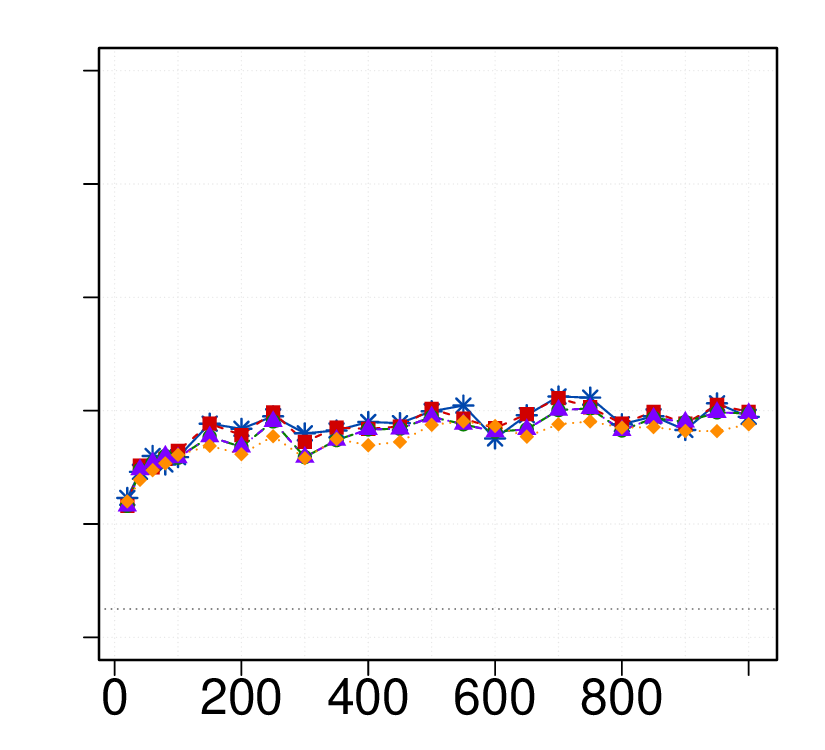}
        \hspace{-1 cm}
            \includegraphics[width=5.5cm, height=5cm]    {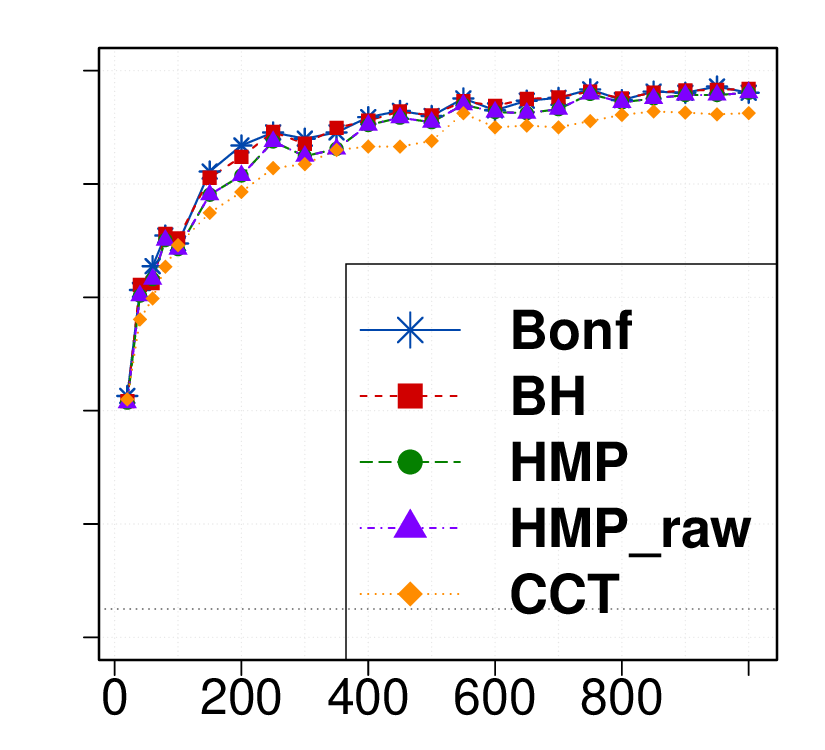}
        \hspace{-1 cm}
        \vspace{-0.2in}
            \caption*{\small{(b) $SNR=0.5$, with $m=1,5,20$  } }
            
        \hspace{-1 cm}    
             \includegraphics[width=5.5cm, height=5cm]{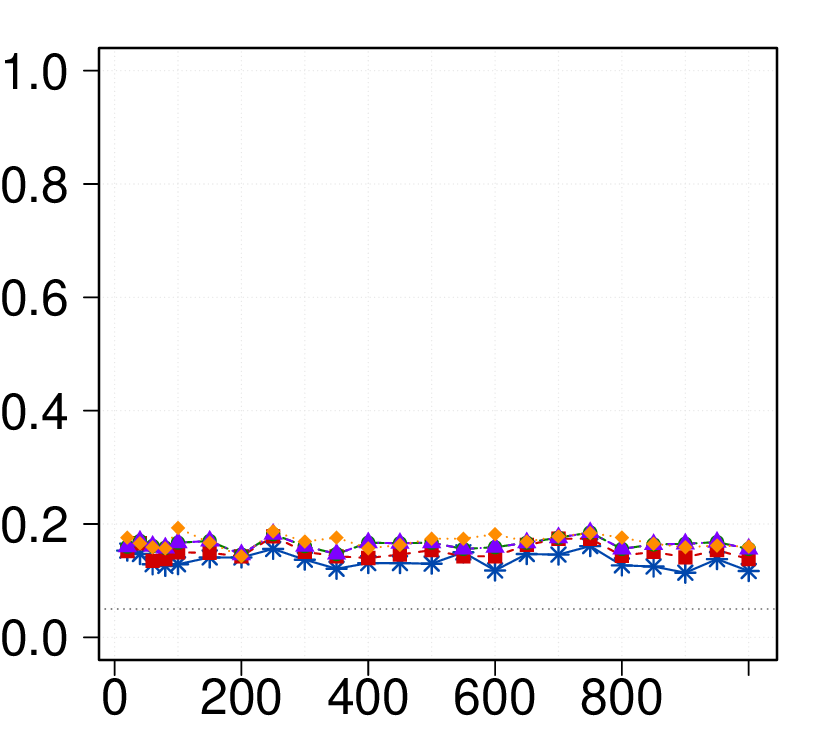}
        \hspace{-1 cm}
            \includegraphics[width=5.5cm, height=5cm]{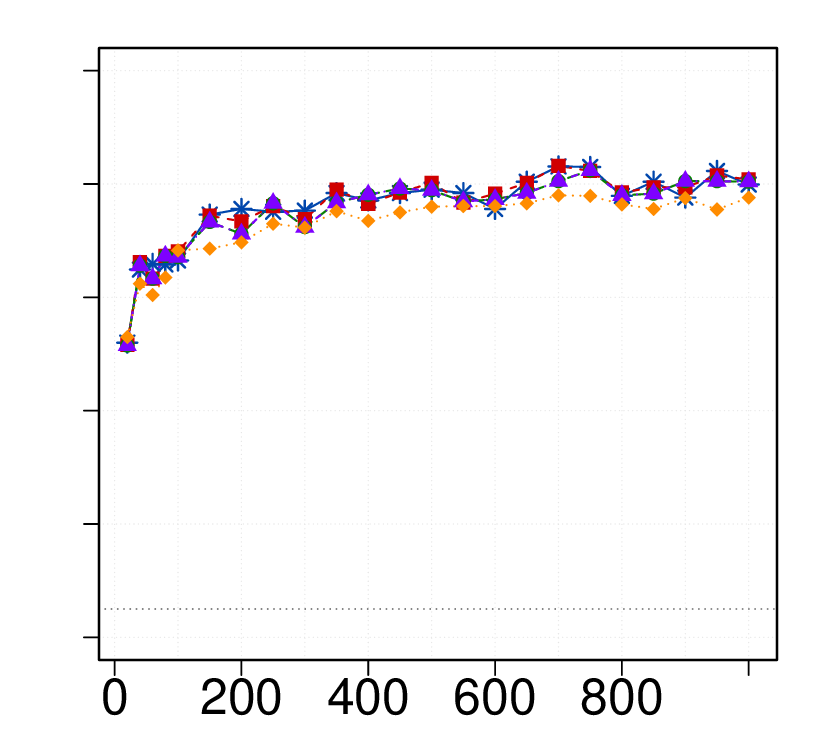}
        \hspace{-1 cm}
            \includegraphics[width=5.5cm, height=5cm]    {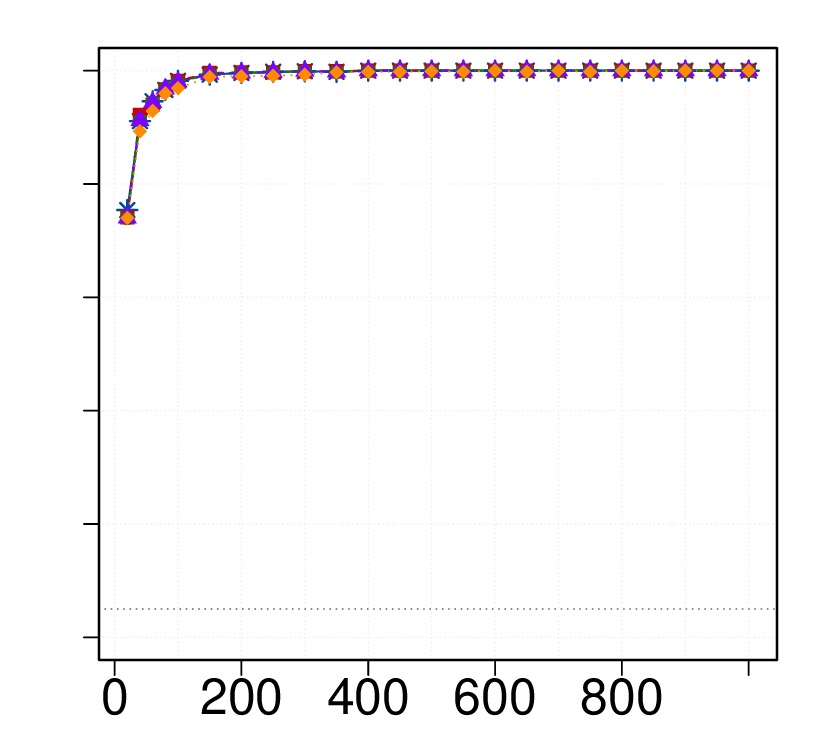}
        \hspace{-1 cm}
            \vspace{-0.2in}
            \caption*{\small{(c) $SNR=1.5$, with $m=1,5,20$  } }
            
            \caption{\small{
             Adjusted empirical rejection rates of the RP-Bonf, RP-BH, RP-HMP, RP-HMP-raw, and RP-CCT methods with the standard CUSUM test for various choices of number $k$ of random projections in the x-axis. 
            The data-generating process follows (\ref{eq:model for fd}), (\ref{eq:data generating process}), and (\ref{break function}), where the standard deviation $\sigma_{g}$ follows \textit{Setting 3}.
            The change point location is set at $\theta=0.25$.
            The empirical rejection rate is based on 1000 simulations.
            }}
            \label{fig:  tuning num_rps_sigma3(adj)}
    \end{figure}

\clearpage 
\section{Comparison of different methods using weighted CUSUM}
\label{subsec: weightedCUSUM}
\begin{figure} [H]
            \centering
            \hspace{-1 cm} 
            \includegraphics[width=5.5cm, height=5cm]{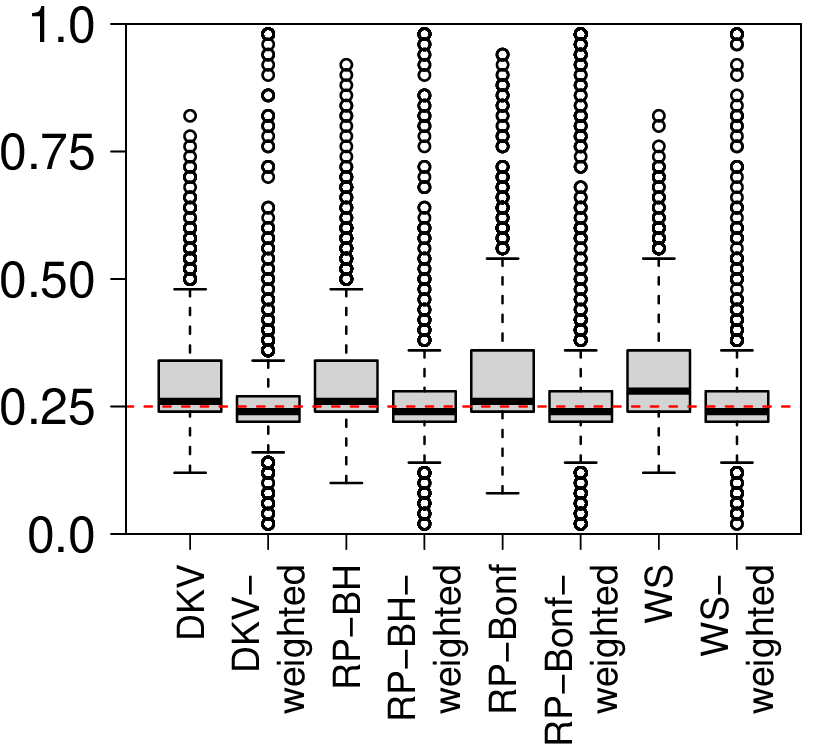}
            \hspace{-1 cm} 
            \includegraphics[width=5.5cm, height=5cm]{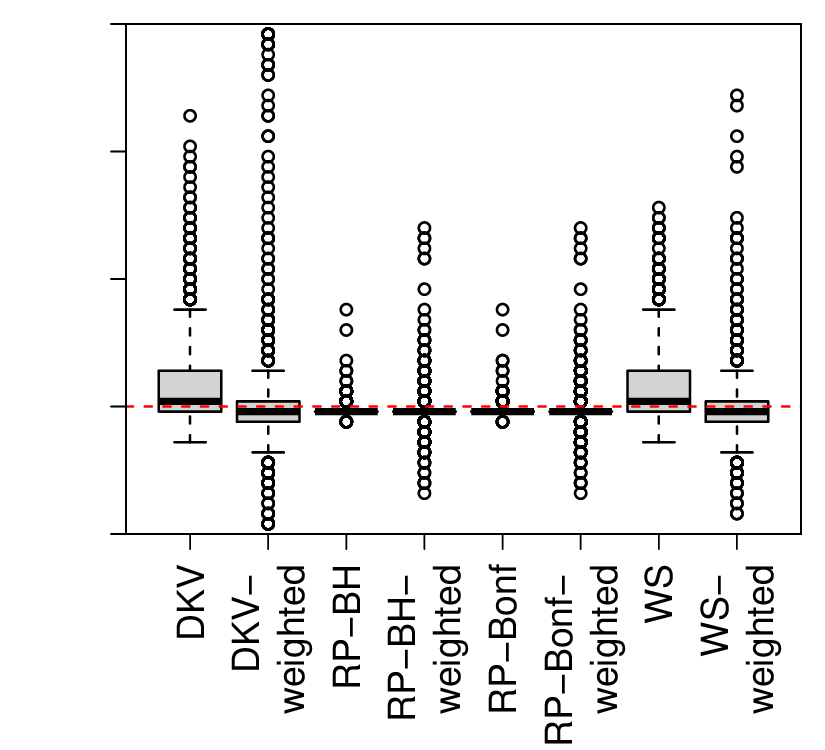}
            \hspace{-1 cm} 
            \includegraphics[width=5.5cm, height=5cm]{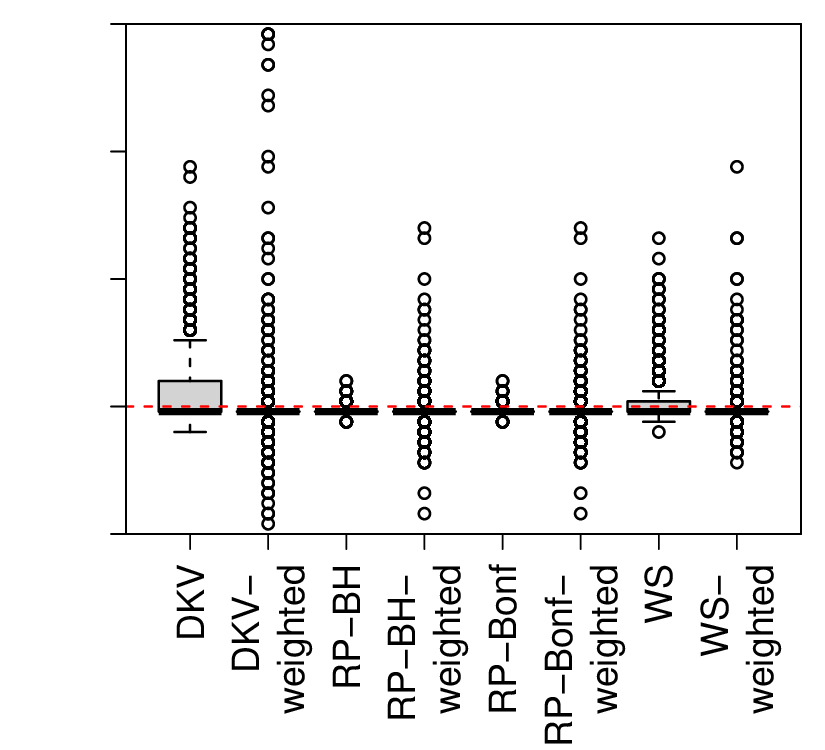}
            \hspace{-1 cm} 
            \vspace{-0.2in}
            \caption*{\small{(a) \textit{Setting 1}, with $m=1,5,20$ } }

            \centering
            \hspace{-1 cm} 
            \includegraphics[width=5.5cm, height=5cm]{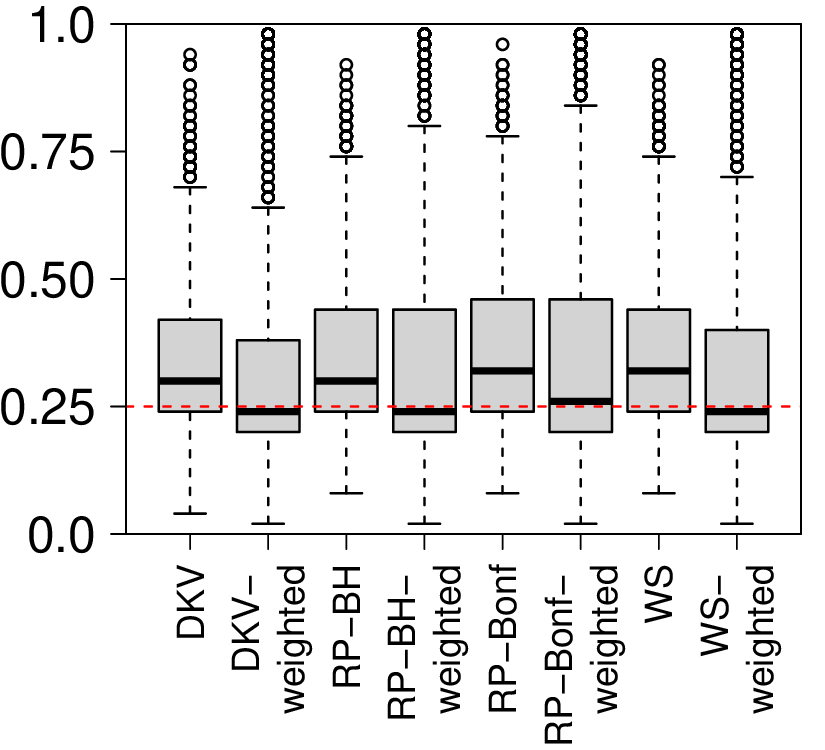}
            \hspace{-1 cm} 
            \includegraphics[width=5.5cm, height=5cm]{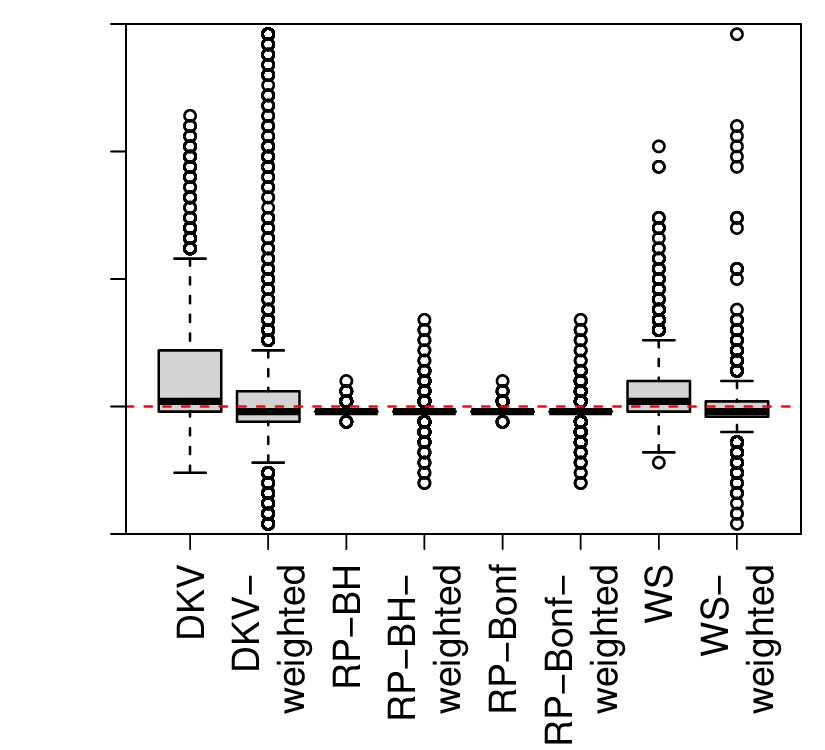}
            \hspace{-1 cm} 
            \includegraphics[width=5.5cm, height=5cm]{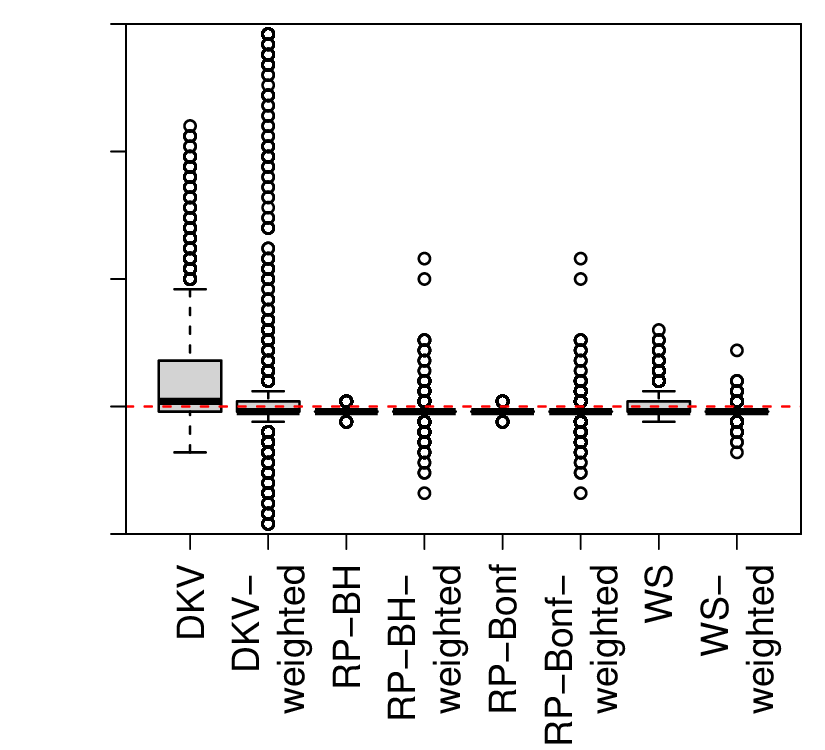}
            \hspace{-1 cm} 
            \vspace{-0.2in}
            \caption*{\small{(b) \textit{Setting 2}, with $m=1,5,20$ } }

            \centering
            \hspace{-1 cm} 
            \includegraphics[width=5.5cm, height=5cm]{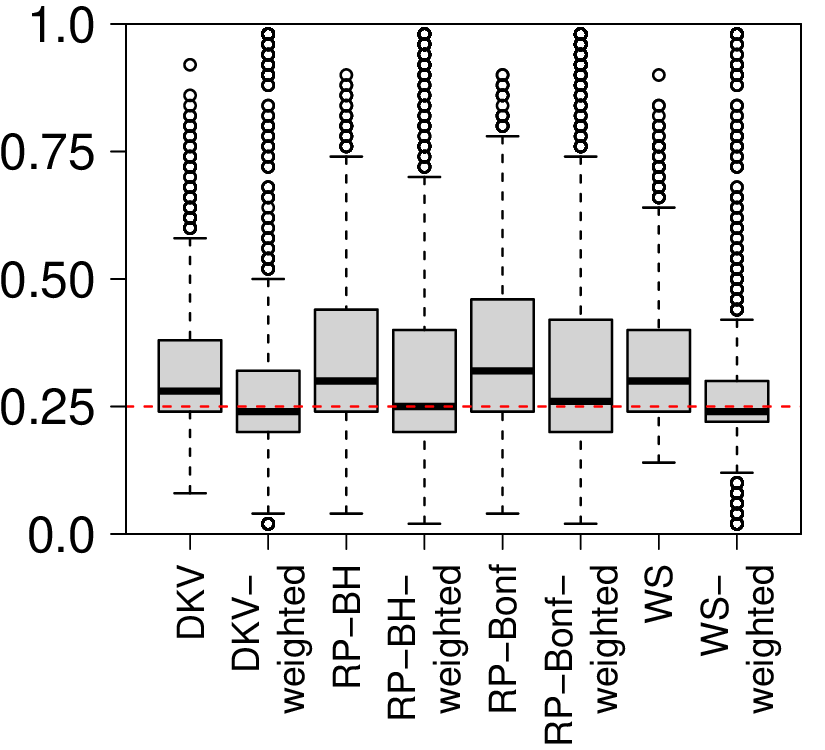}
            \hspace{-1 cm} 
            \includegraphics[width=5.5cm, height=5cm]{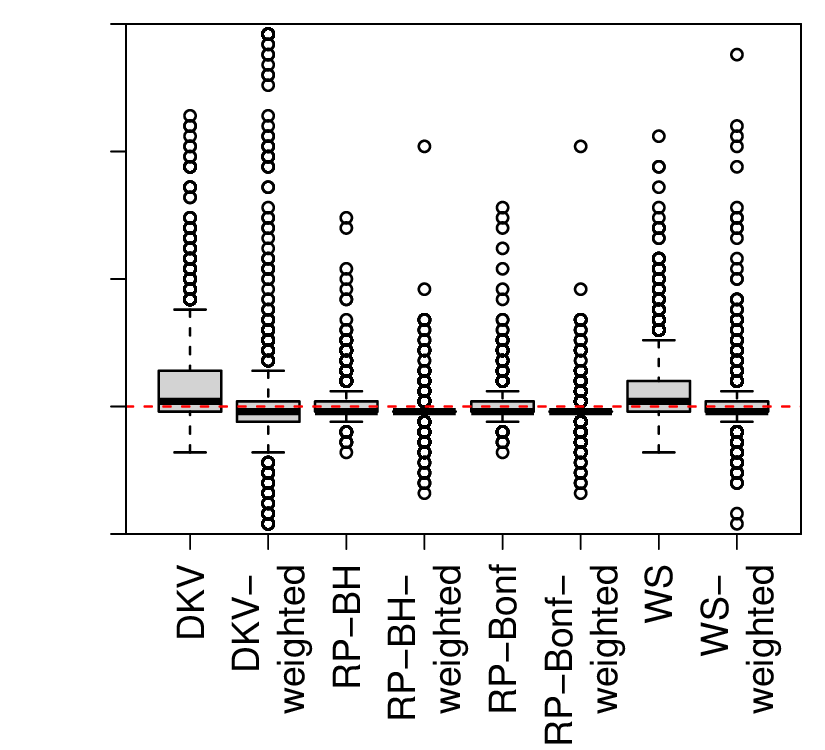}
            \hspace{-1 cm} 
            \includegraphics[width=5.5cm, height=5cm]{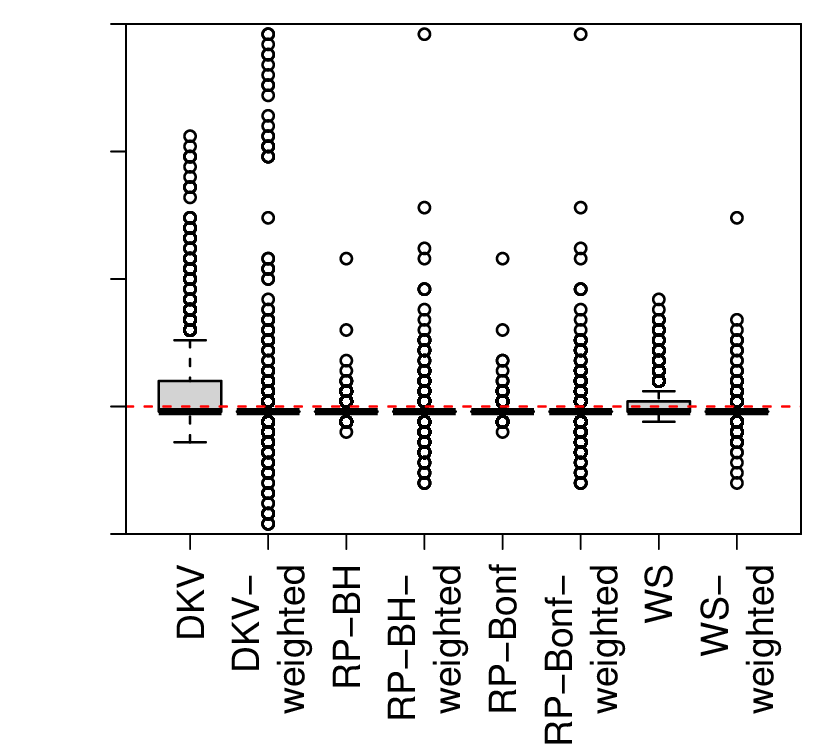}
            \hspace{-1 cm} 
            \vspace{-0.2in}
            \caption*{\small{(c) \textit{Setting 3}, with $m=1,5,20$ } }
            
            \caption{Estimated change point locations detected by different methods on 1000 simulations, using the standard CUSUM approaches or weighted CUSUM approaches with trimming choice of $\lfloor 
            n\tau \rfloor=1$. The data-generating process follows (\ref{eq:model for fd}), (\ref{eq:data generating process}), and (\ref{break function}), where the standard deviation $\sigma_{g}$ follows \textit{Settings 1--3}.  The change point location is set at $\theta = 0.25$. The magnitude of the break function is scaled by $SNR=0.5$.}
            \label{fig:  CP locations Auedata}
    \end{figure}

\section{The procedure in Subsection \ref{subsec:repeat RP method on one data} using RP-BH}\label{subsec:Repeat methods for data1-5}

\begin{figure}[H]
    \centering
    \includegraphics[width=\textwidth]{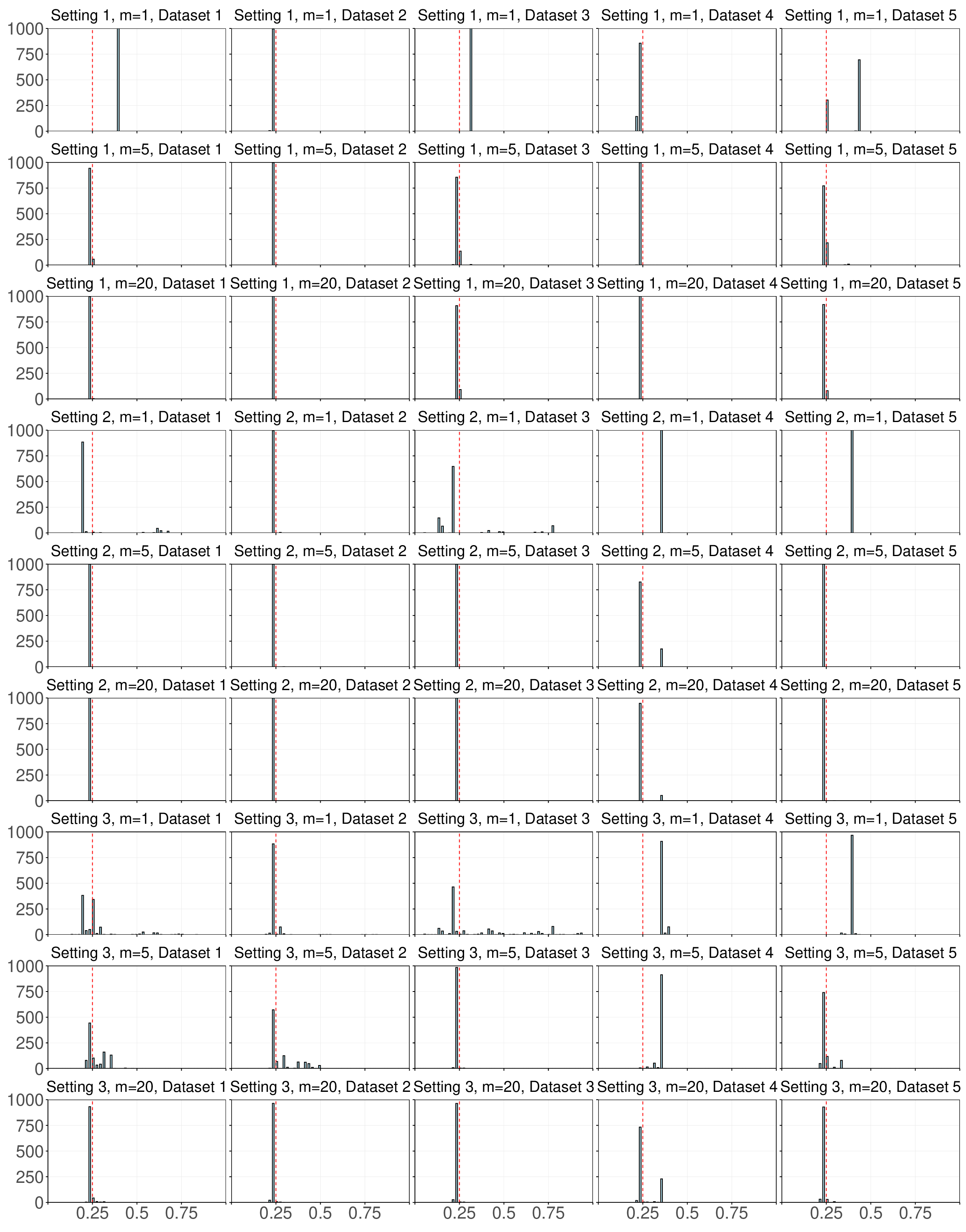}
    \caption{Frequency of estimated locations from repeating the RP-BH method on one dataset (Dataset 1 through Dataset 5) for 1000 times.            
            The data-generating process follows (\ref{eq:model for fd}), (\ref{eq:data generating process}), and (\ref{break function}), where the standard deviation $\sigma_{g}$ follows \textit{Settings 1--3}.  The change point location is set at $\theta = 0.25$. The magnitude of the break function is scaled by $SNR=0.5$. }
    \label{fig:hist_bh:data0}
\end{figure}

\begin{figure} [h!]
            \centering
            \hspace{-1 cm}  
            \includegraphics[width=5.5cm, height=5cm]{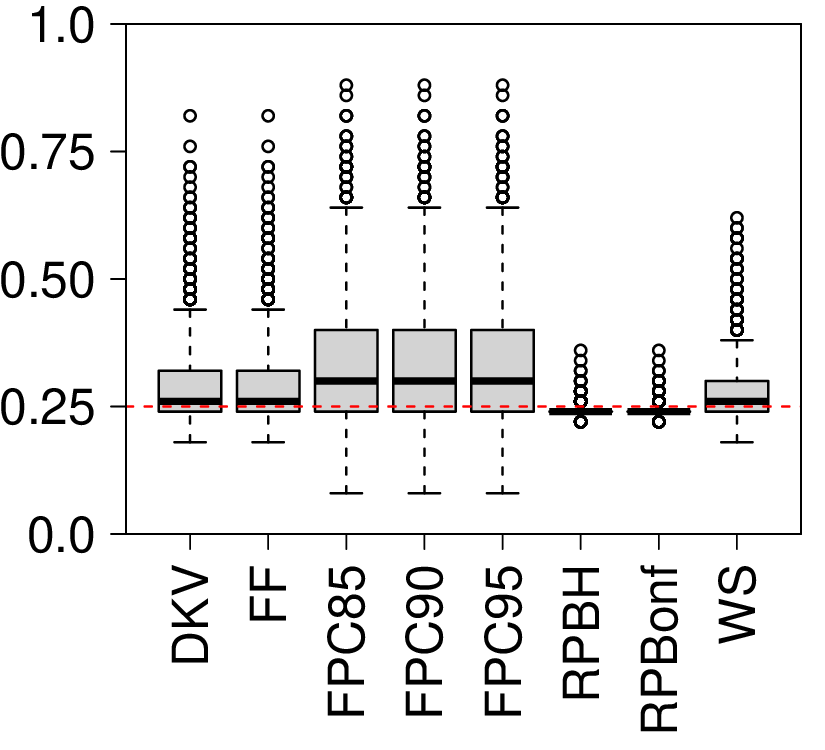}
            \hspace{-1 cm}  
            \includegraphics[width=5.5cm, height=5cm]{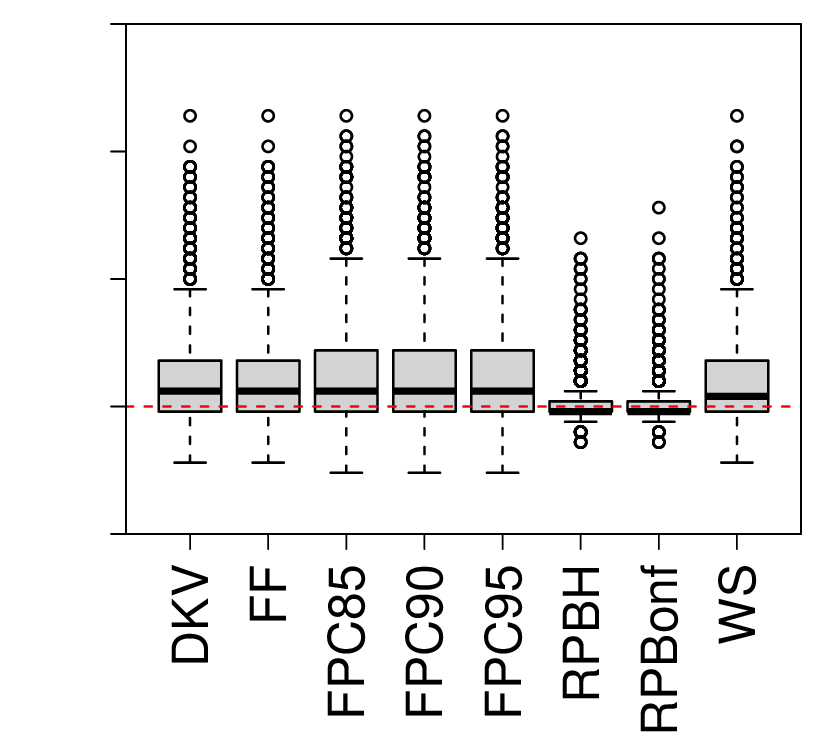}
            \hspace{-1 cm}  
            \includegraphics[width=5.5cm, height=5cm]{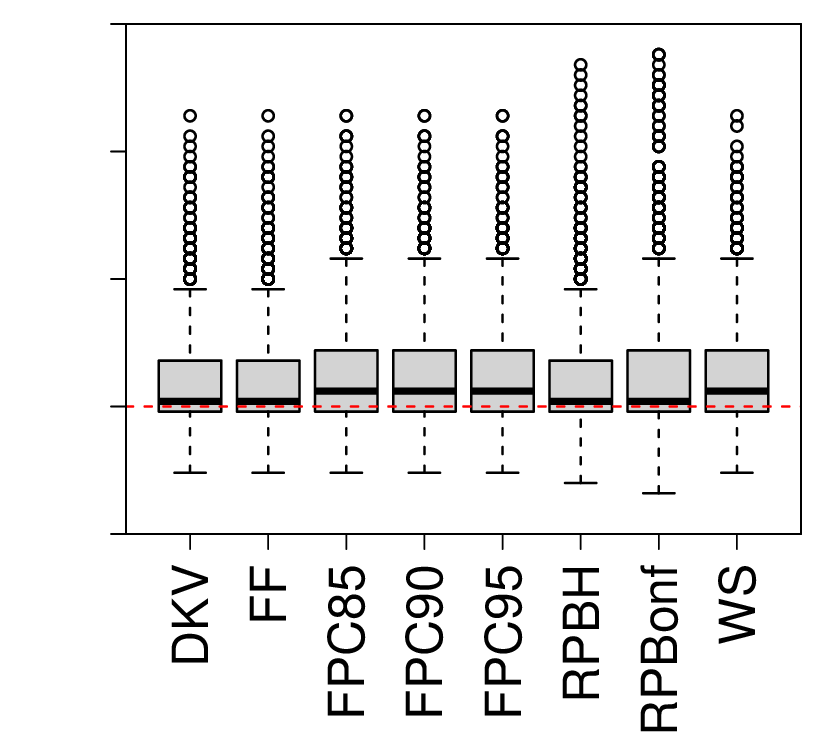}
            \hspace{-1 cm}  
            \vspace{-0.2in}
            \caption*{\small{(a) \textit{Setting 1}, with $m=5,10,23$ } }

            \centering
            \hspace{-1 cm}  
            \includegraphics[width=5.5cm, height=5cm]{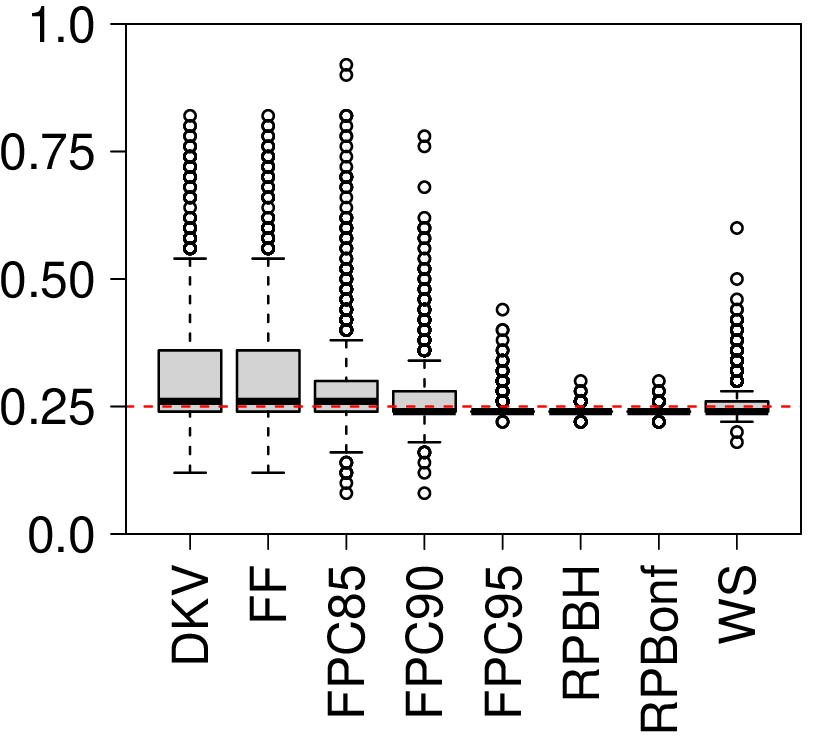}
            \hspace{-1 cm}  
            \includegraphics[width=5.5cm, height=5cm]{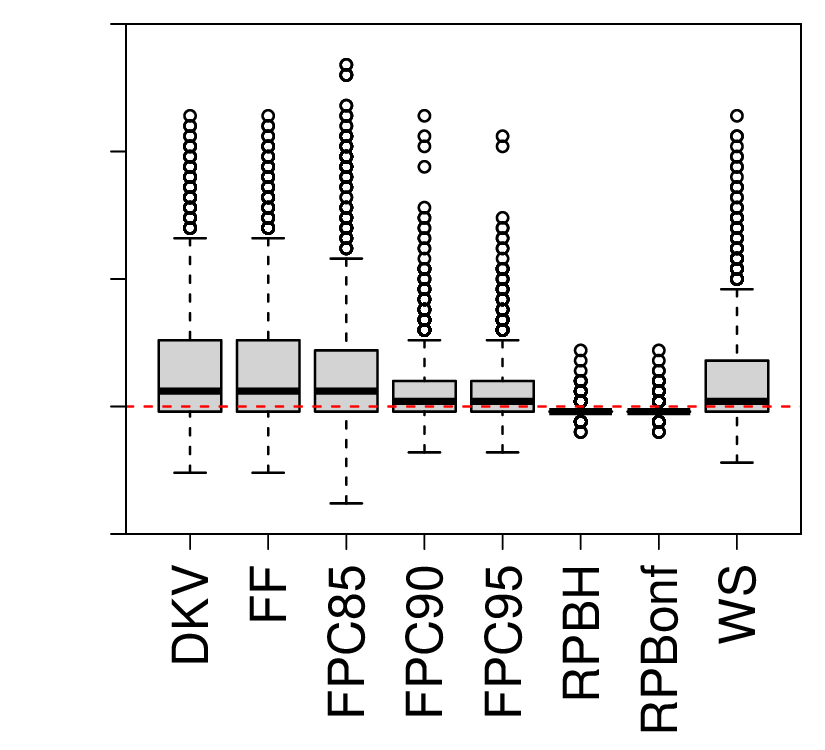}
            \hspace{-1 cm}  
            \includegraphics[width=5.5cm, height=5cm]{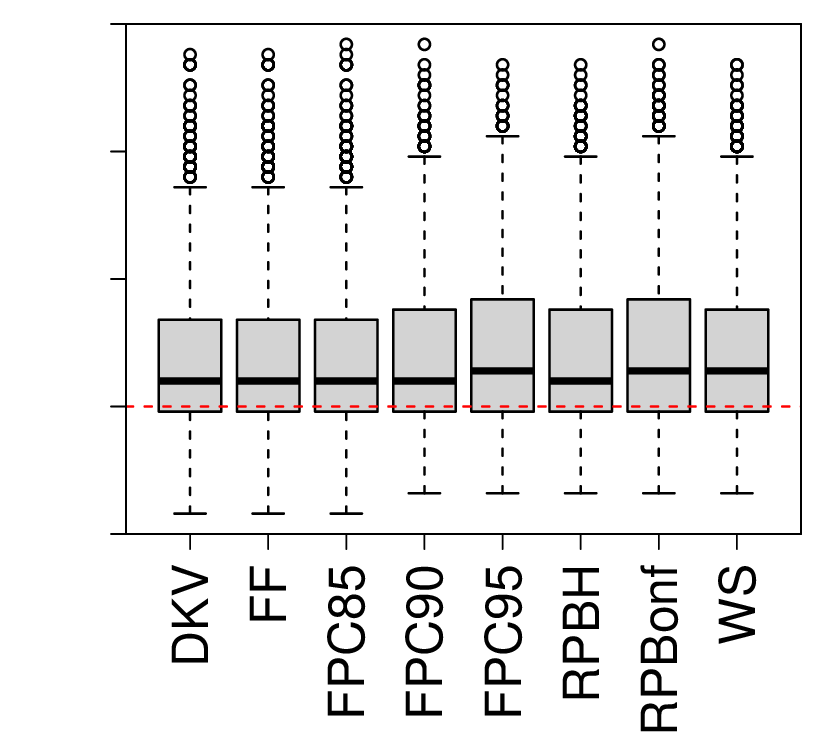}
            \hspace{-1 cm}  
            \vspace{-0.2in}

            \caption*{\small{(b) \textit{Setting 2}, with $m=5,10,23$ } }

            \centering
            \hspace{-1 cm}  
            \includegraphics[width=5.5cm, height=5cm]{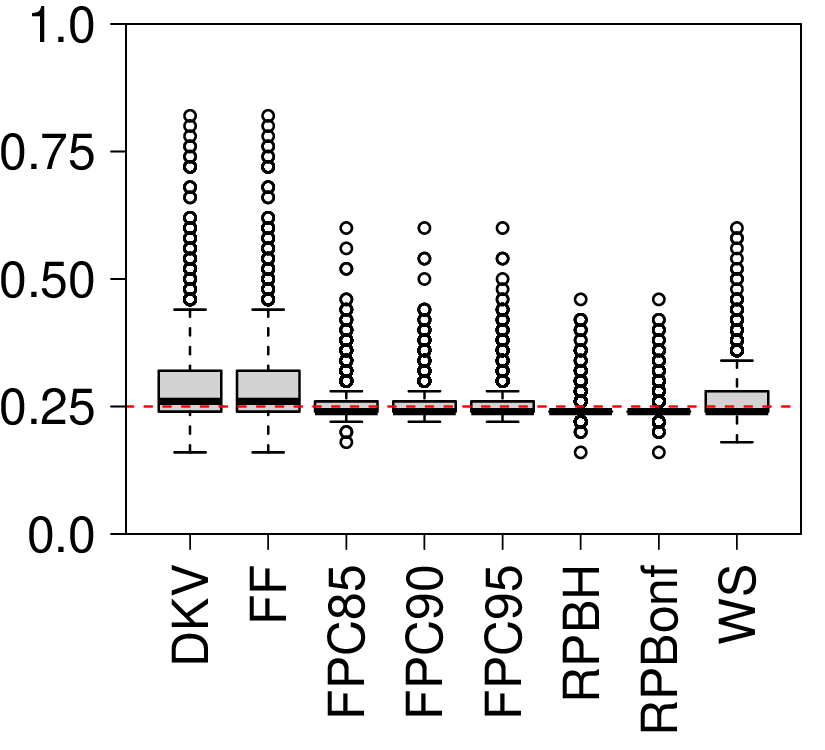}
            \hspace{-1 cm}  
            \includegraphics[width=5.5cm, height=5cm]{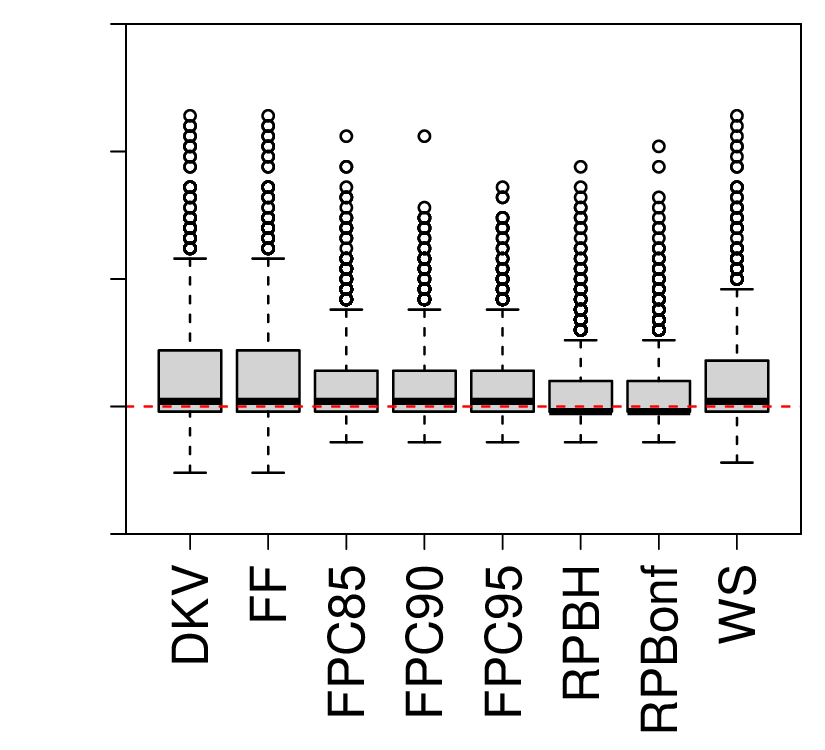}
            \hspace{-1 cm}  
            \includegraphics[width=5.5cm, height=5cm]{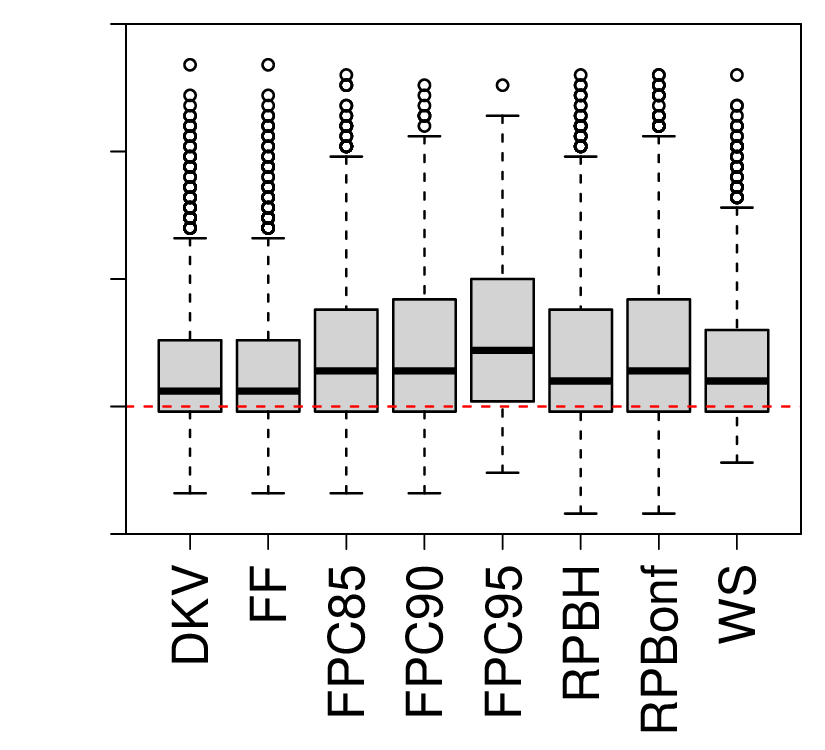}            
            \hspace{-1 cm}  
            \vspace{-0.2in}

            \caption*{\small{(c) \textit{Setting 3}, with $m=5,10,23$ } }
            
            \caption{Sparse functional setting in Section \ref{subsec: sparse functional setting}.   Estimated change point locations detected by different methods on 1000 simulations. The data-generating process follows (\ref{eq:model for fd}), (\ref{eq:data generating process}), and (\ref{bspline break}), where the standard deviation $\sigma_{g}$ follows \textit{Settings 1--3}.  The change point location is set at $\theta = 0.25$. The magnitude of the break function is scaled by $SNR=0.5$.}
            \label{fig:  CP locations Auedata standcusum: data0 bspline}
    \end{figure}

\begin{figure}[H]
    \centering
    \includegraphics[width=\textwidth]{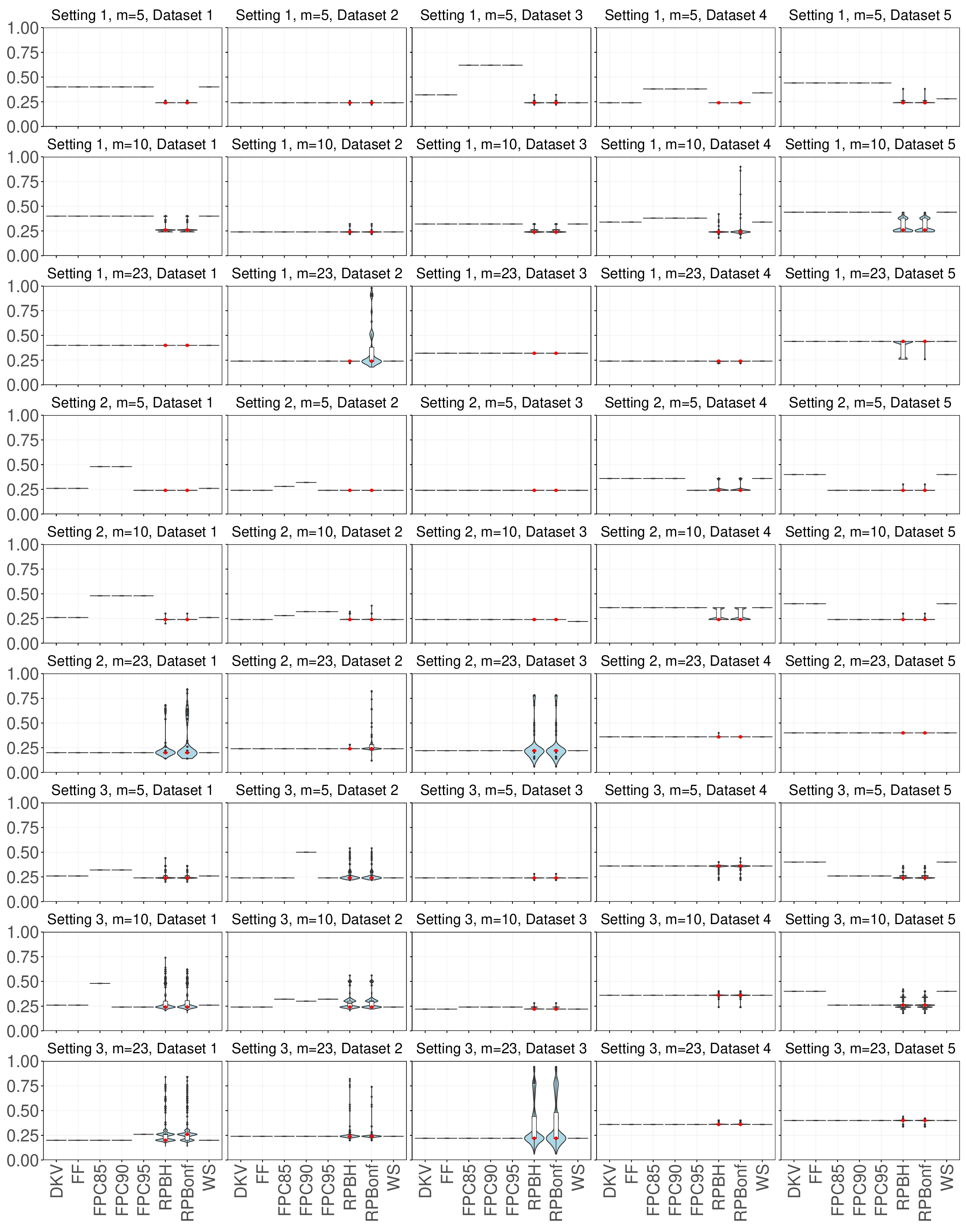}
    \caption{Sparse functional setting in Section \ref{subsec: sparse functional setting}.  Estimated change point locations detected by repeating the RP-based methods on one dataset (Dataset 1 through Dataset 5) for 1000 times. For the RP methods, the mode of the estimated locations across the 1000 repetitions is marked by a red dot. 
            The data-generating process follows (\ref{eq:model for fd}), (\ref{eq:data generating process}), and (\ref{bspline break}), where the standard deviation $\sigma_{g}$ follows \textit{Settings 1--3}.  The change point location is set at $\theta = 0.25$. The magnitude of the break function is scaled by $SNR=0.5$. }
    \label{fig:violin_grid:data0 bspline}
\end{figure}

\begin{figure}[H]
    \centering
    \includegraphics[width=\textwidth]{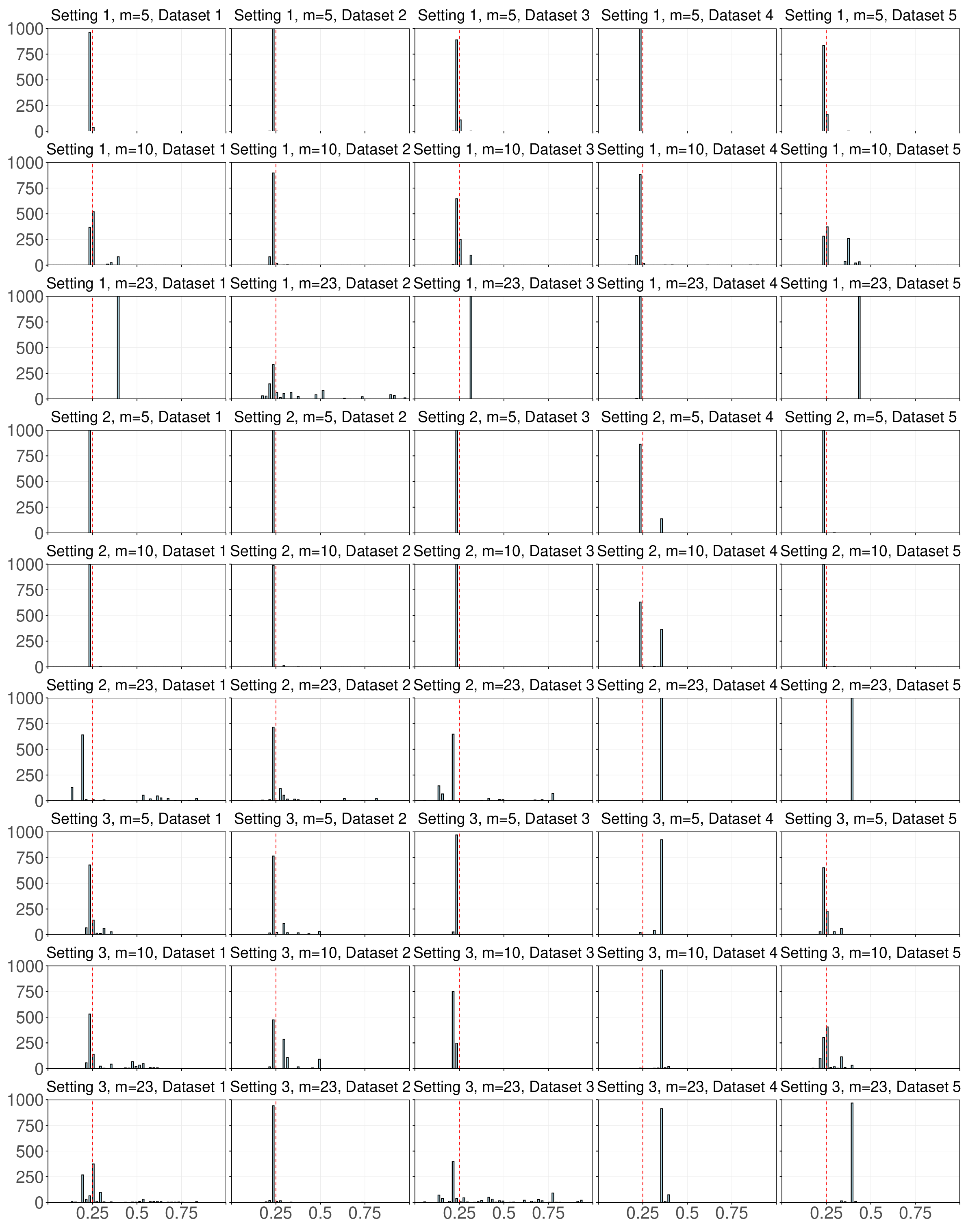}
    \caption{Sparse functional setting in Section \ref{subsec: sparse functional setting}.  Frequency of estimated locations from repeating the RP-Bonf method on one dataset (Dataset 1 through Dataset 5) for 1000 times.            
            The data-generating process follows (\ref{eq:model for fd}), (\ref{eq:data generating process}), and (\ref{bspline break}), where the standard deviation $\sigma_{g}$ follows \textit{Settings 1--3}.  The change point location is set at $\theta = 0.25$. The magnitude of the break function is scaled by $SNR=0.5$. }
    \label{fig:hist_bonf:data0 bspline}
\end{figure}

\begin{figure}[H]
    \centering
    \includegraphics[width=\textwidth]{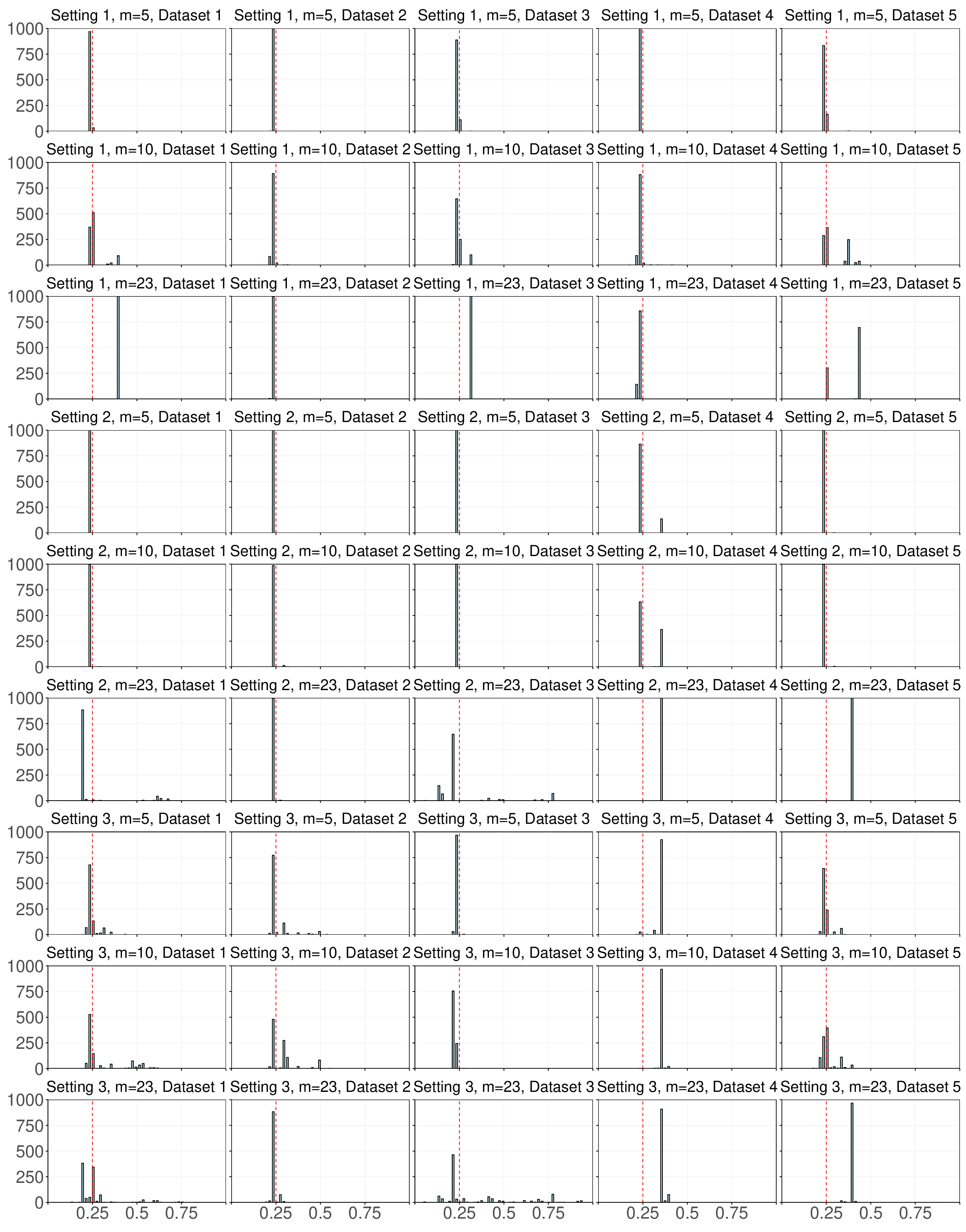}
    \caption{Sparse functional setting in Section \ref{subsec: sparse functional setting}.  Frequency of estimated locations from repeating the RP-BH method on one dataset (Dataset 1 through Dataset 5) for 1000 times.            
            The data-generating process follows (\ref{eq:model for fd}), (\ref{eq:data generating process}), and (\ref{bspline break}), where the standard deviation $\sigma_{g}$ follows \textit{Settings 1--3}.  The change point location is set at $\theta = 0.25$. The magnitude of the break function is scaled by $SNR=0.5$. }
    \label{fig:hist_bh:data0 bspline}
\end{figure}

\clearpage\newpage
{
\section{Additional results for Section \ref{subsec: sparse functional setting}}\label{supp:sec:sparse functional}
Figure \ref{fig:  CP locations Auedata standcusum: data0 bspline} presents location boxplots of 1000 Monte Carlo replications.
For a fixed dataset, we repeat the RP procedure 1000 times, as introduced in Section \ref{subsec:repeat RP method on one data}. 
For five randomly generated datasets, the violin plots of location estimates of RP-based methods are presented along with other methods' location estimates in Figure \ref{fig:violin_grid:data0 bspline}. 
Histograms of the 1000 location estimates of the RP-based methods for each dataset are also presented: RP-Bonf in Figure \ref{fig:hist_bonf:data0 bspline} and RP-BH in Figure \ref{fig:hist_bh:data0 bspline}.
}

\begin{figure} [h!]
            
            \centering
            \hspace{-1 cm}
            \includegraphics[width=5.5cm, height=5cm]{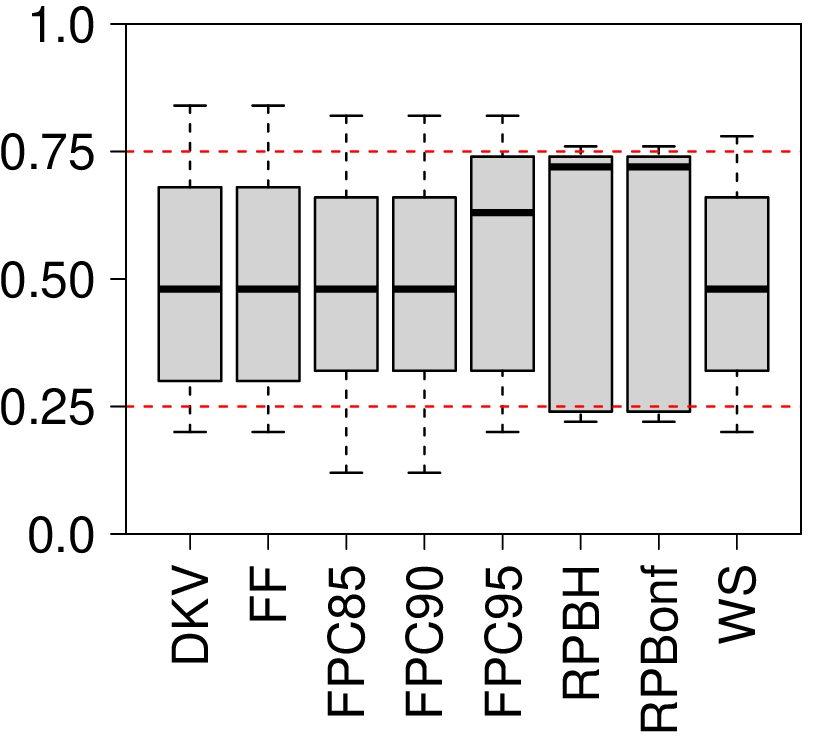}
            \hspace{-1 cm}
            \includegraphics[width=5.5cm, height=5cm]{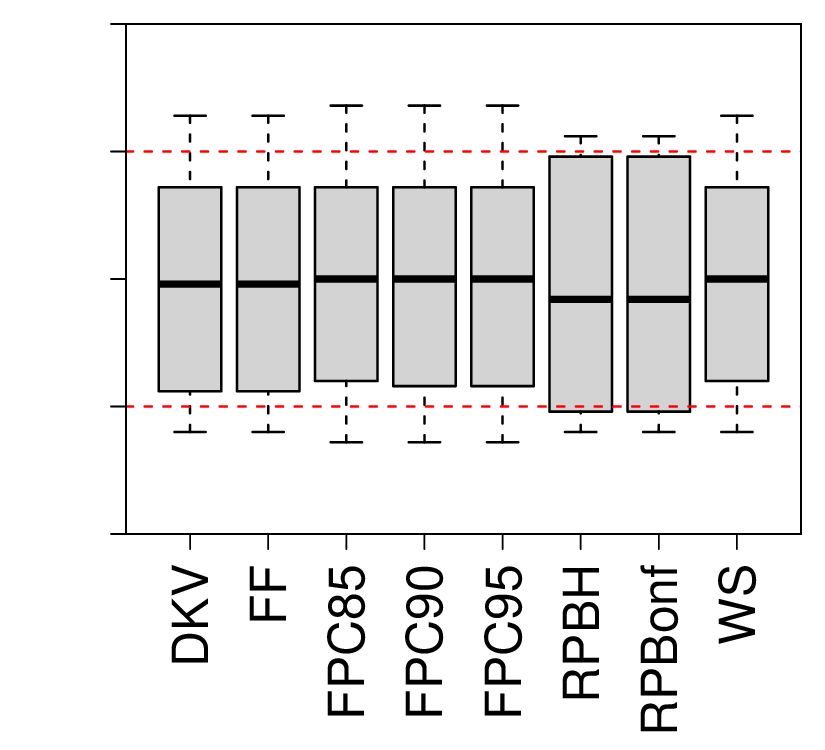}
            \hspace{-1 cm}
            \includegraphics[width=5.5cm, height=5cm]{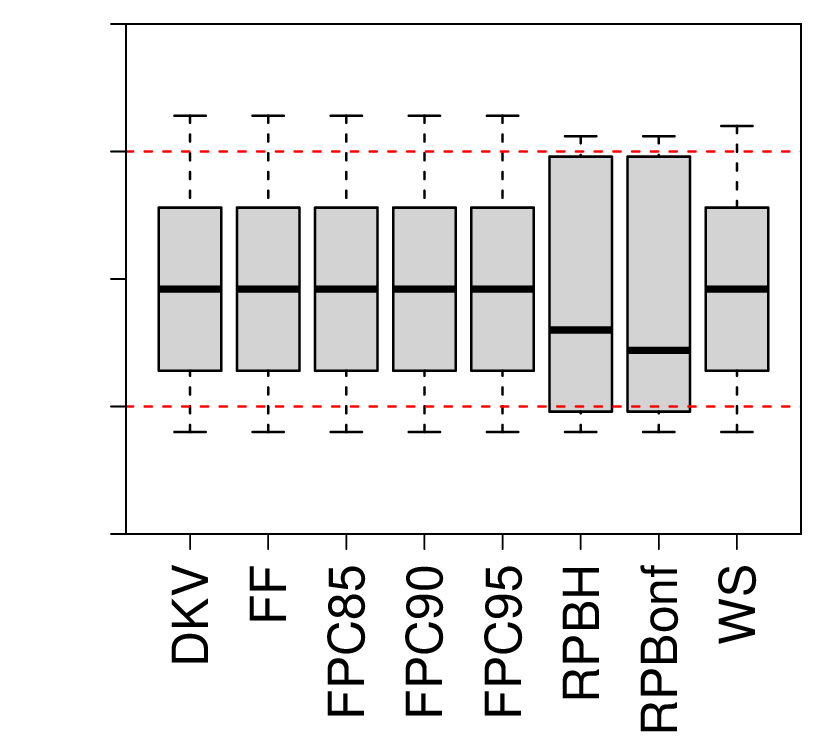}
            \hspace{-1 cm}
            \vspace{-0.2in}
            \caption*{\small{(a) \textit{Setting 1}, with $m=5,10,15$ } }

            \centering
            \hspace{-1 cm}
            \includegraphics[width=5.5cm, height=5cm]{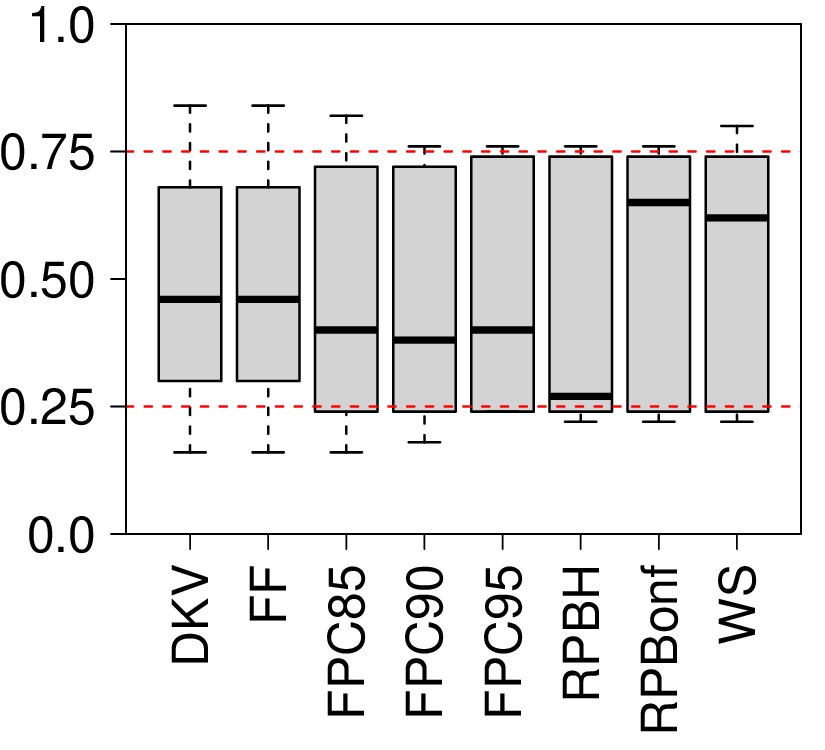}
            \hspace{-1 cm}
            \includegraphics[width=5.5cm, height=5cm]{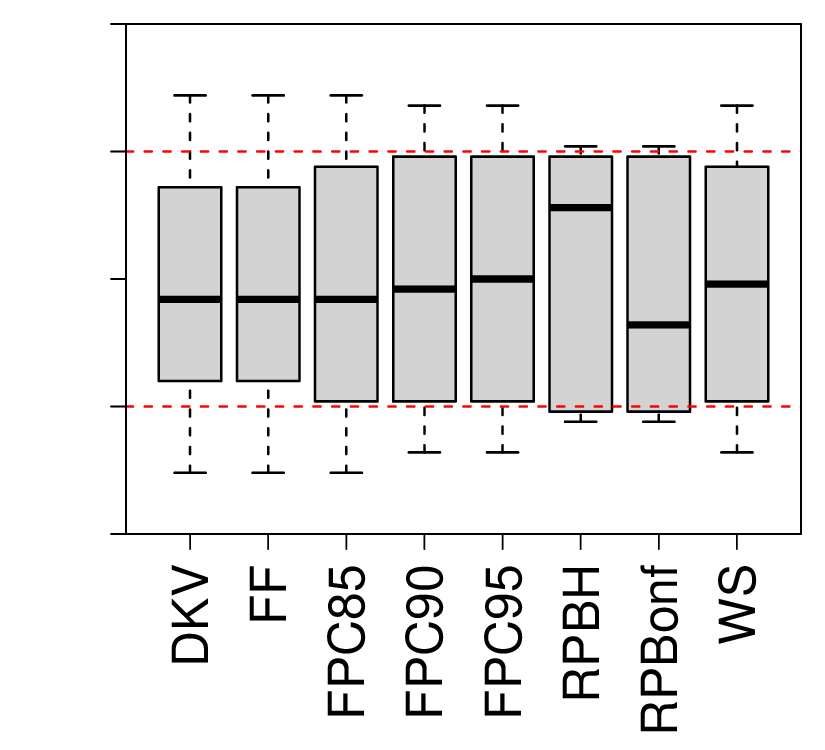}
            \hspace{-1 cm}
            \includegraphics[width=5.5cm, height=5cm]{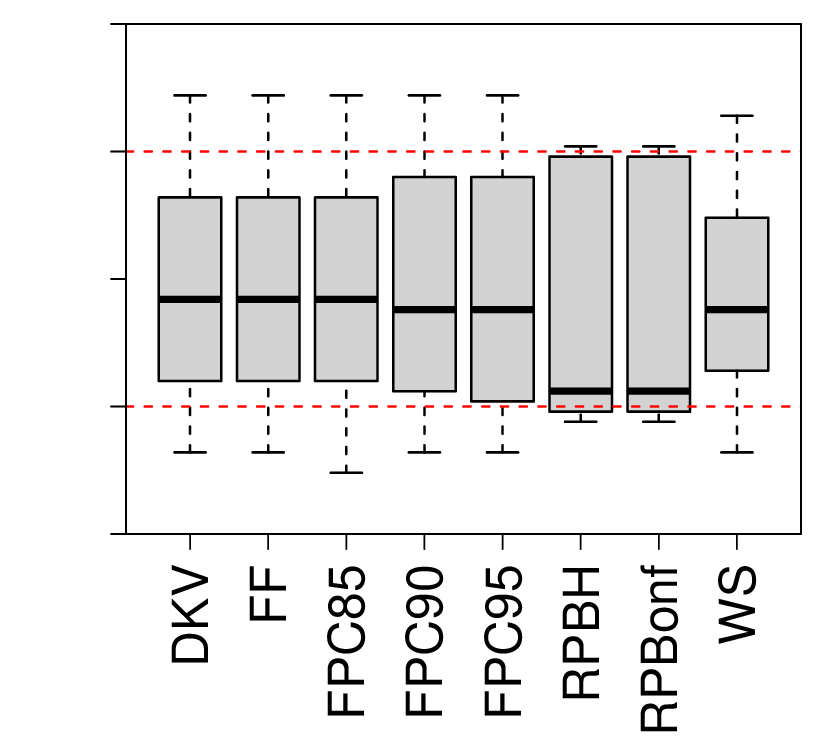}
            \hspace{-1 cm}
            \vspace{-0.2in}
            \caption*{\small{(b) \textit{Setting 2}, with $m=5,10,15$ } }

            \centering
            \hspace{-1 cm}
            \includegraphics[width=5.5cm, height=5cm]{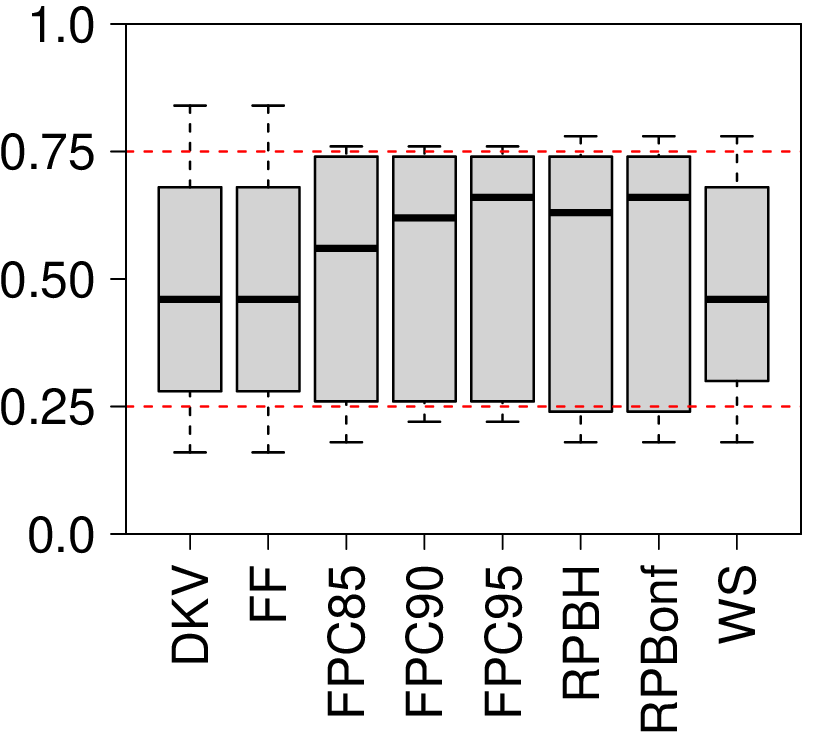}
            \hspace{-1 cm}
            \includegraphics[width=5.5cm, height=5cm]{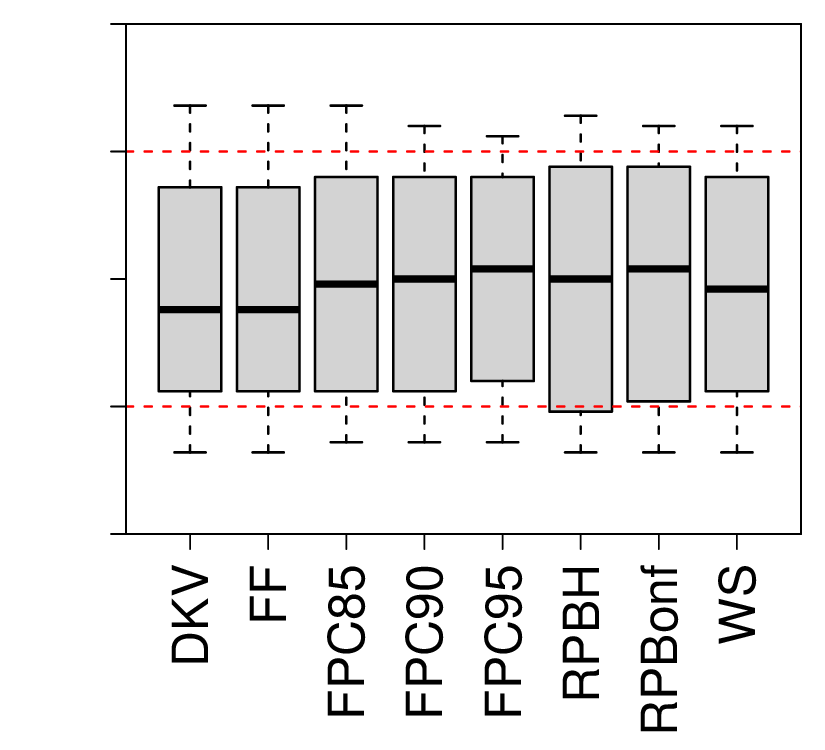}
            \hspace{-1 cm}
            \includegraphics[width=5.5cm, height=5cm]{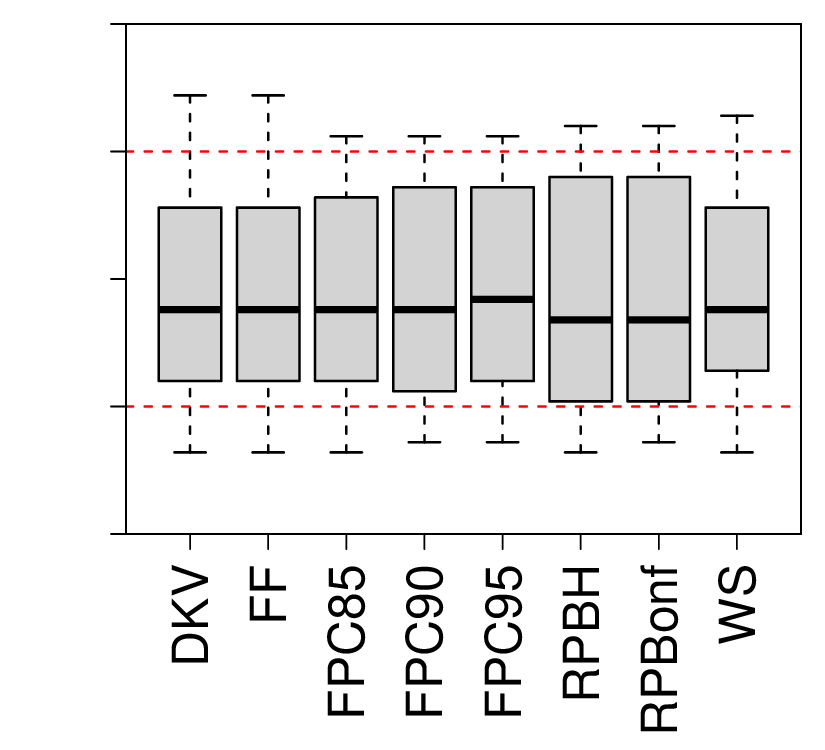}
            \hspace{-1 cm}
            \vspace{-0.2in}
            \caption*{\small{(c) \textit{Setting 3}, with $m=5,10,15$ } }
            
            \caption{Departures from AMOC setting in Section \ref{subsec: deviations from AMOC}. Estimated change point locations detected by different methods on 1000 simulations. The data-generating process follows (\ref{eq:data generating process}), (\ref{eq:dgp-2cpts}), and (\ref{bspline break}), where the standard deviation $\sigma_{g}$ follows \textit{Settings 1--3}.  The change point locations are set at $\theta_1 = 0.25$ and $\theta_2=0.75$. The magnitude of the break function is scaled by $SNR=0.5$.}
            \label{fig:  CP locations Auedata standcusum: data 2cpts}
    \end{figure}

\begin{figure}[H]
    \centering
    \includegraphics[width=\textwidth]{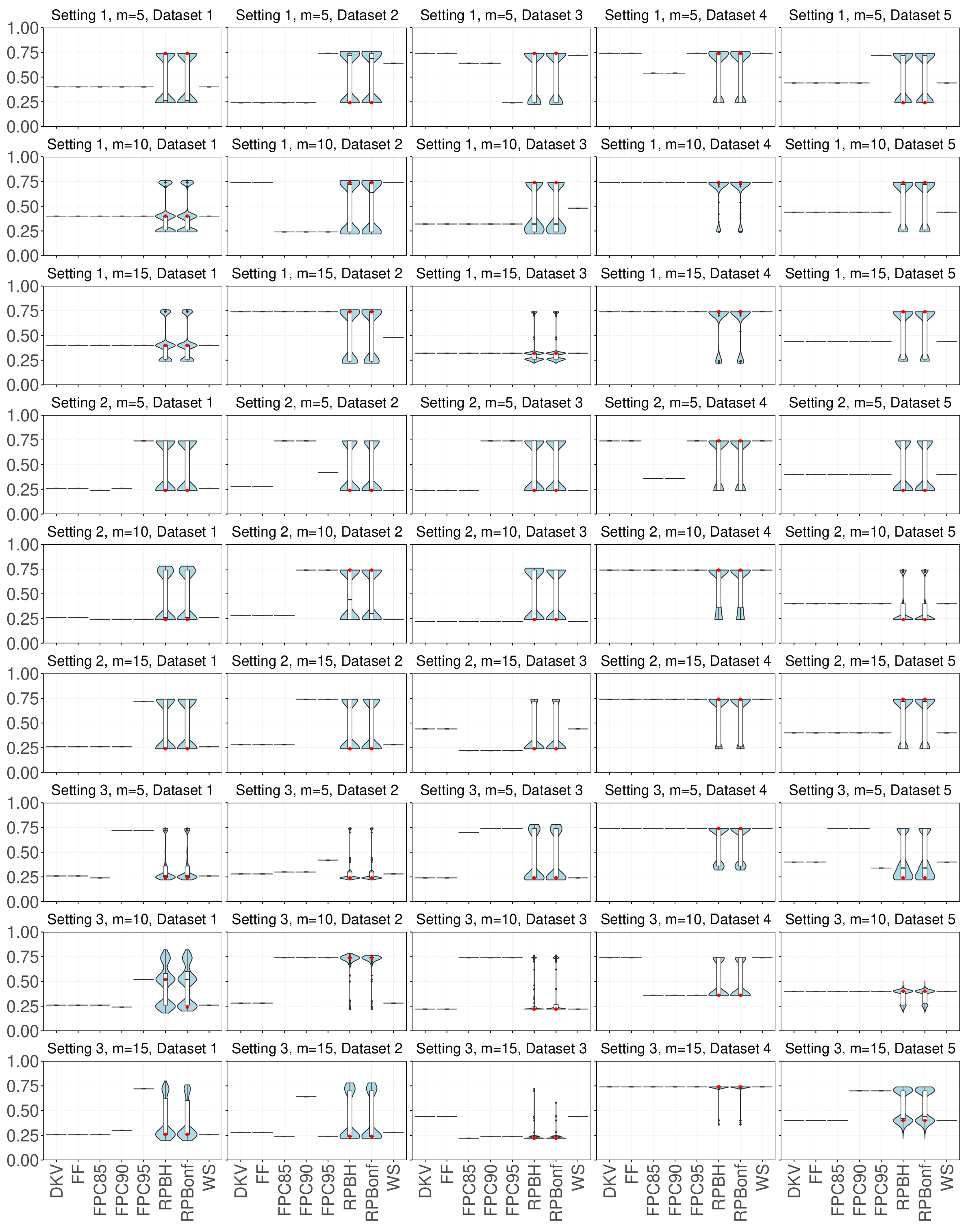}
    \caption{Departures from AMOC setting in Section \ref{subsec: deviations from AMOC}. Estimated change point locations detected by repeating the RP-based methods on one dataset (Dataset 1 through Dataset 5) for 1000 times. For the RP methods, the mode of the estimated locations across the 1000 repetitions is marked by a red dot. 
            The data-generating process follows (\ref{eq:data generating process}), (\ref{eq:dgp-2cpts}), and (\ref{bspline break}), where the standard deviation $\sigma_{g}$ follows \textit{Settings 1--3}.  The change point locations are set at $\theta_1 = 0.25$ and $\theta_2=0.75$. The magnitude of the break function is scaled by $SNR=0.5$. }
    \label{fig:violin_grid:data 2cpts}
\end{figure}

\begin{figure}[H]
    \centering
    \includegraphics[width=\textwidth]{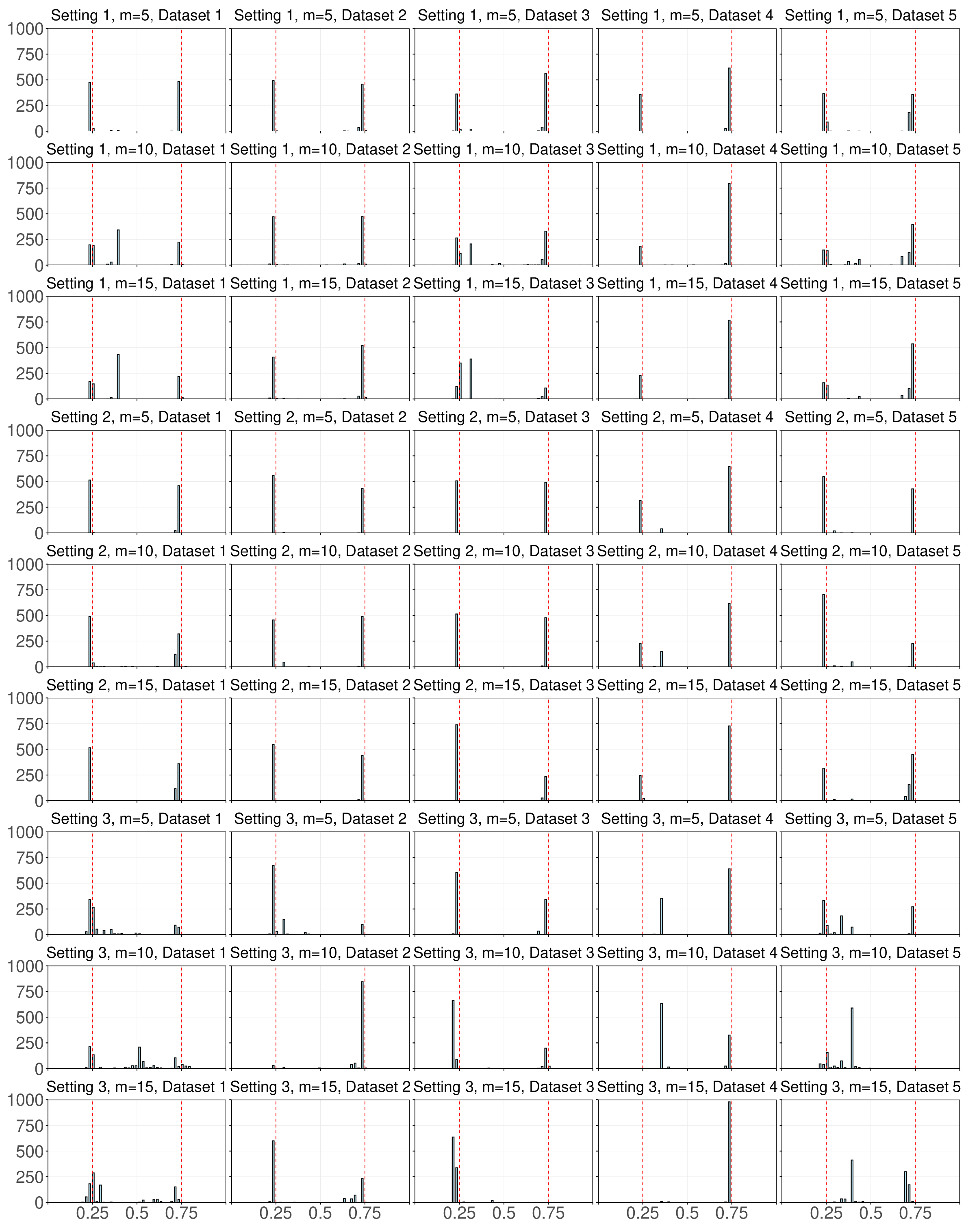}
    \caption{Departures from AMOC setting in Section \ref{subsec: deviations from AMOC}. Frequency of estimated locations from repeating the RP-Bonf method on one dataset (Dataset 1 through Dataset 5) for 1000 times.            
            The data-generating process follows (\ref{eq:data generating process}), (\ref{eq:dgp-2cpts}), and (\ref{bspline break}), where the standard deviation $\sigma_{g}$ follows \textit{Settings 1--3}.  The change point locations are set at $\theta_1 = 0.25$ and $\theta_2=0.75$. The magnitude of the break function is scaled by $SNR=0.5$. }
    \label{fig:hist_bonf:data 2cpts}
\end{figure}

\begin{figure}[H]
    \centering
    \includegraphics[width=\textwidth]{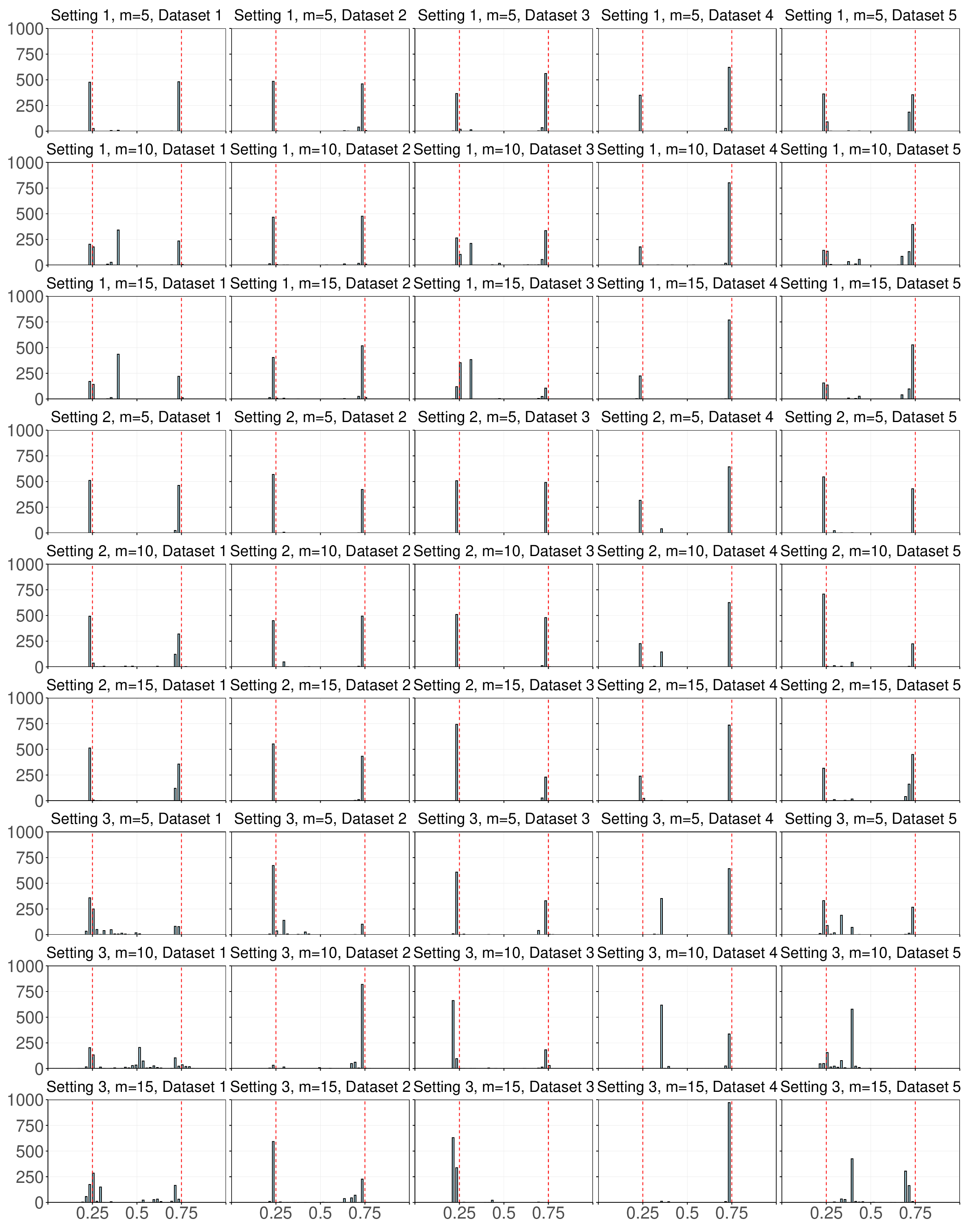}
    \caption{Departures from AMOC setting in Section \ref{subsec: deviations from AMOC}. Frequency of estimated locations from repeating the RP-BH method on one dataset (Dataset 1 through Dataset 5) for 1000 times.            
            The data-generating process follows (\ref{eq:data generating process}), (\ref{eq:dgp-2cpts}), and (\ref{bspline break}), where the standard deviation $\sigma_{g}$ follows \textit{Settings 1--3}.  The change point locations are set at $\theta_1 = 0.25$ and $\theta_2=0.75$. The magnitude of the break function is scaled by $SNR=0.5$.}
    \label{fig:hist_bh:data 2cpts}
\end{figure}

\clearpage\newpage
{
\section{Additional results for Section \ref{subsec: deviations from AMOC}}\label{supp:sec:AMOC}
Figure \ref{fig:  CP locations Auedata standcusum: data 2cpts} presents location boxplots of 1000 Monte Carlo replications.
For a fixed dataset, we repeat the RP procedure 1000 times, as introduced in Section \ref{subsec:repeat RP method on one data}. 
For five randomly generated datasets, the violin plots of location estimates of RP-based methods are presented along with other methods' location estimates in Figure \ref{fig:violin_grid:data 2cpts}. 
Histograms of the 1000 location estimates of the RP-based methods for each dataset are also presented: RP-Bonf in Figure \ref{fig:hist_bonf:data 2cpts} and RP-BH in Figure \ref{fig:hist_bh:data 2cpts}.
}

\begin{figure}[H]
    \centering
    \includegraphics[width=\textwidth]{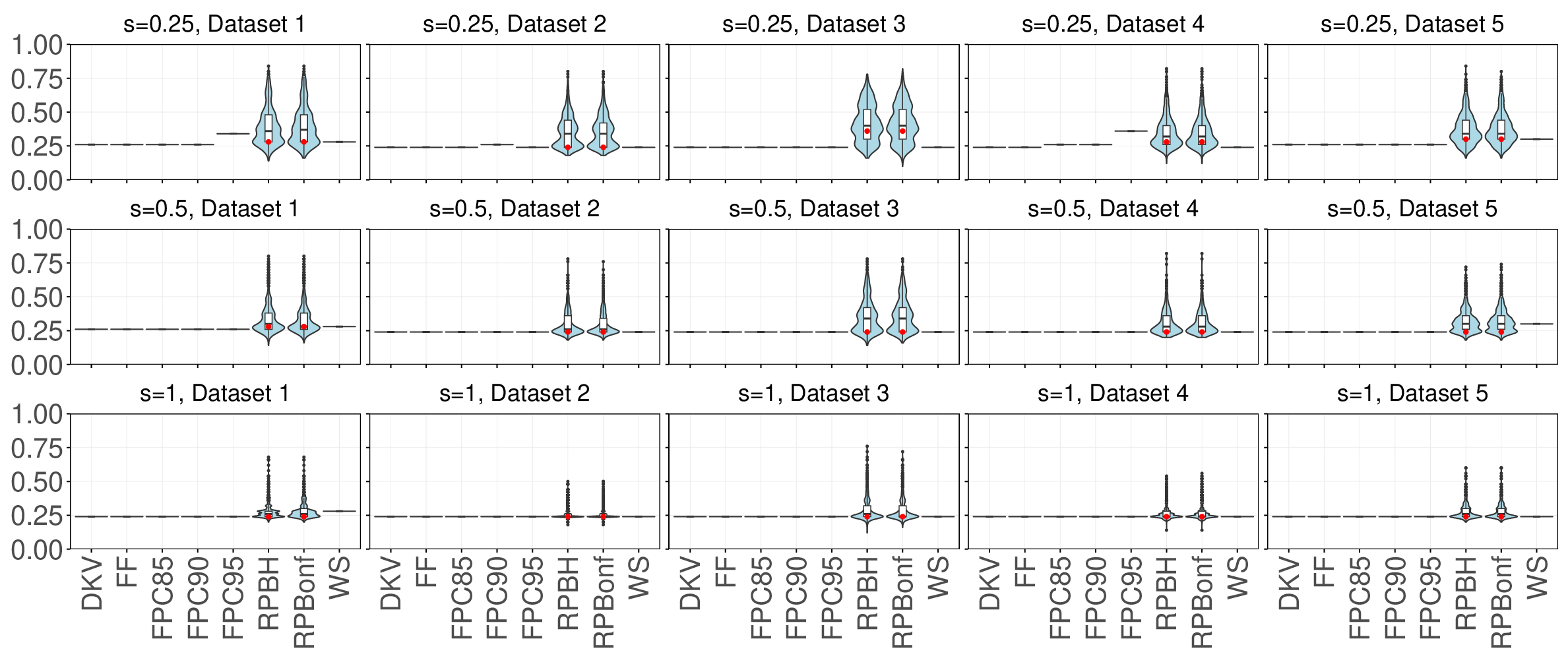}
    \caption{High-dimensional vector setting in Section \ref{subsec: high dimensional vector setting}. Estimated change point locations detected by repeating the RP-based methods on one dataset (Dataset 1 through Dataset 5) for 1000 times. For the RP methods, the mode of the estimated locations across the 1000 repetitions is marked by a red dot. 
            The data-generating process follows 
            (\ref{eq: model}) and (\ref{wang break}) with iid $\varepsilon_{tj}$. The change point location is set at $\theta=0.25$. The magnitude of the break function is scaled by $\vartheta=0.6$.
            }
    \label{fig:violin_grid:data0 wang}
\end{figure}

\begin{figure}[H]
    \centering
    \includegraphics[width=\textwidth]{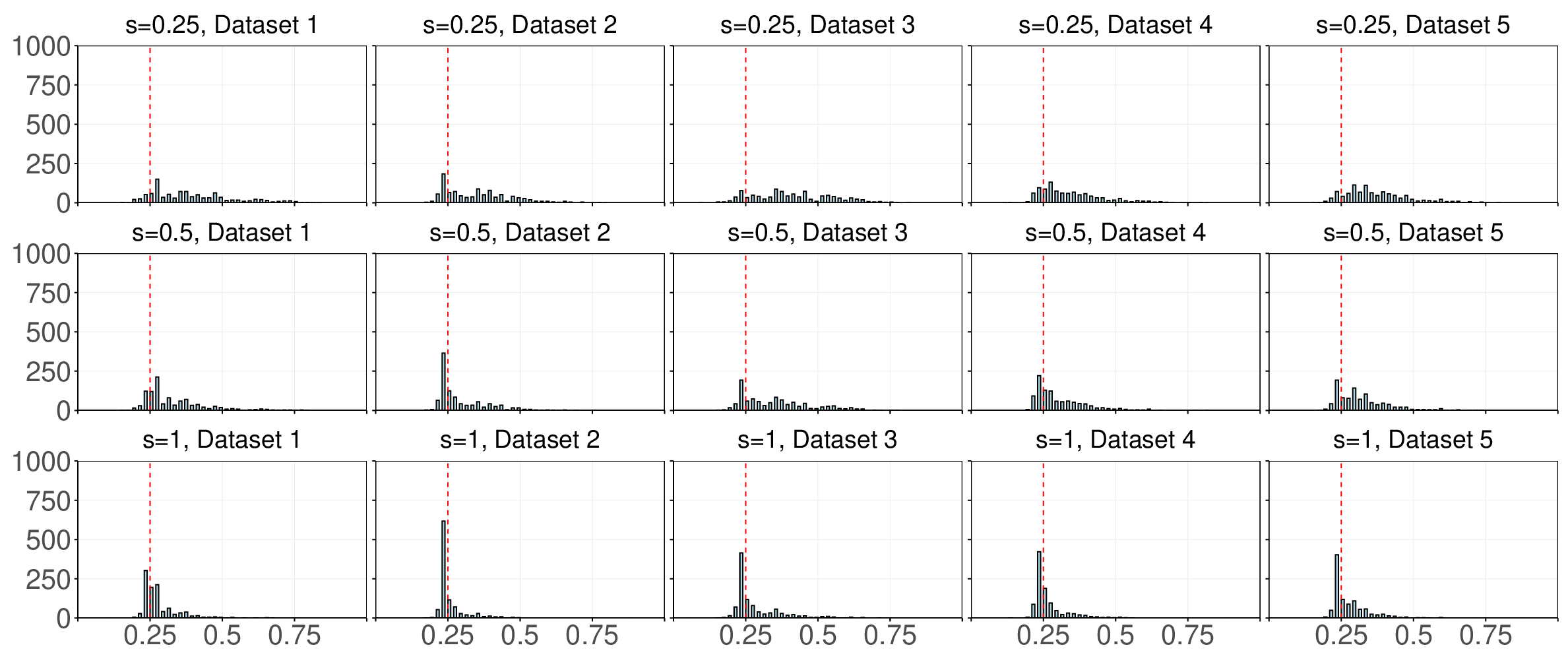}
    \caption{High-dimensional vector setting in Section \ref{subsec: high dimensional vector setting}. Frequency of estimated locations from repeating the RP-Bonf method on one dataset (Dataset 1 through Dataset 5) for 1000 times.          
            The data-generating process follows 
            (\ref{eq: model}) and (\ref{wang break}) with iid $\varepsilon_{tj}$. The change point location is set at $\theta=0.25$. The magnitude of the break function is scaled by $\vartheta=0.6$.}
    \label{fig:hist_bonf:data0 wang}
\end{figure}

\begin{figure}[H]
    \centering
    \includegraphics[width=\textwidth]{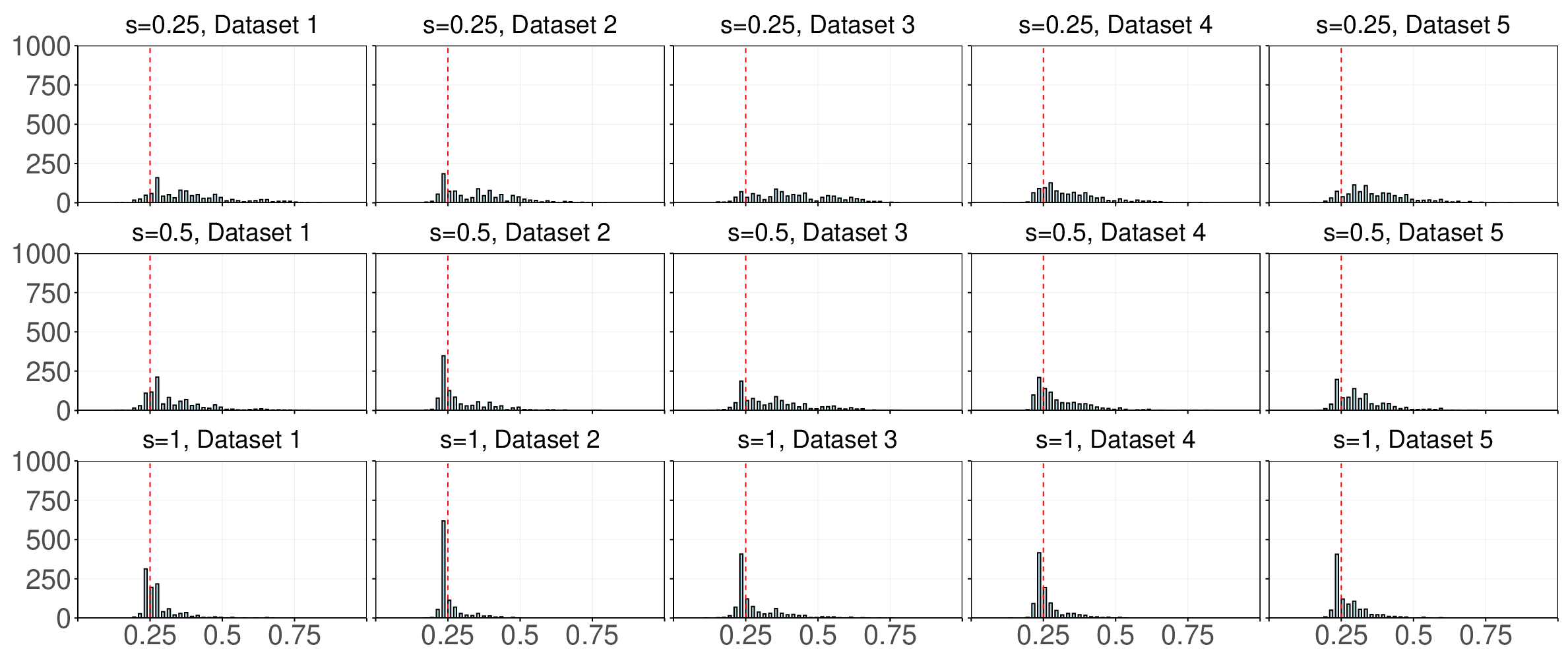}
    \caption{High-dimensional vector setting in Section \ref{subsec: high dimensional vector setting}. Frequency of estimated locations from repeating the RP-BH method on one dataset (Dataset 1 through Dataset 5) for 1000 times.            
            The data-generating process follows 
            (\ref{eq: model}) and (\ref{wang break}) with iid $\varepsilon_{tj}$. The change point location is set at $\theta=0.25$. The magnitude of the break function is scaled by $\vartheta=0.6$.}
    \label{fig:hist_bh:data0 wang}
\end{figure}

\clearpage\newpage
{
\section{Additional results for Section \ref{subsec: high dimensional vector setting}}\label{supp:sec:Wang}
This section presents additional results for the high-dimensional vector setting presented in \ref{subsec: high dimensional vector setting}. 
For a fixed dataset, we repeat the RP procedure 1000 times, as introduced in Section \ref{subsec:repeat RP method on one data}. 
For five randomly generated datasets, the violin plots of location estimates of RP-based methods are presented along with other methods' location estimates in Figure \ref{fig:violin_grid:data0 wang}. 
Histograms of the 1000 location estimates of the RP-based methods for each dataset are also presented: RP-Bonf in Figure \ref{fig:hist_bonf:data0 wang} and RP-BH in Figure \ref{fig:hist_bh:data0 wang}.
}

{
\section{Additional results for application in Section \ref{sec:application}}
\label{subsec:temp app supp}

This section presents the frequency histograms for the RP-Bonf and RP-BH methods corresponding to Figure \ref{fig:app_temp_diff_comp}.
}

\begin{figure} [H]
        \centering
            \hspace{-1.2 cm}
            \includegraphics[width=5cm, height=5cm]       {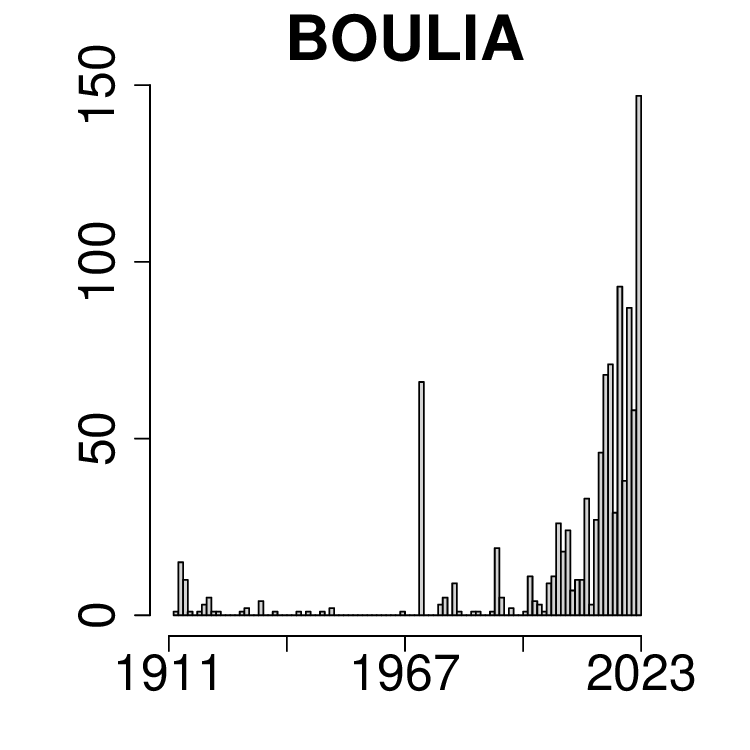}
             \hspace{-1.2 cm}
            \includegraphics[width=5cm, height=5cm]   {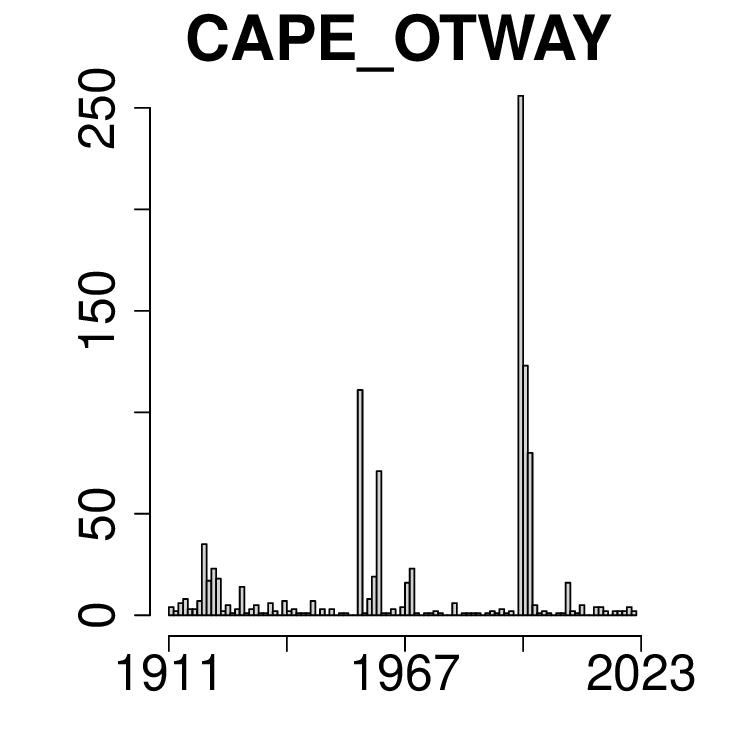}
             \hspace{-1.2 cm}
            \includegraphics[width=5cm, height=5cm]       {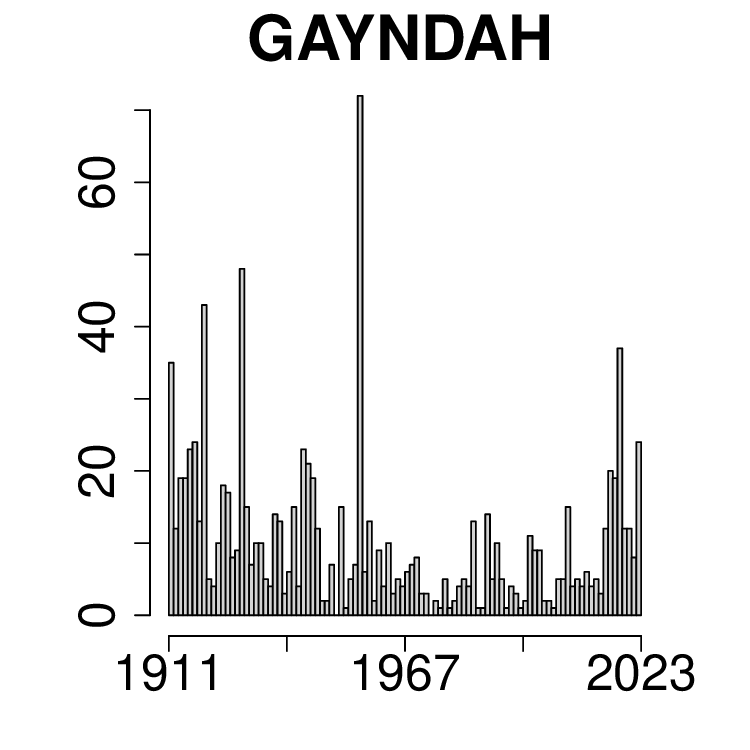}
             \hspace{-1.2 cm}
            \includegraphics[width=5cm, height=5cm]       {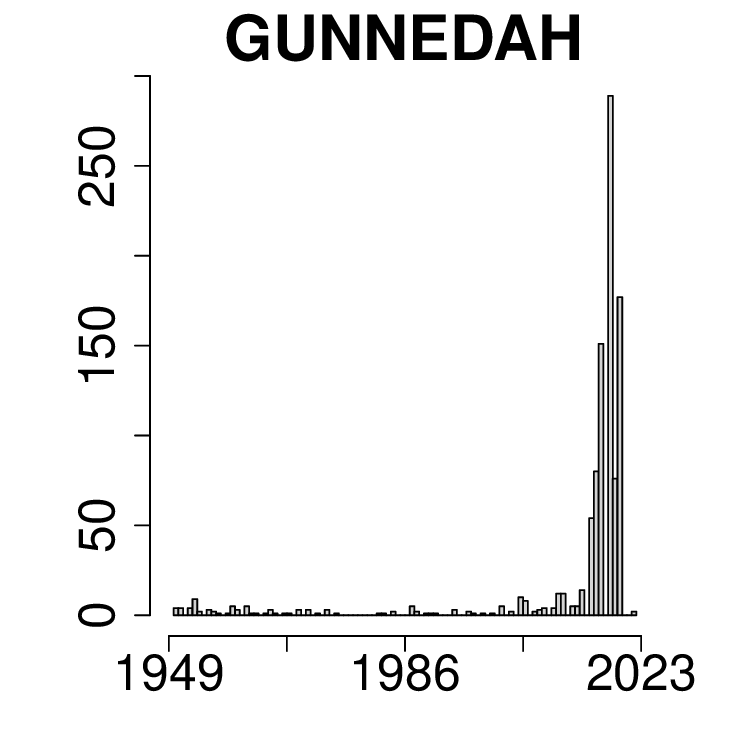}
            \hspace{-1.2 cm}

            \centering              
           \hspace{-1.2 cm}
           \includegraphics[width=5cm, height=5cm]        {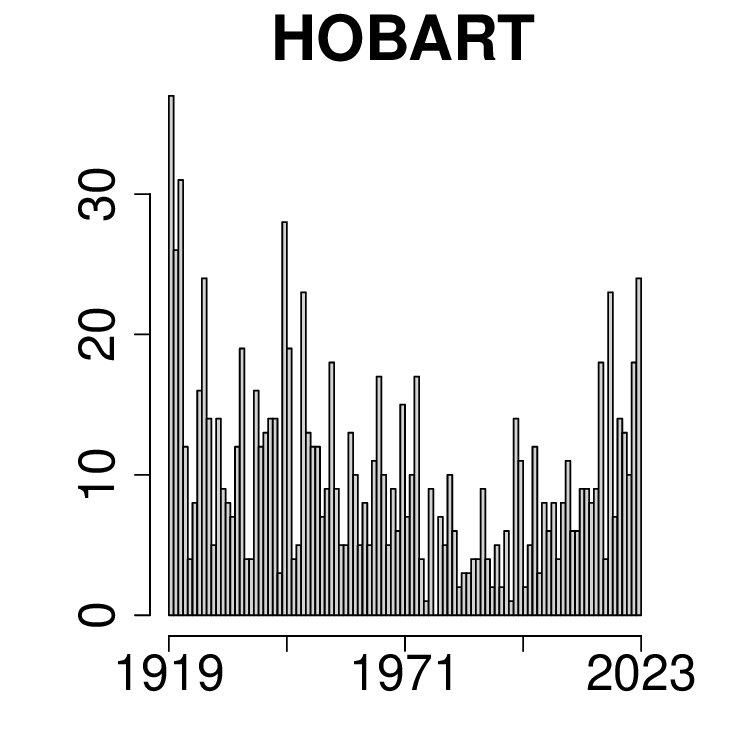}
            \hspace{-1.2 cm}
            \includegraphics[width=5cm, height=5cm]        {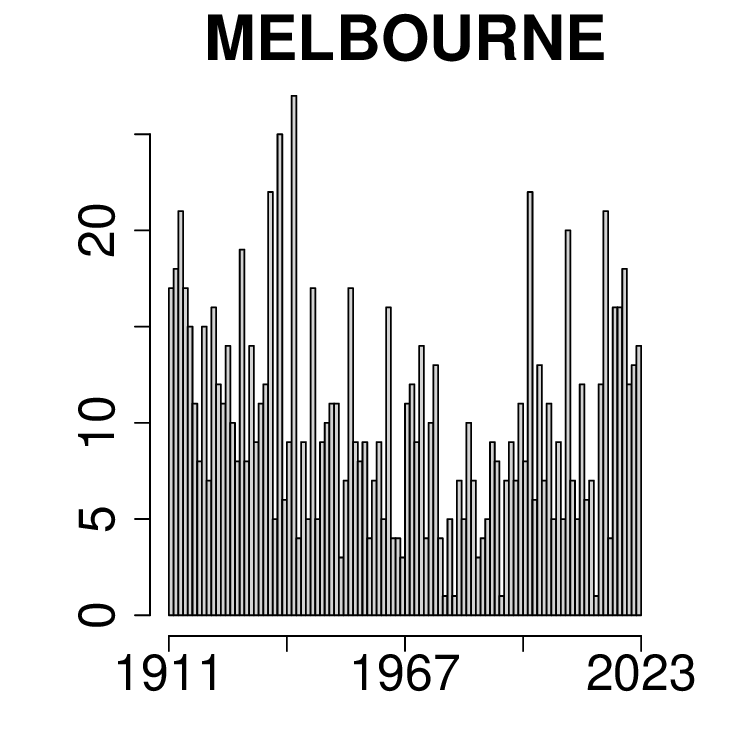}
             \hspace{-1.2 cm}
            \includegraphics[width=5cm, height=5cm]        {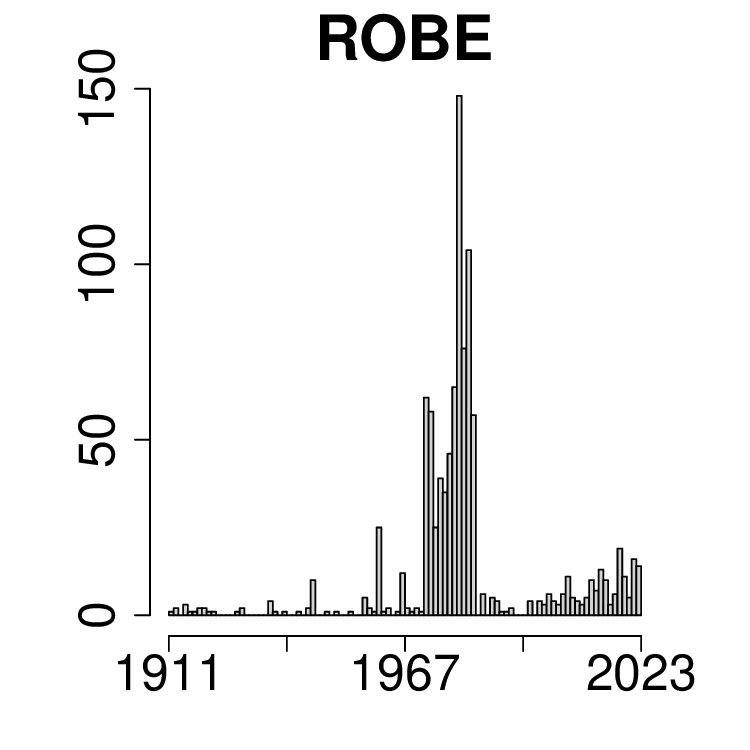}
            \hspace{-1.2 cm}
            \includegraphics[width=5cm, height=5cm]        {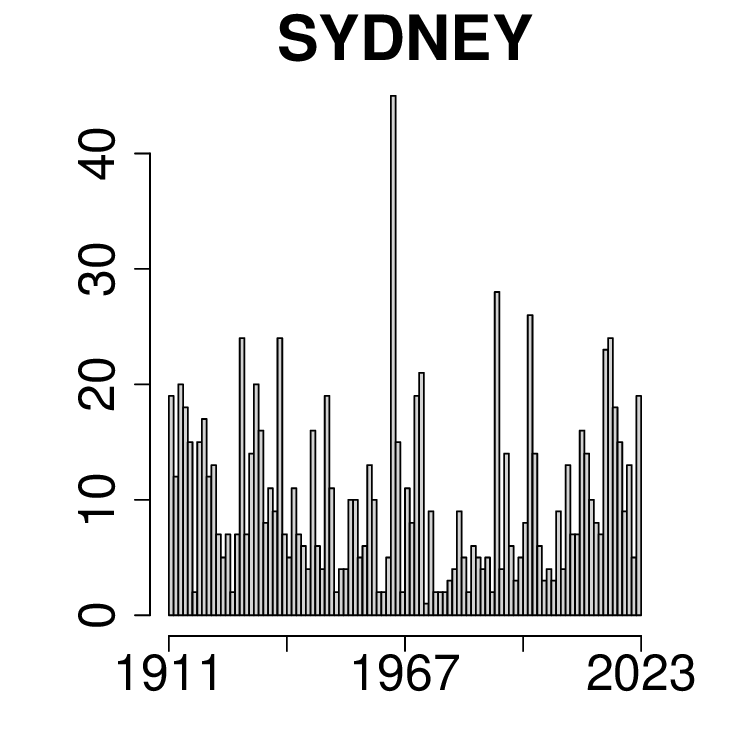}
            \hspace{-1.2 cm}

              \caption{\small{Empirical frequency of estimated change-point locations obtained from applying the RP-Bonf method to the detrended temperature dataset across 1,000 replications. 
            }}
            \label{fig:app_diff_temp_hac_hist}

    \end{figure} 
\begin{figure} [H]
        \centering
            \hspace{-1 cm}
            \includegraphics[width=5cm, height=5cm]       {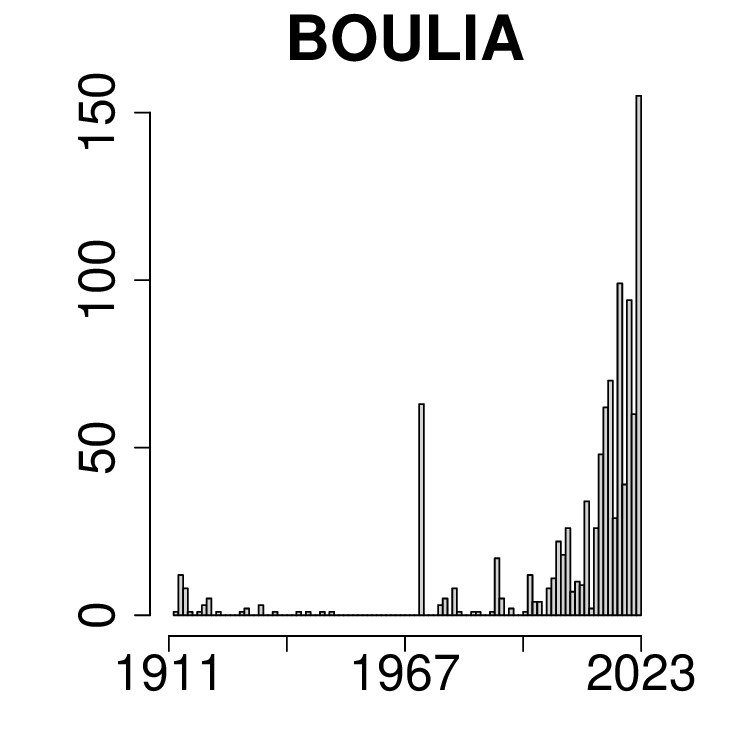}
             \hspace{-1.2 cm}
            \includegraphics[width=5cm, height=5cm]   {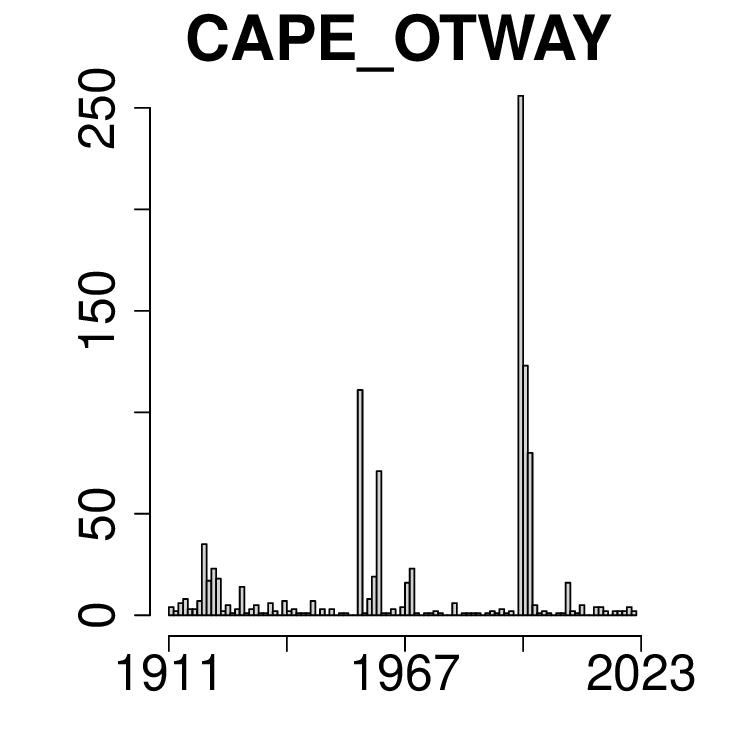}
            \hspace{-1.2 cm}
            \includegraphics[width=5cm, height=5cm]       {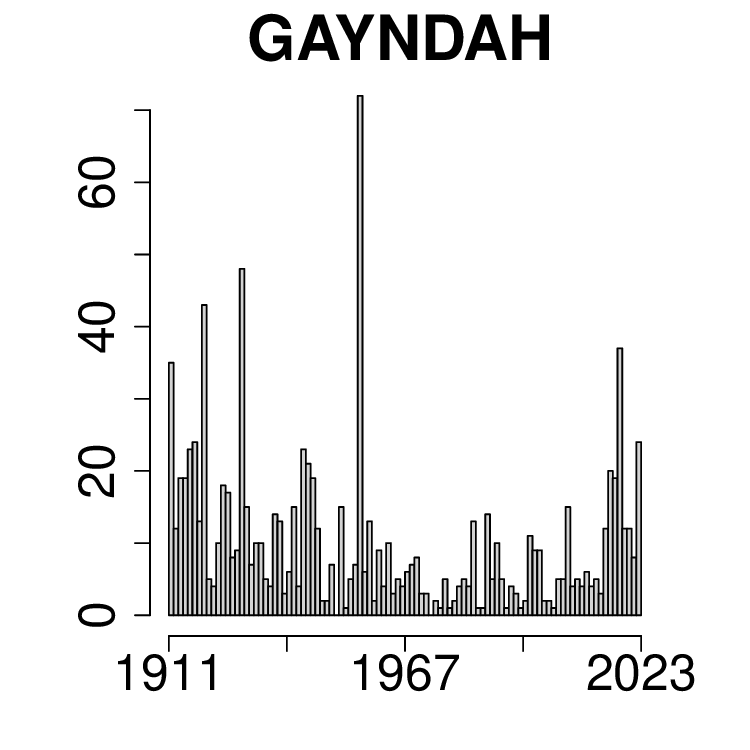}
            \hspace{-1.2 cm}
            \includegraphics[width=5cm, height=5cm]       {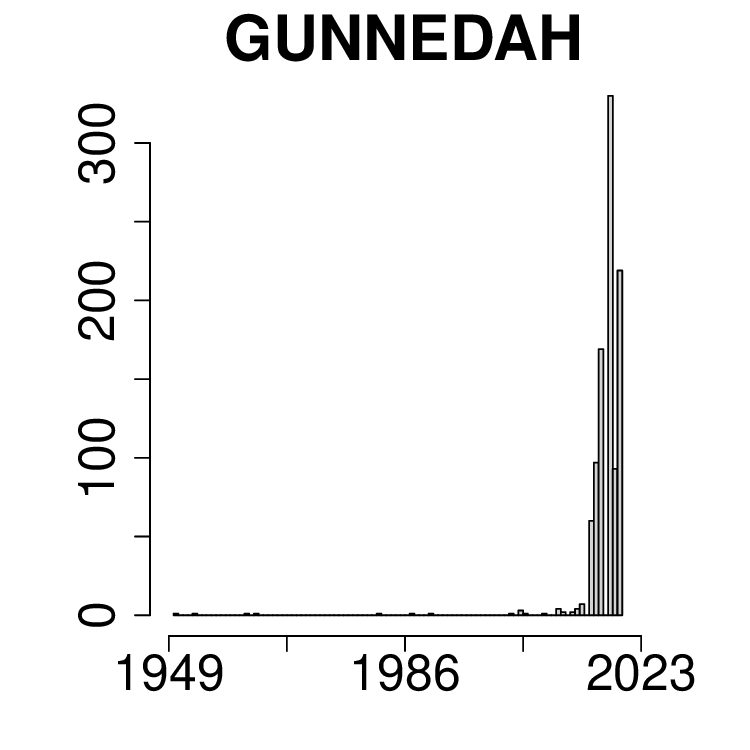}
            \hspace{-1cm}

            \centering              
           \hspace{-1 cm}
           \includegraphics[width=5cm, height=5cm]        {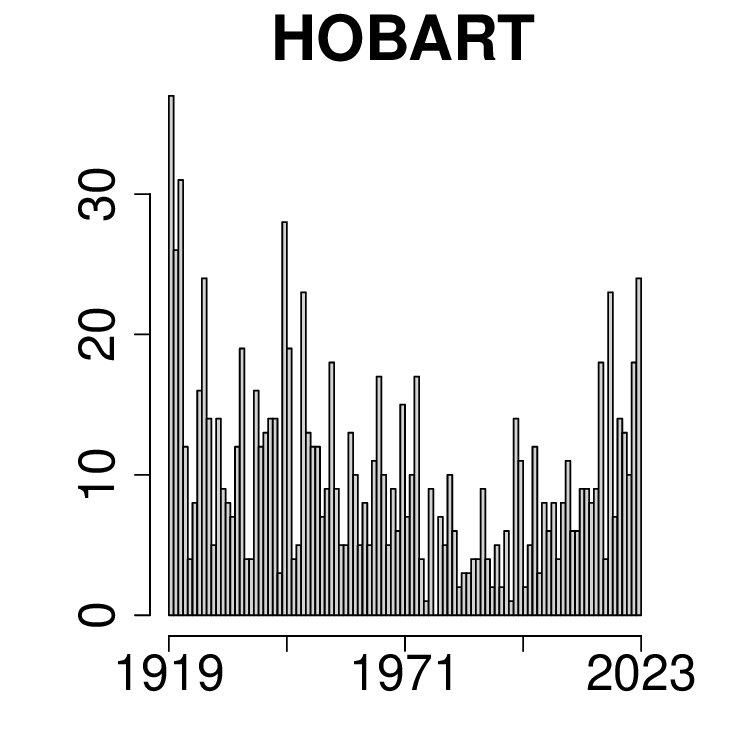}
           \hspace{-1.2 cm}
            \includegraphics[width=5cm, height=5cm]        {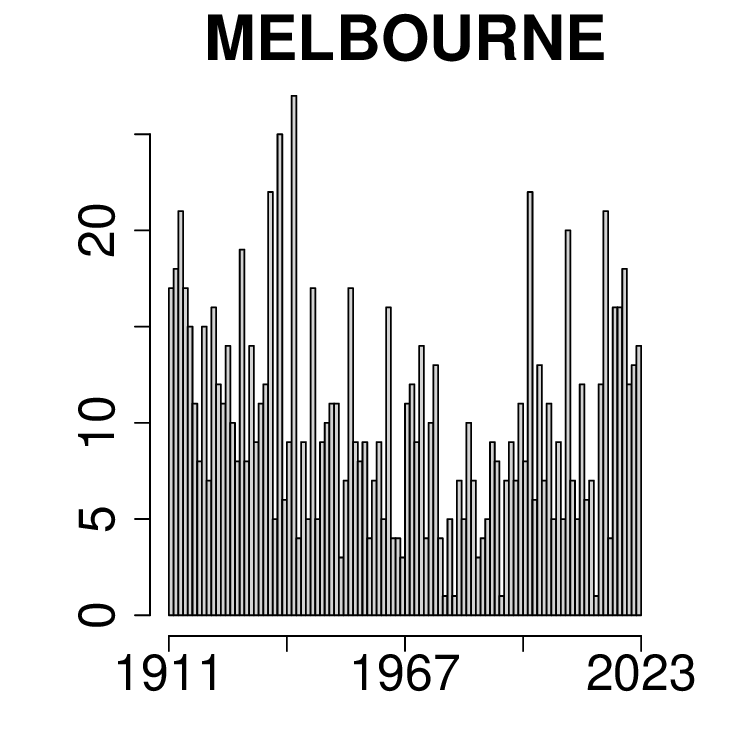}
            \hspace{-1.2 cm}
            \includegraphics[width=5cm, height=5cm]        {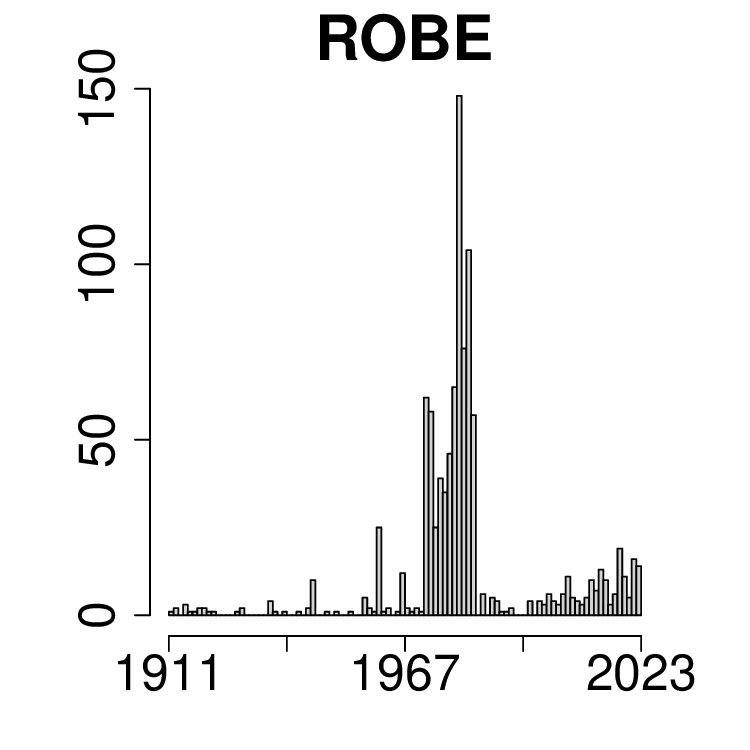}
            \hspace{-1.2 cm}
            \includegraphics[width=5cm, height=5cm]        {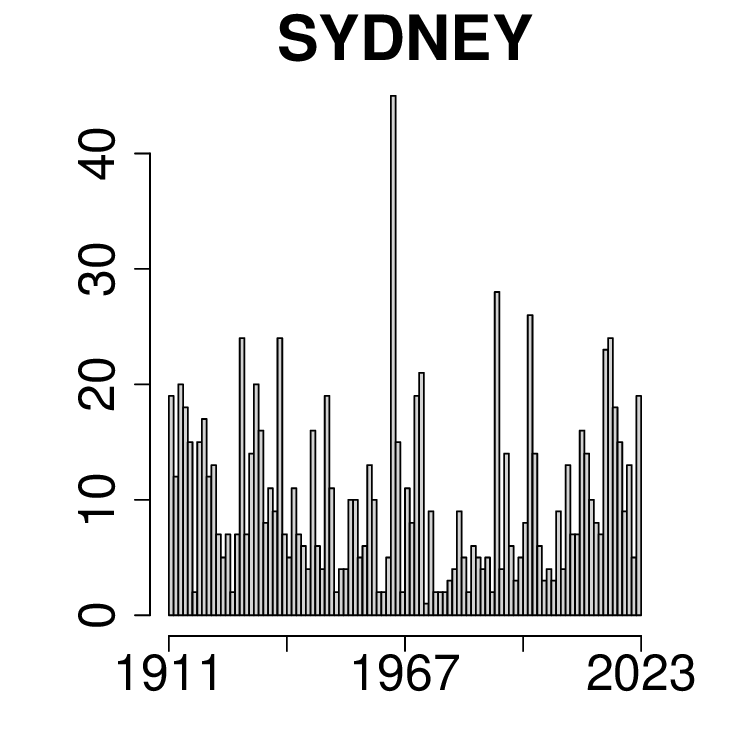}
            \hspace{-1 cm}
        
              \caption{\small{Empirical frequency of estimated change-point locations obtained from applying the RP-BH method to the detrended temperature dataset across 1,000 replications.  
            }}
            \label{fig:app_diff_temp_hac_hist_BH}

    \end{figure}

\end{document}